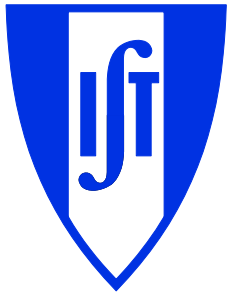

**INSTITUTO SUPERIOR TÉCNICO**

**UNIVERSIDADE TÉCNICA DE LISBOA**

**INSTITUTO SUPERIOR TÉCNICO**

SUPERSYMMETRIC MODELS

AND

NEUTRINO MASSES

João Miguel Barreto Nestal Esteves

Ph.D. Thesis

**January 2011**

# Abstract

Lepton flavour violation and neutrino masses are a signal for new Physics beyond the Standard Model and are deeply related. The minimal extension of the Standard Model to make it include neutrino masses is not satisfactory from a conceptual point of view, since it requires a severe fine-tuning of Yukawa couplings. Seesaw models provide a consistent and natural mechanism to generate neutrino masses and require Physics beyond the Standard Model. In this thesis, the connections between models for neutrino masses and processes that violate lepton flavour are explored. The reconstruction of high energy parameters from neutrino data is partially possible within the framework of these seesaw models and it is enhanced by the knowledge of phenomena outside the neutrino sector, as it is the case of lepton number violation processes. Grand Unification SUSY models offer a consistent theoretical framework for Type I, Type II and Type III seesaw models. On the other hand, Type II seesaw has the attractive feature of producing a lepton asymmetry through triplet decays which can be converted into a baryon asymmetry with leptogenesis. Thus being, we studied a model to explain the baryonic density of the Universe constrained by the phenomenological neutrino data.

**Keywords:**
Neutrino Masses; Lepton Flavour Violation; Baryogenesis; Leptogenesis; Majoron; Seesaw Models.

# Acknowledgments

This thesis is a teamwork. A team made of people to which I had close contact with or I just know from books and scientific articles. From this second group, I select three particular sets of authors to whom I would like to pay tribute. The first is Steven Weinberg and his "Quantum Field Theory" in three volumes from where I took inspiration and learned much of the fundamentals of the Physics I think I know today. It was particularly useful his discussion on anomalies in [1] and on supersymmetry in [2]. The second set includes Carlo Giunti and Chung W. Kim, authors of "Fundamentals of Neutrino Physics and Astrophysics" [3] that was invaluable in setting the details straight and also as a source of experimental and bibliographical data. The third is comprises the authors of "Theory and Phenomenology of Sparticles" [4], which I believe found just the right balance between theoretical formalism and phenomenological implications of SUSY.

From the first group of people, I begin with my Ph.D. supervisor, J. C. Romão, who, despite the multiple requests he has due to his academic responsibilities, was always available to discuss scientific matters with me and to answer all questions (even the unexpected) I put to him, thus making this thesis also a consequence of his perseverance. Next, the AHEP group in València and Werner Porod in Würzburg also had a major role in the research that is presented here: the published paper [5], which is a joint work with J. C. Romao, A. Villanova del Moral, M. Hirsch and W. Porod, had its genesis on an idea by Martin and Werner, if I remember correctly, like it did also the article [6] published with S. Kaneko, J. C. Romao, M. Hirsch and W. Porod. In these two papers the publicly available Fortran code SPheno [7], which is mostly a one man's job, played a crucial role. I decided to reproduce here the contents of these published works because I find hard, if not impossible, to select the parts to which I gave a contribution even if a modest one and also as not to adulterate the scientific content and language style. The director of AHEP, José W. F. Valle, was also invaluable in the work presented in Chapter 4 as this was mainly his idea and also in an invitation to visit València for 2 weeks, where I deeply appreciated the scientific and human environment I found there. My gratitude also to Filipe Joaquim, now at CERN, who pointed out an inconsistency in the work "Leptogenesis in an $A_4$ model" and that is described in detail in Chapter 3. I also did benefit from discussions with Mariam Tórtola, Albert Villanova del Moral, Avelino Vicente and Stefano Morisi.

I also thank Gustavo Castelo Branco, director of CFTP, for accepting me at this research center where I had exceptional working conditions and a very friendly atmosphere.

Last but not least I wish to thank to my Mother, for obvious reasons and others not so obvious, as without her this thesis would not have been possible and also to Maria dos Anjos Carapau for a very professional work in reviewing the text and pointing out its inconsistencies in the English language. Of course, I'm the sole responsible for any spelling or grammar error that remains, or for some lack of proficiency. It's not an easy task to write in a language other than the native one.

This thesis was done under the grant SFRH/BD/29642/2006 from "Fundação para a Ciência e Tecnologia".

João Esteves

# Contents

# List of Figures

# List of Tables

# Chapter 1

# Supersymmetry and Neutrino Masses

## 1.1 The Minimal Supersymmetric Standard Model

Here we give a brief review of the Minimal Supersymmetric Standard Model (MSSM), following mainly [4] and [2]. The left chiral matter and Higgs superfields on one hand, and vector gauge superfields on the other, of the MSSM are shown on tables 1.1 and 1.2, respectively. The Lagrangian density will be a sum of a term $\mathcal{L}_{\text{SUSY}}$ that is fully supersymmetric and a term $\mathcal{L}_{\text{SOFT}}$ that breaks supersymmetry softly:

$$\mathcal{L}_{\text{MSSM}} = \mathcal{L}_{\text{SUSY}} + \mathcal{L}_{\text{SOFT}}. \tag{1.1}$$

Moreover, $\mathcal{L}_{SUSY}$ is by itself a sum of a pure gauge part $\mathcal{L}_g$, a matter part $\mathcal{L}_M$ and a Higgs part $\mathcal{L}_H$:

$$\mathcal{L}_{\text{SUSY}} = \mathcal{L}_g + \mathcal{L}_M + \mathcal{L}_H. \tag{1.2}$$

The pure gauge Lagrangian contains the spinorial gauge field strengths associated with $SU(3)$, $SU(2)$ and $U(1)$ that define the kinetic and self interaction terms for the gauge bosons and gauginos:

$$\mathcal{L}_g = \frac{1}{4} \int d^2\theta \left( W_g^{a\alpha} W_{\alpha g}^a + \vec{W}_W^\alpha \cdot \vec{W}_{W\alpha} + W_Y^\alpha W_{Y\alpha} \right) + \text{h.c.}, \tag{1.3}$$

**Table 1.1:** Matter superfields in the MSSM

| Lepton Doublets | $SU(3) \otimes SU(2) \otimes U(1)$ | Quark doublets | $SU(3) \otimes SU(2) \otimes U(1)$ |
|---|---|---|---|
| $L_1 = \begin{pmatrix} L_{\nu_e} \\ L_e \end{pmatrix}$ | $(\mathbf{1}, \mathbf{2}, -1)$ | $Q_1 = \begin{pmatrix} Q_u \\ Q_d \end{pmatrix}$ | $(\mathbf{3}, \mathbf{2}, \frac{1}{3})$ |
| $L_2 = \begin{pmatrix} L_{\nu_\mu} \\ L_\mu \end{pmatrix}$ | $(\mathbf{1}, \mathbf{2}, -1)$ | $Q_2 = \begin{pmatrix} Q_c \\ Q_s \end{pmatrix}$ | $(\mathbf{3}, \mathbf{2}, \frac{1}{3})$ |
| $L_3 = \begin{pmatrix} L_{\nu_\tau} \\ L_\tau \end{pmatrix}$ | $(\mathbf{1}, \mathbf{2}, -1)$ | $Q_3 = \begin{pmatrix} Q_t \\ Q_b \end{pmatrix}$ | $(\mathbf{3}, \mathbf{2}, \frac{1}{3})$ |
| Anti-lepton singlets | $SU(3) \otimes SU(2) \otimes U(1)$ | Anti-quarks singlets | $SU(3) \otimes SU(2) \otimes U(1)$ |
| $\overline{E}_e$ | $(\mathbf{1}, \mathbf{1}, 2)$ | $\overline{U}_1, \overline{D}_1$ | $(\overline{\mathbf{3}}, \mathbf{1}, -\frac{4}{3}), (\overline{\mathbf{3}}, \mathbf{1}, \frac{2}{3})$ |
| $\overline{E}_\mu$ | $(\mathbf{1}, \mathbf{1}, 2)$ | $\overline{U}_2, \overline{D}_2$ | $(\overline{\mathbf{3}}, \mathbf{1}, -\frac{4}{3}), (\overline{\mathbf{3}}, \mathbf{1}, \frac{2}{3})$ |
| $\overline{E}_\tau$ | $(\mathbf{1}, \mathbf{1}, 2)$ | $\overline{U}_3, \overline{D}_3$ | $(\overline{\mathbf{3}}, \mathbf{1}, -\frac{4}{3}), (\overline{\mathbf{3}}, \mathbf{1}, \frac{2}{3})$ |

**Table 1.2:** Higgs and gauge superfields in the MSSM

| Vector Gauge Superfields | | Left Chiral Higgs Superfields | |
| --- | --- | --- | --- |
| $V^Y$ | Hypercharge | $H_1 = \begin{pmatrix} H_1^0 \\ H_1^- \end{pmatrix}$ | $Y = -1$ |
| $\vec{V}^W$ | Weak Isospin | | |
| $V_g^a$ | Color | $H_2 = \begin{pmatrix} H_2^+ \\ H_2^0 \end{pmatrix}$ | $Y = 1$ |

The gauge field strengths are obtained from the vector superfields through the relations

$$W_\alpha = -\frac{1}{4} \bar{\mathcal{D}} \bar{\mathcal{D}} e^{-V} \mathcal{D}_\alpha e^V, \tag{1.4}$$

$$\bar{W}^{\dot{\alpha}} = -\frac{1}{4} \mathcal{D} \mathcal{D} e^V \bar{\mathcal{D}}^{\dot{\alpha}} e^{-V}, \tag{1.5}$$

with $V = 2gV^a T^a$ and $W_\alpha = 2gW_\alpha^a T^a$, where $T^a$ are matrices from the adjoint representation of the gauge group with coupling constant $g$. The vector superfields $V$ are expanded as functions of the Grassmann variables $\theta$, $\bar{\theta}$, in the Wess-Zumino gauge,

$$V_{WZ}^a(x, \theta, \bar{\theta}) = \theta \sigma^\mu \bar{\theta} A_\mu^a(x) + \theta\theta \bar{\theta} \bar{\lambda}^a(x) + \bar{\theta}\bar{\theta}\theta \lambda^a(x) + \frac{1}{2} \theta\theta \bar{\theta}\bar{\theta} D^a(x), \tag{1.6}$$

which give the expansions for the field strengths $W^a$, expressed in the chiral coordinates $y^\mu = x^\mu + i\theta \sigma^\mu \bar{\theta}$,

$$W_\alpha^a = \lambda_\alpha^a + D^a(y)\theta_\alpha - (\sigma^{\mu\nu}\theta)_\alpha F_{\mu\nu}^a(y) + i\theta\theta \sigma_{\alpha\dot{\beta}}^\mu \nabla_\mu \bar{\lambda}^{a\dot{\beta}}(y). \tag{1.7}$$

As for the matter Lagrangian density, it is a D-term, which means that it can be expressed as a 4-integration in the Grassmann variables:

$$\mathcal{L}_M = \int d^4\theta \left[ L_i^\dagger e^{(g_2 \vec{V}^W \cdot \vec{\sigma} + g_1 V^Y Y)} L_i + \bar{E}_i^\dagger e^{g_1 V^Y Y} \bar{E}_i + \bar{U}_i^\dagger e^{(g_3 V_g^a \bar{\lambda}^a + g_1 V^Y Y)} \bar{U}_i \right. \tag{1.8}$$

$$\left. + \bar{D}_i^\dagger e^{(g_3 V_g^a \lambda^a + g_1 V^Y Y)} \bar{D}_i + Q_i^\dagger e^{(g_3 V_g^a \lambda^a + g_2 \vec{V}^W \cdot \vec{\sigma} + g_1 V^Y Y)} Q_i \right],$$

with $\vec{\sigma}$ the Pauli Matrices, $\lambda^a$ the Gell-Mann matrices, as usual, and an implicit sum over $i = 1, 2, 3$. It gives the kinetic terms for the matter fields (SM particles and superpartners) and also the couplings between these and the gauge fields (gauge bosons and gauginos). Assuming the expansion of the left chiral superfield $\Phi$ in terms of the chiral coordinate $y$ and the Grassmann variable $\theta$

$$\Phi(y, \theta) = \phi(x) - i\theta \sigma^\mu \bar{\theta} \partial_\mu \phi(x) - \frac{1}{4} \theta\theta \bar{\theta}\bar{\theta} \partial^\mu \partial_\mu \phi(x)$$

$$+ \sqrt{2}\theta \xi(x) + \frac{i}{\sqrt{2}} \theta\theta \partial_\mu \xi \sigma^\mu \bar{\theta} + \theta\theta F(x), \tag{1.9}$$

with $\phi$ a scalar field, $\xi$ a spinor field and $F$ an auxiliary field, the component Lagrangian can be obtained. Finally, the Higgs Lagrangian establishes, in the D-term, the kinetic parts for the 2 Higgs superfields as also the couplings between them and the gauge superfields, and in the F-term, the couplings of the Higgs superfields and the matter ones, through the superpotential $\mathcal{W}_{MSSM}$:

$$\mathcal{L}_H = \int d^4\theta \left[ H_p^\dagger e^{(g_2 \vec{V}_W \cdot \vec{\sigma} + g_1 V^Y Y)} H_p + \mathcal{W}_{MSSM} \delta^2(\bar{\theta}) + \mathcal{W}_{MSSM}^\dagger \delta^2(\theta) \right], \qquad (1.10)$$

$$\mathcal{W}_{\mathrm{MSSM}} = \epsilon_{ab}(\mu H_1^a H_2^b - f_{ij}^E H_1^a L_i^b \bar{E}_j - f_{ij}^D H_1^a Q_i^b \bar{D}_j - f_{ij}^U Q_i^a H_2^b \bar{U}_j), \qquad (1.11)$$

with an implicit sum over $p = 1, 2$. The dimension four Yukawa couplings from the Standard Model are obtained from the F-terms of the products of three left chiral superfields. That is why in the previous expression there are 2 delta functions in the Grassmann variables. Also, the need for at least two Higgs superfields can be understood from the fact that the superpotential must have a definite chirality (holomorphy) and so it is impossible to define the couplings from a superfield and its complex conjugate in the same functional. This is also related with the cancelation of anomalies. In the previous Lagrangian $\mathcal{L}_{MSSM}$, there are auxiliary fields that do not have a kinematic term. In the general expansion of a left chiral superfield, there is an F-term that is associated with a 2-power of Grassmann variables, and in the general expansion of a vector superfield, there is a D-term that is associated with a 4-power in these variables. The functions of $x$ that are the coefficients of these terms do not have a kinetic part, and so can be eliminated with the Lagrange equations. For a generic superpotential of the type

$$\mathcal{W} = f_i \Phi_i + \frac{1}{2} f_{ij} \Phi_i \Phi_j + \frac{1}{3!} f_{ijk} \Phi_i \Phi_j \Phi_k, \qquad (1.12)$$

where the $\Phi_i$ are generic left-chiral superfields and a sum over repeated indexes is assumed, the equations of motion for the coefficient $F_i^\dagger$ of the F-term of $\Phi_i^\dagger$ give

$$F_i^\dagger = -f_i - f_{ij}\phi_j - \frac{1}{2} f_{ijk} \phi_j \phi_k, \qquad (1.13)$$

with $\phi_i$ the scalar components of $\Phi_i$ and where it is assumed that the Lagrangian contains both $W$ and $W^\dagger$. This is the same as taking the derivative of $W$ with respect to $\Phi_i$ and projecting it onto the scalar component:

$$F_i^\dagger = \left. \frac{\partial W}{\partial \Phi_i} \right|_{\theta=0}. \qquad (1.14)$$

In the same way, we have

$$F_i = \left. \frac{\partial W^\dagger}{\partial \Phi_i^\dagger} \right|_{\bar{\theta}=0}. \qquad (1.15)$$

Also, as can be seen by eqs. (1.3), (1.6) and (1.7), the gauge part of the Lagrangian contains a $D^a$ field, where $a$ is a gauge index, that can be eliminated, giving the relation

$$D^a = -g\phi_i^\dagger T_{ij}^a \phi_j, \tag{1.16}$$

assuming that the chiral superfields $\Phi_i$ belong to the fundamental representation of the gauge group, labeled by the multi-index $i$. In the end, we will get a scalar potential $V(\phi_i, \phi_j^*)$ given by

$$V(\phi_i, \phi_j^*) = F_i^\dagger F_i + \frac{1}{2} D^a D^a. \tag{1.17}$$

The Higgs sector is specially important in what follows, so lets concentrate on the part of the scalar potential that is responsible for EW symmetry breaking and the Higgs bosons masses. Its origin is in the D-terms (1.16), associated with $SU(2)$ and $U(1)$:

$$
\begin{aligned}
V_D =& \frac{1}{2} \vec{D}^2 + \frac{1}{2} D_Y^2 \\
=& \frac{g_2^2}{8} \left[ \left( h_1^\dagger \vec{\sigma} h_1 \right) + \left( h_2^\dagger \vec{\sigma} h_2 \right) \right]^2 + \frac{g_1^2}{8} \left[ \left( h_1^\dagger h_1 \right) - \left( h_2^\dagger h_2 \right) \right]^2,
\end{aligned} \tag{1.18}
$$

where $h_i$ are the scalar components of the superfields $H_i$, $i = 1, 2$. With the relation

$$(\vec{\sigma})_{il} \cdot (\vec{\sigma})_{kj} = 2\delta_{ij}\delta_{kl} - \delta_{il}\delta_{kj} \tag{1.19}$$

the previous expression for $V_D$ can be written as

$$V_D = \frac{g_2^2}{2} \left| \left( h_1^\dagger h_2 \right) \right|^2 + \frac{g_1^2 + g_2^2}{8} \left| \left( h_1^\dagger h_1 \right) - \left( h_2^\dagger h_2 \right) \right|^2. \tag{1.20}$$

Also, as we can see by expression (1.10), there is an interaction term between $H_1$ and $H_2$ that originates the F-terms

$$F_{H_1}^* = -\left.\frac{\partial W}{\partial h_1}\right|_{\theta=0} = -\mu h_2 \tag{1.21}$$

$$F_{H_2}^* = -\left.\frac{\partial W}{\partial h_2}\right|_{\theta=0} = \mu h_1, \tag{1.22}$$

and gives the contribution to the scalar potential

$$V_\mu = |\mu|^2 \left[ (h_1^\dagger h_1 + h_2^\dagger h_2) \right]. \tag{1.23}$$

We must add also the part from the supersymmetry breaking Lagrangian $\mathcal{L}_{SOFT}$

$$V_{\text{SOFT}} = m_1^2 h_1^\dagger h_1 + m_2^2 h_2^\dagger h_2 + \epsilon_{ab} B\mu \text{Re} \left( h_1^a h_2^b \right). \tag{1.24}$$

This Lagrangian may be obtained from first principles in a theory of supergravity, where there are superfields in a hidden "world" that communicate with the MSSM superfields only via gravitation. We shall not go into the details here, but merely state that such a Lagrangian is soft, in the sense that it does not introduce quadratic divergences. Also, the explicit and soft supersymmetry breaking for the effective Lagrangian of the MSSM is the most well motivated process of breaking supersymmetry, since breaking at tree level with an F-term or a D-term is phenomenologically excluded, and it is known that, if supersymmetry is not broken at tree level, then it cannot be broken to all orders in perturbation theory[1].

### 1.1.1 Effective Potential and Higgs Bosons Masses

The total tree level Higgs scalar potential is then

$$
\begin{aligned}
V =& V_D + V_\mu + V_{\text{SOFT}} \\
=& \frac{g_2^2}{2} \left| \left( h_1^\dagger h_2 \right) \right|^2 + \frac{g_1^2 + g_2^2}{8} \left| \left( h_1^\dagger h_1 \right) - \left( h_2^\dagger h_2 \right) \right|^2 \\
& + \left( m_1^2 + |\mu|^2 \right) \left( h_1^\dagger h_1 \right) + \left( m_2^2 + |\mu|^2 \right) \left( h_2^\dagger h_2 \right) \\
& + \epsilon_{ab} B\mu \text{Re}(h_1^a h_2^b).
\end{aligned}
\tag{1.25}
$$

The condition for this potential to be bounded below can most be seen by making the charged components of $h_1$ and $h_2$ to vanish and the neutral components to be equal. After doing this, one gets the relation

$$
2|\mu|^2 + m_1^2 + m_2^2 > B\mu,
\tag{1.26}
$$

assuming, as usual, that $B\mu$ is positive. Also, to look for a minimum of the scalar potential with electromagnetic gauge invariance, we must set the charged components in $V$ also equal to 0, so that $V$ becomes

$$
\begin{aligned}
V^0 =& \frac{g_1^2 + g_2^2}{8} \left| \left| h_1^0 \right|^2 - \left| h_2^0 \right|^2 \right|^2 + \left( m_1^2 + |\mu|^2 \right) \left| h_1^0 \right|^2 \\
& + \left( m_2^2 + |\mu|^2 \right) \left| h_2^0 \right|^2 - B\mu \text{Re}(h_1^0 h_2^0).
\end{aligned}
\tag{1.27}
$$

With the vacuum expectation values $v_i$ of $h_i$ that break $SU(2) \otimes U(1)$, the fields can be expanded around it[2]

$$
h_i^0 = v_i + \phi_i
\tag{1.28}
$$

---

[1] Details can be found in [2].

[2] With this convention we have $\sqrt{v_1^2 + v_2^2} = 174$ GeV.

which gives, up to second order in $\phi_i$,

$$V^0 = \frac{g_1^2 + g_2^2}{4} \left( |v_1|^2 - |v_2|^2 \right) \left[ 2\mathrm{Re}\left( v_1^* \phi_1 - v_2^* \phi_2 \right) \right] \tag{1.29}$$

$$+ \frac{g_1^2 + g_2^2}{2} \left[ 2\mathrm{Re}\left( v_1^* \phi_1 - v_2^* \phi_2 \right) \right] + \left( m_1^2 + |\mu|^2 \right) \left[ 2\mathrm{Re}(v_1^* \phi_1 + |\phi_1|^2) \right]$$

$$+ \left( m_2^2 + |\mu|^2 \right) \left[ 2\mathrm{Re}(v_2^* \phi_2 + |\phi_2|^2) \right] \tag{1.30}$$

$$- B\mu \mathrm{Re}(v_1 \phi_2 + v_2 \phi_1 + \phi_1 \phi_2) + \text{constant}.$$

The minimization of this expression in order to the fields $\phi_i$ forces the linear terms to vanish:

$$\left( m_1^2 + |\mu|^2 \right) v_1 + \frac{g_1^2 + g_2^2}{4} \left( v_1^2 - v_2^2 \right) v_1 - \frac{1}{2} B\mu v_2 = 0, \tag{1.31}$$

$$\left( m_2^2 + |\mu|^2 \right) v_2 - \frac{g_1^2 + g_2^2}{4} \left( v_1^2 - v_2^2 \right) v_2 - \frac{1}{2} B\mu v_1 = 0. \tag{1.32}$$

where, without loss of generality, we have assumed that the vacuum expectation values are real. It is convenient to define the parameters

$$\tan\beta = \frac{v_2}{v_1}, \tag{1.33}$$

$$m_A^2 = 2|\mu|^2 + m_1^2 + m_2^2, \tag{1.34}$$

and also to consider the $Z$ boson mass

$$m_Z^2 = \frac{1}{2}(g_1^2 + g_2^2)(v_1^2 + v_2^2). \tag{1.35}$$

Multiplying (1.31) and (1.32) by $v_2$ and $v_1$, respectively, and summing and subtracting, gives

$$B\mu = m_A^2 \sin 2\beta, \tag{1.36}$$

$$m_1^2 - m_2^2 = -\left( m_A^2 + m_Z^2 \right) \cos 2\beta. \tag{1.37}$$

Also, the following relations will be useful:

$$m_1^2 + |\mu|^2 = \frac{1}{2} m_A^2 - \frac{1}{2} (m_A^2 + m_Z^2) \cos 2\beta \tag{1.38}$$

$$m_2^2 + |\mu|^2 = \frac{1}{2} m_A^2 + \frac{1}{2} (m_A^2 + m_Z^2) \cos 2\beta \tag{1.39}$$

From (1.34) and (1.37) we see that the square of the $\mu$ parameter can be fixed by these equations:

$$|\mu|^2 = \frac{1}{1 - \tan^2 \beta} \left( m_2^2 \tan^2 \beta - m_1^2 \right) - \frac{1}{2} m_Z^2. \tag{1.40}$$

There is an important feature in this equation. We see that for $\tan\beta > 1$ $m_2^2$ must be negative. This behavior is typical from spontaneous symmetry breaking where usually it is necessary to put by hand a negative mass of the scalar field. However in the MSSM is possible to drive $m_2^2$ negative through the running of RGEs from some unification scale. This fact is known as Radiative Electroweak Symmetry Breaking. This equation will be important in the discussion of neutralino masses.

We have three equations[3] for five quantities, $m_1, m_2, B, \mu$ and $A$, so only two of them are independent, modulus the signs. These can be chosen to be $m_1$ and $m_2$. In the next section it will be clarified why, in fact, only one parameter $m_0$ is taken as arbitrary, so that $m_1 = m_2 = m_0$ at some high scale.

With the vacuum expectation values that are the solutions of (1.31) and (1.32), the quadratic scalar potential (1.30) is

$$
\begin{aligned}
V^0 = {} & \frac{1}{2} m_Z^2 \cos 2\beta \left[ |\phi_1|^2 - |\phi_2|^2 \right] + m_Z^2 \left[ \mathrm{Re}(\cos\beta\phi_1 - \sin\beta\phi_2) \right]^2 \\
& + \frac{1}{2} m_A^2 \left( |\phi_1|^2 + |\phi_2|^2 \right) - \frac{1}{2}(m_A^2 + m_Z^2) \cos 2\beta \left[ |\phi_1|^2 - |\phi_2|^2 \right] \\
& - m_A^2 \sin 2\beta \, \mathrm{Re} \left( \phi_1\phi_2 \right) + \text{constant.}
\end{aligned}
\tag{1.41}
$$

From this, it is possible to obtain the mass matrices of the neutral Higgs bosons. First, for the imaginary components of $\phi_i$, we have

$$
M_{\mathrm{Im}}^2 = \begin{pmatrix} \frac{1}{2} m_A^2 (1 - \cos 2\beta) & \frac{1}{2} m_A^2 \sin 2\beta \\ \frac{1}{2} m_A^2 \sin 2\beta & \frac{1}{2} m_A^2 (1 + \cos 2\beta) \end{pmatrix}.
\tag{1.42}
$$

This has a zero eigenvalue, which is identified with one of the Goldstone bosons from electroweak symmetry breaking. The other eigenvalue is equal to the trace, which is just $m_A^2$. This will be the CP-odd neutral Higgs boson $A$. For the real parts of the $\phi_i$, we get for the elements of the mass matrix

$$
(M_{\mathrm{Re}}^2)_{11} = \frac{1}{2} m_A^2 (1 - \cos 2\beta) + \frac{1}{2} m_Z^2 (1 + \cos 2\beta)
\tag{1.43}
$$

$$
(M_{\mathrm{Re}}^2)_{12} = (M_{\mathrm{Re}}^2)_{21} = -\frac{1}{2}(m_A^2 + m_Z^2) \sin 2\beta
\tag{1.44}
$$

$$
(M_{\mathrm{Re}}^2)_{22} = \frac{1}{2} m_A^2 (1 + \cos 2\beta) + \frac{1}{2} m_Z^2 (1 - \cos 2\beta).
\tag{1.45}
$$

---

[3]The Z boson mass is, of course, given by its phenomenological value and $\tan\beta$ is assumed to be an independent parameter of the theory, only barely constrained.

The eigenvalues equation gives

$$m_H^2 = \frac{1}{2}\left[m_A^2 + m_Z^2 + \sqrt{(m_A^2 + m_Z^2)^2 - 4m_A^2 m_Z^2 \cos^2 2\beta}\right] \tag{1.46}$$

$$m_h^2 = \frac{1}{2}\left[m_A^2 + m_Z^2 - \sqrt{(m_A^2 + m_Z^2)^2 - 4m_A^2 m_Z^2 \cos^2 2\beta}\right]. \tag{1.47}$$

With the natural assumption $m_A > m_Z$, we can expand the previous expressions in powers of $m_Z^2/m_A^2$ to conclude that

$$m_H^2 > m_A^2 \tag{1.48}$$

$$m_h^2 < m_Z^2. \tag{1.49}$$

The second inequality is obviously excluded by LEP data, but it should be emphasized that these are tree level results, and radiative corrections can change this picture in a significant way, as we shall see.

Finally, the masses of the charged scalars can be obtained by taking in (1.25) the neutral scalars equal to their vev and the charged components to be $h_1^-$ and $h_2^+$. Thus, the quadratic part of the potential is

$$V^C = \frac{g_2^2}{2}\left|v_2(h_1^-)^* + v_1 h_2^+\right|^2 + \frac{g_1^2 + g_2^2}{4}(v_1^2 - v_2^2)\left(|h_1^-|^2 - |h_2^+|^2\right)$$
$$+ (m_1^2 + |\mu|^2)|h_1^-|^2 + (m_2^2 + |\mu|^2)|h_2^+|^2 + \text{Re}(B\mu h_1^- h_2^+). \tag{1.50}$$

With formulas (2.11), (1.38), (1.39) and the $W$ boson mass

$$m_W^2 = \frac{1}{2}g_2^2(v_1^2 + v_2^2) \tag{1.51}$$

$V^C$ can be written as

$$V^C = (m_A^2 + m_Z^2)\left(\sin^2\beta|h_1^-|^2 + \cos\beta^2|h_2^+|^2 + \frac{1}{2}\sin 2\beta\left(h_1^- h_2^+ + (h_1^-)^*(h_2^+)^*\right)\right) \tag{1.52}$$

which, in matrix form, is

$$V^C = (m_A^2 + m_Z^2)\begin{pmatrix} h_1^- \\ (h_2^+)^* \end{pmatrix}^\dagger \begin{pmatrix} \sin^2\beta & \frac{1}{2}\sin 2\beta \\ \frac{1}{2}\sin 2\beta & \cos^2\beta \end{pmatrix}\begin{pmatrix} h_1^- \\ (h_2^+)^* \end{pmatrix}, \tag{1.53}$$

which gives a mass matrix with a zero eigenvalue (the other Goldstone boson) and the other eigenvalue equal to the trace, $m_W^2 + m_A^2$. This is the squared mass of the charged scalar.

To see how quantum corrections affect the mass of the Higgs neutral scalars, we must compute the effective potential [17], [1]. Lets start to consider a simple $\phi^4$ model with

Lagrangian

$$L = \frac{1}{2}(\partial_\mu \phi)^2 + \frac{1}{2}m^2\phi^2 - \frac{1}{4!}g\phi^4 + \text{c.t.}, \tag{1.54}$$

where $\phi$ is a real scalar field, $m$ and $g$ are the physical constants and c.t. denote counterterms necessary to renormalize the theory. Assuming that the scalar field acquires a vev $v$, expanding $\phi$ at $v$, with

$$\phi \longrightarrow \phi + v, \tag{1.55}$$

we obtain for the physical action

$$I[\phi + v] = -\mathcal{V}_4 \left( -\frac{1}{2}m^2 v + \frac{1}{24}gv^4 \right) + \left( m^2 v - \frac{1}{6}v^3 \right) \int d^4x \phi$$
$$+ \int d^4x \left( \frac{1}{2}\partial_\mu \phi \partial^\mu \phi - \frac{1}{2}\mu^2 \phi^2 \right) - \int d^4x \left( \frac{1}{6}gv\phi^3 + \frac{1}{24}g\phi^4 \right), \tag{1.56}$$

Here, $\mathcal{V}_4$ is the four-volume that we can regularize in some way and $\mu^2$ is the field dependent mass

$$\mu^2 = -m^2 + \frac{1}{2}gv^2. \tag{1.57}$$

The effective action $\Gamma$ and the effective potential $V_{eff}$, at $v$, are related by

$$\Gamma[v] = -\mathcal{V}_4 V_{eff}(v). \tag{1.58}$$

Also, the exponential of effective action is given by the functional integral of the exponential of the action, with the integration restricted to one particle irreducible diagrams [1]. This allows us to conclude that the zero-loop term is just the constant in (1.56)

$$i\Gamma^{(0)}(v) = -i\mathcal{V}_4 \left( -\frac{1}{2}m^2 v^2 + \frac{g}{24}v^4 \right) \tag{1.59}$$

and the one-loop term is

$$\exp\left( i\Gamma^{(1)}(v) \right) = \int \prod_x d\phi(x) \exp\left[ \frac{1}{2}i \int d^4x \left( \partial_\mu \phi(x) \right)^2 - \mu^2 \phi(x)^2 \right]. \tag{1.60}$$

This can be put in a Gaussian form with an integration by parts, so that one needs to compute

$$\ln \text{Det}\left( \frac{iK}{\pi} \right), \tag{1.61}$$

in order to have

$$i\Gamma^{(1)}[v] = -\frac{1}{2}\ln \text{Det}\left( \frac{iK}{\pi} \right), \tag{1.62}$$

where $K$ is the Klein-Gordon operator:

$$K_{xy} = \left( -\frac{\partial}{\partial x_\mu} \frac{\partial}{\partial y^\mu} + \mu^2 \right) \delta^4(x-y). \tag{1.63}$$

It is useful to consider the identity

$$\mathrm{Det}A = \exp\left( \ln \mathrm{Tr} A \right) \tag{1.64}$$

and to Fourier transform the operator K

$$\begin{aligned} K_{p,q} &= \int \frac{d^4x}{(2\pi)^2} e^{-ip\cdot x} \frac{d^4y}{(2\pi)^2} e^{iq\cdot y} K_{x,y} \\ &= (-p^2 + \mu^2) \delta^4(p-q), \end{aligned} \tag{1.65}$$

so that

$$\begin{aligned} i\Gamma^{(1)}[v] &= -\frac{1}{2} \int d^4p \left[ \ln\left( \frac{iK}{\pi} \right)_{p,p} \right] \\ &= -\frac{\mathcal{V}_4}{2(2\pi)^4} \int d^4p \ln\left( \frac{i}{\pi} \left( -p^2 + \mu^2 \right) \right), \end{aligned} \tag{1.66}$$

where the four-volume factor arrives from the delta function in $K(p,p)$, as shown by its Fourier representation. This integral is obviously ill-defined, but it can be made convergent with dimensional regularization [18]. Also, performing a Wick rotation, we have

$$\begin{aligned} \int \frac{d^4p}{(2\pi)^4} \ln\left( \frac{i}{\pi}\left( -p^2 + \mu^2 \right) \right) &\longrightarrow \int \frac{d^dp}{(2\pi)^d} \ln\left( \frac{i}{\pi}\left( -p^2 + \mu^2 \right) \right) \\ &= i \int \frac{d^dp_E}{(2\pi)^d} \ln\left( \frac{i}{\pi}(p_E^2 + \mu^2) \right) \\ &= -i\frac{\partial}{\partial \alpha} \int \frac{d^dp_E}{(2\pi)^d} \left( \frac{-i\pi}{p_E^2 + \mu^2} \right)^\alpha \bigg|_{\alpha=0} \\ &= -i\frac{\partial}{\partial \alpha} \left( \frac{(-i\pi)^\alpha}{(4\pi)^{d/2}} \frac{\Gamma(\alpha - \frac{d}{2})}{\Gamma(\alpha)} \frac{1}{(\mu^2)^{\alpha - d/2}} \right) \bigg|_{\alpha=0} \\ &= -i\frac{\Gamma(-\frac{d}{2})}{(4\pi)^{d/2}} \frac{1}{(\mu^2)^{-d/2}}, \end{aligned} \tag{1.67}$$

noting that $\Gamma'(\alpha - d/2)$ is finite and $\Gamma(\alpha) \longrightarrow 1/\alpha$ as $\alpha \longrightarrow 0$. Then, the effective potential up to one-loop is

$$V_{\mathrm{eff}} = -\frac{1}{2}m^2 v^2 + \frac{g}{24}v^4 - \frac{1}{2}\frac{\Gamma(-\frac{d}{2})}{(4\pi)^{d/2}}(\mu^2)^{d/2} + \mathrm{c.t.}. \tag{1.68}$$

The one-loop term is divergent in the limit $\epsilon = 4 - d \longrightarrow 0$, but the counter-terms allow us to eliminate the divergences [17]. Using the well known properties of the $\Gamma$ function

$$\Gamma(z+1) = z\Gamma(z), \tag{1.69}$$

$$\Gamma(\epsilon) = \frac{1}{\epsilon} - \gamma + O(\epsilon), \tag{1.70}$$

where $\gamma$ is the Euler-Mascheroni constant, we see that

$$\frac{\Gamma(-\frac{d}{2})}{(4\pi)^{d/2}}(\mu^2)^{d/2} = \frac{1}{\frac{d}{2}(\frac{d}{2}-1)}\frac{\Gamma(2-\frac{d}{2})}{(4\pi)^{d/2}}(\mu^2)^{d/2} \tag{1.71}$$

$$= \frac{\mu^4}{2(4\pi)^2}\left(\frac{2}{\epsilon} - \gamma + \ln(4\pi) - \ln(\mu^2) + \frac{3}{2}\right). \tag{1.72}$$

In the $\overline{MS}$ renormalization scheme, the divergent part as the $\ln(4\pi)$ and the $\gamma$ terms are subtracted using the counter-terms. In the end, we get the following expression for the effective potential up to one-loop:

$$V_{\text{eff}} = -\frac{1}{2}m^2v^2 + \frac{g}{24}v^4 - \frac{1}{64\pi^2}\mu^4\left(\ln\frac{\mu^2}{Q^2} - \frac{3}{2}\right), \tag{1.73}$$

where an arbitrary mass scale $Q$ was introduced for the expression to be dimensionally correct. This mass scale can be understood in the framework of the Renormalization Group equations.

When one has, in addition, Dirac fermions that couple to the scalars through the typical Yukawa term $\lambda\overline{\psi}\psi\phi$, one has to consider the Gaussian of some operator on fermionic variables $\chi$:

$$\text{Det}A = \exp\left(\int d\overline{\chi}d\chi\,\overline{\chi}A\chi\right). \tag{1.74}$$

On the other hand, it is known that the square of the Dirac operator is the Klein-Gordon operator, with 4 multiplicity, so we can reproduce the computation performed above, with some extra-factors of 2 and 1/2. Putting all together, we get for the scalar potential in a model with a Dirac fermion and a real scalar boson,

$$V_{\text{eff}} = -\frac{1}{2}m^2v^2 + \frac{g}{24}v^4 - \frac{1}{64\pi^2}\mu^4\left(\ln\frac{\mu^2}{Q^2} - \frac{3}{2}\right)$$
$$+ \frac{1}{16\pi^2}m_D^4\left(\ln\frac{m_D^2}{Q^2} - \frac{3}{2}\right), \tag{1.75}$$

with the Dirac mass $m_D = \lambda v$.

In a supersymmetric theory, we have the same degrees of freedom for super-partners, so if there is a Dirac fermion with 4 degrees of freedom, 2 complex scalar bosons must correspond. This means that we must multiply the one-loop scalar part by 4, assuming a full boson mass degeneracy:

$$V_{\text{SUSY}}^{1-\text{loop}} = -\frac{1}{16\pi^2}\left(m_B^4\left(\ln\frac{m_B^2}{Q^2}-\frac{3}{2}\right)-m_f^4\left(\ln\frac{m_f^2}{Q^2}-\frac{3}{2}\right)\right), \qquad (1.76)$$

where $m_B$ and $m_f$ are the boson and the fermion masses. Remarkably, if they are equal, the two terms cancel and there is no one-loop correction. This is one more evidence for the need of breaking supersymmetry. We have seen that the tree level mass value for the lightest neutral Higgs boson is too low to be correct. This means that higher order corrections are essential to bring it to acceptable values, and for that there must be an unbalance between fermion and boson masses.

Higher order corrections to the Higgs boson masses are dominated by loops with the top and the stop, so that a color factor of 3 must be included. On the other hand, the top mass is given by the Yukawa coupling from the last term in (1.10) and the stop mass comes from the SUSY soft breaking Lagrangian with a mass parameter $\widetilde{m}$ and also from the F-terms in the superpotential. Neglecting generation mixing and assuming Left-Right mass degeneracy, the field dependent top and stop masses are

$$m_t^2 = (f_{33}^U)^2|h_2^0|^2, \qquad (1.77)$$

$$m_{\tilde{t}}^2 = \widetilde{m}^2 + (f_{33}^U)^2|h_2^0|^2. \qquad (1.78)$$

The one-loop effective potential will give a contribution to the total scalar potential equal to

$$\Delta V^{1-\text{Loop}} = \frac{3}{16\pi^2}\left((\widetilde{m}^2+(f_{33}^U)^2|h_2^0|^2)\left(\log\frac{\widetilde{m}^2+(f_{33}^U)^2|h_2^0|^2}{Q^2}-\frac{3}{2}\right)\right.$$
$$\left.-(f_{33}^U)^2|h_2^0|^2\left(\log\frac{(f_{33}^U)^2|h_2^0|^2}{Q^2}-\frac{3}{2}\right)\right). \qquad (1.79)$$

This will change the minimization equation (1.32), introducing the extra term

$$\frac{3(f_{33}^U)^2}{8\pi^2}\left(m_{\tilde{t}}^2\left(\ln\frac{m_{\tilde{t}}^2}{Q^2}-1\right)-m_t^2\left(\ln\frac{m_t^2}{Q^2}-1\right)\right), \qquad (1.80)$$

which gives an extra contribution to (1.39) and, as a consequence, also to (1.41) and (1.45). However, this contribution partially cancels the second derivative of (1.79), leaving us with the positive correction

$$\Delta_{h^0} = \frac{3(f_{33}^U)^2 m_t^2}{4\pi^2}\ln\frac{m_{\tilde{t}}^2}{m_t^2} \equiv \frac{\epsilon_h}{\sin^2\beta}, \qquad (1.81)$$

where

$$\epsilon_h = \frac{3G_F m_t^4}{\sqrt{2}\pi^2}\ln\frac{m_{\tilde{t}}^2}{m_t^2}. \qquad (1.82)$$

Then the matrix element $M_{\text{Re}}^2)_{22}$ in (1.45) will have this additional contribution which changes the eigenvalues to

$$
\begin{aligned}
m_{h,H}^2 =& \frac{1}{2}\left(m_A^2 + m_Z^2 + \frac{\epsilon_h}{\sin^2 \beta}\right) \\
& \pm \frac{1}{2}\left((m_A^2 + m_Z^2)^2 \sin^2 2\beta + \left((m_Z^2 - m_A^2)\cos 2\beta + \frac{\epsilon_h}{\sin^2\beta}\right)^2\right)^{1/2}.
\end{aligned} \tag{1.83}
$$

### 1.1.2 Neutralino and Chargino Masses

The gauginos and higgsinos mass eigenstates are known as neutralinos and charginos. They are a superposition of the neutral gauginos and neutral higgsinos in the case of neutralinos and the charged ones for charginos. To obtain the relevant mass matrices, one has to consider the Higgs Lagrangian (1.10) and the general expansion (1.9) applied to the Higgs superfields $H_1$ and $H_2$. This generates terms that generically are of the type

$$
-\sqrt{2}g(T^a)_{ij}\lambda^a \xi_j \phi_i^* + \text{h.c.}, \tag{1.84}
$$

where $\lambda^a$, $\phi_i$ and $\xi^j$ are a gaugino, a Higgs and a Higgsino fields respectively. When the neutral components of the Higgs fields acquire vevs, a mixed Higgsino/gaugino mass term appears. Thus, for the charged fermions,

$$
\mathcal{L}_{\text{MASS}}^c = -\frac{g_2}{\sqrt{2}}\left(v_1\lambda^+\tilde{h}_1^2 + v_2\lambda^-\tilde{h}_2^1 + \text{h.c.}\right) - \left(M_2\lambda^+\lambda^- + \mu\tilde{h}_1^2\tilde{h}_2^1 + \text{h.c.}\right) \tag{1.85}
$$

where the gaugino states are $\lambda^\pm = 1/\sqrt{2}(\lambda_1 \mp i\lambda_2)$. Defining two component fermionic fields as

$$
\psi^+ = \begin{pmatrix} \lambda^+ \\ \tilde{h}_2^1 \end{pmatrix}, \tag{1.86}
$$

$$
\psi^- = \begin{pmatrix} \lambda^- \\ \tilde{h}_1^2 \end{pmatrix}, \tag{1.87}
$$

the chargino mass Lagrangian is

$$
\mathcal{L}_{MASS}^c = -\left(\psi^-\right)^T M_c \psi^+ \tag{1.88}
$$

with the chargino mass matrix

$$
M_c = \begin{pmatrix} M_2 & \sqrt{2}M_W\sin\beta \\ \sqrt{2}M_W\cos\beta & \mu \end{pmatrix} \tag{1.89}
$$

Upon diagonalization of this matrix, we obtain the mass eigenstates which are linear combinations of gauginos and higgsinos.

As for the neutralinos, the mass Lagrangian is

$$\mathcal{L}^n_{MASS} = -\frac{g_2}{\sqrt{2}}\lambda_3 \left(v_1\tilde{h}^1_1 - v_2\tilde{h}^2_2\right) + \frac{g_1}{\sqrt{2}}\lambda_0 \left(v_1\tilde{h}^1_1 - v_2\tilde{h}^2_2\right)$$
$$- M_2\lambda_3\lambda_3 - M_1\lambda_0\lambda_0 + \mu\tilde{h}^1_1\tilde{h}^2_2 + \text{h.c.} \tag{1.90}$$

Defining a four component fermionic field as

$$\psi^0 = \begin{pmatrix} \lambda_0 \\ \lambda_3 \\ \tilde{h}^1_1 \\ \tilde{h}^2_2 \end{pmatrix}, \tag{1.91}$$

the neutralino mass Lagrangian can be written in the compact way

$$\mathcal{L}^n_{MASS} = -\frac{1}{2}(\psi^0)^T M_n \psi^0 + \text{h.c.} \tag{1.92}$$

with the neutralino mass matrix

$$M_n = \begin{pmatrix} M_1 & 0 & -M_Z c_\beta s_W & M_Z s_\beta s_W \\ 0 & M_2 & M_Z c_\beta c_W & -M_Z s_\beta c_W \\ -M_Z c_\beta s_W & M_Z c_\beta c_W & 0 & -\mu \\ M_Z s_\beta s_W & -M_Z s_\beta c_W & -\mu & 0 \end{pmatrix} \tag{1.93}$$

with $s_W \equiv \sin\theta_W$, $c_W \equiv \cos\theta_W$, $s_\beta \equiv \sin\beta$ and $c_\beta \equiv \cos\beta$. Upon diagonalization, we obtain four mass eigenstates that are linear combinations of the neutral gauginos and higgsinos.

We see that the neutralino and chargino masses are strongly dependent on the soft breaking mass parameters $M_1$ and $M_2$ and also on the $\mu$ parameter. Since this last one is fixed by the electroweak breaking conditions (1.36) and (1.37) and assuming universal boundary conditions at the GUT scale, as it is the case in mSUGRA, in strict sense the neutralinos and charginos masses depend on the scalar $m_0$ and fermionic $M_{1/2}$ soft breaking mass parameters and also, of course, on $\tan\beta$. This allows us to obtain some limit cases for the neutralinos. Since the $\mu$ parameter belongs to the Higgsino part of the neutralino mass matrix, we see that for $\mu \ll M_1, M_2$, the lightest neutralino state will be mainly a Higgsino and for the opposite condition it will be mainly a gaugino. This can happen in some regions of the mSUGRA parameter space, as we shall see further on.

### 1.1.3 Lepton and Slepton Masses

To obtain the slepton masses we need to consider the relevant part of the soft SUSY breaking Lagrangian:

$$-\mathcal{L}^{\tilde{l}}_{SOFT} = \tilde{l}^*_{iL}(M^2_{\tilde{l}})_{ij}\tilde{l}_{jL} + \tilde{e}^*_{iR}(M^2_{\tilde{e}})_{ij}\tilde{e}_{jR} + \left[\epsilon_{ab}h^a_1\tilde{l}^b_i(f^E A^E)_{ij}\tilde{e}^*_{jR} + \text{h.c.}\right] \tag{1.94}$$

There are also $F$ and $D$ terms that arrive from (1.11) and (1.8) through (1.14) and (1.16), respectively, and that contribute to the scalar potential (1.17). The $F$ terms are

$$-\mathcal{L}^{\tilde{l}}_F = \left|\mu^* h^-_2 - \tilde{\nu}_i f^{E*}_{ij}\tilde{e}_{jR}\right|^2 + \left|\mu^* h^{0*}_2 - \tilde{e}^*_{iL}f^{E*}_{ij}\tilde{e}_{jR}\right|^2$$
$$+ \sum_i \left|f^E_{ji}\epsilon_{ab}h^a_1\tilde{l}^b_{jL}\right| + f^E_{ij}f^{E*}_{ij'}\tilde{e}^*_{jR}\tilde{e}_{j'R}\left(|h^0_1|^2 + h^+_1 h^-_1\right) \tag{1.95}$$

and the $D$ terms are

$$-\mathcal{L}^{\tilde{l}}_D = \frac{1}{4}g^2_1\left(|h_1|^2 - |h_2|^2\right)\sum_i\left(|\tilde{l}_{iL}|^2 - 2|\tilde{e}_{iR}|^2\right) + \frac{1}{4}g^2_2\left(h^\dagger_1\vec{\sigma}h_1 + h^\dagger_2\vec{\sigma}h_2\right)\tilde{l}^\dagger_{iL}\vec{\sigma}\tilde{l}_{iL}. \tag{1.96}$$

When the neutral Higgs fields acquire vevs, the previous expressions lead to the following mass terms

$$-\mathcal{L}^{\tilde{l}}_M = \tilde{\nu}^*_i\left((M^2_{\tilde{l}})_{ij} + M^2_Z\cos 2\beta\delta_{ij}\right)\tilde{\nu}_j$$
$$+ \tilde{e}^*_{iL}\left((M^2_{\tilde{l}})_{ij} - M^2_Z\cos 2\beta(1/2 - \sin^2\theta_W)\delta_{ij} + m^2_{e_i}\delta_{ij}\right)\tilde{e}_{jL}$$
$$+ \tilde{e}^*_{iR}\left((M^2_{\tilde{e}})_{ij} - M^2_Z\cos 2\beta\sin^2\theta_W\delta_{ij} + m^2_{e_i}\delta_{ij}\right)\tilde{e}_{jR}$$
$$- \left(\tilde{e}^*_{iL}(A^{E*}_{ij} + m_{e_i}\delta_{ij}\mu\tan\beta)\tilde{e}_{jR} + \text{h.c.}\right). \tag{1.97}$$

Defining a six-component slepton field $\tilde{\mathbf{f}}$

$$\tilde{\mathbf{f}} = \begin{pmatrix} \tilde{f}_L \\ \tilde{f}_R \end{pmatrix} \tag{1.98}$$

where each $\tilde{f}_{L,R}$ carries a generation index $i$, we can write the sfermion mass Lagrangian in a compact way,

$$-\mathcal{L}^{\tilde{l}}_M = \sum_{f=\nu,e}\tilde{\mathbf{f}}^\dagger\mathbf{M}_{\tilde{\mathbf{f}}}\tilde{\mathbf{f}}, \tag{1.99}$$

with $\mathbf{M}_{\tilde{\mathbf{f}}}$ a $2\times 2$ block matrix, each block being a $3\times 3$ matrix:

$$\mathbf{M}^2_{\tilde{\mathbf{f}}} = \begin{pmatrix} \mathbf{M}^2_{\tilde{f}LL} & \mathbf{M}^2_{\tilde{f}LR} \\ \mathbf{M}^2_{\tilde{f}RL} & \mathbf{M}^2_{\tilde{f}RR} \end{pmatrix} \tag{1.100}$$

Explicitly, for sneutrinos,

$$\mathbf{M}_{\tilde{\nu}}^2 = \begin{pmatrix} \mathbf{M}_{\tilde{l}}^2 + 1/2 M_Z^2 \cos 2\beta \, \mathbf{1} & 0 \\ 0 & 0 \end{pmatrix} \tag{1.101}$$

and for selectrons, smuons and staus,

$$\mathbf{M}_{\tilde{\mathbf{e}}}^2 = \begin{pmatrix} \mathbf{M}_{\tilde{l}}^2 - M_Z^2(1/2 - \sin^2\theta_W)\cos 2\beta \, \mathbf{1} + \mathbf{m}_{\mathbf{e}}^\dagger \mathbf{m}_{\mathbf{e}} & -\mathbf{m}_{\mathbf{e}}(\mathbf{A}^{E*} + \mu\tan\beta \, \mathbf{1}) \\ -(\mathbf{A}^{ET} + \mu^*\tan\beta \, \mathbf{1})\mathbf{m}_{\mathbf{e}}^\dagger & \mathbf{M}_{\tilde{e}}^2 - M_Z^2 \sin^2\theta_W \cos 2\beta \, \mathbf{1} + \mathbf{m}_{\mathbf{e}}^\dagger \mathbf{m}_{\mathbf{e}} \end{pmatrix} \tag{1.102}$$

with $(\mathbf{m}_{\mathbf{e}})_{ij} = m_{e_i}\delta_{ij}$. In the case of a significative mixing, either by large values of $\mathbf{A}^E$ or of the $\mu$ parameter, the mass of the lightest slepton can be reduced considerably.

### 1.1.4 Renormalization Group Equations

It is well known that the 1-loop Renormalization Group Equation (RGE) for the gauge coupling in Yang-Mills theories is [17]

$$\frac{dg^2}{dt} = \frac{1}{8\pi^2}\left(-\frac{11}{3}C_2(G) + \frac{2}{3}\sum_i T_i(R) + \frac{1}{3}\sum_\alpha T_\alpha(R)\right)g^4, \tag{1.103}$$

where $C_2(G)$ is the quadratic Casimir factor of the gauge group $G$ ($N$ for $SU(N)$), $T_i(R)$ refers to the representation constant to which the $i$ chiral fermions belong and $\alpha$ refers to the complex scalars in the loop, with the same meaning for $T_\alpha(R)$. In a supersymmetric theory, for each chiral fermion in some representation, there is a complex scalar in the same representation. In this way for the MSSM we must add the second and the third terms in the previous expression for each $i$ and with $\alpha = i$. Also, for each gauge boson in the adjoint representation there will be a gaugino in the same representation and this means that we must add $-11/3C_2(G)$ with $2/3C_2(G)$. Then for SUSY Yang-Mills theories we have

$$\frac{dg^2}{dt} = \frac{1}{8\pi^2}\left(-3C_2(G) + \sum_i T_i(R)\right)g^4. \tag{1.104}$$

To get the explicit evolution of the three gauge couplings in the MSSM we must apply this formula to its gauge, Higgs and matter content as shown in table 1.1. For the $U(1)_Y$ gauge group we must add $Y^2/4$ for the matter and Higgs fields which gives 11 so that, for $g_1$,

$$\frac{dg_1^2}{dt} = \frac{11}{8\pi^2}g_1^4. \tag{1.105}$$

For $SU(2), C_2(G) = 2$ and the gauge contribution is $-6$. The matter and Higgs fields contribute with $14/2$ as there are 14 $SU(2)$ doublets in the MSSM (counting color) and

$T_i(G) = 1/2$. So

$$\frac{dg_2^2}{dt} = \frac{1}{8\pi^2}g_2^4. \tag{1.106}$$

Finally for $SU(3)$ the gauge part contributes with $-9$ and there are 12 color triplets and anti-triplets that contribute to $\sum_i T_i(R)$ with 6. Then

$$\frac{dg_3^2}{dt} = -\frac{3}{8\pi^2}g_3^4. \tag{1.107}$$

Writing generically

$$\frac{dg_i^2}{dt} = -\frac{1}{8\pi^2}\beta_i g_i^4. \tag{1.108}$$

we can integrate this equations to give at 1-loop

$$g_i^2 = \frac{g_i^2(\mu_0)}{\left(1 - \frac{1}{8\pi^2}\beta_i g_i^2(\mu)\ln(\mu/\mu_0)\right)}, \tag{1.109}$$

with $t = \ln(\mu/\mu_0)$. The evolution is shown on figure 1.1 with the usual $SU(5)$ normalization for the $U_1$ gauge coupling, $g_1^2 \to 3/5g_1^2$. Note the unification of the three gauge couplings on an energy scale around $10^{16}$ GeV. For the sake of completeness we present here other RGEs in the MSSM. We list only the ones relative to the third generation. Details can be found on [4].

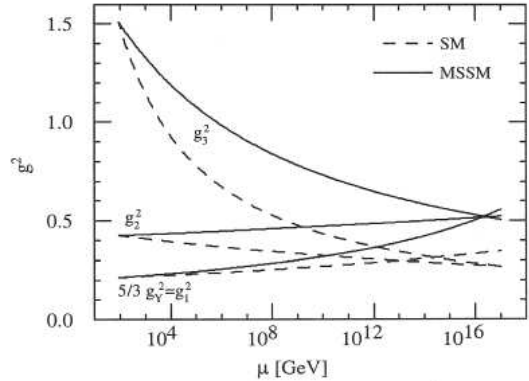

**Figure 1.1:** Evolution of the SM and MSSM gauge couplings with energy scale. Taken from [4].

### Yukawa Couplings

$$\frac{df_t}{dt} = \frac{f_t}{16\pi^2}\left(6f_t^2 + f_b^2 - \frac{16}{3}g_3^2 - 3g_2^2 - \frac{13}{9}g_1^2\right) \tag{1.110}$$

$$\frac{df_b}{dt} = \frac{f_b}{16\pi^2}\left(6f_b^2 + f_t^2 + f_\tau^2 - \frac{16}{3}g_3^2 - 3g_2^2 - \frac{7}{9}g_1^2\right) \tag{1.111}$$

$$\frac{df_\tau}{dt} = \frac{f_\tau}{16\pi^2}\left(3f_b^2 + 4f_\tau^2 - 3g_2^2 - \frac{7}{9}g_1^2\right) \tag{1.112}$$

**$\mu$ parameter**

$$\frac{d\mu}{dt} = \frac{\mu}{16\pi^2}(3f_t^2 + 3f_b^2 + f_\tau^2 - 3g_2^2 - g_1^2) \qquad (1.113)$$

**Trilinear soft SUSY breaking parameters**

$$\frac{dA^t}{dt} = \frac{1}{8\pi^2}\left(6f_t^2 A^t + f_b^2 A^b - \frac{16}{3}g_3^2 M_3 - 3g_2^2 M_2 - \frac{13}{9}g_1^2 M_1\right) \qquad (1.114)$$

$$\frac{dA^b}{dt} = \frac{1}{8\pi^2}\left(6f_b^2 A^b + f_t^2 A^t + f_\tau^2 A^\tau - \frac{16}{3}g_3^2 M_3 - 3g_2^2 M_2 - \frac{7}{9}g_1^2 M_1\right) \qquad (1.115)$$

$$\frac{dA^\tau}{dt} = \frac{1}{8\pi^2}\left(3f_b^2 A^b + 4f_\tau^2 A^\tau - 3g_2^2 M_2 - 3g_1^2 M_1\right) \qquad (1.116)$$

**Bilinear soft SUSY breaking parameter**

$$\frac{dB}{dt} = \frac{1}{8\pi^2}\left(-3f_t^2 A^t - 3f_b^2 A^b - f_\tau^2 A^\tau + 3g_2^2 M_2 + g_1^2 M_1\right) \qquad (1.117)$$

**Soft SUSY breaking Higgs boson masses**

$$\frac{dm_1^2}{dt} = \frac{1}{8\pi^2}\left(3f_b^2 S_b + f_\tau^2 S_\tau - 3g_2^2|M_2|^2 - g_1^2|M_1|^2 - \frac{1}{2}g_1^2 S_1^2\right) \qquad (1.118)$$

$$\frac{dm_2^2}{dt} = \frac{1}{8\pi^2}\left(3f_t^2 S_t - 3g_2^2|M_2|^2 - g_1^2|M_1|^2 + \frac{1}{2}g_1^2 S_1^2\right) \qquad (1.119)$$

**Soft SUSY breaking third generation slepton masses**

$$\frac{dm_{\tilde{L}_3}^2}{dt} = \frac{1}{8\pi^2}\left(f_\tau^2 S_\tau - 3g_2^2|M_2|^2 - g_1^2|M_1|^2 - \frac{1}{2}g_1^2 S_1^2\right) \qquad (1.120)$$

$$\frac{dm_{\tilde{\tau}_R}^2}{dt} = \frac{1}{8\pi^2}\left(2f_t^2 S_t - 4g_1^2|M_1|^2 + g_1^2 S_1^2\right) \qquad (1.121)$$

**Soft SUSY breaking gaugino masses**

$$\frac{dM_i}{dt} = \frac{dg_i^2}{dt}, \; i = 1, 2, 3 \qquad (1.122)$$

## 1.2 Neutrino Masses and Mixing

### 1.2.1 Neutrino oscillations

Lepton flavour changing currents are absent in the Standard Model. This happens because neutrinos are massless and it is always possible to choose a physical basis where the leptonic Yukawa couplings are diagonal. However, the evidence of neutrino oscillations is an indirect proof for neutrino masses. In fact, lets assume that a neutrino of a specific flavour $\nu_i$ is created at $t = 0$ and propagates as a superposition of mass eigenstates $\nu_a$.

We would like to know the probability of finding the flavour $j$ at a time $t$. The flavour and the mass states are related through a change of basis performed by a unitary matrix $U$:

$$|\nu_i\rangle = \sum_a U_{ia} |\nu_a\rangle \ , \ \ i = e, \mu\tau. \tag{1.123}$$

As a simplifying assumption, we take a beam of definite momentum $\vec{p}$, related to the energy of each neutrino with a definite mass through the familiar relation

$$E_a = \sqrt{|\vec{p}|^2 + m_a^2}. \tag{1.124}$$

For quasi relativistic neutrinos, we can expand this in powers of $m^2/|\vec{p}|^2$, obtaining

$$E_a \sim |\vec{p}|(1 + \frac{m_a^2}{2|p|^2}). \tag{1.125}$$

The time evolution is given by the usual exponential factor and the flavour state at a time $t$ is

$$|\nu_i(t)\rangle = \sum_a e^{-iE_a t} U_{ia} |\nu_a\rangle \ , \tag{1.126}$$

so that the amplitude for finding a flavour $j$ at time $t$ is given by

$$\langle \nu_j \, | \, \nu_i \rangle \, (t) = \sum_a e^{-iE_a t} U_{ja}^* U_{ia} \tag{1.127}$$

and the conversion probability is

$$
\begin{aligned}
P_{\nu_i \to \nu_j}(t) &= |\langle \nu_j \, | \, \nu_i \rangle \, (t)|^2 \\
&= \sum_{ab} \left| U_{ja}^* U_{ia} U_{ib}^* U_{jb} \right| \cos((E_b - E_a)t + \phi_{abij}),
\end{aligned}
\tag{1.128}
$$

with

$$\phi_{abij} = \mathrm{Arg}(U_{ja}^* U_{ia} U_{ib}^* U_{jb}). \tag{1.129}$$

Using now equation (1.125), we obtain the result

$$P_{\nu_i \to \nu_j}(x) = \sum_{ab} \left| U_{ja}^* U_{ia} U_{ib}^* U_{jb} \right| \cos(\frac{2\pi x}{L_{ba}} + \phi_{abij}), \tag{1.130}$$

where $t$ has been replaced by $x$ and the oscillation length $L_{ab}$ is defined by the formula

$$L_{ab} = \frac{4\pi |\vec{p}|}{m_{ab}^2} \tag{1.131}$$

$$\sim \frac{4\pi E}{m_{ab}^2} \tag{1.132}$$

with $E$ some common energy scale and $m_{ab}^2$ the square masses difference

$$m_{ab}^2 = m_a^2 - m_b^2. \tag{1.133}$$

It is now obvious that, in the limit of zero masses, the conversion probability is zero. Also, neutrino masses cannot be fully degenerate, because in that case we also get a null oscillation probability.

It is possible to give some variations of formula (1.130) that are useful in some applications. From (1.128) we can write

$$P_{\nu_i \to \nu_j}(t) = \sum_{ab} U_{ja}^* U_{ia} U_{ib}^* U_{jb} \exp(i(E_b - E_a)t), \tag{1.134}$$

From the unitarity relation

$$\sum_{ab} U_{ja}^* U_{ia} U_{ib}^* U_{jb} = \delta_{ij} \tag{1.135}$$

and the following fact

$$\sum_{ab} U_{ja}^* U_{ia} U_{ib}^* U_{jb} = \sum_{a=b} U_{ja}^* U_{ia} U_{ib}^* U_{jb} + \sum_{b>a} U_{ja}^* U_{ia} U_{ib}^* U_{jb} + \sum_{b<a} U_{ja}^* U_{ia} U_{ib}^* U_{jb} \tag{1.136}$$

we know that[4]

$$\sum_a |U_{ia}|^2 |U_{ja}|^2 = \delta_{ij} - 2\text{Re}\left(\sum_{b>a} U_{ja}^* U_{ia} U_{ib}^* U_{jb}\right) \tag{1.137}$$

and then

$$P_{\nu_i \to \nu_j}(t) = \text{Re}\left(\sum_{ab} U_{ja}^* U_{ia} U_{ib}^* U_{jb}\right) \cos\left(\frac{m_{ba}^2 x}{2E}\right) - \text{Im}\left(\sum_{ab} U_{ja}^* U_{ia} U_{ib}^* U_{jb}\right) \sin\left(\frac{m_{ba}^2 x}{2E}\right)$$

$$= \sum_a |U_{ia}|^2 |U_{ja}|^2 + 2\text{Re}\left(\sum_{b>a} U_{ja}^* U_{ia} U_{ib}^* U_{jb}\right) \cos\left(\frac{m_{ba}^2 x}{2E}\right)$$

$$- 2\text{Im}\left(\sum_{b>a} U_{ja}^* U_{ia} U_{ib}^* U_{jb}\right) \sin\left(\frac{m_{ba}^2 x}{2E}\right)$$

$$= \delta_{ij} - 2\text{Re}\left(\sum_{b>a} U_{ja}^* U_{ia} U_{ib}^* U_{jb}\right) \left(1 - \cos\left(\frac{m_{ba}^2 x}{2E}\right)\right)$$

$$- 2\text{Im}\left(\sum_{b>a} U_{ja}^* U_{ia} U_{ib}^* U_{jb}\right) \sin\left(\frac{m_{ba}^2 x}{2E}\right) \tag{1.138}$$

---

[4]In the last term just change dummy indices $(a, b) \to (b, a)$

or else

$$P_{\nu_i \to \nu_j}(t) = \delta_{ij} - 4\text{Re}\left(\sum_{b>a} U_{ja}^* U_{ia} U_{ib}^* U_{jb}\right) \sin^2\left(\frac{m_{ba}^2 x}{4E}\right)$$

$$- 2\text{Im}\left(\sum_{b>a} U_{ja}^* U_{ia} U_{ib}^* U_{jb}\right) \sin\left(\frac{m_{ba}^2 x}{2E}\right). \tag{1.139}$$

If we want to compare this result with the one from oscillations for anti-neutrinos, we must start by taking the complex conjugate of formula (1.123):

$$|\overline{\nu}_i\rangle = \sum_a U_{ia}^* |\overline{\nu}_a\rangle \ , \ i = e, \mu\tau. \tag{1.140}$$

Then we can reproduce the same computations but with $U_{ia}$ replaced by $U_{ia}^*$ and vice-versa. In the end we get for anti-neutrinos

$$P_{\overline{\nu}_i \to \overline{\nu}_j} = \delta_{ij} - 4\text{Re}\left(\sum_{b>a} U_{ja}^* U_{ia} U_{ib}^* U_{jb}\right)\left(\sin^2\left(\frac{m_{ba}^2 x}{4E}\right)\right)$$

$$+ 2\text{Im}\left(\sum_{b>a} U_{ja}^* U_{ia} U_{ib}^* U_{jb}\right) \sin\left(\frac{m_{ba}^2 x}{2E}\right). \tag{1.141}$$

This means that a way to test $CP$ violation is to measure the asymmetry

$$A_{ij} = P_{\overline{\nu}_i \to \overline{\nu}_j} - P_{\nu_i \to \nu_j}(t)$$

$$= 4\text{Im}\left(\sum_{b>a} U_{ja}^* U_{ia} U_{ib}^* U_{jb}\right) \sin\left(\frac{m_{ba}^2 x}{2E}\right). \tag{1.142}$$

We see that for CP violation to occur the imaginary part must be non-zero. It can be shown [3] that the quantities

$$s_{ab;ji} J = \text{Im}\left(U_{ja}^* U_{ia} U_{ib}^* U_{jb}\right) \tag{1.143}$$

are all equal up to a sign $s_{ab;ji}$ and $J$, given by

$$J = \text{Im}\left(U_{\mu3} U_{e2} U_{\mu2}^* U_{e3}^*\right), \tag{1.144}$$

is known as the Jarlskog invariant, because it is invariant by a re-phasing of the neutrino fields or equivalently by a re-parametrization of the mixing matrix. A convenient parametrization of this matrix is given by

$$U = \begin{pmatrix} c_{12}c_{13} & s_{12}c_{13} & s_{13}e^{-i\delta_{13}} \\ -s_{12}c_{23} - c_{12}s_{23}s_{13}e^{i\delta_{13}} & c_{12}c_{23} - s_{12}s_{23}s_{13}e^{i\delta_{13}} & s_{23}c_{13} \\ s_{12}s_{23} - c_{12}c_{23}s_{13}e^{i\delta_{13}} & -c_{12}s_{23} - s_{12}c_{23}s_{13}e^{i\delta_{13}} & c_{23}c_{13} \end{pmatrix} \qquad (1.145)$$

where $c_{ab} = \cos\theta_{ab}$, $s_{ab} = \sin\theta_{ab}$, with the mixing angles varying in the range $0 < \theta_{ab} < \pi/2$ and the Dirac CP-phase varying in $0 < \delta_{13} < 2\pi$. In this case the Jarlskog invariant has the expression

$$J = \frac{1}{8}\sin 2\theta_{12}\sin 2\theta_{23}\cos\theta_{13}\sin 2\theta_{13}\sin\delta_{13}. \qquad (1.146)$$

In the approximation of only two neutrino flavours the mixing matrix takes the simpler form

$$U = \begin{pmatrix} \cos\theta & \sin\theta \\ -\sin\theta & \cos\theta \end{pmatrix} \qquad (1.147)$$

and we can give also a simpler formula for the transition probability from (1.134):

$$P_{\nu_i \to \nu_j} = \frac{1}{2}\sin^2 2\theta\left(1 - \cos\frac{m_{ba}^2 x}{2E}\right), \ i \neq j \qquad (1.148)$$

or

$$P_{\nu_i \to \nu_j} = \frac{1}{2}\sin^2 2\theta\sin^2\left(\frac{m_{ba}^2 x}{4E}\right), \ i \neq j. \qquad (1.149)$$

The survival probability is obtained in the obvious way:

$$P_{\nu_i \to \nu_i} = 1 - P_{\nu_i \to \nu_j \ i \neq j} \qquad (1.150)$$

$$= 1 - \frac{1}{2}\sin^2 2\theta\sin^2\left(\frac{m_{ba}^2 x}{4E}\right), \ i \neq j. \qquad (1.151)$$

Noting that the average in $x$ of the cos function is zero, we get

$$\langle P_{\nu_i \to \nu_j}\rangle = \frac{1}{2}\sin^2 2\theta \qquad (1.152)$$

These results can be useful if some experiment is only sensitive to the mass differences between two states, as it is already known that $\Delta m_{\rm Sol}^2 \ll \Delta m_{\rm Atm}^2$.

In reactor oscillation experiments neutrino energies are of the order of 1 MeV and it is convenient to write the transition probability in these specific units:

$$P_{\nu_i \to \nu_j} = \sin^2\theta\sin^2\left(\frac{1.27 m_{ba}^2(\text{eV}^2)x(\text{m})}{E(\text{MeV})}\right) \qquad (1.153)$$

In real experiments it is not possible to determine $E$ and $x$ with 100% accuracy, so that it is necessary to average the probability formula as

$$\langle P_{\nu_i \to \nu_j} \rangle = \frac{1}{2} \sin^2 \theta \left( 1 - \left\langle \cos\left( \frac{m_{ba}^2 x}{2E} \right) \right\rangle \right), \ i \neq j \tag{1.154}$$

with

$$\left\langle \cos\left( \frac{m_{ba}^2 x}{2E} \right) \right\rangle = \int \cos\left( \frac{m_{ba}^2 x}{2E} \right) \phi\left( \frac{x}{E} \right) d\frac{x}{E}. \tag{1.155}$$

This average can be computed analytically if the density $\phi\left( \frac{x}{E} \right)$ is gaussian with an average value $\langle \frac{x}{E} \rangle$ and a standard deviation $\sigma_{x/E}$. In other cases it must be computed numerically.

From formula (1.154) we get an upper bound for $\sin^2 \theta$ in terms of the maximum oscillation probability and the cos average:

$$\sin^2 \theta \leq \frac{2 P_{\nu_i \to \nu_j}^{\max}}{1 - \left\langle \cos\left( \frac{m_{ba}^2 x}{2E} \right) \right\rangle}. \tag{1.156}$$

This allows to make exclusion plots for $\Delta m^2 \equiv m_{ab}^2$ vs $\sin^2 2\theta$, like the one in figure 1.16 if the maximum probability is known.

Two limiting cases are worth mentioning. The first one is

$$\frac{m_{ba}^2}{2} \langle \frac{x}{E} \rangle \simeq \pi \Rightarrow \sin^2 2\theta \leq P_{\nu_i \to \nu_j}^{\max} \tag{1.157}$$

which gives the most stringent bound on $\sin^2 2\theta$. The other is

$$\frac{m_{ba}^2}{2} \langle \frac{x}{E} \rangle \gg \pi \Rightarrow \sin^2 2\theta \leq 2 P_{\nu_i \to \nu_j}^{\max} \tag{1.158}$$

In this case the argument of the cos function oscillates rapidly which gives a zero average.

### 1.2.2 Lepton flavour violation

In principle we could extend minimally the Standard Model, including an $SU(2)$ singlet for the right handed neutrino and generate neutrino masses through the familiar Higgs mechanism. Even though this process wouldn't be completely satisfactory because it would require a fine tuning of Yukawa couplings, we would have lepton flavour violating (LFV) processes like the one in fig. 1.2. However, these processes would be highly suppressed due to the GIM mechanism, in virtue of the smallness of neutrino masses. To see this, note that the amplitude for the process in fig 1.2 is proportional to

$$\sum_b \frac{U_{ab} U_{bc}^\dagger}{k^2 - m_b^2 + i\epsilon} (\not{k} + m_b), \tag{1.159}$$

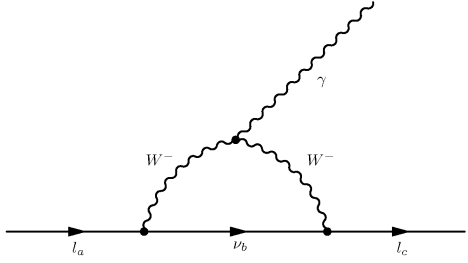

**Figure 1.2:** $l \longrightarrow l^{'}\gamma$ within a minimal extension of the Standard Model with massive neutrinos.

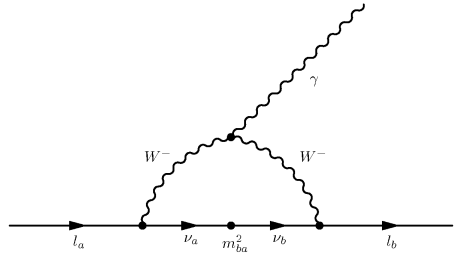

**Figure 1.3:** $l \longrightarrow l^{'}\gamma$ within a minimal extension of the Standard Model with massive neutrinos.

with $\not{k}$ and $m_b$ the momentum and the mass of the loop neutrino and $U$ the matrix that diagonalizes neutrino masses, considering a leptonic weak basis where the charged leptons mass matrix is already diagonal. Because $U$ is unitary, we clearly see that, in the limit $m_b = 0$ or fully degenerate masses, we get no lepton flavour violation. Now, we focus on the first factor of the previous formula and, assuming that neutrino masses deviate little from a common value $m$, expand it in powers of $\Delta m_b^2/(k^2 - m^2)$, with $\Delta m_b^2 = m_b^2 - m^2$, to get an amplitude proportional to

$$\frac{1}{(k^2 - m^2 + i\epsilon)^2} \sum_b U_{ab} \Delta m_b^2 U_{bc}^{\dagger} \qquad (1.160)$$

This illustrates the fact that LFV processes are encoded in the off-diagonal elements of the mass matrices. Usually, one adopts the formalism of mass insertions, where these mass matrix elements are considered as interaction vertices that flip the flavour and propagators are functions of some common mass scale. In this way, we could describe the process with the diagram of fig. 1.3.

In the MSSM the situation is rather different. Fig. 1.4 shows two typical contributions for the LFV process $l \longrightarrow l^{'}\gamma$, mediated by neutralinos and charged sleptons and charginos and sneutrinos, respectively.

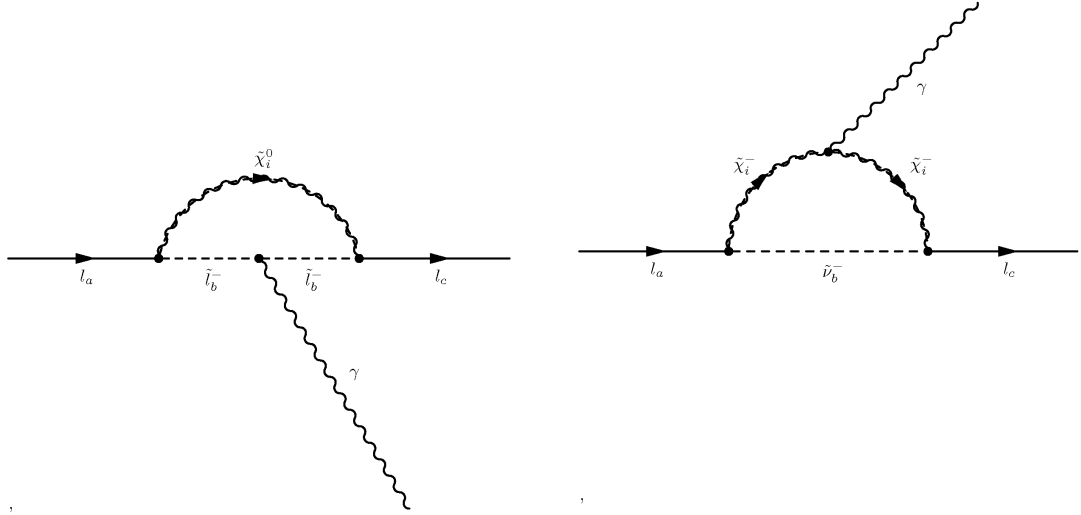

**Figure 1.4:** $l \longrightarrow l'\gamma$ within the MSSM.

### 1.2.3 Seesaw Models

The Dirac Lagrangian

$$\mathcal{L}_D = i\bar{\psi}\gamma^\mu\partial_\mu\psi - m\bar{\psi}\psi \tag{1.161}$$

with the decomposition of the Dirac spinor $\psi$ in left and right chiral states $\psi_L$ and $\psi_R$ can be written as

$$\mathcal{L}_D = i\overline{\psi_L}\gamma^\mu\partial_\mu\psi_L + i\overline{\psi_R}\gamma^\mu\partial_\mu\psi_R - m\overline{\psi_L}\psi_R - m\overline{\psi_R}\psi_L. \tag{1.162}$$

The Lagrange equations for $\psi_L$ and $\psi_R$ give a set of two coupled equations

$$i\gamma^\mu\partial_\mu\psi_L = m\psi_R \tag{1.163}$$

$$i\gamma^\mu\partial_\mu\psi_R = m\psi_L \tag{1.164}$$

which decouple in the limit $m = 0$, giving origin to what are known as Weyl spinors. However, even in the massive case it is not mandatory to have a Dirac spinor with four degrees of freedom. With the Majorana condition

$$\psi_R = \mathcal{C}\overline{\psi_L}^T, \tag{1.165}$$

where $\mathcal{C}$ is the charge conjugation operator, it is possible to show that the equations (1.163) and (1.164) are equivalent. The decomposition of the Dirac spinor

$$\psi = \psi_L + \psi_R \tag{1.166}$$

translates into

$$\psi = \psi_L + \mathcal{C}\overline{\psi_L}^T, \tag{1.167}$$

which means that a Majorana fermion is self conjugate:

$$\psi^c \equiv \mathcal{C}\bar{\psi}^T = \psi. \tag{1.168}$$

Thus only neutral particles can be of Majorana type, as it could be the case of neutrinos. The Dirac mass term

$$\mathcal{L}_D^m = -m\overline{\psi_L}\psi_R + \text{H.c.} \tag{1.169}$$

in the case of a Majorana particle gets transformed into the more economical way

$$\mathcal{L}_M^m = -\frac{1}{2}m\left(\overline{\psi_L}\psi_L^c + \overline{\psi_L^c}\psi_L\right). \tag{1.170}$$

The $1/2$ factor accounts for the fact that $\psi_L^c$ and $\overline{\psi_L}$ are not independent. Noting the relation

$$\overline{\psi_L^c} = -\psi_L^T C^{-1}, \tag{1.171}$$

and the properties of the charge conjugation operator, the Majorana mass term is usually written as

$$\mathcal{L}_M^m = -\frac{1}{2}\psi_L^T \mathcal{C}\psi_L + \text{H.c.} \tag{1.172}$$

The full Lagrangian for a Majorana particle is then

$$\mathcal{L}_M = \frac{1}{2}\left[i\overline{\psi_L}\gamma^\mu\partial_\mu\psi_L + i\psi_L^T\gamma^{\mu T}\partial_\mu\overline{\psi_L}^T - m\left(\psi_L^T\mathcal{C}\psi_L + \overline{\psi_L}\mathcal{C}\overline{\psi_L}^T\right)\right] \tag{1.173}$$

which can be written in a more compact way:

$$\mathcal{L}_M = \frac{1}{2}\bar{\psi}\left(i\gamma^\mu\partial_\mu - m\right)\psi \tag{1.174}$$

where $\psi$ is the Majorana fermion (1.167):

$$\psi = \psi_L + \psi_L^c. \tag{1.175}$$

We have chosen to express $\psi_R$ in terms of $\psi_L$ and obtain the Lagrangian as a funtion of $\psi_L$, but we could equally well have chosen the opposite procedure and obtain a Lagrangian with respect to $\psi_R$. This is more suited for models that include Majorana right handed neutrinos that are singlets under the gauge group of the Standard Model. This is one particular possibility for the implementation of the effective dimension five operator that

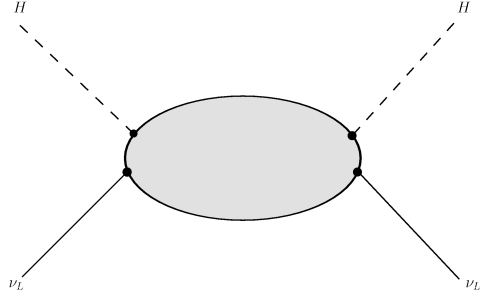

**Figure 1.5:** Dimension 5 operator responsible for neutrino masses.

violates lepton number and generates neutrino masses (fig. 1.5),

$$O = \frac{f}{\Lambda} LHLH, \tag{1.176}$$

long ago proposed by Weinberg [19]. For these theories, after electroweak symmetry breaking, a Dirac mass matrix $m_D$ that couples $\overline{\nu_L}$ and $\nu_R$ is present with generation indexes implicit, as well as a Majorana mass matrix $M$ for $\nu_R$. In the static limit for $\nu_R$ the Lagrange equation gives

$$\nu_R = -M^{-1} m_D^T \mathcal{C} \overline{\nu_L}^T, \tag{1.177}$$

which replaced in the Dirac equation for $\nu_L$

$$i\gamma^\mu \partial_\mu \nu_L - m_D \nu_R = 0 \tag{1.178}$$

gives

$$i\gamma^\mu \partial_\mu \nu_L - M_I^\nu \mathcal{C} \overline{\nu_L}^T = 0, \tag{1.179}$$

that it is the equation for Majorana neutrinos with a mass matrix $M_I^\nu$ given by

$$M_I^\nu = -m_D M^{-1} m_D^T. \tag{1.180}$$

This is the celebrated seesaw mechanism, in this case type I seesaw (fig. 1.6). There are other possibilities for generating the operator (1.176). In Type-II seesaw (fig. 1.7) an $SU(2)$ triplet $T$ mediates through a t-channel the LH scattering which, after electroweak symmetry breaking, generates neutrino masses. The relevant couplings are

$$-\mathcal{L}_{II} = \lambda_1 L^T \mathcal{C} \left( i\tau_2 T \right) L + \lambda_2 H^T \left( i\tau_2 T^\dagger \right) H + \text{H.c.} \tag{1.181}$$

For light neutrino masses of the order of 0.05 eV, the scale of the new physics $\Lambda$ in (1.176) can easily be estimated. In Type-I, $m_D$ is of the order of electroweak symmetry breaking,

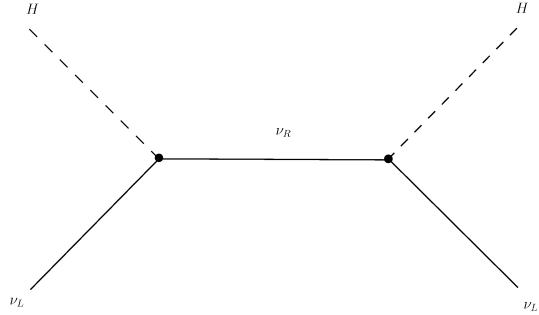

**Figure 1.6:** Type-I seesaw mechanism.

so

$$\Lambda \sim \frac{m_D^2}{m_\nu} \sim \frac{10^4}{0.05} \times 10^9 \sim 10^{14} \text{ GeV}. \tag{1.182}$$

Of course, this depends on the magnitude of the Yukawa couplings in the interaction terms between the Higgs doublet and the left leptons, so there is some freedom in the $\Lambda$ scale. On the other hand the right neutrinos are singlets under the Standard Model gauge group, which means that there is no symmetry that prevents their mass from being high. So this upper scale provides a nice justification for the smallness of light neutrino masses, which otherwise would require a severe fine tuning of Yukawa couplings if it were to be generated by the usual Higgs mechanism.

The estimate for Type-II seesaw models shows that

$$M_T^2 \sim \frac{\lambda_1 \lambda_2 v^2}{0.05} \times 10^9 \text{ GeV}. \tag{1.183}$$

Supersymmetric models require at least 2 triplets because of the holomorphy of the superpotential and the cancelation of anomalies. In that case $\lambda_2$ is proportional to one of the triplets mass $M_T$, leaving us with roughly the same estimate as in Type-I. But in non supersymmetric models $M_T$ can be as low as $10^8$ GeV (note however that $\lambda_2$ has the dimension of mass).

The seesaw models predict that light neutrinos are of Majorana type. The neutrinoless double beta decay $2\beta_{0\nu}$ (fig.1.8), if observed, would be the final answer to the Dirac or Majorana nature of neutrinos. In fact, this process is equivalent to neutrinos being Majorana particles, as it was first proved in [20]. Even if $2\beta_{0\nu}$ is an effect of some new physics beyond the Standard Model, like supersymmetry, Left-Right symmetric models, or other, the same physics would imply a Majorana mass term for neutrinos.

The counting of independent parameters depends on the seesaw type one is considering. For Type-I, from equation (1.180), we know that $M_I^\nu$ is the mass matrix of Majorana neutrinos so that it can be diagonalized via an orthogonal transformation with a unitary

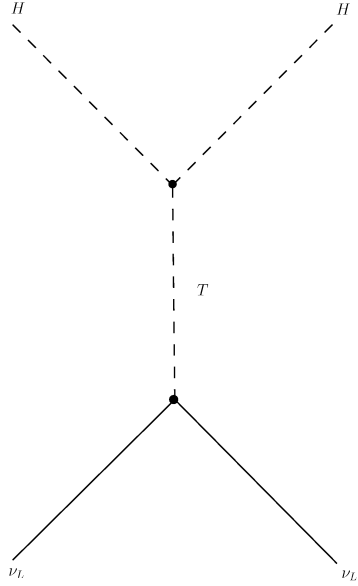

**Figure 1.7:** Type-II seesaw mechanism.

matrix $U$

$$D_I^\nu \equiv \text{diag}(m_1, m_2, m_3) = U^T M_I^\nu U \tag{1.184}$$

with $m_i$, $i = 1, 2, 3$ in general complex. An unitary $n \times n$ matrix has $\frac{n(n-1)}{2}$ independent modulus and $\frac{n(n+1)}{2}$ independent phases. This means that $U$ will have 3 mixing angles and 6 phases. Five of these phases can be absorbed by the charged lepton and neutrino fields redefinitions. Not all phases can be absorbed because there is an overall phase that is related with the lepton number and that leaves the leptonic charged current invariant. On the other hand to change to the physical masses we must multiply $D_I^\nu$ by a diagonal matrix of phases. This means that

$$V = U \cdot \text{diag}(e^{-i\phi_1/2}, e^{-i\phi_2/2}, 1) \tag{1.185}$$

where we have factored out one phase and $U$ is of CKM form

$$U = \begin{pmatrix} c_{12}c_{13} & s_{12}c_{13} & s_{13}e^{-i\delta_{13}} \\ -s_{12}c_{23} - c_{12}s_{23}s_{13}e^{i\delta_{13}} & c_{12}c_{23} - s_{12}s_{23}s_{13}e^{i\delta_{13}} & s_{23}c_{13} \\ s_{12}s_{23} - c_{12}c_{23}s_{13}e^{i\delta_{13}} & -c_{12}s_{23} - s_{12}c_{23}s_{13}e^{i\delta_{13}} & c_{23}c_{13} \end{pmatrix}. \tag{1.186}$$

Then we have

$$\text{diag}(m_1, m_2, m_3) = V^T M_I^\nu V, \ m_i > 0. \tag{1.187}$$

Now we must repeat the procedure for the right-handed neutrinos. Then the overall counting is, for the light neutrino sector, 3 positive eigenvalues for the mass matrix, plus three mixing angles, plus 3 CP phases (1 Dirac and 2 Majorana), which gives a total of 9 independent parameters. In the end we have 18 independent parameters for Type-I seesaw. This can be put in a more formal way. In [21] it is shown that the leptonic Yukawa couplings $Y_\nu$ responsible for the Dirac mass term $m_D$ are given by

$$Y_\nu = D_{\sqrt{M}} R D_{\sqrt{m}} V^\dagger \qquad (1.188)$$

where $D_{\sqrt{M}}$ is the diagonal mass matrix with positive eigenvalues for the right-handed neutrinos, $D_{\sqrt{m}}$ is the same but for the light neutrinos, $R$ is a generic $3 \times 3$ complex orthogonal matrix and $V$ is given by (1.185). Of course, this is the most general situation; in many applications it is possible to reduce drastically the number of parameters, imposing certain assumptions. For instance we can assume that the $R$ matrix is the identity, which amounts to say that the right-handed neutrinos are already in the mass base. This was done in [5, 6], as it is mentioned in Chapters 5 and 6. In [22] we made an even more drastic assumption, that the right-handed neutrinos are fully degenerated because of the symmetry group, which reduces the parameters in the right-handed sector from 9 to just 1.

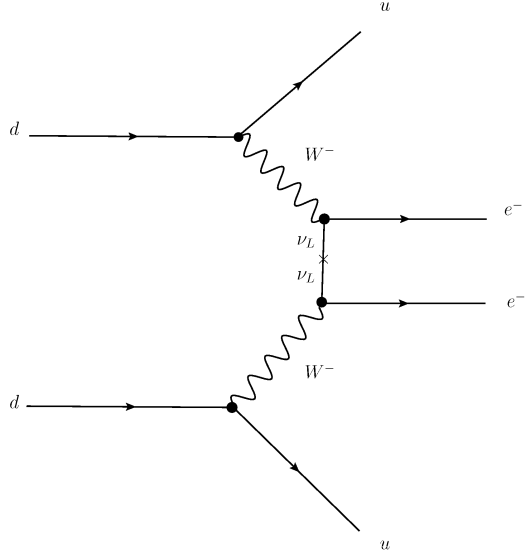

**Figure 1.8:** Neutrinoless double beta decay

With respect to Type-II the number of independent parameters is smaller. We have also the 3 positive eigenvalues from the light neutrinos mass matrix, plus the 3 mixing angles and the 3 phases. Then we have the complex coupling $\lambda_2$ and the triplet mass $M_T$. But two phases can still be removed by field redefinitions. This means that the Yukawa couplings $\lambda_1$ in (1.181) will be a function of 9 parameters from the light neutrinos as in Type-I plus two real parameters, which gives a total of 11 real parameters. Both Type I and Type II seesaw mechanisms have too much degrees of freedom to be fully reconstructed purely from neutrino data. In fact at the moment we only have rigorous data from the neutrino square mass differences (3 parameters), 2 mixing angles and an upper bound on the third mixing angle. For now, no information about the CP phases is available. This means that some data outside the neutrino sector (like lepton flavour violation constrains) will always be necessary to make the models more predictive.

## 1.3 Solar, atmospheric, reactor and accelerator neutrinos.

In this section we present a brief survey on neutrino properties including the most relevant experiments that lead to the discovery of neutrino oscillations. This follows closely [3], where full references can be found for the experimental results presented here. We end with a review of neutrino parameters obtained from a global analysis of the data from the major neutrino oscillation experiments.

### 1.3.1 Solar neutrinos

The sun is a powerful source of neutrinos through the various thermonuclear reactions that occur in its core. It is estimated that the overall neutrino flux from the sun is

$$\Phi_\nu^\odot = 6.54 \times 10^{10} \text{ cm}^{-2} \text{ s}^{-1}.$$  (1.189)

There are two major groups of thermonuclear reactions responsible for the emission of neutrinos and radiation: the pp-chain (fig. 1.9) and the CNO cycle (fig. 1.10). The net result of these reactions is the conversion of four protons and two electrons into a $^4$He nucleus plus two electron neutrinos

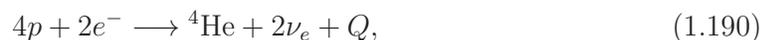

$$4p + 2e^- \longrightarrow {}^4\text{He} + 2\nu_e + Q,$$  (1.190)

where $Q$ is equal to 26.731 MeV and is released as radiation or neutrinos kinetic energy.

Solar neutrinos were first observed during the 60's in the Homestake experiment, a radiochemical apparatus based on inverse $\beta$ decay:

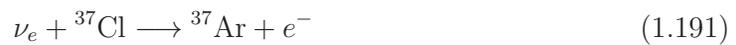

$$\nu_e + {}^{37}\text{Cl} \longrightarrow {}^{37}\text{Ar} + e^-$$  (1.191)

with a neutrino threshold energy $E_\nu = 0.814$ MeV. Therefore this experiment was only able to detect intermediate and high-energy neutrinos (see fig. 1.11).

The Homestake experiment, which was proposed in 1964 and built in the period 1965-1967, released its first data in 1968, declaring a solar neutrino flux less then 3 SNU[5], well below the rate predicted by the Standard Solar Models (SSM) (see tables 1.3 and 1.4). Since the experiment was based on the chemical extraction of the radioactive isotope $^{37}$Ar which is produced in very small quantities, the uncertainties with the data were large and dominated by statistical fluctuations. However the accumulation of data over 23 years of operation allowed to reduce the statistical uncertainty to an average rate of 6%. The solar neutrino flux measured at Homestake is about 1/3 of the predicted by the SSM, with a discrepancy of more than $3\sigma$.

---

[5]1 SNU = $10^{-36}$ events atom$^{-1}$s$^{-1}$

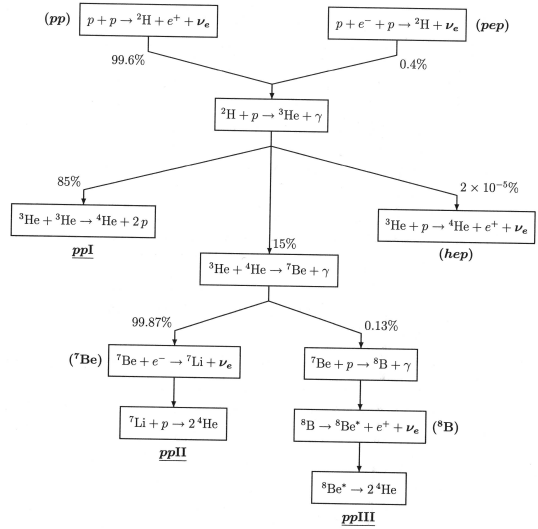

**Figure 1.9:** The pp chain of stellar thermonuclear reactions. Taken from [3]

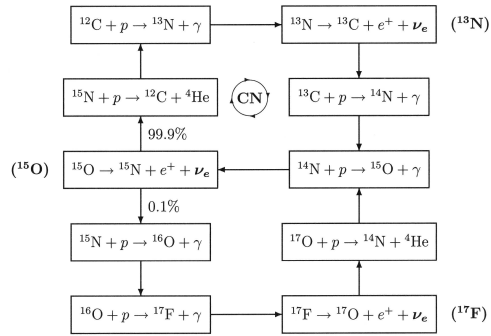

**Figure 1.10:** The CNO cycle of stellar thermonuclear reactions. Taken from [3]

**Table 1.3:** Standard Model Predictions (BP2000): solar neutrino fluxes and neutrino capture rates, with $1\sigma$ uncertainties from all sources (combined quadratically). Taken from [16].

| Source | Flux $(10^{10}\ \mathrm{cm^{-2}s^{-1}})$ | Cl (SNU) | Ga (SNU) | Li (SNU) |
|---|---|---|---|---|
| pp | $5.95\left(1.00^{+0.01}_{-0.01}\right)$ | 0.0 | 69.7 | 0.0 |
| pep | $1.40\times10^{-2}\left(1.00^{+0.015}_{-0.015}\right)$ | 0.22 | 2.8 | 9.2 |
| hep | $9.3\times10^{-7}$ | 0.04 | 0.1 | 0.1 |
| $^7$Be | $4.77\times10^{-1}\left(1.00^{+0.10}_{-0.10}\right)$ | 1.15 | 34.2 | 9.1 |
| $^8$B | $5.05\times10^{-4}\left(1.00^{+0.20}_{-0.16}\right)$ | 5.76 | 12.1 | 19.7 |
| $^{13}$N | $5.48\times10^{-2}\left(1.00^{+0.21}_{-0.17}\right)$ | 0.09 | 3.4 | 2.3 |
| $^{15}$O | $4.80\times10^{-2}\left(1.00^{+0.25}_{-0.19}\right)$ | 0.33 | 5.5 | 11.8 |
| $^{17}$F | $5.63\times10^{-4}\left(1.00^{+0.25}_{-0.25}\right)$ | 0.0 | 0.1 | 0.1 |
| Total | | $7.6^{+1.3}_{-1.1}$ | $128^{+9}_{-7}$ | $52.3^{+6.5}_{-6.0}$ |

**Table 1.4:** Solar Neutrino Rates (units are in SNU for Chlorine, Gallex+GNO and SAGE; for $^8$B and hep are respectively $10^6 \mathrm{cm}^{-2}\mathrm{s}^{-1}$ and $10^3 \mathrm{cm}^{-2}\mathrm{s}^{-1}$): Theory versus Experiment. Taken from [16] with data published up to the year 2000 - see the references therein.

| Experiment | BP2000 | Measured | Measured/BP2000 |
|---|---|---|---|
| Chlorine | $7.6^{+1.3}_{-1.1}$ | $2.56 \pm 0.23$ | $0.34 \pm 0.06$ |
| GALLEX + GNO | $128^{+9}_{-7}$ | $74.1^{+6.7}_{-7.8}$ | $0.58 \pm 0.07$ |
| SAGE | $128^{+9}_{-7}$ | $75.4^{+7.8}_{-7.4}$ | $0.59 \pm 0.07$ |
| $^8$B-Kamiokande | $5.05 \left[1.00 +^{+0.20}_{-0.16}\right]$ | $2.80 \left[1.00 \pm 0.14\right]$ | $0.55 \pm 0.13$ |
| $^8$B-Super-Kamiokande | $5.05 \left[1.00 +^{+0.20}_{-0.16}\right]$ | $2.40 \left[1.00 +^{+0.04}_{-0.03}\right]$ | $0.48 \pm 0.09$ |
| hep-Super-Kamiokande | $9.3$ | $11.3(1 \pm 0.8)$ | $\sim 1$ |

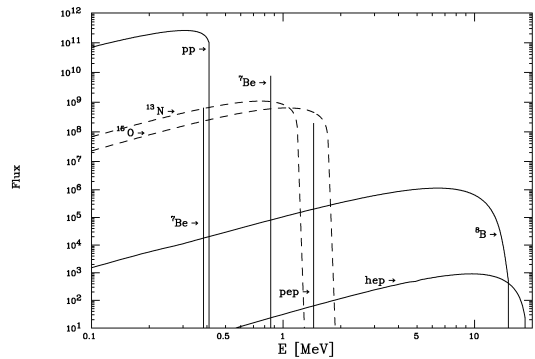

**Figure 1.11:** Energy spectra of neutrino fluxes from the pp and CNO chains as predicted by the SSM. Taken from [8]

In the 90's other experiments (Gallex/GNO and Sage) appeared based on the process

$$\nu_e + {}^{71}\mathrm{Ga} \longrightarrow {}^{71}\mathrm{Ge} + e^- \qquad (1.192)$$

that has an energy threshold $E^\nu = 0.233$ MeV. This allowed to detect solar neutrinos from all sources (see fig. 1.11). The GALLium EXperiment (GALLEX) started operating in May 1991, in Gran Sasso laboratory, Italy and it used a detector with 101 tons of liquid gallium chloride, GaCl$_3$-HCl solution. This experiment was followed by Gallium Neutrino Observatory (GNO) which operated from May 1998 until April 2003. The total collected data of the two experiments from May 1991 until April 2003 resulted in an average solar neutrino capture rate [23]:

$$R^{\mathrm{Gallex/GNO}}_{^{71}\mathrm{Ga}} = 69.3 \pm 5.5 \text{ SNU} \qquad (1.193)$$

which is about one half of the predicted by the SSM (see table 1.3).

Another Gallium experiment is SAGE (Soviet American Gallium Experiment) that started collecting data in 1990. It is located in the Baksan Neutrino Observatory of the Russian Academy of Sciences, in the northern Caucasus mountains. The average neutrino capture rate obtained from the data collected in the period of January 1990 to December 2001 is [24]

$$R_{71\text{Ga}}^{\text{Sage}} = 70.8_{-6.1}^{+6.5} \text{ SNU} \tag{1.194}$$

and it is in agreement with GALLEX/GNO experiments. It is about one half of the predicted by the SSM.

Water Cerenkov detectors provide another way of detecting indirectly solar neutrinos. If a charged particle travels with velocity $v > 1/n$ in a medium with a refraction index $n$ then it emits a cone of light that can be detected by arrays of photomultipliers, allowing a precise determination of the arrival time which is to say of the interaction point. Examples of water Cerenkov detectors are Kamiokande, Super-Kamiokande and SNO.

Kamiokande (Kamioka nucleon detector) was built initially to search for proton decay. It started operating in 1983 and was upgraded in 1986 to Kamiokande-II in order to be able to observe $^8$B solar neutrinos, which generate events in the detector with an energy around 10 MeV. In 1990 a new upgrade allowed to reduce the electrons threshold energy down to 7.0 MeV by the replacement of 100 photomultipliers, which was the start of Kamiokande-III.

The solar neutrino flux is measured by the elastic reaction

$$\nu_\alpha + e^- \longrightarrow \nu_\alpha + e^- \tag{1.195}$$

that is sensitive mostly to electron neutrinos because the cross section is about six times larger then for muon and tau neutrinos. The average $^8$B neutrino flux measured in Kamiokande from January 1987 to February 1995 is [25]

$$\Phi_{8\text{B}}^{\text{Kam}} = 2.80 \pm 0.38 \times 10^6 \text{ cm}^{-2}\text{s}^{-1}. \tag{1.196}$$

It shows a discrepancy with the SSM of about one-half (see table 1.4).

In April 1996 a new experiment in the Kamioka mine started operating, about 500 m from the cavity of the Kamiokande detector that contains now the Kamland experiment. This was Super-Kamiokande, still a water Cerenkov experiment, that went through its first phase from April 1996 until July 2001. It collected solar neutrino data with an electron recoil threshold energy of $E_e^{\text{th}} = 6.5$ MeV for the first 280 days and $E_e^{\text{th}} = 5.0$ MeV in the remaining 1216 days. The corresponding neutrino threshold energies are $E_\nu^{\text{th}} = 6.2$ MeV and $E_e^{\text{th}} = 4.7$ MeV and this means that Super-Kamiokande was also only sensitive to $^8$B solar neutrinos. The measured flux is [26]

$$\Phi_{8B}^{SK} = 2.35 \pm 0.08 \times 10^6 \text{ cm}^{-2}\text{s}^{-1}. \tag{1.197}$$

which once again correspond to one-half of the SSM flux. It was possible to compare day and night fluxes which could be a sign for matter effects, since night neutrinos must cross the Earth before entering upwards in the detector. The results are [26]

$$\Phi_{8B}^{SK,day} = (2.32 \pm 0.03)^{+0.08}_{-0.07} \times 10^6 \text{ cm}^{-2}\text{s}^{-1} \tag{1.198}$$

$$\Phi_{8B}^{SK,night} = (2.37 \pm 0.03)^{+0.08}_{-0.08} \times 10^6 \text{ cm}^{-2}\text{s}^{-1} \tag{1.199}$$

with an asymmetry

$$\mathcal{A}_{day,night}^{SK} = \frac{\Phi_{8B}^{SK,day} - \Phi_{8B}^{SK,night}}{\frac{1}{2}\left(\Phi_{8B}^{SK,day} + \Phi_{8B}^{SK,night}\right)} \tag{1.200}$$

$$= -0.021 \pm 0.020^{+0.013}_{-0.012}$$

that is consistent with a null result.

A variation of the water Cerenkov technique was pursued by the Sudbury Neutrino Observatory (SNO) in Ontario, Canada. In this apparatus neutrino interact with heavy water ($D_2O$), which allows to detect neutrinos with three different processes:

CC:      $\nu_e + d \longrightarrow p + p + e^-$

NC:      $\nu_\alpha + d \longrightarrow p + n + \nu_\alpha$

ES:      $\nu_\alpha + e^- \longrightarrow \nu_\alpha + e^-$

The charged current reaction has a neutrino threshold energy $E_\nu^{th,CC} = 1.442$ MeV but because of the high background at such low energies the effective neutrino threshold energy is about 7 MeV, which means that the charged current is only sensitive to $^8$B neutrinos. The neutral current reaction allows to detect the three neutrino flavours and it revealed to be extremely useful in solving the observed deficit of solar neutrino fluxes, known as the Solar Neutrino Problem. It has a neutrino threshold energy $E_\nu^{th,NC} = 2.224$ MeV and so it is also only sensitive to $^8$B flux. The elastic scattering is the same process that is used in $H_2O$ Cerenkov detectors and it has a neutrino threshold energy $E_\nu^{th,NC} = 5.7$ MeV and once more it is only sensitive to the $^8$B channel.

The SNO experiment is divided in three phases:

**$D_2O$ Phase**: the final neutron in NC is detected through the reaction

$$n + d \longrightarrow {}^3H + \gamma \text{ (6.25 MeV)} \tag{1.201}$$

This phase operated during 306.4 live days, from 2 November 1999 until 28 May 2001.

**NaCl Phase**: in the so-called salt phase 2 tons of NaCl were added to the heavy water in order to detect the neutron from NC, through the reaction

$$n + {}^{35}\text{Cl} \longrightarrow {}^{36}\text{Cl} + \text{ several } \gamma \text{ (8.57 MeV)} \tag{1.202}$$

This improved significantly the efficiency since the photon distribution from this process is very different from the one in Cerenkov radiation, which allows to distinguish between NC and CC reactions with good accuracy.

**Third Phase**: in this phase, a grid with three hundred ${}^3$He proportional counter tubes was introduced in the heavy water tank. This isotope has a large cross-section for the capture of thermal neutrons, producing an energetic pair proton-triton that triggers an electric pulse. This process, that started in January 2005, improved further the sensibility of the NC detection.

In the SNO salt phase were observed $2176 \pm 78$ CC, $2010 \pm 85$ NC and $279 \pm 26$ ES events. This corresponds to the following ${}^8$B neutrino fluxes [27]:

$$\Phi_{\text{CC}}^{\text{SNO}} = \left(1.68 \pm 0.06^{+0.08}_{-0.09}\right) \times 10^6 \text{ cm}^{-2}\text{s}^{-1} \tag{1.203}$$

$$\Phi_{\text{NC}}^{\text{SNO}} = \left(4.94 \pm 0.21^{+0.38}_{-0.34}\right) \times 10^6 \text{ cm}^{-2}\text{s}^{-1} \tag{1.204}$$

$$\Phi_{\text{ES}}^{\text{SNO}} = (2.35 \pm 0.22 \pm 0.15) \times 10^6 \text{ cm}^{-2}\text{s}^{-1} \tag{1.205}$$

We see immediately that these fluxes are not compatible with each other. More specifically, the fact that the recorded flux for NC (that is sensitive to the three neutrino flavours) is larger than the CC one (only sensitive to electron neutrinos) indicates that some electron neutrinos were converted to muon or tau neutrinos in their travel from the core of the sun to its surface. From the experimental data the fluxes of muon and tau neutrinos can be estimated:

$$\Phi_{\nu_\mu,\nu_\tau}^{\text{SNO,NC}} = \left(3.26 \pm 0.25^{+0.4}_{-0.35}\right) \times 10^6 \text{ cm}^{-2}\text{s}^{-1} \tag{1.206}$$

$$\Phi_{\nu_\mu,\nu_\tau}^{\text{SNO,ES}} = \left(4.36 \pm 1.52^{+0.9}_{-0.87}\right) \times 10^6 \text{ cm}^{-2}\text{s}^{-1} \tag{1.207}$$

which are in good agreement with each other. Thus, the neutral current channel used in SNO was able to suggest a solution to the Solar Neutrino Problem, providing the first evidence that neutrinos oscillate and as so have mass.

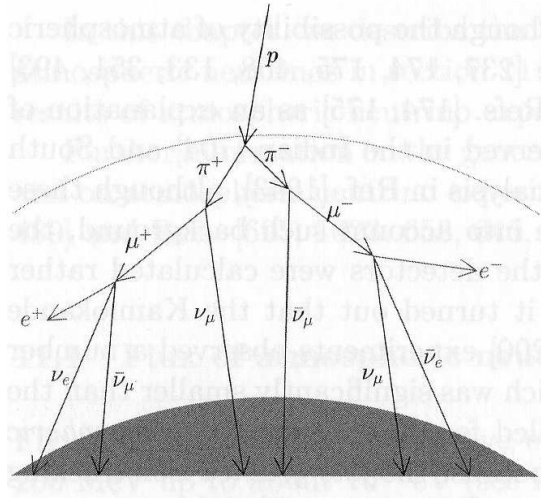

**Figure 1.12:** Neutrinos from cosmic rays. Taken from [3]

### 1.3.2 Atmospheric neutrinos

Cosmic rays that hit the Earth are mainly made of protons, with some small component of heavy nuclei. These primary cosmic rays interact in the atmosphere and produce all the hadrons and mesons, in particular pions, that decay into leptons and respective neutrinos. These lepton are mainly muons because the ratio of the decay widths is

$$R_{e/\mu} = \frac{\Gamma_{\pi^{\pm} \to e^{\pm}\nu_e}}{\Gamma_{\pi^{\pm} \to \mu^{\pm}\nu_\mu}} \tag{1.208}$$

$$= \frac{m_e^2 \left(1 - m_e^2/m_\pi^2\right)}{m_\mu^2 \left(1 - m_\mu^2/m_\pi^2\right)}$$

$$\simeq 1.28 \times 10^{-4}.$$

The muons will decay afterwards into electrons and neutrinos (see fig. 1.12). Thus, it is estimated that the neutrino flavours ratio is of the order of 2:

$$\frac{\Phi_{\nu_\mu} + \Phi_{\overline{\nu}_\mu}}{\Phi_{\nu_e} + \Phi_{\overline{\nu}_e}} \simeq 2 \tag{1.209}$$

in the absence of neutrino oscillations and at low and moderate energies, roughly below 5 GeV. At higher energies the production and decay of Kaons will contribute to the emission of neutrinos. Also, the flavour ratio cannot be measured directly in real experiments because what is observed is the Cerenkov light produced by charged leptons. Since electron and muon neutrinos have different cross-sections for the interaction with matter and the efficiencies and detection criteria are different for e-like and mu-like events, it is preferred to report the experimental data in terms of e-like and mu-like events and to reveal the

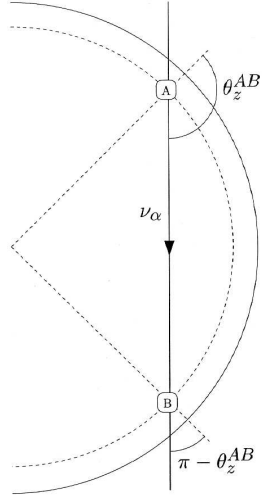

**Figure 1.13:** Symmetry in neutrino flux distribution. Taken from [3]

anomaly through a ratio of ratios:

$$R_{\mu/e} = \frac{(N_{\mu\text{-like}}/N_{e\text{-like}})_{\text{data}}}{(N_{\mu\text{-like}}/N_{e\text{-like}})_{\text{MC}}} \tag{1.210}$$

where $N_e$, $N_\mu$ are the number of events, the numerator is the measured ration and the denominator is the one calculated with Monte Carlo methods.

For high energetic cosmic rays, roughly above 1 GeV, the muons may hit the ground before they decay and this can suppress the ratio between the fluxes of muon and electron neutrinos. On the other hand, low energy cosmic rays may be severely constrained by the geomagnetic field. Below some energies it is expected an overabundance of cosmic rays in the geomagnetic poles and a suppression in the equator. This asymmetry is expected to vanish above few GeV: in this case, if a neutrino is observed in some underground laboratory A (fig. 1.13) with an azimuthal angle $\theta_z^{AB}$, it will hit another laboratory B with an azimuthal angle $\pi - \theta_z^{AB}$. So, for high energies we have

$$\Phi_B^{\nu_\alpha}(\pi - \theta_z^{AB}) = \Phi_A^{\nu_\alpha}(\theta_z^{AB}). \tag{1.211}$$

Since the flux distribution is assumed to be independent of the location,

$$\Phi_A^{\nu_\alpha}(\theta_z) = \Phi_B^{\nu_\alpha}(\theta_z), \tag{1.212}$$

we conclude that it has a symmetry

$$\Phi_A^{\nu_\alpha}(\theta_z) = \Phi_A^{\nu_\alpha}(\pi - \theta_z). \tag{1.213}$$

This means that it is expected an equality between upwards and downwards neutrino fluxes. Now, upwards neutrinos observed at some underground facility have crossed the Earth and are exposed to matter effects. This means that an asymmetry in the downward and upward fluxes is an indication that interaction with matter has induced neutrino oscillations. This can be quantified by the formula

$$A_\alpha^{\text{up-down}} = \left( \frac{U-D}{U+D} \right)_\alpha, \tag{1.214}$$

where $U$ and $D$ are respectively the up and down neutrino fluxes integrated in the ranges $0.2 < \cos\theta_z < 1$ and $-1 < \cos\theta_z < -0.2$ with $\alpha = e, \mu$. Figure 1.14 shows Monte Carlo calculations of the up-down asymmetry in the absence of neutrino oscillations for two locations, Kamioka and Soudan. Note the cut-off on the asymmetry for energies bigger than 3 GeV: the geomagnetic field is no longer strong enough to deviate the cosmic rays which results in a equality between up and down fluxes. The different signs for the asymmetry are due to the different geographic locations. Kamioka is near the geomagnetic equator and Soudan is near the geomagnetic pole.

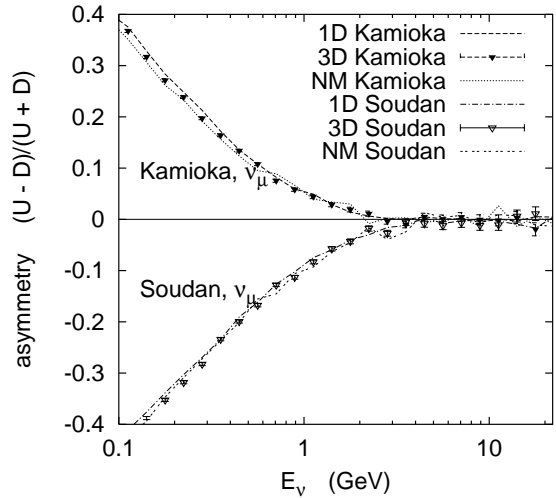

**Figure 1.14:** Asymmetry in atmospheric muon neutrinos as computed with Monte Carlo methods. Taken from [9]

In 1988 Kamiokande published data from sub-GeV events that indicated clearly an anomaly in atmospheric neutrino fluxes. In terms of ratio of ratios the final results of the Kamiokande sub-GeV and multi-GeV events are [3]

$$R_{\mu/e}^{\text{sub-GeV}} = 0.60_{-0.06}^{+0.07} \pm 0.05 \tag{1.215}$$

$$R_{\mu/e}^{\text{multi-GeV}} = 0.57_{-0.07}^{+0.08} \pm 0.07 \tag{1.216}$$

These anomalies are well explained by neutrino oscillations, even though Kamiokande was unable to distinguish between $\mu \leftrightarrow \tau$ and $\mu \leftrightarrow e$ transitions. However the results by the CHOOZ experiment in 1997 excluded $\mu \leftrightarrow e$ transitions.

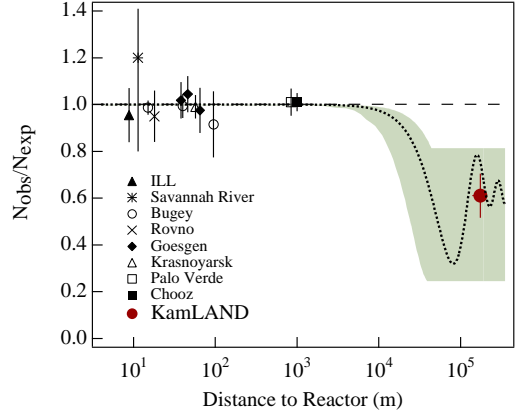

**Figure 1.15:** Ratio of measured to expected $\overline{\nu}_e$ flux of different reactor experiments. The dotted curve corresponds to the best-fit values $\Delta m_{\mathrm{SOL}}^2 = 5.5 \times 10^{-5}$ eV$^2$ and $\sin^2 2\theta_{\mathrm{SOL}} = 0.83$. Taken from [3] and [10].

### 1.3.3 Reactor neutrinos

The fission process in nuclear reactors produces electron anti-neutrinos in abundance from $\beta$-decay in neutron-rich nucleus. However the energy of these anti-neutrinos is small, in the order of a few MeV and this excludes the possibility of detecting $\mu$ and $\tau$ neutrinos from charged current interactions. So in these experiments only the disappearance of $\overline{\nu}_e$ can be investigated. Figure 1.15 shows the ratio of expected to observed $\overline{\nu}_e$ flux for various experiments. The experiments with $L \sim 10 - 100$ m are short-baseline (SBL), the ones with $L \sim 1$ Km are long-baseline (LBL) and finally the KamLAND experiment with $L \sim 200$ km is very-long-baseline (VLBL). It is clear that only KamLAND found evidence of neutrino oscillations. However the absence of a signal for the other experiments leads to exclusions curves in the $\Delta m_{\mathrm{SOL}}^2$, $\sin^2 2\theta_{\mathrm{SOL}}$ plane and to upper bounds for $\Delta m_{\mathrm{SOL}}^2$. Recall that the sensitive to $\Delta m_{\mathrm{SOL}}^2$ is determined by the value of $x/E$ for which we have

$$\frac{\Delta m^2 x}{2E} \sim 1. \tag{1.217}$$

It is now established that $\Delta m_{\mathrm{SOL}}^2$ is of order $10^{-5}$ eV$^2$ and for an energy of few MeV this requires a source-detector distance around 100 km in order to get an evidence of neutrino oscillations, as it was the case with the KamLAND experiment.

**CHOOZ Experiment:** The CHOOZ experiment was located near the CHOOZ power plant in France which is composed by two nuclear reactors. The distance between the detector and the two reactors was 998 m and 1115 m which, by (1.217), means that the experiment was sensible to $\Delta m^2$ of the order of $\times 10^{-3}$ eV$^2$. The CHOOZ experiment gathered data from April 1997 to July 1998 and the ratio of measured to expected $\overline{\nu}_e$ events was [3]

$$R = 1.01 \pm 0.028 \pm 0.027 \tag{1.218}$$

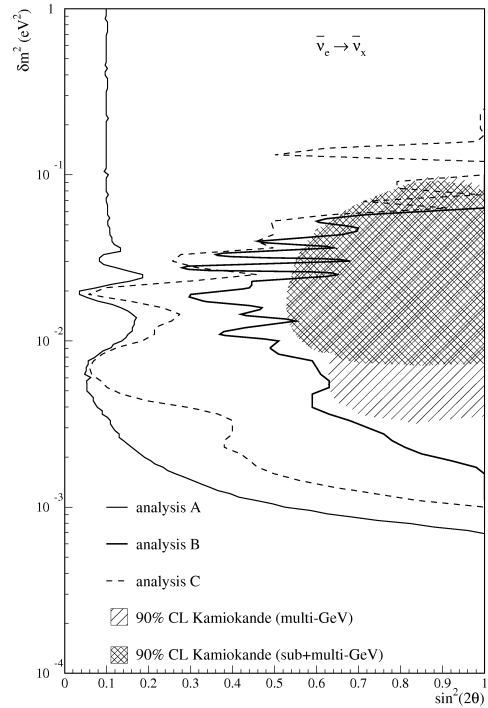

**Figure 1.16:** Exclusion curves for the CHOOZ experiment. Also shown is the allowed region for $\nu_\mu \to \nu_e$ from the Kamiokande experiment. Taken from [3] and [11].

The exclusion curve is shown in figure 1.16 together with the allowed range from Kamiokande for $\nu_\mu \to \nu_e$ transitions as the solution for the atmospheric neutrino deficit. The data implies that [3]

$$\sin^2 2\theta \leq 0.1, \quad \text{for } \Delta m^2 \geq 2 \times 10^{-3} \text{ eV}^2, \tag{1.219}$$

$$\Delta m^2 < 7 \times 10^{-4} \text{ eV}, \quad \text{for } \sin^2 2\theta = 1. \tag{1.220}$$

and excludes the Kamiokande solution for the atmospheric neutrino anomaly.

**KamLAND Experiment:** this experiment [3] was designed to measure $\overline{\nu}_e$ fluxes from nuclear reactors of 53 power plants in Japan. The detector is located in the Kamioka mine where the Kamiokande experiment was previously installed. It consists of 1200 m³ of liquid scintillator that occupies a spherical balloon with a diameter of 13 m. Neutrinos are detected through the inverse $\beta$ decay

$$\overline{\nu}_e + p \longrightarrow e^+ + n \tag{1.221}$$

where the delayed coincidence from the light signals emitted by the positron and the neutron capture allows an efficient reduction of the background. The ratio of measured

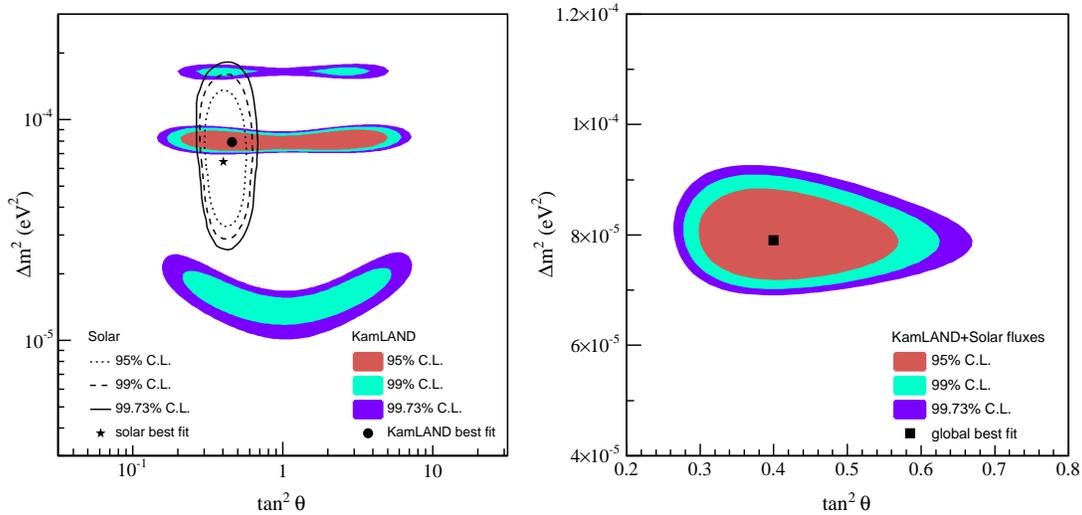

**Figure 1.17:** (a) Allowed region in the $\tan^2\theta, \Delta m^2$ plane obtained in the KamLAND experiment. The lines show the regions allowed by solar neutrino data. (b) Combined analysis of two-neutrino oscillations with KamLAND and solar neutrino data. Taken from [3] and [12].

to expected $\overline{\nu}_e$ events from March 2002 to January 2004 is [3]

$$R = 0.658 \pm 0.044 \pm 0.047 \qquad (1.222)$$

which deviates from unity by $5\sigma$. The best fit of the data is obtained with

$$\Delta m^2 = 7.9^{+0.6}_{-0.5} \times 10^{-5} \text{ eV}^2. \qquad (1.223)$$

Figure 1.17 shows the data in the $\Delta m^2, \tan\theta$ plane and it is clear that there is a big uncertainty in the mixing angle. But a global analysis with the solar data allows a much better resolution for $\theta$.

### 1.3.4 Accelerator neutrinos

These experiments detect muon neutrinos from the decays of muon and kaons produced by the collision of a proton beam with a target. We will discuss only the K2K experiment. **K2K Experiment:** the accelerator is located in the KEK laboratory in Japan and the neutrino flux is detected at the Superkamiokande experiment. The two facilities dist around 250 km. The neutrino beam is an almost pure $\nu_\mu$ beam with roughly 1% of muon anti-neutrinos and 1% of electron neutrinos. The events for K2K at Superkamiokande are selected using GPS synchronization with the proton beam at KEK and the initial neutrino flux is monitored by a near detector similar to the one of Superkamiokande. K2K has been divided in two phases. The first one, K2K-I, lasted from June 1999 to July 2001 and the second one, K2K-II, started in January 2003 and ended in February 2004. In total the experiment observed 107 fully contained $\mu-$like events with an expected number without oscillations of $151^{+12}_{-10}$. Figure 1.18 shows the energy distribution of K2K events plotted with the best fit curve with oscillations (solid) and the curve without oscillations (dashed). The best fit values are [3]

$$\sin^2 2\theta = 1.0 \qquad \Delta m^2 = 2.8 \times 10^{-3} \text{ eV}^2. \qquad (1.224)$$

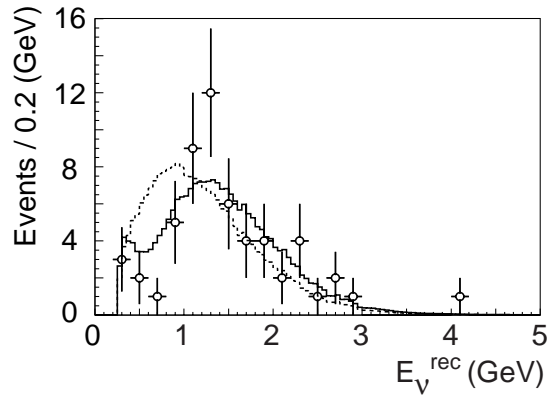

**Figure 1.18:** Number of events as a function of the reconstructed neutrino energy. Points with error bars correspond to data, the solid curve corresponds to best fit with oscillations and the dashed line refers to no oscillations. Taken from [3] and [13].

### 1.3.5 Neutrino parameters

In [28] a global analysis of neutrino oscillations was done, taking into account the latest data at the time of KamLAND and K2K, as well as state-of-the-art solar and atmospheric neutrino fluxes. In [14] the discussion was updated taking also into account the data released in 2008 by the MINOS collaboration, the Sudbury Neutrino Observatory (SNO), KamLAND and Borexino. We present here the major results of these references. In figure 1.19 the curves for solar, Kamland and global analysis are shown. For this result new data from the SNO experiment was taken into account, in particular the data from CC and NC fluxes, that quote a lower value for the flux ratio, $\phi_{\text{CC}}/\phi_{\text{NC}} = 0.301 \pm 0.033$ which leads to a stronger upper bound on $\sin^2 \theta_{12}$. Also included are direct measurements of the $^4$B solar neutrino rate performed by the Borexino collaboration. For the global study a recent re-analysis of the Gallex data was also taken into account. More details can be found on [14]. At $1\sigma$ the best fit points are [14]

$$\sin^2 \theta_{12} = 0.304^{+0.022}_{-0.016}, \qquad \Delta m^2_{21} = 7.65^{+0.23}_{-0.20} \times 10^{-5} \text{ eV}^2. \qquad (1.225)$$

For the atmospheric neutrino parameters, long-baseline accelerator and atmospheric neutrino measurements from Superkamiokande were combined. The results are shown in figure 1.20. In [14] were found the following best fit points for atmospheric parameters,

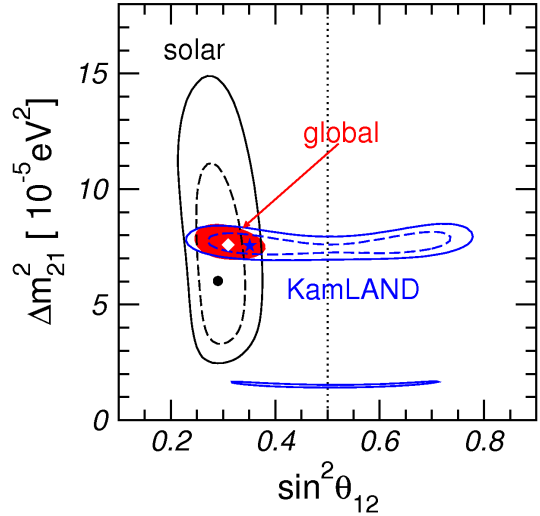

**Figure 1.19:** Determination of the leading solar oscillation parameters from artificial and natural neutrino sources. Curves for data from solar, KamLAND and global analysis are shown. Taken from [14].

at $1\sigma$:

$$\sin^2\theta_{23} = 0.50^{+0.07}_{-0.06}, \qquad |\Delta m^2_{31}| = 2.40^{+0.12}_{-0.11} \times 10^{-3} \, \text{eV}^2 \,. \qquad (1.226)$$

We note that the data is consistent with maximal mixing $\theta_{23} = \pi/4$ and both normal $m_3 \gg m_1 \sim m_2$ or inverted $m_3 \ll m_1 \sim m_2$ hierarchy. The resolution of this last ambiguity lies in a direct determination of neutrino masses, as oscillation experiments are only sensitive to the squares of mass differences. As for $\theta_{13}$ we see from (1.142), (1.144) and (1.146) that a non-zero value for this mixing angle is a necessary condition to observe CP violation. Figure 1.21 on the left summarizes the information on $\theta_{13}$ from the present data, where the different experiments are complementary and all contribute to give a more rigorous bound. At 90% confidence level ($3\sigma$) [14] reports the following limits:

$$\sin^2\theta_{13} \leq \begin{cases} 0.060 \ (0.089) & (\text{solar+KamLAND}) \\ 0.027 \ (0.058) & (\text{CHOOZ+atm+K2K+MINOS}) \\ 0.035 \ (0.056) & (\text{global data}) \end{cases} \qquad (1.227)$$

The interplay of solar and KamLAND is shown on the right of figure 1.21 where the slightly smaller ration CC/NC for SNO mentioned above gives a hint for a non-zero value of $\theta_{13}$, even though is still compatible with a null value for this mixing angle. Details can be found on [14].

In summary, we see that neutrino data is consistent with the Tri-Bi-Maximal mixing

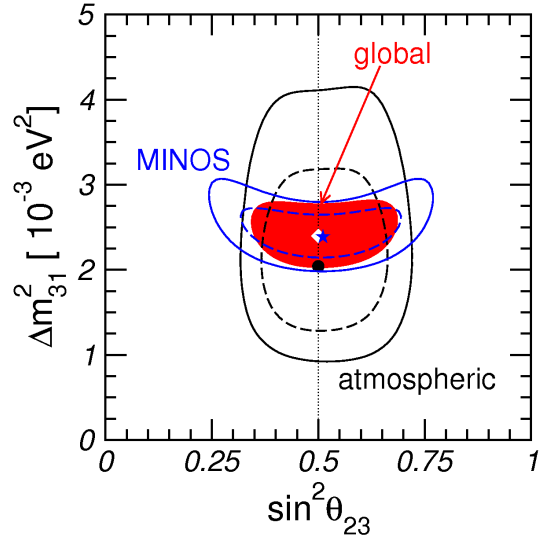

**Figure 1.20:** Determination of the leading atmospheric oscillation parameters from artificial and natural neutrino sources. Curves for data from atmospheric, MINOS and global analysis are shown. The dot, star and diamons indicate the best fit points of atmospheric, MINOS and global data, respectively. Taken from [14].

simplifying hypothesis [29] that leads to the HPS mixing matrix U (see 1.145):

$$
U = \begin{pmatrix} \sqrt{\frac{2}{3}} & \frac{1}{\sqrt{3}} & 0 \\ -\frac{1}{\sqrt{6}} & \frac{1}{\sqrt{3}} & \frac{1}{\sqrt{2}} \\ -\frac{1}{\sqrt{6}} & \frac{1}{\sqrt{3}} & -\frac{1}{\sqrt{2}} \end{pmatrix}. \tag{1.228}
$$

In Chapters 5 and 6 we will assume that the leptonic mixing matrix has this form.

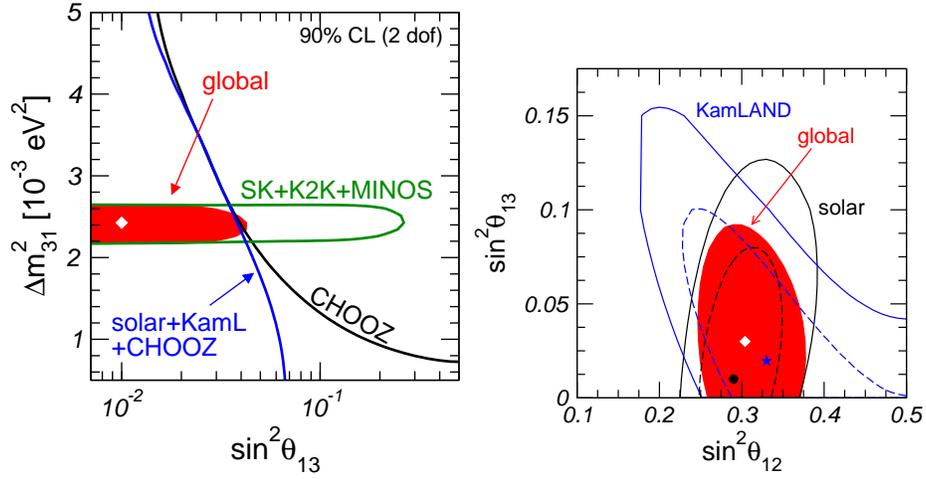

**Figure 1.21:** Left: constraints on $\sin^2 \theta_{13}$ from different parts of the global data. Right: allowed regions in the $(\theta_{12} - \theta_{13})$ plane at 90 and 99.73% CL as well as the 99.73% region for the combined analysis. The dot, star and diamond indicate the best fit points of solar, KamLAND and combined data, respectively. Taken from [14].

## Chapter 2

# Baryon Asymmetry, Leptogenesis and Dark Matter

## 2.1  Anomalies, Baryonic and Leptonic Currents

Anomalies play a central role in the Standard Model, where its absence or total cancelation is essential to the renormalizability of the theory. The anomaly associated with the chiral gauge transformations is related with the non invariance of the fermionic measure in the path integral formulation of the theory. This was first pointed out by Fujikawa [30] and it is in connection with a deep result in Mathematics, the Atiyah-Singer theorem [31], as it is shown in [1].

The explicit computation of the divergence of the gauge 3-vertex function with fermion loops (fig. 2.1) shows that the anomaly is given by

$$[\partial_\mu J_\alpha^\mu]_{\text{anom}} = -\frac{1}{32\pi^2} D_{\alpha\beta\gamma} \epsilon^{\kappa\nu\lambda\rho} F_{\kappa\nu}^\beta F_{\lambda\rho}^\gamma, \qquad (2.1)$$

where $F_{\mu\nu}^\alpha$ is the non-Abelian field strenght associated with the gauge bosons $A_\alpha^\mu$ that belong to the adjoint representation $T_\alpha$ of the gauge group, $D_{\alpha\beta\gamma}$ is the totally symmetric quantity

$$D_{\alpha\beta\gamma} = \frac{1}{2}\text{Tr}(\{T_\alpha, T_\beta\}T_\gamma) \qquad (2.2)$$

and $J_\alpha^\mu$ is the fermionic classically conserved current

$$J_\alpha^\mu = \overline{\chi}\gamma^\mu T_\alpha \chi. \qquad (2.3)$$

Here $\chi$ is a fermionic field that unifies all left-handed fermions and anti-fermions that transform nontrivially under the representations of the gauge group.

There is always some freedom in choosing which of the 3 fermionic currents on the triangle diagram is anomalous, even though it is not possible to remove simultaneously the anomaly. In this way, we could as well have written the anomaly (2.1) for $J_\beta^\nu$ or $J_\gamma^\rho$. This is particularly useful when one has a global symmetry that it is not gauged, coupled to 2 gauge fields. In this case, one can choose the anomaly to be in the non-gauged current. That is what happens with the baryon number current,

$$J_B^\mu = B \overline{\chi}_B \gamma^\mu \chi_B \tag{2.4}$$

where B is the baryon number, which is conserved at the classical level but gets anomalous with the triangle diagram (fig. 2.2). To compute the anomaly, we must calculate the coefficient (2.2) for all the possible couplings with the gauge fields of the Standard Model. It is easy to see that only pairs of fields in the same gauge group contribute, since the traces of the $SU(3)$ and $SU(2)$ generators are zero. The relevant quantum numbers are shown in table 2.1 (see also sec. 2.3). The normalization condition adopted here for $SU(N)$ generators is

$$\text{Tr} \left( T_\alpha T_\beta \right) = \frac{N_G}{2} \delta_{\alpha\beta}, \tag{2.5}$$

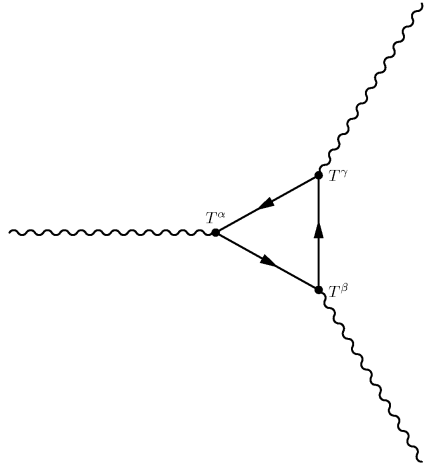

**Figure 2.1:** One loop gauge 3-vertex that contributes to the triangle anomaly.

**Table 2.1:** Standard Model quantum numbers

| Particle | $\widetilde{g}$ | B | L | $SU(3)$ | $SU(2)$ | $U(1)$ |
|---|---|---|---|---|---|---|
| $\begin{pmatrix} u_{Li} \\ d_{Li} \end{pmatrix}$ | 3 | $\frac{1}{3}$ | 0 | **3** | **2** | $\frac{1}{3}$ |
| $u_{Li}^c$ | 3 | $-\frac{1}{3}$ | 0 | $\overline{\mathbf{3}}$ | **1** | $-\frac{4}{3}$ |
| $d_{Li}^c$ | 3 | $-\frac{1}{3}$ | 0 | $\overline{\mathbf{3}}$ | **1** | $\frac{2}{3}$ |
| $\begin{pmatrix} \nu_{Li} \\ e_{Li} \end{pmatrix}$ | 1 | 0 | 1 | **1** | **2** | $-1$ |
| $e_{Li}^c$ | 1 | 0 | $-1$ | **1** | **1** | $+2$ |
| $W^+$ | 4 | 0 | 0 | **1** | **3** | 0 |
| $\begin{pmatrix} \phi^+ \\ \phi^0 \end{pmatrix}$ | 2 | 0 | 0 | **1** | **2** | 1 |
| gluons | 4 | 0 | 0 | **3** | **1** | 0 |

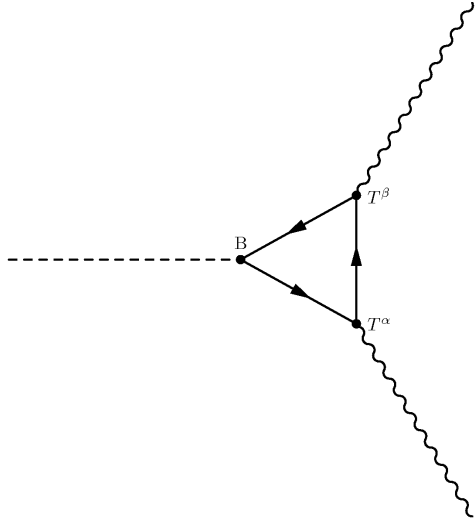

**Figure 2.2:** Anomaly in the baryonic current.

where $N_G$ is a constant that depends on the representation of the group $G \equiv SU(N)$. For the baryonic current coupled to 2 $SU(3)$ fields there is no anomaly, because

$$\sum_{\mathbf{3},\overline{\mathbf{3}}} B = 0. \tag{2.6}$$

In the case of $SU(2)$, the coefficient $D_{\alpha\beta\gamma}$ is

$$\sum_{\mathbf{2}} B = N_2, \tag{2.7}$$

for one generation, where a color factor of 3 was taken into account and $N_2$ is the factor $N_G$ above for $SU(2)$, and for $U(1)$, the same coefficient is

$$\sum BY^2 = -2. \tag{2.8}$$

This shows that, in the Standard Model, the baryonic current has a total anomaly

$$[\partial_\mu J^\mu_B]_{\text{anom}} = \frac{N_g}{16\pi^2} \epsilon^{\alpha\beta\mu\nu} \left( F_{1\alpha\beta} F_{1\mu\nu} - \frac{N_2}{2} F^\rho_{2\alpha\beta} F^\rho_{2\mu\nu} \right), \tag{2.9}$$

with $N_g$ the number of generations.

In exactly the same way one can compute the anomaly for the leptonic current and find that it is equal, assuming the same number of generations for quarks and leptons, as it is required by the cancelation of gauge anomalies in the Standard Model. So, even if the leptonic and baryonic currents are anomalous, their difference $J^\mu_B - J^\mu_L$ is conserved, which makes $B - L$ a good quantum number. This may have serious cosmological implications in the way the asymmetry between matter and anti-matter is obtained, as we will see. We finish noting that the differences between lepton flavour currents are also conserved, so that $L_i - L_j$ are also good quantum numbers.

## 2.2 Instantons and Sphalerons

The variation of macroscopic charge $Q$ associated with a given classically conserved Noether current $J^\mu$ is usually given by

$$\dot{Q} = \int d^3x \, \partial_i J^i = 0. \tag{2.10}$$

In the case of the anomalous baryonic current, the variation of baryon number is

$$\dot{B} = \frac{N_g}{16\pi^2} \epsilon^{\alpha\beta\mu\nu} \int d^3x \left( F_{1\alpha\beta} F_{1\mu\nu} - \frac{N_2}{2} F_{2\alpha\beta}^\rho F_{2\mu\nu}^\rho \right), \tag{2.11}$$

so that

$$B_f = B_0 + \frac{N_g}{16\pi^2} \epsilon^{\alpha\beta\mu\nu} \int d^4x \left( F_{1\alpha\beta} F_{1\mu\nu} - \frac{N_2}{2} F_{2\alpha\beta}^\rho F_{2\mu\nu}^\rho \right). \tag{2.12}$$

Now, it can be seen that the current

$$K_l = \epsilon_{lijk} \frac{2}{N_G} \text{Tr} \left( A_i F_{jk} + \frac{2i}{3} A_i A_j A_k \right), \tag{2.13}$$

where we have changed to Euclidean space through a Wick rotation, gives

$$\partial_m K_m = \frac{1}{2} \epsilon_{ijkl} F_{\alpha ij} F_{\alpha kl}, \tag{2.14}$$

so that the variation of baryon number can be given as a surface integral of a total baryon current density. Assuming that the field strengths $F_{ij}$ and $F_{\alpha ij}$ vanish at infinity, the $U(1)$ gauge field does not contribute and then

$$\Delta B = -\frac{N_g i}{24\pi^2} \epsilon_{ijk} \int_{S^3} \text{Tr} \left( A_i A_j A_k \right), \tag{2.15}$$

with $S^3$ a three dimensional sphere with a radius approaching infinity. Naively, we would be tempted to consider that this integral is zero for gauge fields vanishing sufficiently fast. However, there are configurations for $A_i^\alpha$ that do not vanish at infinity, but give there zero field strength. These are known as Instantons, after Belavin, Polyakov, Schwarz and Tyupkin [32] and have asymptotically a pure gauge configuration:

$$i A_i \longrightarrow g(\hat{x})^{-1} \partial_i g(\hat{x}), \tag{2.16}$$

where $A_i \equiv A_i^{\alpha} t^{\alpha}$ and $g(\hat{x})$ is a group element of $SU(2)$. Belavin *et al.* [32] found a solution:

$$iA_i(x) = \left( \frac{r^2}{r^2 + R^2} \right) g^{-1}(\hat{x}) \partial_i g(\hat{x}),$$ (2.17)

where, explicitly,

$$g(\hat{x}) = \left( \frac{x_4 + 2i\mathbf{x} \cdot \mathbf{t}}{r} \right),$$ (2.18)

and $\mathbf{t}$ are the usual $SU(2)$ generators. This allows us to define the quantity

$$\mathcal{I} = -i \lim_{r \to \infty} r^3 \int d\theta^1 d\theta^2 d\theta^3 \epsilon^{abc} \frac{\partial \hat{x}_i}{\partial \theta^a} \frac{\partial \hat{x}_j}{\partial \theta^b} \frac{\partial \hat{x}_k}{\partial \theta^c} \text{Tr} \left( A_i A_j A_k \right)$$

$$= \int d\theta^1 d\theta^2 d\theta^3 \epsilon^{abc} \text{Tr} \left\{ g^{-1}(\theta) \frac{\partial g(\theta)}{\partial \theta^a} g^{-1}(\theta) \frac{\partial g(\theta)}{\partial \theta^b} g^{-1}(\theta) \frac{\partial g(\theta)}{\partial \theta^c} \right\}$$ (2.19)

in such a way that the variation of baryon number is

$$\Delta B = \frac{N_g}{24\pi^2} \mathcal{I}.$$ (2.20)

The quantity $\mathcal{I}$ is well known by mathematicians and it is called the Maurer-Cartan form. It classifies homotopy classes of maps[1] of $S^3$ into a Lie group. A computation of $\mathcal{I}$ can be found in [1], where it is shown that

$$\mathcal{I} = 24\pi^2,$$ (2.21)

because $g(\hat{x})$ belongs to the homotopy class of the identity. Since the homotopy group $\pi_3(G)$ of a compact connected simple Lie group $G$ that classifies homotopy classes of maps from $S^3$ into $G$ is[2] $\mathbf{Z}$, a general homotopy class is identified by an integer $\nu$ called the winding number and can be obtained by superimposing several solutions that have winding number equal to 1 or -1. So, in general,

$$\mathcal{I} = 24\pi^2 \nu.$$ (2.22)

This shows that Instantons change the baryon number by an integral value:

$$\Delta B = N_g \nu.$$ (2.23)

In virtue of the inequality

$$0 \leq \frac{1}{8} \int \left( F_{\alpha ij} \mp \frac{1}{2} \epsilon_{ijkl} F_{\alpha kl} \right)^2 d^4 x,$$ (2.24)

---

[1] An homotopy class of maps on a Lie group is a set of maps that can be continuously deformed into each other.

[2] $S^3$ is homeomorphic to $SU(2)$, so this is the same as considering maps from $SU(2)$ into $G$. Moreover, it is possible to restrict these maps to a $SU(2)$ subgroup of $G$

the Euclidean Yang-Mills action

$$S \equiv -\frac{1}{4g_2^2} \int F_{\alpha ij} F_{\alpha ij} d^4 x, \tag{2.25}$$

will satisfy the inequality

$$-S \geq \frac{1}{8g_2^2} \left| \epsilon_{ijkl} \int F_{ij}^{\alpha} F_{kl}^{\alpha} d^4 x \right|. \tag{2.26}$$

Because the field strengths $F_{ij}^{\alpha}$ from (2.17) are self-dual

$$F_{ij}^{\alpha} = \frac{1}{2} \epsilon_{ijkl} F_{kl}^{\alpha}, \tag{2.27}$$

equation (2.26) is in fact an equality for Instantons, which allows us to write

$$S = 8\pi^2 |\nu| / g_2^2. \tag{2.28}$$

In the Standard Model, $g_2 = e/\sin\theta_W$ and, for $e^2/4\pi \simeq 1/129$ and $\sin^2\theta_W \simeq 0.23$ evaluated at $M_Z$, equation (2.28) gives a suppression factor of $\exp(-186)$, which indicates that Instanton processes are unlikely to be observed nowadays. However, the potential barrier could be surpassed if thermal fluctuations are important, as it could be the case in the early Universe.

Until now the discussion has been somehow oversimplified because we are considering a pure gauge theory. The inclusion of matter and Higgs fields can change the picture in a significant way. This issue was first addressed in two papers by Klinkhammer [33] and Manton [33] [34], where the name Sphaleron appeared to describe a saddle point for the energy on the Weinberg-Salam model configuration space. Here, we present a short summary, following [35]. In the limit of vanishing mixing angle $\sin\theta_W \to 0$, the Sphaleron has the following form, in the $A_0^a = 0$ gauge:

$$A_i = f(g_2 v r) U^{\infty} \partial_i (U^{\infty})^{-1} \tag{2.29}$$

$$\Phi = \frac{iv}{\sqrt{2}} h(g_2 v r) U^{\infty} \begin{pmatrix} 0 \\ 1 \end{pmatrix}, \tag{2.30}$$

with $U^{\infty} = g(\hat{x}_0 = 0, \vec{\hat{x}})$, where $g(\hat{x})$ is given by (2.18), and the functions $f$ and $h$ have the asymptotic behavior

$$f(\xi) \longrightarrow \begin{cases} \sim \xi^2, & \xi \to 0 \\ 1, & \xi \to \infty \end{cases} \tag{2.31}$$

$$h(\xi) \longrightarrow \begin{cases} \sim \xi, & \xi \to 0 \\ 1, & \xi \to \infty \end{cases} \tag{2.32}$$

The Euclidean energy functional

$$E = \int d^3x \left[ \frac{1}{4g_2^2} F_{ij}^a F_{ij}^a + (D_i\Phi)^\dagger(D_i\Phi) + \lambda \left( \Phi^\dagger\Phi - \frac{1}{2}v^2 \right)^2 \right] \qquad (2.33)$$

can be reexpressed as

$$E = \frac{4\pi v}{g_2} \int_0^\infty d\xi \left[ 4 \left( \frac{df}{d\xi} \right)^2 + \frac{8}{\xi^2}[f(1-f)]^2 + \frac{1}{2}\xi^2 \left( \frac{dh}{d\xi} \right)^2 + [h(1-f)]^2 + \frac{\lambda}{g_2^2}\xi^2(h^2-1)^2 \right]. \qquad (2.34)$$

The Sphaleron solution (2.29) and (2.30) interpolates between a vacuum state for $\xi \to \infty$ and the maximum of the Higgs potential ($\Phi^a = 0$) for $\xi \to 0$. The vacuum solution for the gauge fields is the same as the Instanton with winding number 1, and superimposing the solution $U^\infty$ n times we get a solution with winding number n. So, the Sphaleron is a saddle point in field configuration space, the lowest barrier between two topological distinct vacua and transitions from one vacuum state to another will induce violations of baryon and lepton number. Using dimensional analysis, one estimates roughly the Sphaleron energy:

$$E(A_i) \sim \frac{4\pi}{g_2^2 l}, \qquad (2.35)$$

$$E(\Phi) \sim 4\pi v^2 l, \qquad (2.36)$$

where $l$ is a typical length scale for the Sphaleron solution. Minimizing $E(A_i) + E(\Phi)$, one gets

$$l_{sp} \sim \frac{1}{g_2 v} \sim 10^{-16} \text{ cm} \qquad (2.37)$$

and

$$E_{sp} \sim \frac{8\pi v}{g_2} \sim 10 \text{ TeV}, \qquad (2.38)$$

which sets the order of magnitude for the Sphaleron potential barrier. A more accurate result was found in [33] by means of variational methods:

$$E_{sp} = \frac{4\pi v}{g_2} B(\frac{\lambda}{g_2}), \qquad (2.39)$$

where $B$ is a function that depends weakly on $\lambda/g_2$: $B(0) \simeq 1.52$ and $B(\infty) \simeq 2.72$. The effects of a mixing angle $\theta_w$ different from zero are small.

In [36] the rate of Sphaleron processes was computed:

$$\Gamma_{sph} = TC \exp\left( -\frac{2M_W(T)}{Tg_2(T)} B(\lambda(T)/g_2(T)) \right), \qquad (2.40)$$

where $C$ is a constant. This should be compared with the expansion rate of the Universe

$$H \sim N_{eff}^{1/2} T^2 / m_{pl},$$
(2.41)

where $N_{eff} \sim 100$ are the effective relativistic degrees of freedom of the Standard Model and $m_{pl}$ is the Planck Mass. One sees that the Sphaleron rate exceeds $H$ for $T > T^*$ (see the comments in [36]), where

$$T^* = (2M_W(T)/g_2(T)\ln(m_{pl}/T^*))\,B(\lambda/g_2).$$
(2.42)

This will depend on the value of $\lambda(T)$, but for $\lambda(T) \sim g_2^2$, one finds that $B = 2.1$ and $T^* \sim 0.6T_c$, where $T_c$ is the temperature for the electroweak symmetry breaking transition. Since Sphalerons are only active below $T_c$, we see that there is a temperature range where they are in thermal equilibrium. On the other hand, the trivial relation

$$B = \frac{1}{2}(B+L) + \frac{1}{2}(B-L)$$
(2.43)

and the fact that $B - L$ is a good quantum number, suggest that any previous baryon asymmetry generated by any process that gives $B - L = 0$ will be erased by Sphalerons when in thermal equilibrium, since they violate $B + L$. However, if some process creates a non-zero value of $B - L$ at a high scale, for instance, by means of the decaying of a heavy particle into leptons or anti-leptons, then it could survive through the period of thermal equilibrium, even though the values of $B$ and $L$ could change. This was first proposed by Fukugita and Yanagida [37] and it is the basis of Leptogenesis.

## 2.3 Baryogenesis and Leptogenesis

In order to estimate the baryon number that survives the period of thermal equilibrium, one must express it as a function of the conserved quantum numbers. This was done in [38], but the argument presented here is from [39].

From a set of conserved quantities $Q_a$, like isospin, hypercharge or $B - L$ and their quantum numbers $q_{ai}$ for each particle specie $i$, we can express the chemical potentials as

$$\mu_i = \sum_a q_{ai}\mu_a,$$
(2.44)

with some linear coefficients $\mu_a$. The relativistic particle density is

$$n_i = \frac{g_i}{(2\pi)^3} \int \frac{d^3p}{e^{(p-\mu_i)/T} \mp 1}$$
(2.45)

$$= 4\pi g_i \left(\frac{T}{2\pi}\right)^3 \int_0^\infty \frac{x^2 dx}{e^{x-\mu_i/T} \mp 1}$$
(2.46)

where $g_i$ are the effective degrees of freedom for the particle specie $i$. The antiparticle density will be given by the same formula, but with $\mu_i$ replaced by $-\mu_i$, so the difference is

$$n_i - \overline{n}_i = 8\pi g_i \left(\frac{T}{2\pi}\right)^3 \frac{\mu_i}{T} \int_0^\infty \frac{x^2 dx}{(e^x \mp 1)^2}, \qquad (2.47)$$

assuming that the imbalance between particles and antiparticles is small, that is, $|\mu_i| \ll 1$. The integral has the value $\pi^2/3$ for bosons and $\pi^2/6$ for fermions. This means that

$$n_i - \overline{n}_i = \frac{T^2}{6} \widetilde{g}_i \mu_i, \qquad (2.48)$$

where $\widetilde{g}_i$ are the effective degrees of freedom, with an additional factor of 2 for bosons. So, from equation (2.44) we have

$$n_i - \overline{n}_i = \frac{T^2}{6} \widetilde{g}_i \sum_a q_{ai} \mu_a. \qquad (2.49)$$

The density of the conserved quantity $Q_a$ is

$$n_a = \sum_i q_{ai}(n_i - \overline{n}_i) = \frac{T^2}{6} \sum_b M_{ab} \mu_b, \qquad (2.50)$$

where $M$ is the positive definite matrix

$$M_{ab} = \sum_i \widetilde{g}_i q_{ai} q_{bi}. \qquad (2.51)$$

Since $M$ has an inverse, we can invert equation (2.50) and use this in equation (2.49) to obtain

$$n_i - \overline{n}_i = \sum_{ab} \widetilde{g}_i q_{ai} M_{ab}^{-1} n_b. \qquad (2.52)$$

From this, we can obtain the baryon number density as

$$n_B = \sum_i B_i(n_i - \overline{n}_i). \qquad (2.53)$$

It is a simple exercise to compute the matrix elements $M_{ab}$ from table 2.1:

$$M_{B-L,B-L} = \frac{13N_g}{3}, \qquad (2.54)$$

$$M_{B-L,Y} = -\frac{8N_g}{3}, \qquad (2.55)$$

$$M_{Y,Y} = \frac{10N_g}{3} + N_d, \qquad (2.56)$$

where $N_d$ is the number of scalar doublets, from which we can obtain $M_{ab}^{-1}$:

$$M_{B-L,B-L}^{-1} = \frac{10N_g}{3D} + \frac{N_d}{D}, \qquad (2.57)$$

$$M_{B-L,Y}^{-1} = -\frac{8N_g}{3D}, \qquad (2.58)$$

$$M_{Y,Y}^{-1} = \frac{13N_g}{3D}, \qquad (2.59)$$

with the determinant

$$D = \frac{22N_g^2}{3} + \frac{13N_g N_d}{3}. \qquad (2.60)$$

This allows to compute the baryon number density from equation (2.53):

$$n_B = \sum_i \widetilde{g}_i B_i \left( (B-L)_i M_{B-L,B-L}^{-1} + Y_i M_{Y,B-L} \right) n_{B-L} \qquad (2.61)$$

$$= \left( \frac{8N_g + 4N_d}{22N_g + 13N_d} \right) n_{B-L}. \qquad (2.62)$$

taking note that the hypercharge density is zero, from gauge invariance. For $N_g = 3$ and $N_d = 1$ this gives $n_B = (28/79)n_{B-L}$. In any case, $n_B$ is of the same order of magnitude as $n_{B-L}$, as it would be expected from the naive relation (2.43).

## 2.4 Cosmology and Dark Matter

In this section we review briefly the Standard Cosmological Model, based on the Firedmann-Robertson-Walker metric, then we discuss in detail the Boltzmann equation that describes the out of equilibrium processes in the early Universe and we finish with the applications of this equation to the cosmological bound on neutrino masses and to the mass estimate of WIMPs as the major constituents of Dark Matter. Major references are [3, 40–42].

### 2.4.1 The Standard Cosmological Model

Modern Cosmology is based on the General Theory of Gravitation, first proposed by Albert Einstein in 1915. The dynamics of space-time are described by Einstein's equations

$$R^{\mu\nu} - \frac{1}{2}Rg^{\mu\nu} = 8\pi G_N T^{\mu\nu} + \Lambda g^{\mu\nu} \qquad (2.63)$$

where $g^{\mu\nu}$ is the space-time metric, $R^{\mu\nu}$ is the Ricci tensor, $R$ the scalar curvature, $T^{\mu\nu}$ the energy-momentum tensor and $\Lambda$ the cosmological constant. The Ricci tensor and the scalar curvature are obtained from the full curvature tensor

$$R_{\nu\rho\sigma}^{\mu} = \frac{\partial \Gamma_{\nu\sigma}^{\mu}}{\partial x^\rho} - \frac{\partial \Gamma_{\nu\rho}^{\mu}}{\partial x^\sigma} + \Gamma_{\eta\rho}^{\mu}\Gamma_{\nu\sigma}^{\eta} - \Gamma_{\eta\sigma}^{\mu}\Gamma_{\nu\rho}^{\eta} \qquad (2.64)$$

with the Christoffel symbols being given by the metric

$$\Gamma^{\alpha}_{\beta\gamma} = \frac{1}{2} g^{\alpha\mu} \left( \frac{\partial g_{\gamma\mu}}{\partial x^{\beta}} + \frac{\partial g_{\beta\mu}}{\partial x^{\gamma}} - \frac{\partial g_{\beta\gamma}}{\partial x^{\mu}} \right).$$ (2.65)

More specifically we have

$$R^{\mu\nu} = g_{\rho\sigma} R^{\mu\rho\nu\sigma}$$ (2.66)

and

$$R = g_{\mu\nu} R^{\mu\nu}.$$ (2.67)

The free falling motion is determined by the geodesics equation

$$\frac{d^2 x^{\mu}}{d\tau^2} + \Gamma^{\mu}_{\rho\sigma} \frac{dx^{\rho}}{d\tau} \frac{dx^{\sigma}}{d\tau} = 0$$ (2.68)

with the proper time given by

$$d\tau^2 = g_{\alpha\beta} dx^{\alpha} dx^{\beta}.$$ (2.69)

The energy momentum tensor describes the matter content of the model. For a perfect fluid we have

$$T^{\mu\nu} = (\rho + p) u^{\mu} u^{\nu} - p g^{\mu\nu}$$ (2.70)

with $u$ the proper velocity, $\rho$ the energy density and $p$ the pressure.

Observations of the large scale structure of the Universe suggest that it is spatially homogeneous and isotropic at distances roughly bigger than 10 Mpc. A solution for Einstein equations for such an Universe is given by the Friedmann-Robertson-Walker metric

$$d\tau^2 = dt^2 - R(t)^2 \left( \frac{dr^2}{1 - kr^2} + r^2 (d\theta^2 + \sin^2 \theta d\phi^2) \right)$$ (2.71)

which describes a perfect fluid with the energy momentum tensor given by (2.70) in a co-moving frame. The scale factor describes the expansion of the Universe and it has the dimension of length; $k$ is the spacial curvature normalized to

$$k = \begin{cases} -1 & \text{open Universe} \\ 0 & \text{flat Universe} \\ 1 & \text{closed Universe} \end{cases}$$ (2.72)

The fact that $R(t)$ depends on time implies an apparent cosmological doppler effect known as redshift, even though its origin is in the expansion of spacetime. In fact, since photons are obviously light-like ($d\tau = 0$), from (2.71) we have for a particular wave front

$$\int_t^{t_0} \frac{dt}{R(t)} = \int_r^{r_0} \frac{dr}{\sqrt{1 - kr^2}},$$ (2.73)

where $(t, r)$ are relative to the emission and $(t_0, r_0)$ are relative to the reception. If we consider the next wave front differing from the previous in space and time by the wavelength and the period we have that

$$\int_{t+T}^{t_0+T'} \frac{dt}{R(t)} = \int_r^{r_0} \frac{dr}{\sqrt{1-kr^2}} = \int_t^{t_0} \frac{dt}{R(t)} \tag{2.74}$$

so that

$$\int_{t_0}^{t_0+T'} \frac{dt}{R(t)} = \int_t^{t+T} \frac{dt}{R(t)} \tag{2.75}$$

which means that

$$\frac{\lambda_0}{R(t_0)} = \frac{\lambda}{R(t)} \Leftrightarrow$$
$$\frac{\lambda_0}{\lambda} = \frac{R(t_0)}{R(t)}$$
$$= 1 + z, \tag{2.76}$$

where $z$ is the redshift:

$$z = \frac{\lambda_0 - \lambda}{\lambda}. \tag{2.77}$$

The redshift can be directly measured observing the electromagnetic spectrum from stars, nebulae and galaxies. Observational data shows irrefutably that $z > 1$ and as so the Universe is expanding.

When the FRW metric is used in Einstein's equations with the perfect fluid energy-momentum tensor, we obtain Friedmann equations

$$H^2 = \frac{8\pi G_N}{3} \rho - \frac{k}{R^2} \tag{2.78}$$

with $H$, the Hubble parameter, given by

$$H = \frac{\dot{R}}{R}. \tag{2.79}$$

This definition has its origin in the expansion

$$\frac{1}{1+z} = \frac{R(t)}{R(t_0)} \tag{2.80}$$

$$= 1 + \frac{\dot{R}(t_0)}{R(t_0)}(t - t_0) + O((t - t_0)^2) \tag{2.81}$$

which can be seen to give the Hubble law for $z \ll 1$,

$$z = H_0 d_L \tag{2.82}$$

where $d_L$ is the luminosity distance.

From the Friedmann equation we see that for a flat Universe ($k = 0$) the density is equal to the critical density

$$\rho_c = \frac{3H^2}{8\pi G_N}.$$ (2.83)

Its present value is [3]

$$\rho_c = \frac{3H_0^2}{8\pi G_N} = (10.5369 \pm 0.0016)\, h^2 \text{ keV cm}^{-3}$$ (2.84)

where the present value of the Hubble parameter is $H_0 = 100h \text{ km s}^{-1}\text{Mpc}^{-1}$. The Particle Data Group [43] quotes $h = 0.72(3)$.

The energy density is usually expressed in terms of a relative density

$$\Omega = \frac{\rho}{\rho_c},$$ (2.85)

which allows to write Friedmann equation as

$$\Omega - 1 = \frac{k}{H^2 R^2}.$$ (2.86)

Matter, radiation and vacuum have different evolutions. Matter is understood as non-relativistic particles with vanishing pressure. Radiation obeys the equation of state

$$p_R = \frac{1}{3}\rho_R$$ (2.87)

and vacuum has a negative pressure which is a consequence of the energy conservation equation. As the Universe evolves matter density is suppressed by a factor $R^{-3}$

$$\rho_M \propto R^{-3} \propto (1 + z)^3$$ (2.88)

because the volume expands by a factor $R^3$. As for the energy density, it is suppressed by an additional factor of $R^{-1}$ that takes into account the redshift

$$\rho_R \propto R^{-4} \propto (1 + z)^4.$$ (2.89)

Finally the vacuum energy density is constant. These different evolution rates suggest that the Universe has three distinct epochs: radiation dominated, its early epoch just after inflation; matter dominated, an intermediate stage; vacuum dominated, its latest era. It will be instructive to obtain an estimate of the Universe age, taking into account these different contributions by radiation, matter and vacuum. Friedmann equation (2.86) allows to obtain the curvature constant as a function of current parameters

$$k = (\Omega_0 - 1) H_0^2 R_0^2 \tag{2.90}$$

and replacing back in (2.86) we get

$$
\begin{aligned}
H^2 &= H_0^2 \frac{R_0^2}{R^2} \frac{\Omega_0 - 1}{\Omega - 1} \\
&= H_0^2 (1+z)^2 \frac{\Omega_0 - 1}{\Omega - 1}.
\end{aligned}
\tag{2.91}
$$

or

$$\Omega - 1 = \frac{H_0^2}{H^2}(1+z)^2(\Omega_0 - 1). \tag{2.92}$$

Formula (2.91) gives the evolution of the Hubble parameter in terms of the relative density and the redshift and it will be useful in a moment. The last formula can give the relative density as a function of present values and the redshift $z$: combining the evolutions of the three states

$$\frac{\rho_{0M}}{\rho_M} = \left(\frac{R}{R_0}\right)^3 = (1+z)^{-3} \tag{2.93}$$

$$\frac{\rho_{0\Lambda}}{\rho_\Lambda} = 1 \tag{2.94}$$

$$\frac{\rho_{0R}}{\rho_R} = \left(\frac{R}{R_0}\right)^4 = (1+z)^{-4}, \tag{2.95}$$

with the observation

$$
\begin{aligned}
\Omega &= \frac{\rho}{\rho_c} \\
&= \frac{\rho}{\rho_0} \Omega_0 \left(\frac{H_0}{H}\right)^2,
\end{aligned}
\tag{2.96}
$$

we can get the three relative densities as

$$\Omega_R = \Omega_{0R}(1+z)^4 \left(\frac{H_0}{H}\right)^2 \tag{2.97}$$

$$\Omega_M = \Omega_{0M}(1+z)^3 \left(\frac{H_0}{H}\right)^2 \tag{2.98}$$

$$\Omega_\Lambda = \Omega_{0\Lambda} \left(\frac{H_0}{H}\right)^2 \tag{2.99}$$

and the total $\Omega$ as

$$\Omega = \Omega_R + \Omega_M + \Omega_\Lambda$$
$$= \left((1+z)^3\Omega_{0M} + (1+z)^4\Omega_{0R} + \Omega_{0\Lambda}\right)\left(\frac{H_0}{H}\right)^2. \tag{2.100}$$

Obtaining the Hubble parameters ratio from this equation and replacing in (2.92) we get after some trivial algebra

$$\Omega - 1 = \frac{\Omega_0 - 1}{1 - \Omega_0 + (1+z)\Omega_{0M} + (1+z)^2\Omega_{0R} + (1+z)^{-2}\Omega_{0\Lambda}}. \tag{2.101}$$

This in turn allows to express the Hubble parameter solely in terms of present values and the redshift. Replacing in (2.91) we have

$$H = H_0(1+z)\sqrt{1 - \Omega_0 + (1+z)\Omega_{0M} + (1+z)^2\Omega_{0R} + (1+z)^{-2}\Omega_{0\Lambda}}. \tag{2.102}$$

From the definition of redshift (2.76) we can relate the variation of time with the variation of $z$

$$dt = -\frac{1}{H}\frac{dz}{1+z} \tag{2.103}$$

so that finally we can integrate with equation (2.102) to obtain

$$t_0 = \frac{1}{H_0}\int_0^1 \frac{dx}{\sqrt{1 - \Omega_0 + x^{-1}\Omega_{0M} + x^{-2}\Omega_{0R} + x^2\Omega_{0\Lambda}}}. \tag{2.104}$$

with $x = (1+z)^{-1}$. Neglecting the radiation contribution and for $k = 0$ the integral can be performed analytically to give

$$t_0 = \frac{2}{3}H_0^{-1}\frac{1}{\sqrt{\Omega_{0\Lambda}}}\ln\frac{1 + \sqrt{\Omega_{0\Lambda}}}{\sqrt{1 - \Omega_{0\Lambda}}} \tag{2.105}$$

The energy density for a dilute, weakly-interacting gas of particles with $g$ internal degrees of freedom is given in terms of its distribution function $f(\mathbf{p})$:

$$\rho = \frac{g}{(2\pi)^3}\int E(\mathbf{p})f(\mathbf{p})d^3p. \tag{2.106}$$

If the specie is in thermal equilibrium then the phase space distribution is given by the Fermi-Dirac or Bose-Einstein distribution

$$f(\mathbf{p}) = \frac{1}{\exp\left((E(\mathbf{p}) - \mu)/T \pm 1\right)} \tag{2.107}$$

with $+1$ for fermions and $-1$ for bosons and $\mu$ the chemical potential. Moreover, if the particles are relativistic then this expression can be easily integrated to give

$$\rho = \begin{cases} \frac{\pi^2}{30} g T^4, & \text{bosons} \\ \frac{7}{8} \frac{\pi^2}{30} g T^4, & \text{fermions} \end{cases} \tag{2.108}$$

Also, for relativistic particles

$$p = \frac{1}{3}\rho. \tag{2.109}$$

Replacing the relativistic energy density in Friedmann equation with $k = 0$ gives the evolution of the Hubble parameter as a function of the temperature in the radiation dominated era:

$$H = \left( \frac{8\pi G g_* \pi^2}{90} \right)^{1/2} T^2, \tag{2.110}$$

where $g_* = (1, 7/8) \leftrightarrow (\text{bosons}, \text{fermions})$. Also, from the fact that $\rho$ decreases as $R^{-4}$ Friedmann equation shows once again that

$$H = \frac{1}{2t} \tag{2.111}$$

which allows to obtain a relation between time and temperature:

$$t = \frac{1}{2} \left( \frac{90}{8\pi^3 g_*} \right)^{1/2} \frac{M_{\text{Pl}}}{T^2}, \tag{2.112}$$

with $M_{\text{Pl}}$ the Planck mass in natural units. This will be useful in the discussion of Boltzmann equation.

### 2.4.2 Boltzmann equation

Following closely [40] and [41] we now turn to the derivation of Boltzmann equation describing the out-of equilibrium processes in the early Universe.

For an affine parameter $\tau$ we have

$$d\tau^2 = g_{\mu\nu} dx^\mu dx^\nu, \tag{2.113}$$

such that the four-velocity $u^\mu$ has norm 1:

$$u^\mu = \frac{dx^\mu}{d\tau}$$
$$\Rightarrow g_{\mu\nu} u^\mu u^\nu = 1. \tag{2.114}$$

It is convenient to reparametrize $\tau$ [39] as $\lambda = \tau/m$ where $m$ is the particle mass in a way that

$$m^2 d\lambda^2 = g_{\mu\nu} dx^\mu dx^\nu \tag{2.115}$$

and that the four-momentum $p^\mu$ is

$$
\begin{aligned}
p^\mu &= m\frac{dx^\mu}{d\tau} \\
&= \frac{dx^\mu}{d\lambda}.
\end{aligned}
\tag{2.116}
$$

In this case the geodesics equation (2.68) becomes

$$
\frac{dp^\mu}{d\lambda} + \Gamma^\mu_{\rho\sigma}p^\rho p^\sigma = 0.
\tag{2.117}
$$

From the FRW metric (2.71) we can obtain the evolution equations for $p^\mu$:

$$
\frac{dp^0}{d\lambda} = -\mathbf{p}^2 R\dot{R}
\tag{2.118}
$$

$$
\frac{dp^i}{d\lambda} = -2\frac{\dot{R}}{R}p^0 p^i.
\tag{2.119}
$$

To get the evolution of some scalar density function $f$ with variables $x(\lambda)$ and $p(\lambda)$, we note that

$$
\begin{aligned}
\frac{df}{d\lambda} &= \frac{dx^\mu}{d\lambda}\frac{\partial f}{\partial x^\mu} + \frac{dp^\mu}{d\lambda}\frac{\partial f}{\partial p^\mu} \\
&= p^0\frac{\partial f}{\partial t} + \mathbf{p}\cdot\nabla f - 2\frac{\dot{R}}{R}p^0 p^i\frac{\partial f}{\partial p^i} - \mathbf{p}^2 R\dot{R}\frac{\partial f}{\partial p^0}.
\end{aligned}
\tag{2.120}
$$

The variation in $x^i$ can be dropped on the basis of the FRW metric isotropy. Also, we are interested in the variation on mass-shell for a spatially flat metric. In this way we consider the quantity

$$
\tilde{f}(p,t) = \int f(p^0,\mathbf{p},t)\delta\left(p_0 - (\mathbf{p}^2 R^2 + m^2)^{\frac{1}{2}}\right)dp^0.
\tag{2.121}
$$

Then, integrating (2.120), we have

$$
\begin{aligned}
\int \frac{df}{d\lambda}\frac{1}{p^0}&\delta\left(p_0 - (\mathbf{p}^2 R^2 + m^2)^{\frac{1}{2}}\right)dp^0 \\
&= \int \frac{\partial f}{\partial t}\delta\left(p_0 - (\mathbf{p}^2 R^2 + m^2)^{\frac{1}{2}}\right)dp^0 \\
&\quad - 2\frac{\dot{R}}{R}p^i\frac{\partial \tilde{f}}{\partial p^i} \\
&\quad + \frac{\mathbf{p}^2}{p^0}R\dot{R}\frac{\partial f}{\partial p^0}\bigg|_{p^0 = (\mathbf{p}^2 R^2 + m^2)^{1/2}}.
\end{aligned}
\tag{2.122}
$$

Observe that in obtaining this formula we must differentiate the delta function with respect to $p^i$ (see below). That is why the last sign is symmetric from (2.120). Differentiating

(2.121) with respect to $t$

$$\frac{\partial \tilde{f}(p,t)}{\partial t} = \int \frac{\partial f(p^0, \mathbf{p}, t)}{\partial t} \delta \left( p_0 - (\mathbf{p}^2 R^2 + m^2)^{\frac{1}{2}} \right) dp^0$$
$$+ \int f(p^0, \mathbf{p}, t) \frac{\partial}{\partial t} \delta \left( p_0 - (\mathbf{p}^2 R^2 + m^2)^{\frac{1}{2}} \right) dp^0. \qquad (2.123)$$

Noting that

$$\int f(p^0, \mathbf{p}, t) \frac{\partial}{\partial t} \delta \left( p_0 - (\mathbf{p}^2 R^2 + m^2)^{\frac{1}{2}} \right) dp^0$$
$$= \int f(p^0, \mathbf{p}, t) \frac{\partial}{\partial t} \frac{1}{2\pi} e^{ix(p^0 - (\mathbf{p}^2 R^2 + m^2)^{\frac{1}{2}})} dx dp^0$$
$$= -\int f(p^0, \mathbf{p}, t) R\dot{R} \frac{\mathbf{p}^2}{(\mathbf{p}^2 R^2 + m^2)^{\frac{1}{2}}} \frac{ix}{2\pi} e^{ix(p^0 - (\mathbf{p}^2 R^2 + m^2)^{\frac{1}{2}})} dp^0 dx$$
$$= R\dot{R} \int \frac{\mathbf{p}^2}{(\mathbf{p}^2 R^2 + m^2)^{\frac{1}{2}}} \frac{\partial}{\partial p^0} f(p^0, \mathbf{p}, t) \delta \left( p_0 - (\mathbf{p}^2 R^2 + m^2)^{\frac{1}{2}} \right) dp^0$$
$$= \frac{\mathbf{p}^2}{p^0} R\dot{R} \frac{\partial f}{\partial p^0} \bigg|_{p^0 = (\mathbf{p}^2 R^2 + m^2)^{1/2}} \qquad (2.124)$$

we see that

$$\frac{\partial \tilde{f}(p,t)}{\partial t} = \int \frac{\partial f(p^0, \mathbf{p}, t)}{\partial t} \delta \left( p_0 - (\mathbf{p}^2 R^2 + m^2)^{\frac{1}{2}} \right) dp^0$$
$$+ \frac{\mathbf{p}^2}{p^0} R\dot{R} \frac{\partial f}{\partial p^0} \bigg|_{p^0 = (\mathbf{p}^2 R^2 + m^2)^{1/2}} \qquad (2.125)$$

which allows to write (2.122) as

$$\int \frac{df}{d\lambda} \frac{1}{p^0} \delta \left( p_0 - (\mathbf{p}^2 R^2 + m^2)^{\frac{1}{2}} \right) dp^0 = \frac{\partial \tilde{f}(p,t)}{\partial t} - 2\frac{\dot{R}}{R} p^i \frac{\partial \tilde{f}}{\partial p^i}$$

Dropping the tildes, using the isotropy of FRW metric and putting $p = |\mathbf{p}|$ we can define the operator

$$L(f) = \frac{\partial f(p,t)}{\partial t} - 2\frac{\dot{R}}{R} p \frac{\partial f}{\partial p}. \qquad (2.126)$$

To see that this is the pertinent operator in studying processes with a variable number of particles, we start by defining the local momentum [41] $\overline{\mathbf{p}} = R\mathbf{p}$. Using the zero component of (2.116) it is easy to show that

$$\frac{d\overline{p}^i}{d\lambda} = -\frac{\dot{R}}{R} \overline{p}^0 \overline{p}^i, \qquad (2.127)$$

and repeating the computation that led to (2.126) gives the operator $L(f)$ in terms of

local momentum:

$$L(f) = \frac{\partial f(\overline{p}, t)}{\partial t} - \frac{\dot{R}}{R}\overline{p}_i\frac{\partial f}{\partial \overline{p}_i}. \tag{2.128}$$

From now on we will drop the bars and consider only local momentum. The particle current density is

$$N^\mu = g \int \frac{d^3p}{(2\pi)^3} f \frac{p^\mu}{p^0} \tag{2.129}$$

where $f$ is the momentum distribution function and $g$ the effective degrees of freedom. Because of spatial symmetry only the zero component does not vanish, which gives the particle number density

$$n \equiv N^0 = g \int \frac{d^3p}{(2\pi)^3} f. \tag{2.130}$$

As it is well known the covariant divergence of a four-vector is

$$A^\mu_{;\mu} = \frac{1}{g^{1/2}} \frac{\partial}{\partial x^\mu}(g^{1/2} A^\mu), \tag{2.131}$$

where here and only here $g$ denotes the metric determinant. So, for the flat FRW metric,

$$\begin{aligned} N^\mu_{;\mu} &= \frac{g}{R^3}\frac{\partial}{\partial t}\left(R^3 \int f\frac{d^3p}{(2\pi)^3}\right) \\ &= 3g\frac{\dot{R}}{R}\int f\frac{d^3p}{(2\pi)^3} + g\int\frac{\partial f}{\partial t}\frac{d^3p}{(2\pi)^3} \\ &= 3g\frac{\dot{R}}{R}\int f\frac{d^3p}{(2\pi)^3} + g\frac{\dot{R}}{R}\int p_i\frac{\partial}{\partial p_i}f\frac{d^3p}{(2\pi)^3} + g\int L(f)\frac{d^3p}{(2\pi)^3} \end{aligned} \tag{2.132}$$

where we used (2.128). Integrating by parts cancels the first and the second terms, so we end up with

$$N^\mu_{;\mu} = g \int L(f)\frac{d^3p}{(2\pi)^3} = \frac{g}{R^3}\frac{\partial}{\partial t}\left(R^3 \int f\frac{d^3p}{(2\pi)^3}\right). \tag{2.133}$$

The operator $L$ is the relativistic Liouville operator and a necessary and sufficient condition for the conservation of the number of particles is $L(f) = 0$. Boltzmann equation can be written in a generic way as

$$L(f) = C(E)/E, \tag{2.134}$$

where $C(E)$ is the collision term and encodes all the specificities of the processes that change the number of particles. The evolution of the particle density can then be given in the more familiar form

$$\frac{dn}{dt} + 3\frac{\dot{R}}{R}n = g\int\frac{d^3p}{(2\pi)^3E}C(E). \tag{2.135}$$

In particular, $C(E)$ contains the scattering matrix for some specific process. Consider as an example the process that destroys the particles $i, j$ and creates the particles $k, l$. Then the amplitude $A$ for this process must be proportional to the creation operators of $k, l$ and anihilation operators of $i, j$

$$A \propto a_k^\dagger a_l^\dagger a_i a_j. \tag{2.136}$$

On the other hand, the normalization of creation and annihilation operators is

$$a_k^\dagger \left| n_1 n_2 \ldots n_k \ldots \right\rangle = \sqrt{1 \pm n_k} \left| n_1 n_2 \ldots n_k + 1 \ldots \right\rangle \tag{2.137}$$

$$a_i \left| n_1 n_2 \ldots n_i \ldots \right\rangle = \sqrt{n_i} \left| n_1 n_2 \ldots n_i - 1 \ldots \right\rangle. \tag{2.138}$$

To compute the $T$ matrix, which is the nontrivial scattering matrix after factoring out the conservation delta functions, we must take the square of the amplitude and this means that for this process

$$T \propto \left| \langle kl | A | ij \rangle \right|^2 \propto (1 \pm n_k)(1 \pm n_l) n_i n_j. \tag{2.139}$$

Since we are integrating over the phase-space density what counts is in fact the distribution densities $f_i, f_j, \ldots$ related with the number density by (2.130). Following [40] we can write the Boltzmann equation for the specie $i$

$$\frac{dn_i}{dt} + 3Hn_i = -g_i \int \frac{d^3 p_i}{(2\pi)^3 E_i} d\Pi_j d\Pi_k d\Pi_l \times (2\pi)^4 \delta \left( p_k + p_l - p_i - p_j \right)$$
$$\times \left( |A|_{i+j \to k+l}^2 f_i f_j (1 \pm f_k)(1 \pm f_l) - |A|_{k+l \to i+j}^2 f_k f_l (1 \pm f_i)(1 \pm f_j) \right). \tag{2.140}$$

where $d\Pi$ is the relativistic phase space density

$$d\Pi = \frac{g}{(2\pi)^3} \frac{d^3 p}{2p^0}. \tag{2.141}$$

The minus sign in (2.140) reflects the fact that if the forward scattering prevails with respect to the inverse one, then obviously $n_i$ will decrease. Of course, identical equations must be introduced to take into account the evolution of the other species. This means that often we have a set of coupled Boltzmann equations that can only be solved numerically. But in some limit cases it is possible to assume that only one specie departs significantly from equilibrium and approximate analytical solutions are possible. When the $i$ specie is heavy and interacts with some relativistic particles, it is possible to obtain an approximate expression for the Boltzmann equation that occurs often in the literature, first stated without proof in [44] and derived in [42]. We will follow closely this last reference. Assuming a two body inelastic scattering between two heavy fermion $L, \overline{L}$ and two light ones $l, \overline{l}$

$$l(p_1) + \overline{l}(p_2) \leftrightarrow L(p_1^{'}) + \overline{L}(p_2^{'}) \tag{2.142}$$

with Fermi-Dirac distribution functions $f$ and $g$, Boltzmann equation assumes the form

$$\frac{dn}{dt} + 3Hn = 2 \int \frac{d^3 p_1^{'}}{(2\pi)^3 E_1^{'}} d\Pi_1 d\Pi_2 d\Pi_2^{'} \times (2\pi)^4 \delta \left( p_1 + p_2 - p_1^{'} - p_2^{'} \right)$$
$$\times |A|^2 \left( g(p_1, t) g(p_2, t)(1 - f(p_1^{'}, t))(1 - f(p_2^{'}, t)) \right.$$
$$\left. - f(p_1^{'}, t) f(p_2^{'}, t)(1 - g(p_1, t))(1 - g(p_2, t)) \right). \tag{2.143}$$

with the CP-invariance hypothesis. Here $n$ is the particle number density of $L$ and $\overline{L}$. The light fermions are kept in thermal equilibrium by other interactions that are very fast in comparison with the expansion rate so that its distribution function is

$$g(p, t) = \frac{1}{\exp(p^0 / T(t)) + 1} \tag{2.144}$$

with $p^0 = |\mathbf{p}|$. As for the heavy fermion, the distribution function can be approximated [42] by

$$f(p^{'}, t) = \frac{1}{\exp\left(\alpha(t) + E(p^{'}) / T(t)\right) + 1}, \tag{2.145}$$

where $\alpha(t)$ has the meaning of a chemical potential. With (2.144) and (2.145) note that

$$(1 - f(p_1^{'}, t))(1 - f(p_2^{'}, t)) = \exp(2\alpha(t)) \exp\left((E_1^{'} + E_2^{'}) / T(t)\right) f(p_1^{'}, t) f(p_2^{'}, t). \tag{2.146}$$

Energy conservation implies that

$$E_1^{'} + E_2^{'} = E_1 + E_2 \tag{2.147}$$

and this allows to write

$$(1 - f(p_1^{'}, t))(1 - f(p_2^{'}, t)) g(p_1, t) g(p_2, t) =$$
$$\exp(2\alpha(t))(1 - g(p_1, t))(1 - g(p_2, t)) f(p_1^{'}, t) f(p_2^{'}, t). \tag{2.148}$$

In this way Boltzmann equation (2.143) becomes

$$\frac{dn}{dt} + 3Hn = 2 \int \frac{d^3 p_1^{'}}{(2\pi)^3 E_1^{'}} d\Pi_1 d\Pi_2 d\Pi_2^{'} \times (2\pi)^4 \delta \left( p_1 + p_2 - p_1^{'} - p_2^{'} \right)$$
$$\times |A|^2 f(p_1^{'}, t) f(p_2^{'}, t)(1 - g(p_1, t))(1 - g(p_2, t)) \left( \exp(2\alpha(t)) - 1 \right) \tag{2.149}$$

In [42] it is argued that the chemical potential $\alpha(t)$ only departs significantly from zero for temperatures well below the mass of the heavy lepton. In this case the Fermi-Dirac distribution can be approximated by the Maxwell-Boltzmann distribution

$$f(p,t) = \exp\left(-\alpha(t) - \frac{E(p)}{T}\right).$$ (2.150)

This means that the actual density $n(t)$ is related with the equilibrium one $n_0$ by

$$n(t) = e^{-\alpha(t)} n_0$$ (2.151)

with

$$n_0 = 2 \int \frac{d^3 p}{(2\pi)^3} e^{-E(p)/T}.$$ (2.152)

Replacing (2.150) in (2.149) and remembering (2.151) we obtain the promised result

$$\frac{dn}{dt} + 3Hn = \langle \sigma v \rangle \left(n_0^2 - n(t)^2\right)$$ (2.153)

with the thermally averaged cross-section identified as

$$\langle \sigma v \rangle = \frac{1}{n_0^2} \int d\Pi_1 d\Pi_2 d\Pi_1' d\Pi_2' \times (2\pi)^4 \delta\left(p_1 + p_2 - p_1' - p_2'\right)$$
$$\times 2e^{-E(p_1')/T} 2e^{-E(p_2')/T} (1 - g(p_1,t))(1 - g(p_2,t)) |A|^2.$$ (2.154)

On an isentropic Universe the entropy density will decrease by a factor $R^{-3}$. This can be used to scale out the Universe expansion effect in Boltzmann equation by defining the quantity

$$Y = \frac{n}{s}.$$ (2.155)

A trivial computation shows that

$$\dot{Y} = \frac{\dot{n}}{s} + 3H\frac{n}{s}$$ (2.156)

and Boltzmann equation in terms of $Y$ is

$$s\dot{Y} = \langle \sigma v \rangle \left(Y_0^2 - Y^2\right)$$ (2.157)

where now $\langle \sigma v \rangle$ is defined with $Y_0$. Usually the evolution of $Y$ is given as a function of the temperature instead of time. From (2.110), (2.112) and the reduced variable $x = m/T$, where $m$ is the mass of the particle in study, we get finally

$$sHx\frac{dY}{dx} = \langle \sigma v \rangle \left(Y_0^2 - Y^2\right).$$ (2.158)

The Boltzmann equation in this form will be used in Chapter 3. With a little rearrangement it can be used to set a rough estimate for a specie to freeze-out. If the interaction rate is much smaller than the expansion rate determined by the Hubble parameter

$$\Gamma = n_0 \langle \sigma v \rangle \ll H, \tag{2.159}$$

where $\langle \sigma v \rangle$ is again given by (2.154), then the relative density $Y$ may stabilize in a value different from the one given by the equilibrium density. In a scattering this just means that the expansion of the Universe makes more and more difficult the bath particles to meet and interact. On a decay it simply says that the particle's half-life is bigger than the age of the Universe.

Relativistic particles contribute the most to the entropy density, so that in a very good approximation

$$s = \frac{2\pi^2}{45} g_{*S} T^3 \tag{2.160}$$

with

$$g_{*S} = \sum_{i=bosons} g_i \left( \frac{T_i}{T} \right)^3 + \frac{7}{8} \sum_{i=fermions} g_i \left( \frac{T_i}{T} \right)^3 . \tag{2.161}$$

When all the Standard Model degrees of freedom are active and the particles are in thermal equilibrium, then $g_{*S} = 106.75$. On the other hand the equilibrium relativistic particle density can be obtained by integration of (2.130) with (2.107) and $E = |\mathbf{p}|$. The result is

$$n = \begin{cases} \frac{\zeta(3)g}{\pi^2} T^3 & \text{bosons} \\ \frac{3\zeta(3)g}{4\pi^2} T^3 & \text{fermions} \end{cases} \tag{2.162}$$

with $\zeta$ the Riemann's zeta function. Then the equilibrium relative density $Y_{eq}$ for relativistic particles will be

$$Y_{eq} = \frac{45\zeta(3)}{2\pi^4} \frac{g_{eff}}{g_{*S}} = 0.278 \left( \frac{g_{eff}}{g_{*S}} \right) \tag{2.163}$$

where $g_{eff} = g$ for bosons and $g_{eff} = 3g/4$ for fermions. This can be used to obtain a cosmological bound on neutrino masses. The present value for the entropy density is [40]

$$s_0 = 2.97 \times 10^3 \text{ cm}^{-3}. \tag{2.164}$$

Neutrinos decouple when they are still relativistic, roughly at temperatures of few MeV [40] when $g_{*S} = 10.75$, so that their relic density is related with $Y_{eq}$ by

$$n_0 = s_0 Y_{eq} = 825 \left( \frac{3}{2g_{*S}} \right) = 115.12 \text{ cm}^{-3} \tag{2.165}$$

and then

$$\Omega_\nu = \frac{m_\nu n_0}{\rho_c} = m_\nu \frac{115.12}{10.54h^2} \times 10^{-3} \tag{2.166}$$

which gives

$$\Omega_\nu h^2 = \frac{m_\nu}{91.5} \text{ eV} \tag{2.167}$$

for just one family of a 2-component neutrino. The value quoted at PDG [43] for the relative matter density is $\Omega_m h^2 = 0.133 \pm 0.006$. Assuming 3 neutrino families we get the cosmological upper bound

$$\sum_{i=1}^{3} m_{\nu i} < 12.2 \text{ eV}. \tag{2.168}$$

### 2.4.3  Dark Matter

If the particle decouples from the thermal bath when it is non-relativistic the precise determination of $Y_\infty$ is more difficult. Following [40] we start by noting that $\sigma v$ can be expanded in powers of $v$

$$\sigma v \propto v^p, \tag{2.169}$$

with $p = 0$ for s-wave annihilation, $p = 2$ for p-wave annihilation, etc... Because $v^2 \propto T$ we have that

$$\langle \sigma v \rangle \equiv \sigma_0 x^{-n}, \tag{2.170}$$

for $n = p/2$, and Boltzmann equation (2.153) expressed in terms of $Y$ becomes

$$\frac{dY}{dx} = -\lambda x^{-n-2} \left( Y^2 - Y_{eq}^2 \right) \tag{2.171}$$

with

$$\lambda = \left( \frac{\langle \sigma v \rangle s(x) m^3}{1.67 g_*^{1/2} m^2 / M_{Pl}} \right)_{x=1} = 0.264 \left( g_{*S} / g_*^{1/2} \right) M_{Pl} m \sigma_0 \tag{2.172}$$

and

$$Y_{eq}(x) = \frac{n_{eq}(x)}{s(x)} = \frac{45}{2\pi^4} \left( \frac{\pi}{8} \right)^{1/2} \frac{g}{g_{*S}} x^{3/2} e^{-x}$$
$$= 0.145 \frac{g}{g_{*S}} x^{3/2} e^{-x}. \tag{2.173}$$

This differential equation can be solved approximately. Start by defining the quantity

$$\Delta = Y - Y_{eq} \tag{2.174}$$

and consider the corresponding differential equation

$$\frac{d\Delta}{dx} = -\frac{dY_{eq}}{dx} - \lambda x^{-n-2} \Delta \left( 2Y_{eq} + \Delta \right). \tag{2.175}$$

At early times, for $1 < x \ll x_f$ where $x_f$ is relative to the decoupling point, $Y$ follows closely $Y_{eq}$ so that $\Delta$ and $\frac{d\Delta}{dx}$ can be safely ignored. In this way

$$\Delta = -\lambda^{-1} x^{n+2} \frac{dY_{eq}}{dx} (2Y_{eq})^{-1}. \tag{2.176}$$

Differentiating (2.173) we get

$$\Delta \simeq \frac{1}{2\lambda} x^{n+2}. \tag{2.177}$$

At late times, for $x \gg x_f$, $Y$ is very different from $Y_{eq}$ and we can take $\Delta \simeq Y \gg Y_{eq}$. Then

$$\frac{d\Delta}{dx} = -\lambda x^{-n-2} \Delta^2 \tag{2.178}$$

which by integration gives

$$\frac{1}{\Delta_\infty} - \frac{1}{\Delta(x_f)} = \frac{\lambda}{n+1} \frac{1}{x^{n+1}}. \tag{2.179}$$

Assuming that $\Delta(x_f) - \Delta_\infty \sim \Delta(x_f)$ we get

$$\Delta_\infty \sim \frac{n+1}{\lambda} x_f^{n+1}. \tag{2.180}$$

Now we must determine the decoupling point $x_f$ that is when $Y$ starts to deviate from $Y_{eq}$. We choose the criteria $\Delta(x_f) = cY_{eq}(x_f)$ where $c$ is a constant of order 1. Then solving the general equation for $\Delta$ (2.175) with the freeze-out criteria we get the approximate transcendental equation

$$\lambda c(2+c)ae^{-x_f} = (n + \frac{1}{2})\ln x_f \tag{2.181}$$

with $a = 0.145 \, (g/g_{*S})$ that can be solved by iteration. The first two iterates give

$$x_f = \ln\left((2+c)\,\lambda ac\right) - (n + \frac{1}{2})\ln\left(\ln\left((2+c)\,\lambda ac\right)\right). \tag{2.182}$$

The choice $c(c+2) = n+1$ gives the best fit to the numerical integration [40]. With this choice we have

$$x_f = \ln\left(0.038\,(n+1)\left(g/g_*^{1/2}\right)M_{Pl}m\sigma_0\right), \tag{2.183}$$

$$Y_\infty = \frac{3.79(n+1)x_f^{n+1}}{(g_{*S}/g_*^{1/2})M_{Pl}m\sigma_0}. \tag{2.184}$$

Now it is easy to get the number and relative densities for a cold relic:

$$n_\chi = s_0 Y_\infty = 1.13 \times 10^4 \frac{(n+1)x_f^{n+1}}{(g_{*S}/g_*^{1/2})M_{Pl}m\sigma_0} \text{ cm}^{-3} \qquad (2.185)$$

$$\Omega_\chi h^2 = \frac{m_\chi n}{\rho_c} = \frac{1.07 \times 10^9 \text{ GeV}^{-1}(n+1)x_f}{(g_{*S}/g_*^{1/2})M_{Pl}\sigma_0}. \qquad (2.186)$$

To get an order of magnitude for the mass of a cold relic let's assume that the interaction strength is of the order of the electroweak coupling and on dimensional grounds we may suppose that

$$\sigma_0 \sim \frac{\alpha^2}{m_\chi^2}. \qquad (2.187)$$

For s-wave and p-wave annihilations and with $\Omega_\chi < 1$ we obtain the order of magnitude

$$m_\chi \sim 1 \text{ TeV}. \qquad (2.188)$$

That is why in the search for Dark Matter candidates, WIMPs (Weakly Interating Massive Particles) are well motivated. In this respect mSugra provides an attractive framework for these searches, not only because in most regions of parameter space the LSP (Lightest Supersymmetric Particle) has a mass on the hundreds of GeVs range but also because there are only 5 independent parameters

$$\tan\beta, \ |\mu|, \ m_0, \ m_{1/2}, \ A_0 \qquad (2.189)$$

from which the mass spectrum depends and this makes the phenomenological study simpler.

Usually in most regions of parameter space the LSP is the neutralino and we will focus on this possibility. The dominant channels for the neutralino annihilation or co-annihilation depend on its nature. As explained in section 1.1.2 the neutralino mass matrix depends on the gaugino masses $M_1$ and $M_2$, that are obtained at the electroweak scale by the running of the RGEs with the boundary value $m_{1/2}$ at the GUT scale and also on the $\mu$ parameter which is fixed by the electroweak breaking condition (1.40) and that depends on $m_1$ and $m_2$. These in turn are the solutions at the electroweak scale of the corresponding RGEs, so they are in fact functions of $m_0$, and also of $m_{1/2}$, since these equations are coupled with the gauge part. This means that what determines the predominant neutralino component are the parameters $m_0$, $m_{1/2}$ and $\tan\beta$. On large regions of mSugra parameter space the $\mu$ parameter is high, which means that the lightest neutralino will have a big gaugino component. But for $\tan\beta$ low (roughly of order 10 or below) and low to moderate values of $m_{1/2}$ the RGEs are relatively insensitive to the value of $m_0$, in the sense that for a wide range of initial values for $m_0$, these equations give virtually the same negative value for $m_2^2$, small in absolute value. This is known as the focus point. Since in this region the value of $m_2^2$ is small in absolute value, the $\mu$ parameter obtained by the electroweak breaking condition (1.40) can also be small and the lightest neutralino for this region has a big higgsino component. In this case the relevant couplings are between neutralinos and the $Z$ boson and neutralino, charginos and the $W$ boson. The annihilation proceeds through s and t channels with the exchange of another neutralino or a chargino. Since the mass differences between $\chi_1^0$ and $\chi_2^0$ and also between $\chi_1^0$ and $\chi_1^+$ are small, co-annihilation must also be taken into account. These processes can be efficient enough to reduce the neutralino density down to values compatible with $\Omega_{DM} = 0.105 \pm 0.008$ [43].

In the case where $\chi_1^0$ is mainly a bino, then the only gauge coupling is by $U(1)_Y$ and the annihilation cross-section will be suppressed by a factor $\tan^4 \theta_W$. However, for high values of $m_{1/2}$ roughly of order 1 TeV and low values of $m_0$, the sleptons are relatively light and co-annihilation between the lightest $\tilde{\tau}$ and $\chi_1^0$ is very active and can also reduce the neutralino density to acceptable values.

For high values of $\tan \beta$ there is another process that should be taken into account. For these values of $\tan \beta$ the CP-odd Higgs boson mass is lowered to a few hundreds of GeVs. The amplitude $A$ for the process in figure 2.3 goes as

$$A \sim \frac{1}{4 - (m_A/m_\chi)^2 + i\Gamma_A m_A/m_\chi^2}, \tag{2.190}$$

where $\Gamma_A$ is the decay width of the $A$ scalar, and it is resonant for the condition

$$m_\chi = \frac{m_A}{2}, \tag{2.191}$$

which can be fulfilled in some regions of the $(m_0, m_{1/2})$ plane and makes the annihilation of neutralinos very effective.

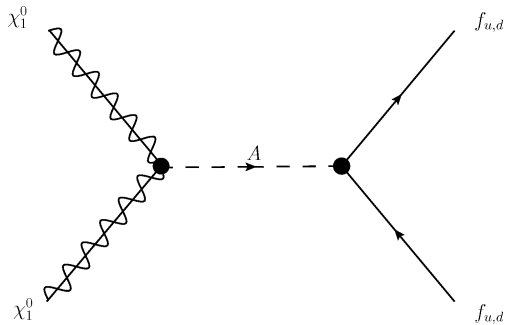

**Figure 2.3:** Higgs Funnel

Some explicit examples of these processes will be given in Chapter 6.

### 2.4.4  The Majoron as a Dark Matter candidate

In [45] the interplay between spontaneous breaking of lepton number and neutrinos masses was fully addressed. We review here the discussion for completeness. The model consists of an $SU(2) \times U(1)$ gauge theory under which there are $n$ species of neutrinos that are doublets for this gauge group and another $n$ species which are singlets. These neutrinos are massive. We assume the most general situation where the masses are given by a mixed Type-I/Type-II seesaw mechanism and where the heavy neutrino mass matrix is obtained by spontaneous breaking of lepton number. In this way the model contains 2 Higgs fields, a $SU(2)$ triplet $T$ with vev $u$ and a doublet $h$ with vev $v$. Additionally it contains a scalar field $\phi$ with vev $\sigma$ that it is a singlet under this gauge group and that carries lepton number $l = -2$. The vev of this field is responsible for the heavy neutrino mass matrix. The full neutrino mass matrix is

$$M = \begin{pmatrix} M_{II} & m_D \\ m_D^T & M_R \end{pmatrix} \tag{2.192}$$

and the corresponding mass term in the lagrangian is

$$\mathcal{L}_{\text{mass}} = -\frac{1}{2}\rho^T i\sigma_2 M\rho + \text{ h.c..} \tag{2.193}$$

Here $\rho$ refers to a $2n$ column vector of Weyl spinors, where the first $n$ are $SU(2)$ doublets and the last $n$ are singlets. The term $M_{II}$ is typically generated by a Type-II seesaw mechanism, whereas $m_D$ refers to Dirac terms. The physical fields $\nu$ are obtained by a diagonalization of the mass matrix through a unitary matrix such that

$$U^T M U = \begin{pmatrix} m & 0 \\ 0 & m_R \end{pmatrix} = \text{real positive, diagonal,} \tag{2.194}$$

$$\rho = U\nu. \tag{2.195}$$

The full diagonalization procedure has been described in the *op. cit.* In the process the expansion parameter

$$\epsilon = O\left(\frac{m_D}{M_R}\right) \tag{2.196}$$

is of relevance. For Type-I seesaw $M_R$ can be at most at the energy scale of $10^{15}$ GeV and because $m_D$ is of the order of the electroweak scale, $\epsilon$ can be as low as $10^{-13}$. In diagonalizing (2.192) we make the *ansatz*

$$U = (\exp iH)V,$$

$$H = \begin{pmatrix} 0 & S \\ S^\dagger & 0 \end{pmatrix}, \quad V = \begin{pmatrix} V_1 & 0 \\ 0 & V_2 \end{pmatrix}, \tag{2.197}$$

where $S$ is of order $\epsilon$ and $V_1$ and $V_2$ are unitary and of order 1. Replacing (2.197) and (2.192) in (2.194) gives

$$S = -im_D^* \, (M_R^*)^{-1} \tag{2.198}$$

and

$$V_1^T \left( -m_D M_R^{-1} m_D^T + M_{II} \right) V_1 = m = \text{real positive, diagonal}, \tag{2.199}$$

$$V_2^T \left( M_R + \frac{1}{2}(M_R^*)^{-1} m_D^\dagger m_D + \frac{1}{2} m_D^T m_D^* (M_R^*)^{-1} \right) V_2 = m_R = \text{real positive, diagonal} \tag{2.200}$$

The transformation matrix $U$ is then

$$\begin{aligned}
U &= \begin{pmatrix} U_a & U_b \\ U_c & U_d \end{pmatrix} \\
&= \begin{pmatrix} \left(1 - \frac{1}{2} m_D^* (M_R^*)^{-1} M_R^{-1} m_D^T \right) V_1 & m_D^* (M_R^*)^{-1} V_2 \\ -M_R^{-1} m_D^T V_1 & \left(1 - \frac{1}{2} M_R^{-1} m_D^T m_D^* (M_R^*)^{-1}\right) V_2 \end{pmatrix} \\
&\quad + O(\epsilon^2).
\end{aligned} \tag{2.201}$$

We now turn to the discussion of the Yukawa couplings between the neutrinos and the scalar fields. As the neutrino masses are obtained by a Type-I/Type-II seesaw mechanism and by the spontaneous breaking of lepton number, the Yukawa couplings are just the full neutrino mass matrix divided by the corresponding vevs:

$$\mathcal{L}_{\text{int}} = -\frac{1}{2} \rho^T i \sigma_2 \begin{pmatrix} M_{II} \frac{T^0}{u} & m_D \frac{h^0}{v} \\ m_D^T \frac{h^0}{v} & M_R \frac{\phi^*}{\sigma} \end{pmatrix} \rho + \text{ h.c.} \tag{2.202}$$

Due to the breaking of a global lepton number symmetry the physical model will contain a Goldstone boson, the Majoron. In order to identity the coupling of the Majoron with the light neutrinos we must first address the question of the minimization of the Higgs potential and identify the massless components of the Higgs fields. To do that we note that the scalar potential is invariant under the hyper-charge and lepton number gauge transformations. For example, in the first case we have the infinitesimal transformations

$$\delta(T_r) = -2\alpha_Y(T_i)$$
$$\delta(T_i) = 2\alpha_Y(T_r) \tag{2.203}$$

where $r, i$ refer to the real and imaginary components, $\alpha_Y$ is the abelian gauge parameter and noting that $Y(T) = 2$. There is a similar expression for $h$ with $Y(h) = 1$. Recall that

$\phi$ is a singlet under $SU(2)$. In this way, for hypercharge, we get

$$\frac{\partial V}{\partial \alpha_Y} = 2\frac{\partial V}{\partial T_r}T_i - 2\frac{\partial V}{\partial T_i}T_r + \frac{\partial V}{\partial h_r}h_i - \frac{\partial V}{\partial h_i}h_r = 0 \qquad (2.204)$$

and for lepton number we have

$$\frac{\partial V}{\partial \alpha_l} = \frac{\partial V}{\partial \phi_r}\phi_i - \frac{\partial V}{\partial \phi_i}\phi_r + \frac{\partial V}{\partial T_r^0}T_i^0 - \frac{\partial V}{\partial T_i^0}T_r^0 = 0 \qquad (2.205)$$

assuming that the triplet has $l = -2$. Differentiating this second relation with respect to $\phi_i$ and taking the vacuum expectation value we get, with the hypothesis of non spontaneous breaking of CP,

$$\left\langle \frac{\partial V}{\partial \phi_r} \right\rangle - \langle \phi_r \rangle \left\langle \frac{\partial^2 V}{\partial \phi_i^2} \right\rangle - \langle T_r^0 \rangle \left\langle \frac{\partial^2 V}{\partial \phi_i \partial T_i^0} \right\rangle = 0 \qquad (2.206)$$

and using the minimization conditions we obtain

$$\sigma \left\langle \frac{\partial^2 V}{\partial \phi_i^2} \right\rangle + u \left\langle \frac{\partial^2 V}{\partial \phi_i \partial T_i^0} \right\rangle = 0. \qquad (2.207)$$

Proceeding in exactly the same way we can derive four more relations:

$$\sigma \left\langle \frac{\partial^2 V}{\partial T_i \partial \phi_i} \right\rangle + u \left\langle \frac{\partial^2 V}{\partial (T_i^0)^2} \right\rangle = 0$$
$$v \left\langle \frac{\partial^2 V}{\partial (h_i^0)^2} \right\rangle + 2u \left\langle \frac{\partial^2 V}{\partial h_i^0 \partial T_i^0} \right\rangle = 0$$
$$v \left\langle \frac{\partial^2 V}{\partial h_i^0 \partial T_i^0} \right\rangle + 2u \left\langle \frac{\partial^2 V}{\partial (T_i^0)^2} \right\rangle = 0$$
$$\sigma \left\langle \frac{\partial^2 V}{\partial \phi_i \partial h_i^0} \right\rangle + u \left\langle \frac{\partial^2 V}{\partial T_i^0 \partial h_i} \right\rangle = 0 \qquad (2.208)$$

These five equations are enough to express the 3 by 3 symmetric mass matrix of the imaginary components of scalar fields in terms of just one quantity, which can be chosen to be

$$a = \left\langle \frac{\partial^2 V}{\partial (h_i^0)^2} \right\rangle. \qquad (2.209)$$

This means that in the basis $T_i^0, h_i^0, \phi_i^0$ this matrix is

$$\begin{pmatrix} 1 & \frac{v}{2\sigma} & -\frac{v}{2u} \\ \frac{v}{2\sigma} & \frac{v^2}{4\sigma^2} & -\frac{v^2}{4\sigma u} \\ -\frac{v}{2u} & -\frac{v^2}{4\sigma u} & \frac{v^2}{4u^2} \end{pmatrix} a. \qquad (2.210)$$

The explicit computation of the determinant shows that there is a null eigenvalue. In fact this eigenvalue has double multiplicity which means that there are two Goldstone bosons: one should be identified with the Majoron and the other is absorbed by the $Z$ boson.

Since the $Z$ does not couple to the scalar $\phi$ the second Goldstone boson is proportional to

$$v h_i^0 + 2u T_i^0 \qquad (2.211)$$

while the Majoron is orthogonal to it:

$$J = N \left( -2vu^2 h_i^0 + \sigma(v^2 + 4u^2)\phi_i + uv^2 T_i^0 \right), \qquad (2.212)$$

with $N = (4v^2 u^4 + \sigma^2(v^2 + 4u^2)^2 + u^2 v^4)^{-1/2}$. We see that each of the fields $h^0, T^0$ and $\phi$ contains a piece of the Majoron:

$$\phi^* = -iN\sigma(v^2 + 4u^2)J + \ldots \qquad (2.213)$$

$$h^0 = -2iNvu^2 J + \ldots \qquad (2.214)$$

$$T^0 = iNuv^2 J + \ldots \qquad (2.215)$$

In this way, from (2.202), we see that neutrinos will couple to the Majoron. Specifically for the light neutrinos we have

$$\mathcal{L}_{\text{int}} = -\frac{1}{2} \nu^T i\sigma_2 \left( \frac{T^0}{u} U_a^T M_{II} U_a + \frac{h^0}{v} (U_a^T m_D U_c + U_c^T m_D^T U_a) + \frac{\phi^*}{\sigma} U_c^T M_R U_c \right) \nu + \text{ h.c.} \qquad (2.216)$$

Using the components (2.213), the expressions (2.201) and the definition of $\epsilon$ we find that

$$\mathcal{L}_{\nu J} = -\frac{1}{2} N J v^2 \nu^T i\sigma_2 V_1^T \left( M_{II} - m_D M_R^{-1} m_D^T \right) V_1 \nu + \text{ h.c.} + O(\epsilon^3) \qquad (2.217)$$

Note that in obtaining this result it is important to observe that

$$m_\nu \sim O\left( \frac{m_D^2}{M_R} \right),$$

$$M_{II} \sim O(m_\nu). \qquad (2.218)$$

Using (2.199) we see that the coupling is proportional to the neutrino masses.

As the Higgs doublet has some Majoron component, also charged leptons will couple to the Majoron. It can be seen that this coupling is

$$\mathcal{L}_{\psi J} = \mp \frac{2iu m_\psi}{v^2} J \overline{\psi} \gamma_5 \psi. \qquad (2.219)$$

with $m_\psi$ the fermion mass and $-(+)$ corresponds to a positively (negatively) charged fermion. There is also a one-loop induced coupling between the Majoron and photons that will be explicitly computed on section 4.5.2. This coupling is the most active in the Majoron decay since the Majoron may acquire a mass on the $keV$ range and also because the coupling with neutrinos is proportional to neutrino masses. The present density $n_J(t_0)$

of Majorons is

$$n_J(t_0) = n_J(t_D)e^{-t_0/\tau},$$
(2.220)

where $\tau$ is the mean half-life ant $n(t_D)$ is the density at the decoupling time $t_D$, when Majorons start to depart from thermal equilibrium with photons. Assuming constant total entropy $S$ from $t_D$ to $t_0$, the entropy density varies as $s \propto R^{-3}$. This means that $N = nR^3 \propto n/s$ and so

$$\frac{n_J(t_0)}{s(t_0)} = \frac{n_J(t_D)}{s(t_D)}e^{-t_0/\tau}$$
(2.221)

The photon density is related with the entropy density as

$$s = 1.80g_{*s}n_\gamma$$
(2.222)

which means that (2.221) can be rewritten as

$$\frac{n_J(t_0)}{n_\gamma(t_0)} = \frac{g_{*s}(t_0)}{g_{*s}(t_D)}\frac{n_J(t_D)}{n_\gamma(t_D)}e^{-t_0/\tau}$$
(2.223)

Assuming equal temperatures for the relativistic effective degrees of freedom and for $t_D \gtrsim 170$ GeV, we have $g_*(t_D) = 106.75$ and $g_*(t_0) = 3.91$. Also $n_\gamma(t_0) = 422$ cm$^{-3}$. When Majorons are in thermal equilibrium with photons, the ration $n_J/n_\gamma$ is equal to $1/2$. Putting all this together gives

$$\Omega_J h^2 = \frac{m_J}{1.36 \text{ keV}}e^{-t_0/\tau}$$
(2.224)

for the relic density. We have considered the simplest possibility, that Majorons where in thermal equilibrium when produced. If more complex scenarios are allowed, when can encode our ignorance on the production process by a parameter $\beta$ and write

$$\Omega_J h^2 = \beta \frac{m_J}{1.36 \text{ keV}}e^{-t_0/\tau}$$
(2.225)

Clearly the Majoron must be long-lived ($\tau > t_0$) if it is to explain the Dark Matter density.

# Chapter 3

# Leptogenesis in an $A_4$ Model

## 3.1 The Model

In [46] it was proposed a model for neutrino masses with mixed Type-I and Type-II seesaw, based on the $A_4$ symmetry group. The $A_4$ or alternating group is the rotational symmetry group of the tetrahedron or equivalently the group of even permutations of four objects. Even permutations are obtained from pairs of transpositions, so the generators of $A_4$ can be taken as $S = (21)(43)$ and $T = (312)$ which form a presentation of the group. Note that $(231)$ is conjugate of $(312)$ by the single transposition $(32)$ which means that these two elements belong to different classes in $A_4$, which are the inverse of each other. In summary, $A_4$ has four classes

$$
\begin{aligned}
&C_1 : I = (1234) \\
&C_2 : T = (2314), ST = (4132), TS = (3241), STS = (1423) \\
&C_3 : T^2 = (3124), ST^2 = (4213), T^2S = (2431), TST = (1342) \\
&C_4 : S = (4321), T^2ST = (3412), TST^2 = (2143)
\end{aligned}
\tag{3.1}
$$

Then it has four irreducible representations and a well-known relation for discrete groups

$$
N = \sum_i d_i^2,
\tag{3.2}
$$

where $N$ is the number of group elements and $d_i$ is the dimension of the i-th irreducible representation, shows that it has three one-dimensional and 1 three-dimensional representations. The characters are shown in table 3.1. The procedure to obtain the characters can be seen in [47]. Note that as expected $C_2$ and $C_3$, being inverse classes of each other, have characters that are complex conjugate. Note also the important relation

$$
1 + w + w^2 = 0.
\tag{3.3}
$$

**Table 3.1:** Characters of A4

| Class | $\chi^1$ | $\chi^{1'}$ | $\chi^{1''}$ | $\chi^3$ |
|-------|----------|-------------|--------------|----------|
| $C_1$ | 1 | 1 | 1 | 3 |
| $C_2$ | 1 | $\omega$ | $\omega^2$ | 0 |
| $C_3$ | 1 | $\omega^2$ | $\omega$ | 0 |
| $C_4$ | 1 | 1 | 1 | -1 |

|         | $L_1$ | $L_2$ | $L_3$ | $l_{Ri}$ | $\nu_{Ri}$ | $\Phi_i$ | $\Delta$ |
|---------|-------|-------|-------|----------|------------|----------|----------|
| $SU(2)$ | 2 | 2 | 2 | 1 | 1 | 2 | 3 |
| $U(1)$ | $-1$ | $-1$ | $-1$ | $-2$ | 0 | 1 | 2 |
| $A_4$ | 1 | $1'$ | $1''$ | 3 | 3 | 3 | $1'$ or $1''$ |

**Table 3.2:** Lepton multiplet structure of the model

The model of [46] is an extension of the Standard Model with right-handed neutrinos in the singlet representation of $SU(2)$ and with an extra Higgs boson in the triplet representation. In table 3.2 the quantum numbers for the leptons and the Higgs bosons in the model are indicated. The Higgs potential for the Higgs doublets has been calculated in [48] and we present it here for completeness. Since the Higgs doublets belong to the **3** dimensional representation of $A_4$, we can produce invariants from $\mathbf{3} \otimes \mathbf{3}$:

$$\mathbf{3} \otimes \mathbf{3} = \mathbf{1} \oplus \mathbf{1'} \oplus \mathbf{1''} \oplus \mathbf{3}_s \oplus \mathbf{3}_a \tag{3.4}$$

This means that in condensed form the Higgs potential can be written as

$$
\begin{aligned}
V(\Phi) = {} & m^2 \left(\Phi^\dagger \cdot \Phi\right)_1 + \lambda_1 \left(\Phi^\dagger \cdot \Phi\right)_1^2 + \lambda_2 \left(\Phi^\dagger \cdot \Phi\right)_{1'} \left(\Phi^\dagger \cdot \Phi\right)_{1''} \\
& + \lambda_3 \left(\Phi^\dagger \cdot \Phi\right)_{3_s} \left(\Phi^\dagger \cdot \Phi\right)_{3_s} + \lambda_4 \left(\Phi^\dagger \cdot \Phi\right)_{3_s} \left(\Phi^\dagger \cdot \Phi\right)_{3_a} + \lambda_5 \left(\Phi^\dagger \cdot \Phi\right)_{3_a} \left(\Phi^\dagger \cdot \Phi\right)_{3_a}
\end{aligned}
\tag{3.5}
$$

By a redefinition of the coupling constants we obtain the expression in [48]:

$$
\begin{aligned}
V = {} & m^2 \sum_i \Phi_i^\dagger \Phi_i + \frac{1}{2}\lambda_1 \left(\sum_i \Phi_i^\dagger \Phi_i\right)^2 \\
& + \lambda_2(\Phi_1^\dagger \Phi_1 + \omega^2 \Phi_2^\dagger \Phi_2 + \omega \Phi_3^\dagger \Phi_3)(\Phi_1^\dagger \Phi_1 + \omega \Phi_2^\dagger \Phi_2 + \omega^2 \Phi_3^\dagger \Phi_3) \\
& + \lambda_3[(\Phi_2^\dagger \Phi_3)(\Phi_3^\dagger \Phi_2) + (\Phi_3^\dagger \Phi_1)(\Phi_1^\dagger \Phi_3) + (\Phi_1^\dagger \Phi_2)(\Phi_2^\dagger \Phi_1)] \\
& + \left\{ \frac{1}{2}\lambda_4[(\Phi_2^\dagger \Phi_3)^2 + (\Phi_3^\dagger \Phi_1)^2 + (\Phi_1^\dagger \Phi_2)^2] + h.c. \right\}
\end{aligned}
\tag{3.6}
$$

It has been shown in [48] that the vacuum alignment

$$\langle \Phi_1^0 \rangle = \langle \Phi_2^0 \rangle = \langle \Phi_3^0 \rangle = \frac{v}{\sqrt{3}} \tag{3.7}$$

is a solution for the minimization of (3.6). The Lagrangian pertinent to neutrino masses is

$$
\begin{aligned}
-\mathcal{L}_L \;=\; & h_{1D}\overline{L}_1\left(\nu_R\widetilde{\Phi}\right)_1 + h_{2D}\overline{L}_2\left(\nu_R\widetilde{\Phi}\right)_1' + h_{3D}\overline{L}_3\left(\nu_R\widetilde{\Phi}\right)_1'' + \frac{M}{2}\left(\nu_R^T C\nu_R\right)_1 \\
& + \lambda\left(L_1^T C i\sigma_2\vec{\sigma}\cdot\vec{\Delta}L_2\right) + \lambda\left(L_2^T C i\sigma_2\vec{\sigma}\cdot\vec{\Delta}L_1\right) \\
& + \lambda'\left(L_3^T C i\sigma_2\vec{\sigma}\cdot\vec{\Delta}L_3\right) + \text{h.c.}
\end{aligned}
\tag{3.8}
$$

where the $A_4$ representations for each term are shown explicitly. Assuming that the triplet acquires a vacuum expectation value we obtain a mixed Type-I/Type-II seesaw mechanism for neutrino masses. The charged lepton and Dirac neutrino masses are

$$
M_l = v \,\text{diag}(h_1, h_2, h_3)U
$$

$$
m_D = v \,\text{diag}(h_{1D}, h_{2D}, h_{3D})U \;,
$$

with

$$
U = \frac{1}{\sqrt{3}}\begin{pmatrix} 1 & 1 & 1 \\ 1 & \omega^2 & \omega \\ 1 & \omega & \omega^2 \end{pmatrix}\;, \quad \omega \equiv e^{\frac{2\pi i}{3}}\;.
\tag{3.9}
$$

As a result, the Type I neutrino mass matrix after electroweak symmetry breaking is

$$
\mathcal{M}_{\nu f}^I = m_D M_R^{-1} m_D^T = \frac{v^2}{M}\begin{pmatrix} h_{1D}^2 & 0 & 0 \\ 0 & 0 & h_{2D}h_{3D} \\ 0 & h_{2D}h_{3D} & 0 \end{pmatrix}\;.
\tag{3.10}
$$

On the other hand, a small vev $u$ for the triplet induces a Type II contribution to the light neutrino mass matrix

$$
\mathcal{M}_\nu^{II} = \begin{pmatrix} 0 & \lambda u & 0 \\ \lambda u & 0 & 0 \\ 0 & 0 & \lambda' u \end{pmatrix}\;,
\tag{3.11}
$$

where $\lambda, \lambda'$ are two Yukawa couplings and for definiteness we are assuming that the triplet belongs to the $\mathbf{1}''$ representation of $A_4$. The total neutrino mass matrix is given by the sum of equation (3.10) and (3.11) and has the form

$$
\mathcal{M}_\nu = \begin{pmatrix} a & x & 0 \\ x & 0 & b \\ 0 & b & y \end{pmatrix}\;,
\tag{3.12}
$$

where $a$, $b$ and $x$, $y$ refer to the type-I and type-II contributions, respectively. This provides a simple derivation of the two-zero texture classified as $B_1$ in Ref. [49]. As it is referred in [46], the main feature of two-zero texture models, such as the ones derived here, is their power in predicting the as yet undetermined neutrino parameters. Current neutrino oscillation experiments determine two mass splittings $\Delta m^2_{\text{atm}}$ and $\Delta m^2_{\text{sol}}$ and the corresponding mixing angles $\theta_{12}$ and $\theta_{23}$, with some sensitivity on $\theta_{13}$ which is bounded [28]. The Dirac CP phase will be probed in future oscillation experiments. Similarly, the absolute neutrino mass scale will be probed by future cosmological observations [50], tritium beta decays [51] and neutrinoless double beta decay experiments [52] with improved sensitivity. The latter will also shed light on the two Majorana CP phases which are hard to test otherwise, as they do not affect lepton number conserving processes. The general $3 \times 3$ light neutrino mass matrix $\mathcal{M}_\nu$ in the flavour basis contains *a priori* nine independent real parameters, once the three unphysical phases associated with the charged lepton fields are removed. In contrast, in the proposed model all the above nine parameters are given in terms of only five unknowns. Hence the number of physical parameters characterizing the charged current weak interaction is reduced with respect to what is expected in the general case [53].

## 3.2   Triplet Decays and Leptogenesis

Now we turn to the study of leptogenesis in this model. As our model has not only right-handed neutrinos but also Higgs boson triplets, the analysis of leptogenesis is more complicated than in the usual type-I seesaw case. This case has been studied, in a generic way, in Ref.[54] and we will use their results. To make contact with their notation we need to fully specify our model, giving the couplings between the $SU(2)$ triplet and doublets. These are

$$-\mathcal{L}_{\Delta/\phi} = \sum_i f_i \left( \Phi_i^T i\sigma_2 \vec{\sigma} \cdot \vec{\Delta}^\dagger \Phi_i \right) + \text{h.c.}, \tag{3.13}$$

where $i = 1, 2, 3$ is an $A_4$ indices and $f_i = f(1, w, w^2)$. These two Lagrangians, Eq. (3.8) and Eq. (3.13), are gauge invariant for the quantum numbers given in Table 3.2. With the identifications $\Delta_1 + i\Delta_2 \equiv \Delta^0$, $\Delta_1 - i\Delta_2 \equiv \Delta^{++}$ and $\Delta_3 \equiv \Delta^+/\sqrt{2}$, and the $\Delta$ in the $1''$ representation, the Lagrangian $\mathcal{L}_L$ is, explicitly,

$$
\begin{aligned}
-\mathcal{L}_L = {} & h_{1D}\ \overline{\nu}_{1L}\nu_{1R}\phi_1^{0*} - h_{1D}\ \overline{l}_{1L}\nu_{1R}\phi_1^- + h_{1D}\ \overline{\nu}_{1L}\nu_{2R}\phi_2^{0*} - h_{1D}\ \overline{l}_{1L}\nu_{2R}\phi_2^- \\
& + h_{1D}\ \overline{\nu}_{1L}\nu_{3R}\phi_3^{0*} - h_{1D}\ \overline{l}_{1L}\nu_{3R}\phi_3^- + h_{2D}\ \overline{\nu}_{2L}\nu_{1R}\phi_1^{0*} - h_{2D}\ \overline{l}_{2L}\nu_{1R}\phi_1^- \\
& + \omega^2 h_{2D}\ \overline{\nu}_{2L}\nu_{2R}\phi_2^{0*} - \omega^2 h_{2D}\ \overline{l}_{2L}\nu_{2R}\phi_2^- + \omega h_{2D}\ \overline{\nu}_{2L}\nu_{3R}\phi_3^{0*} - \omega h_{2D}\ \overline{l}_{2L}\nu_{3R}\phi_3^- \\
& + h_{3D}\ \overline{\nu}_{3L}\nu_{1R}\phi_1^{0*} - h_{3D}\ \overline{l}_{3L}\nu_{1R}\phi_1^- + \omega h_{3D}\ \overline{\nu}_{3L}\nu_{2R}\phi_2^{0*} - \omega h_{3D}\ \overline{l}_{3L}\nu_{2R}\phi_2^- \\
& + \omega^2 h_{3D}\ \overline{\nu}_{3L}\nu_{3R}\phi_3^{0*} - \omega^2 h_{3D}\ \overline{l}_{3L}\nu_{3R}\phi_3^- + \lambda\Delta^0\nu_{2L}^T C\nu_{1L} - \frac{\lambda}{\sqrt{2}}\Delta^+\nu_{2L}^T C l_{1L}
\end{aligned}
$$

$$- \frac{\lambda}{\sqrt{2}} \Delta^+ l_{2L}^T C \nu_{1L} - \lambda \Delta^{++} l_{2L}^T C l_{1L} + \lambda \Delta^0 \nu_{1L}^T C \nu_{2L} - \frac{\lambda}{\sqrt{2}} \Delta^+ \nu_{1L}^T C l_{2L}$$

$$- \frac{\lambda}{\sqrt{2}} \Delta^+ l_{1L}^T C \nu_{2L} - \lambda \Delta^{++} l_{1L}^T C l_{2L} + \lambda' \Delta^0 \nu_{3L}^T C \nu_{3L} - \frac{\lambda'}{\sqrt{2}} \Delta^+ \nu_{3L}^T C l_{3L}$$

$$- \frac{\lambda'}{\sqrt{2}} \Delta^+ l_{3L}^T C \nu_{3L} - \lambda' \Delta^{++} l_{3L}^T C l_{3L} + \text{h.c.} \tag{3.14}$$

On the other hand, $\mathcal{L}_{\Delta/\phi}$ can be written as

$$\begin{aligned}
-\mathcal{L}_{\Delta/\phi} = & f \Delta^{--} \phi_1^+ \phi_1^+ + \omega f \Delta^{--} \phi_2^+ \phi_2^+ + \omega^2 f \Delta^{--} \phi_3^+ \phi_3^+ \\
& - \sqrt{2} f \Delta^- \phi_1^0 \phi_1^+ - \sqrt{2} \omega f \Delta^- \phi_2^0 \phi_2^+ - \sqrt{2} \omega^2 f \Delta^- \phi_3^0 \phi_3^+ \\
& - f \Delta^{0*} \phi_1^0 \phi_1^0 - \omega f \Delta^{0*} \phi_2^0 \phi_2^0 - \omega^2 f \Delta^{0*} \phi_3^0 \phi_3^0 + \text{h.c.}
\end{aligned} \tag{3.15}$$

The expressions for the asymmetries are normally written in the basis where the right-handed neutrinos are diagonal. In our case this already happens because the right handed neutrinos are fully degenerate as a consequence of the $A_4$ symmetry. Translating our notation into the one of [54] we get

$$\mu_i^* = \left( f, \omega f, \omega^2 f \right) \tag{3.16}$$

$$Y_\Delta = \begin{pmatrix} 0 & \lambda & 0 \\ \lambda & 0 & 0 \\ 0 & 0 & \lambda' \end{pmatrix} \tag{3.17}$$

$$Y_N = U^\dagger \text{diag}(h_{1D}^*, h_{2D}^*, h_{3D}^*) \tag{3.18}$$

The diagrams contributing to the CP-asymmetry are shown in Fig. 3.1. The asymmetries for these diagrams have been computed generically in [54]. The first two do not depend on the triplet and give

$$\epsilon_{\nu_{kR}} = \frac{1}{8\pi} \sum_j \frac{\mathcal{I}m\left[(Y_N Y_N^\dagger)_{kj}^2 \delta_{kj}\right]}{\sum_i \mid (Y_N)_{ki} \mid^2} \times f(M), \tag{3.19}$$

where $f(M)$ is some kinematical factor. The third diagram depends on the triplet and is given by

$$\epsilon_{\nu_{kR}}^\Delta = -\frac{1}{2\pi} \frac{\sum_{il} \mathcal{I}m\left[(Y_N)_{ki}(Y_N)_{kl}(Y_\Delta^*)_{il}(\mu_k)\right]}{\sum_i |(Y_N)_{ki}|^2 M} \times \left(1 - \frac{M_\Delta^2}{M^2} \log(1 + M^2/M_{\Delta^2})\right), \tag{3.20}$$

while for the fourth diagram we have

$$\epsilon_\Delta = \frac{1}{8\pi} \sum_k M_{\nu_{kR}} \frac{\sum_{il} \mathcal{I}m\left[(Y_N^*)_{ki}(Y_N^*)_{kl}(Y_\Delta)_{il}(\mu_k^*)\right]}{\sum_{ij} |(Y_\Delta)_{ij}|^2 M_\Delta^2 + \sum_i |\mu_i|^2} \times \log(1 + M_\Delta^2/M^2), \tag{3.21}$$

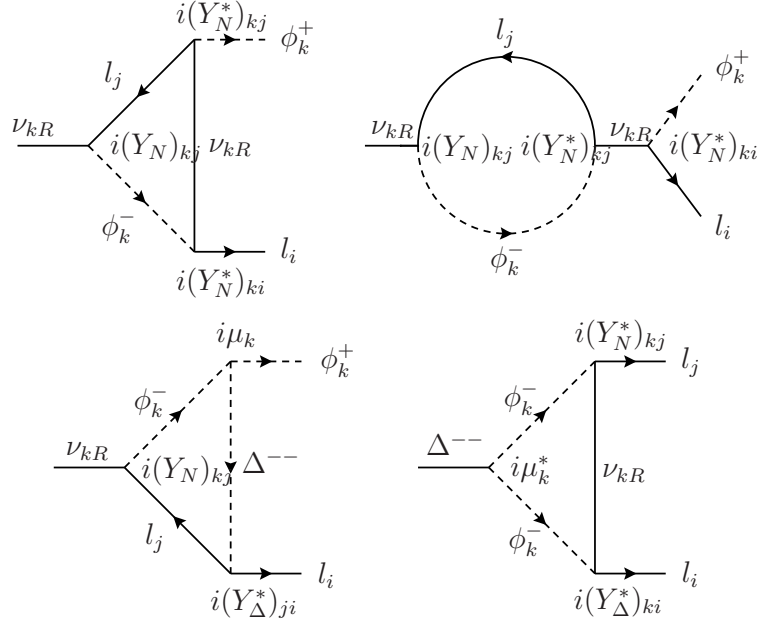

**Figure 3.1:** One Loop diagrams contributing to asymmetry from the $\nu_R$ decay.

With the couplings in Eq. (3.16) we can see immediately that the diagrams with only right-handed neutrinos do not contribute to the asymmetry due to the $A_4$ group properties. The same is not true for the other asymmetries. As we see that without the triplet there is no asymmetry, we will concentrate in the case where all the asymmetry comes from the fourth diagram in Fig. 3.1, given in Eq. (3.21). As it was discussed in Ref. [54] this corresponds to the case where $M_\Delta \ll M$.

We have performed a scan of the parameter space subject to the condition that all the points satisfy the neutrino data [14]. The results for the asymmetry $\epsilon_\Delta$ are shown in Fig. 3.2 as a function of the triplet parameters, the mass $M_\Delta$ and an *effective* coupling defined as

$$\lambda_L = \sqrt{2}\sqrt{\text{Tr}\left(Y_\Delta Y_\Delta^\dagger\right)} \tag{3.22}$$

We see that the asymmetry can easily be high. However, to be able to discuss if the washout effects are significant, we have to solve the Boltzmann equations for our model.

For their study we will also need the decay rates of the triplet into leptons and Higgs bosons. The tree-level decay rates are,

$$\Gamma(\Delta \to LL) \;\; = \;\; \Gamma_L = \frac{M_\Delta}{16\pi}\lambda_L^2$$

$$\Gamma(\Delta \to HH) \;\; = \;\; \Gamma_H = \frac{M_\Delta}{16\pi}\lambda_H^2 \tag{3.23}$$

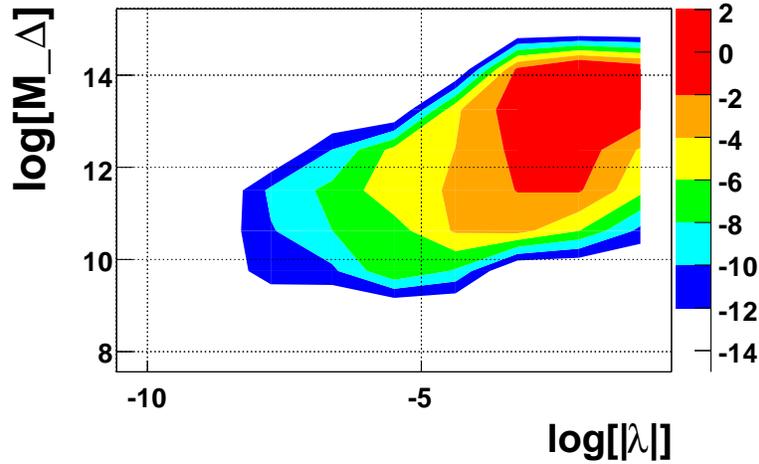

**Figure 3.2:** $\log(\epsilon_\Delta)$ contours as a function of the triplet mass and the triplet-lepton coupling $\lambda$.

where we have defined the dimensionless effective couplings

$$\lambda_L \equiv \sqrt{2} \ \sqrt{\text{Tr}\left(Y_\Delta Y_\Delta^\dagger\right)}, \quad \lambda_H \equiv \sqrt{2} \ \frac{\sqrt{|\mu_1|^2 + |\mu_2|^2 + |\mu_3|^2}}{M_\Delta} \qquad (3.24)$$

In terms of these couplings the branching ratios into leptons and Higgs bosons, $B_L, B_H$, are

$$B_L = \frac{\lambda_L^2}{\lambda_L^2 + \lambda_H^2}, \quad B_H = \frac{\lambda_H^2}{\lambda_L^2 + \lambda_H^2} \ . \qquad (3.25)$$

We solve the Boltzmann equation following the setup of Ref.[55]. Using their notation we recall that the Boltzmann equations describe the evolution as function of $z = M_\Delta/T$ of the total triplet density, $\Sigma_T = (n_T + n_{\overline{T}})/s$, and of the asymmetries $\Delta_p = (n_p - n_{\overline{p}})/s$ for the species $p = T, L, H$, where $n_p$ is the number density of the type $p$ particles, and $s$ is the total entropy density [1]. With this notation the Boltzmann equations are

$$sHz\frac{d\Sigma_T}{dz} = -\left(\frac{\Sigma_T}{\Sigma_T^{\text{eq}}} - 1\right)\gamma_D - 2\left(\frac{\Sigma_T^2}{\Sigma_T^{\text{eq}\,2}} - 1\right)\gamma_A$$

$$sHz\frac{d\Delta_L}{dz} = X - 2\gamma_D B_L\left(\frac{\Delta_L}{Y_L^{\text{eq}}} + \frac{\Delta_T}{\Sigma_T^{\text{eq}}}\right)$$

$$sHz\frac{d\Delta_L}{dz} = X - 2\gamma_D B_H\left(\frac{\Delta_H}{Y_H^{\text{eq}}} - \frac{\Delta_T}{\Sigma_T^{\text{eq}}}\right) \qquad (3.26)$$

$$sHz\frac{d\Delta_T}{dz} = -\gamma_D\left(\frac{\Delta_T}{\Sigma_T^{\text{eq}}} + B_L\frac{\Delta_L}{Y_L^{\text{eq}}} - B_H\frac{\Delta_H}{Y_H^{\text{eq}}}\right)$$

_____________

[1] In order to follow as close as possible the notation of Ref.[55], in this section we denote the triplet fields by $T$ instead of $\Delta$.

where $H$ is the Hubble constant at temperature $T$, $Y_p = n_p/s$, a suffix $^{\text{eq}}$ denotes equilibrium, $\gamma_p$ is the spacetime density of type $p$ computed in thermal equilibrium. The other quantities are

$$X = \gamma_D \epsilon_L \left( \frac{\Sigma_T}{\Sigma_T^{\text{eq}}} - 1 \right) - 2 \left( \frac{\Delta_L}{Y_L^{\text{eq}}} + \frac{\Delta_H}{Y_H^{\text{eq}}} \right) \left( \gamma_{Ts}^{\text{sub}} + \gamma_{Tt} \right) \tag{3.27}$$

where $\gamma_{Ts}^{\text{sub}}$ and $\gamma_{Tt}$, are the spacetime densities for the s-channel (properly subtracted to avoid double counting of the processes already counted on the decay, see below) and t-channel $\Delta L = 2$ scatterings, respectively. We recall that the reaction densities are defined through

$$\gamma = \frac{T}{64\pi^4} \int_{s_{\text{min}}}^{\infty} ds \sqrt{s} K_1 \left( \frac{\sqrt{s}}{T} \right) \hat{\sigma}(s) \tag{3.28}$$

where $K_1$ is a Bessel function and $\hat{\sigma}$ the reduced cross-section defined as

$$\hat{\sigma}(s) = \sum \int dt \frac{|A|^2}{8\pi s} \tag{3.29}$$

where the sum runs over initial and final spins and gauge indices of the amplitude $A$.

The expressions for $\gamma_A$ and $\gamma_D$ are the same as in Ref.[55] (with the obvious exception that we have three Higgs bosons) and we do not reproduce them here. There are computational differences for $\Delta L = 2$ scatterings but the order of magnitude for these processes remains similar.

The baryon asymmetry is then obtained as

$$\eta_B = \frac{n_B}{n_\gamma}\bigg|_{\text{today}} = \frac{n_B}{s} \left. \frac{s}{n_\gamma} \right|_{\text{today}} = -a_{\text{sph}} \left. \frac{s}{n_\gamma} \right|_{\text{today}} \Delta_L \tag{3.30}$$

where

$$\frac{s}{n_\gamma}\bigg|_{\text{today}} = \frac{\pi^4}{45\zeta(3)} \frac{43}{11} \tag{3.31}$$

and $a_{\text{sph}} = 36/105$ is the sphaleron conversion factor for three doublets of Higgs bosons. We therefore have to solve the Boltzmann equations, Eqs. (3.26) to find the lepton asymmetry $\Delta_L$ and then we obtain $\eta_B$ using Eq. (3.30). Due the complexity of Eqs. (3.26), they have to be solved numerically. Although our model differs in the details from that of Ref.[55], it shares many of the same features that are due to the presence of the triplet in the Boltzmann equations. Therefore we recover their result that the efficiency can be high in this type of models.

We have performed a scan of the parameter space with the free parameters in the following ranges

$$M_\Delta \in [10^6, 10^{14}] \text{ GeV}, \quad |f| \in [10^6, 10^{14}] \text{ GeV}, \quad M \in [10, 100] M_\Delta \tag{3.32}$$

The other parameters, including the coupling matrix $Y_\Delta$, were obtained from the requirement that the neutrino masses and mixings were within the present allowed $3\sigma$ range [14]. The results are summarized in Fig. 3.3, where we plot contours of equal baryon asymme-try, $\eta_B$, as a function of the triplet mass $M_\Delta$ and the effective coupling, $\lambda_L$, defined in Eq. (3.24). In Fig. 3.4 the contours are shown as functions of the effective couplings $\lambda_L$ and $\lambda_H$. We conclude that in a large region of parameter space the present value obtained from Big Bang Nucleosynthesis (@ 95% CL) [56],

$$4.7 \times 10^{-10} < \eta_B < 6.5 \times 10^{-10} \tag{3.33}$$

can be easily achieved, accounting at the same time for the present experimental limits on neutrino masses and mixings[14].

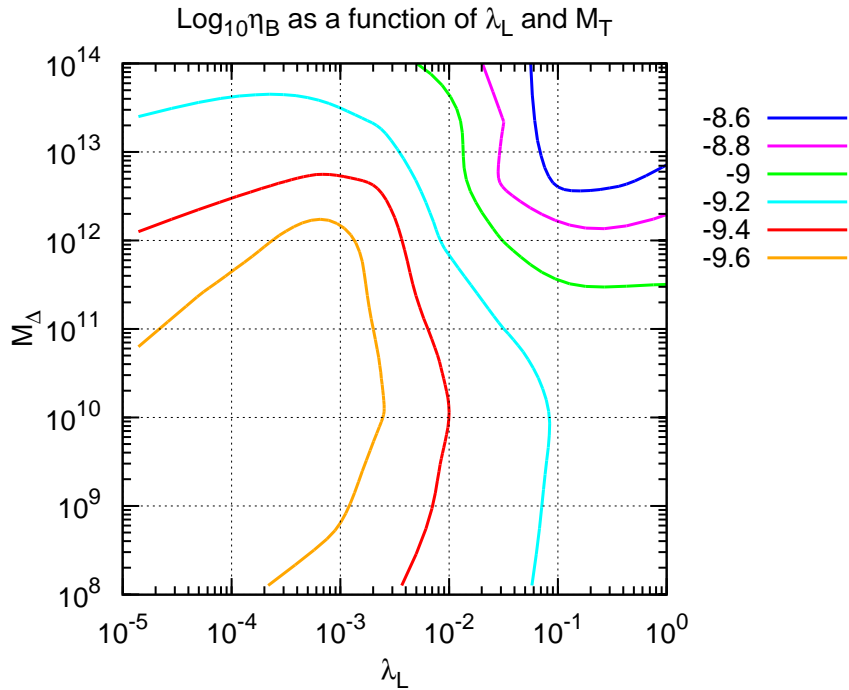

**Figure 3.3:** Contour plots for $\log_{10} \eta_B$ as a function of $M_\Delta$ and $\lambda_L$, after the effect of washout has been taken in account by solving the Boltzmann equations. For comparison we note that (at 95% CL) we have: $-9.35 < \log_{10} \eta_B < -9.18$.

## 3.3 Higgs Potential Minimization

Until now we have just assumed that the triplet acquires a small vev, breaking in this way $A_4$. However, a careful analysis of the Higgs potential in the presence of the triplet shows that this is a non trivial matter[2]. Following a suggestion by A. S. Joshipura[3], let

---

[2]We thank F. Joaquim for pointing this out to us.
[3]A. S. Joshipura, private communication. See also [22].

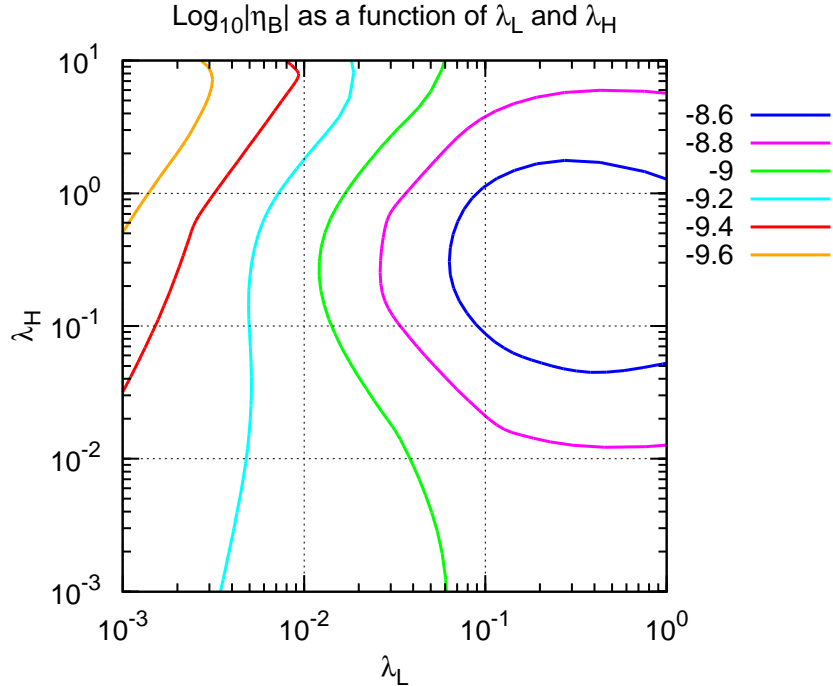

**Figure 3.4:** Contour plots for $\log_{10} \eta_B$ as a function of $\lambda_L$ and $\lambda_H$, after the effect of washout has been taken in account by solving the Boltzmann equations. For comparison we note that (at 95% CL) we have: $-9.35 < \log_{10} \eta_B < -9.18$.

us write the full Higgs potential

$$V = V(\Phi) + V_1(\Phi, T) + V_2(\Phi, T) + V(T) \tag{3.34}$$

with $V(\Phi)$ given by (3.6) and $V_1(\Phi, T)$ the trilinear coupling

$$V_1(\Phi, T) = \lambda T^\dagger \sum_i (c_i \Phi_i^T \Phi_i) + \text{ h.c.} \tag{3.35}$$

in a simplified notation omitting $SU(2)$ indices, with $c_i = (1, w, w^2)$. The terms $V_2(\Phi, T)$ and $V(T)$ refer to quartic couplings and the triplet mass term and its explicit form will not play an essential role in this discussion. The minimum condition with respect to $T$ is

$$\frac{\delta V}{\delta T} = 2\lambda \left( \sum_i c_i \Phi_i^T \Phi_i \right) + \frac{\delta V_2(\Phi, T)}{\delta T} + \frac{\delta V(T)}{\delta T}. \tag{3.36}$$

The first term is essential to give a vev to the triplet, but with the Higgs doublets vacuum alignment (3.7) it vanishes because of the $A_4$ relation

$$1 + w + w^2 = 0. \tag{3.37}$$

There is also a second problem. The minimization with respect to the $\Phi_i$ gives

$$\frac{\delta V}{\delta \Phi_i} = 4\lambda u \frac{v}{\sqrt{3}} c_i + \frac{\delta V_2(\Phi, T)}{\delta \Phi_i} + \frac{\delta V(\Phi)}{\delta \Phi_i}. \quad (3.38)$$

Noting that the last two terms are independent of $i$ at the minimum and summing the three equations for $i = 1, 2, 3$ we get at the minimum, using (3.37),

$$\frac{\delta V}{\delta \Phi_1} + \frac{\delta V}{\delta \Phi_2} + \frac{\delta V}{\delta \Phi_3} = 0$$
$$\Leftrightarrow 3a = 0 \quad (3.39)$$

with

$$a = \frac{\delta V_2(\Phi, T)}{\delta \Phi_i} + \frac{\delta V(\Phi)}{\delta \Phi_i}, \quad i = 1, 2, 3. \quad (3.40)$$

Replacing in (3.38) we conclude that

$$\lambda u v c_i = 0 \quad (3.41)$$

which means that the two vevs cannot be simultaneously non zero.

To solve the first problem A. S. Joshipura proposed to introduce an explicit $A_4$ breaking term

$$\lambda' T^\dagger \sum_i (\Phi_i^T \Phi_i)_{\mathbf{1}} + \text{ h.c.} \quad (3.42)$$

with $\lambda'$ small, where we choose the product of the two Higgs $SU(2)$ doublets $A_4$ triplets to be in the trivial representation of $A_4$. The triplet vev would break $A_4$ anyway, although spontaneously, so this just amounts to break $A_4$ already in the tree level Lagrangian. This avoids the relation (3.37) and it allows the triplet to acquire a vev. For the second problem we propose to introduce a second triplet in the $\mathbf{1}'$ representation and with opposite hypercharge of $T$. Let us call this new triplet $T_2$ and the initial one $T_1$. Then the $A_4$ invariant cubic couplings between the Higgs triplets and doublets are

$$\lambda_1 T_1^\dagger \sum_i (c_i \Phi_i^T \Phi_i) + \lambda_2 T_2 \sum_i (c_i \Phi_i^T \Phi_i) + \text{ h.c.} \quad (3.43)$$

which permits to replace (3.41) by

$$\lambda_1 u_1 v c_i + \lambda_2 u_2 v c_i = 0. \quad (3.44)$$

This no longer requires the vanishing of one of the vevs.

This second solution turns the model less predictive because there are more parameters. But in a supersymmetric framework the second triplet would not couple to the leptons because of the superpotential holomorphy, which means that the Type-II seesaw would still be determined only by one triplet-leptons coupling. On the other hand it is necessary to introduce a second family of $SU(2)$ Higgs doublets for the cancelation of anomalies, which turns the model more elaborated. Alternatively, since the two triplets have different quantum numbers, it is licit to make the assumption that one is much heavier than the other and Type-II seesaw would still be determined mainly by the lightest triplet, assuming small to moderate Yukawa couplings. Of course, leptogenesis in this framework would be much more involved as there are now more processes engaging the two Higgs triplets. This will be the subject of future work. Another possibility of reconciling the $A_4$ symmetry with the $B_1$ zero texture is explored in the next chapter [22].

# Chapter 4

# $A_4$-based neutrino masses with Majoron decaying dark matter

## 4.1 Introduction

As we saw in Chapter 1, the discovery of neutrino oscillations [10, 57–60], now confirmed at reactors and accelerators [61–63], has brought neutrino physics to the center of particle physics research. Global analysis of current oscillation data indicate that the pattern of lepton mixing differs sharply from that characterizing quarks [14]. Understanding the origin of neutrino mass and the pattern of neutrino mixing angles from basic principles constitutes a major challenge [64, 65]. A paradigm framework to generate neutrino masses is provided by the seesaw mechanism, for which several realizations have been proposed [66]. The observed pattern of neutrino mixing may arise from suitable non-abelian flavour symmetries, as those based on the $A_4$ group [48, 67–69].

Elucidating the nature of dark matter constitutes another intriguing problem of modern physics which has so far defied all efforts. It is therefore crucial to build a fundamental particle physics theory of dark matter and, since the Standard Model of elementary particles (SM) fails to provide a dark matter candidate, such theory necessarily requires physics beyond the SM.

In the work [22], that it is described in this Chapter, we suggest a version of the seesaw mechanism containing both type-I [53, 70–76] and type-II contributions [53, 75, 77–80] in which we implement an $A_4$ flavor symmetry with spontaneous violation of lepton number [75, 76]. We study the resulting pattern of vacuum expectation values (vevs) and show that the model reproduces the phenomenologically consistent and predictive two-zero texture proposed in Ref. [46], avoiding the problem of the vanishing of the triplet vev discussed in section 3.3.

In the presence of explicit global symmetry breaking effects, as might follow from gravitational interactions, the resulting pseudo-Goldstone boson - Majoron - may constitute a viable candidate for decaying dark matter if it acquires mass in the keV-MeV range. Indeed, this is not in conflict with the lifetime constraints which follow from current cos-mic microwave background (CMB) observations provided by the Wilkinson Microwave Anisotropy Probe (WMAP) [81]. We also show how the corresponding mono-energetic emission line arising from the sub-leading one-loop induced electromagnetic decay of the Majoron may be observed in future X-ray missions [15].

## 4.2 The Model

Our model is described by the multiplet content specified in Table 4.1 where the transformation properties under the SM and $A_4$ groups are shown (as well as the corresponding lepton number $L$). The $L_i$ and $l_{Ri}$ fields are the usual SM lepton doublets and singlets and $\nu_R$ the right-handed neutrinos. The scalar sector contains an SU(2) triplet $\Delta$, three Higgs doublets $\Phi_i$ (which transform as a triplet of $A_4$) and a scalar singlet $\sigma$. Three additional fermion singlets $S_i$ are also included.

**Table 4.1:** Lepton multiplet structure ($Q = T_3 + Y/2$)

|          | $L_1$ | $L_2$ | $L_3$ | $l_{Ri}$ | $\nu_{iR}$ | $\Phi_i$ | $\Delta$ | $\sigma$ | $S_i$ |
|----------|-------|-------|-------|----------|-----------|----------|----------|----------|-------|
| $SU(2)$  | 2     | 2     | 2     | 1        | 1         | 2        | 3        | 1        | 1     |
| $U(1)_Y$ | $-1$  | $-1$  | $-1$  | $-2$     | 0         | $-1$     | 2        | 0        | 0     |
| $A_4$    | $1'$  | 1     | $1''$ | 3        | 3         | 3        | $1''$    | $1''$    | 3     |
| $L$      | 1     | 1     | 1     | 1        | 1         | 0        | $-2$     | $-2$     | 1     |

Taking into account the information displayed in Table 4.1, and imposing lepton number conservation, the Lagrangian responsible for neutrino masses reads

$$\begin{aligned}
-\mathcal{L}_L = {} & h_1 \overline{L}_1 \left(\nu_R \Phi\right)'_1 + h_2 \overline{L}_2 \left(\nu_R \Phi\right)_1 + h_3 \overline{L}_3 \left(\nu_R \Phi\right)''_1 \\
& + \lambda L_1^T C \Delta L_2 + \lambda L_2^T C \Delta L_1 + \lambda' L_3^T C \Delta L_3 \\
& + M_R \left(\overline{S_L} \nu_R\right)_1 + h \left(S_L^T C S_L\right)'_1 \sigma + \text{h.c.},
\end{aligned} \qquad (4.1)$$

where $h$ and $\lambda$ are adimensional couplings, $M_R$ is a mass scale and

$$\Delta = \begin{pmatrix} \Delta_0 & -\Delta^+/\sqrt{2} \\ -\Delta^+/\sqrt{2} & \Delta^{++} \end{pmatrix}, \; \Phi_i = \begin{pmatrix} \phi_i^0 \\ \phi_i^- \end{pmatrix}. \qquad (4.2)$$

Note that the term $(\nu_R^T C \nu_R)'_1 \sigma$ is allowed by the imposed symmetry. This term however does not contribute to the light neutrino masses to the leading order in the seesaw expansion and we omit it. Alternatively, such term may be forbidden by holomorphy in a supersymmetric framework with the following superpotential terms

$$\mathcal{W} = \cdots + \lambda \epsilon_{ab} h_i^\nu \hat{L}_i^a \hat{\nu}^c \hat{H}_u^b + M_R \hat{\nu}^c \hat{S} + \frac{1}{2} h \hat{S} \hat{S} \hat{\sigma}$$

where the hats denote superfields and the last term replaces the corresponding bilinear employed in Ref. [82, 83]. Assuming that the Higgs bosons $\Phi_i$, $\Delta^0$ and $\sigma$ acquire the following vevs (see section 4.3 below)

$$\langle \phi_1^0 \rangle = \langle \phi_2^0 \rangle = \langle \phi_3^0 \rangle = \frac{v}{\sqrt{3}}, \quad \langle \Delta^0 \rangle = u_\Delta, \quad \langle \sigma \rangle = u_\sigma \,, \tag{4.3}$$

we obtain an extended seesaw neutrino mass matrix $\mathcal{M}$ [82–84] in the $(\nu_L, \nu^c, S)$ basis

$$\mathcal{M} = \begin{pmatrix} 0 & m_D & 0 \\ m_D^T & 0 & M \\ 0 & M^T & \mu \end{pmatrix} \,, \ m_D = v \ \mathrm{diag}(h_1, h_2, h_3) \ U, \quad U = \frac{1}{\sqrt{3}} \begin{pmatrix} 1 & \omega^2 & \omega \\ 1 & 1 & 1 \\ 1 & \omega & \omega^2 \end{pmatrix}, \tag{4.4}$$

with $\omega = e^{2\pi i/3}$, $M = M_R \ \mathrm{diag}(1, 1, 1)$ and $\mu = u_\sigma h \ \mathrm{diag}(1, w^2, w)$. This leads to an effective light neutrino mass matrix $\mathcal{M}_\nu^{\mathrm{I}}$ given by

$$\mathcal{M}_\nu^{\mathrm{I}} = m_D M^{T-1} \mu M^{-1} m_D^T = \frac{h v^2 u_\sigma}{M_R^2} \begin{pmatrix} h_1^2 & 0 & 0 \\ 0 & 0 & h_2 h_3 \\ 0 & h_2 h_3 & 0 \end{pmatrix} \,. \tag{4.5}$$

On the other hand the vev of the triplet, $u_\Delta$, will induce an effective mass matrix for the light neutrinos from type-II seesaw mechanism

$$\mathcal{M}_\nu^{\mathrm{II}} = 2 u_\Delta \begin{pmatrix} 0 & \lambda & 0 \\ \lambda & 0 & 0 \\ 0 & 0 & \lambda' \end{pmatrix} \,, \tag{4.6}$$

and the total effective light neutrino mass matrix will then be

$$\mathcal{M}_\nu = \mathcal{M}_\nu^{\mathrm{I}} + \mathcal{M}_\nu^{\mathrm{II}} \,. \tag{4.7}$$

In Ref.[46] it was shown that the neutrino mass matrix given by Eq. (4.7) could explain the currently available neutrino data. In section 4.4 we will present an update of that analysis taking into account the latest neutrino oscillation data.

## 4.3  *$A_4$ Invariant Higgs Potential*

As in Chapter 3, 3.3, we now address the question of the minimization of the neutral Higgs scalar potential in this new framework, which, as we saw, is a necessary condition to reproduce the structure of the neutrino mass matrix presented in the previous section. With the assignments of Table 4.1, the Higgs potential consistent with gauge and $A_4$ invariance and lepton number conservation reads,

$$V = V(\Phi) + V(\Phi, \Delta, \sigma)\,, \qquad (4.8)$$

where $V(\Phi)$ is given as[1]:

$$\begin{aligned}
V(\Phi) = {} & m_\Phi^2 \left(\Phi^\dagger\Phi\right)_1 + \lambda_1 \left(\Phi^\dagger\Phi\right)_1 \left(\Phi^\dagger\Phi\right)_1 + \lambda_2 \left(\Phi^\dagger\Phi\right)_{1'} \left(\Phi^\dagger\Phi\right)_{1''} \\
& + \lambda_3 \left(\Phi^\dagger\Phi\right)_{3s} \cdot \left(\Phi^\dagger\Phi\right)_{3s} + \lambda_4 \left(\Phi^\dagger\Phi\right)_{3s} \cdot \left(\Phi^\dagger\Phi\right)_{3a} + \lambda_5 \left(\Phi^\dagger\Phi\right)_{3a} \cdot \left(\Phi^\dagger\Phi\right)_{3a}\,,
\end{aligned} \qquad (4.9)$$

and $V(\Phi, \Delta, \sigma)$ contains pure $\Delta$, $\sigma$ terms, together with others involving mixed invariant combinations of the scalar fields. Assuming the so-called seesaw hierarchy $u_\Delta \ll v \ll u_\sigma$ [75] [2], the relevant terms in $V(\Phi, \Delta, \sigma)$ are [3]

$$V(\Phi, \Delta, \sigma) = \left(M_\Delta^2 + \rho|\sigma|^2\right)\operatorname{Tr}(\Delta^\dagger\Delta) + \lambda_\sigma|\sigma|^4 + \left[m_\sigma^2 + \xi\left(\Phi^\dagger\Phi\right)_1\right]|\sigma|^2 - (\delta\Phi^T\Delta\Phi\sigma^* + \text{h.c.}), \qquad (4.10)$$

Taking the vacuum alignment for the Higgs doublets $\Phi_a$ given in equation (4.3) the minimization of the Higgs potential with respect to $\Delta$ gives

$$\frac{\delta V}{\delta\Delta} = 0 \Rightarrow \left(M_\Delta^2 + \rho\,u_\sigma^2\right)u_\Delta - \delta v^2 u_\sigma = 0\,. \qquad (4.11)$$

We stress that the $A_4$ symmetry, together with the doublet vev alignment assumed in Eq. (4.3), requires that the product $\Phi \otimes \Phi \sim \mathbf{1}$ under $A_4$. If $\Phi \otimes \Phi \sim \mathbf{1'}$, $\mathbf{1''}$, then the second term in the above equation would reduce to $2\delta(1+\omega+\omega^2)u_\sigma = 0$ implying $u_\Delta \sim 0$. Moreover, as a direct consequence of the requirement $\Phi \otimes \Phi \sim \mathbf{1}$ under $A_4$, $\Delta$ and $\sigma$ must have the same (singlet) transformation properties under that group.

The above equation leads to the following solution for the triplet vev

$$u_\Delta = \frac{\delta v^2 u_\sigma}{M_\Delta^2 + \rho u_\sigma^2} \simeq \frac{\delta v^2}{\rho u_\sigma}\,, \qquad (4.12)$$

where the last approximation holds for $M_\Delta \ll u_\sigma$. This result shows that the "vev-seesaw" relation $u_\Delta u_\sigma \sim v^2$ is fulfilled. The minimization with respect to the $\Phi_a$ gives

$$\frac{\delta V}{\delta\Phi_a} = 0 \Rightarrow \frac{\delta V(\Phi)}{\delta\Phi_a} + 2\xi v u_\sigma^2 - 4\delta v u_\Delta u_\sigma = 0. \qquad (4.13)$$

---

[1] The decomposition of the tensorial product of two triplets in $A_4$ is shown in (3.4)

[2] In contrast to the inverse seesaw models used in Refs. [83, 84] here we consider large values of $u_\sigma$, $u_\sigma > 10^7$ GeV or so.

[3] Notice that the scalar potential contains other invariant terms such as $\Phi^\dagger\Phi\operatorname{Tr}(\Delta^\dagger\Delta)$, $\operatorname{Tr}(\Delta^\dagger\Delta)|\sigma|^2$, $[\operatorname{Tr}(\Delta^\dagger\Delta)]^2$, etc. Assuming the vev hierarchy $u_\Delta \ll v \ll u_\sigma$ and that the adimensional coefficients of these terms are of the same order of the ones in $V(\Phi, \Delta, \sigma)$, then $V(\Phi, \Delta, \sigma)$ is enough for our purposes.

Finally,

$$\frac{\delta V}{\delta \sigma} = 0 \Rightarrow 2\lambda_\sigma u_\sigma^3 + \left(m_\sigma^2 + \xi v^2 + \rho u_\Delta^2\right) u_\sigma - 2\delta v^2 u_\Delta = 0. \tag{4.14}$$

which, in the limit $u_\Delta, v << u_\sigma$, has the approximate solution

$$u_\sigma = \sqrt{-\frac{m_\sigma^2}{2\lambda_\sigma}}, \tag{4.15}$$

as it is typical from spontaneous symmetry breaking scenarios. In summary, we have shown that in our framework it is possible to achieve a consistent minimization of the scalar potential with non-zero vevs satisfying the "vev-seesaw" relation $u_\Delta u_\sigma \sim v^2$.

## 4.4 Neutrino parameter analysis

Given the two contributions to the light neutrino mass matrix discussed in Eqs. (4.5) and (4.6) one finds that the total neutrino mass matrix has the following structure:

$$\mathcal{M}_\nu = \begin{pmatrix} a & b & 0 \\ b & 0 & c \\ 0 & c & d \end{pmatrix}. \tag{4.16}$$

This matrix with two-zero texture has been classified as B1 in [49]. One can show that considering the $(L_1, L_2, L_3)$ transformation properties under $A_4$ as being $(1', 1'', 1)$ or $(1'', 1', 1)$ an effective neutrino mass matrix with $\mathcal{M}_\nu(1, 2) = \mathcal{M}_\nu(3, 3) = 0$ is obtained (type B2 in [49]). Moreover, by choosing $\Delta, \sigma \sim \mathbf{1'}$ and appropriate transformation properties of the $L_i$ doublets, we could obtain the textures B1 and B2 as well. Still, the configuration $\Delta, \sigma \sim \mathbf{1}$ would lead to textures which are incompatible with neutrino data since, in this case, both type I and type II contributions to the effective neutrino mass matrix would have the same form. Since the textures of the type B1 and B2 are very similar in what concerns to neutrino parameter predictions, we will restrict our analysis to B1, shown in (4.16).

As was already mentioned in section 1.2.3, the neutrino mass matrix is described by nine parameters: three masses, three mixing angles and three phases (one Dirac + two Majorana). From neutrino oscillation experiments we have good determinations for two of the mass parameters (mass squared differences) and for two of the mixing angles ($\theta_{12}$ and $\theta_{23}$) as well as an upper-bound on the third mixing angle $\theta_{13}$. Using the $3\sigma$ allowed ranges for these five parameters and the structure of the mass matrix in Eq. (4.16) we can determine the remaining four parameters. The phenomenological implications of this kind of mass matrix have been analyzed in Refs. [46] and [85]. Here we will update the results in light of the recently determined neutrino oscillation parameters [14].

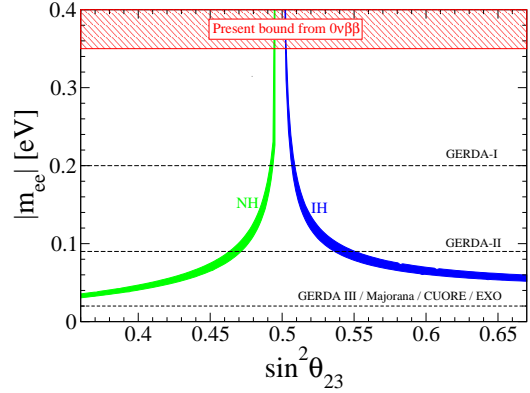

**Figure 4.1:** Correlation between the neutrinoless double beta decay amplitude parameter $|m_{ee}|$ and the atmospheric mixing parameter. Experimental sensitivities are also given for comparison.

The main results are shown in Figs. 4.1 and 4.2. In figure 4.1 we plot the correlation of the mass parameter characterizing the neutrinoless double beta decay amplitude:

$$|m_{ee}| = \left| c_{13}^2 c_{12}^2 m_1 + c_{13}^2 s_{12}^2 m_2 e^{2i\alpha} + s_{13}^2 m_3 e^{2i\beta} \right|, \tag{4.17}$$

with the atmospheric mixing angle $\theta_{23}$. Here $c_{ij}$ and $s_{ij}$ stand for $\cos\theta_{ij}$ and $\sin\theta_{ij}$ respectively. At the zeroth order approximation $m_1/m_3 = \tan^2\theta_{23}$, and therefore $\theta_{23} < 45°$ for normal hierarchy (NH), while $\theta_{23} > 45°$ for inverted hierarchy (IH). The main result from this plot is a lower bound on the effective neutrino mass:$|m_{ee}| > 0.03$ eV. For comparison the range of sensitivities of planned experiments as well as current bounds is also given. Note that the lower bound we obtain lies within reach of the future generation of neutrinoless double beta decay experiments.

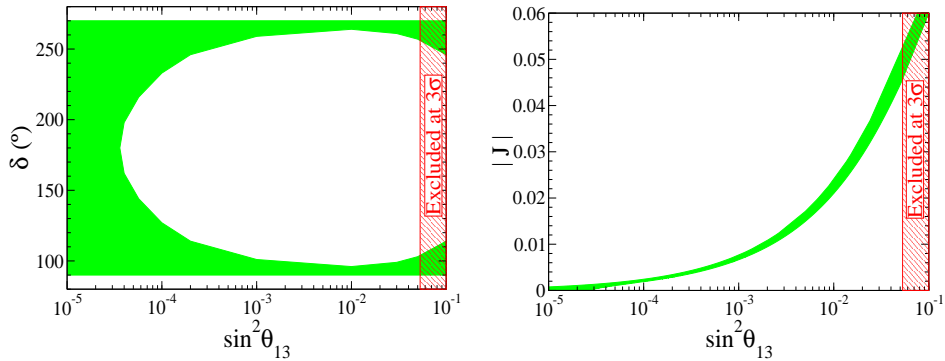

**Figure 4.2:** CP violating phase $\delta$ and CP-invariant $J$ in terms of the reactor mixing parameter. The 3 $\sigma$-excluded range for $\sin^2\theta_{ij}$ is given for comparison.

The panels in Fig. 4.2 show the CP-violating phase $\delta$ and the corresponding CP-violating invariant in neutrino oscillations, already discussed in section 1.2.1 - see formula (1.146):

$$J = s_{12} s_{23} s_{13} c_{12} c_{23} c_{13}^2 \sin\delta, \tag{4.18}$$

versus $\sin^2 \theta_{13}$. Note that these hold both for normal and inverted hierarchy spectra. In the left panel one sees that $\cos \delta < 0$ since, at first order in $\sin^2 \theta_{13}$, $m_1/m_2 = 1 + \frac{\cos \theta_{23}}{\cos \theta_{12} \sin \theta_{12} \sin^2 \theta_{23}} \sin \theta_{13} \cos \delta$, and the ratio of masses should satisfy: $m_1/m_2 < 1$. Moreover, for large $\theta_{13}$ values, where CP violation is likely to be probed in neutrino oscillations, one can see that our model predicts maximal violation of CP. Quantitatively, from the right panel one sees that the $3\sigma$ bound on $\theta_{13}$: $\sin^2 \theta_{13} < 0.053$ implies an upper bound: $|J| \lesssim 0.06$ on the CP-invariant.

In addition, the two-zero texture structure of our neutrino mass matrix may have other implications, for example for the expected pattern of lepton flavor violating decays. In fact, thanks to the strong renormalization effects due to the presence of the triplet states, the latter are quite sizeable in sypersymmetric models [5, 86, 87].

## 4.5 Majoron Dark Matter

In models where neutrinos acquire mass through spontaneous breaking of an ungauged lepton number [75, 76] one expects that, due to non-perturbative effects, the Nambu-Goldstone boson (Majoron) may pick up a mass that we assume to lie in the kilovolt range [88]. This implies that the Majorons will decay, mainly in neutrinos. As the coupling $g_{J\nu\nu}$ is proportional to $\frac{m_\nu}{u_\sigma}$ [75], the corresponding mean lifetime can be extremely long, even longer than the age of the Universe. As a result the Majoron can, in principle, account for the observed cosmological dark matter (DM).

This possibility was explored in Refs. [89, 90] in a general context. Here, we just summarize the results. It was found that the relic Majorons can account for the observed cosmological dark matter abundance provided

$$\Gamma_{J\nu\nu} < 1.3 \times 10^{-19} \text{ s}^{-1} , \ \ 0.12 \text{ keV} < \beta \, m_J < 0.17 \text{ keV} , \tag{4.19}$$

where $\Gamma_{J\nu\nu}$ is the decay width of $J \to \nu\nu$ and $m_J$ is the Majoron mass. The parameter $\beta$ encodes our ignorance about the number density of Majorons, being normalized to $\beta = 1$ if the Majoron was in thermal equilibrium in the early Universe decoupling sufficiently early, when all other degrees of freedom of the standard model were excited [90]. In the following we will follow their choice and will take

$$10^{-5} < \beta < 1, \tag{4.20}$$

and calculate both the width into neutrinos as well as the subleading one-loop induced decay into photons.

### 4.5.1   Decay into neutrinos

We now proceed with the computation of the Majoron decay width into neutrinos, which will be useful to obtain the allowed parameter space for which the Majoron can be a viable DM candidate. In order to calculate the decay amplitude we recall that the coupling $g_{J\nu_i\nu_j}$ is defined through

$$\mathcal{L} = -\frac{1}{2} g_{J\nu_i\nu_j} J \nu_i \nu_j + \text{ h.c.} \tag{4.21}$$

For the evaluation of $g_{J\nu_i\nu_j}$, we follow the steps developed in Ref. [75]. First we notice that with scalar potential defined in section 4.3, the Majoron, in the basis $[\text{Im}(\phi_i^0), \text{Im}(\Delta^0), \text{Im}(\sigma^0)]^T$, is given by

$$J = N_J \left[ 2u_\Delta^2 \frac{v}{\sqrt{3}}, 2u_\Delta^2 \frac{v}{\sqrt{3}}, 2u_\Delta^2 \frac{v}{\sqrt{3}}, u_\Delta v^2, u_\sigma(4u_\Delta^2 + v^2) \right], \tag{4.22}$$

and

$$N_J = \left[ 4v^2 u_\Delta^4 + v^4 u_\Delta^2 + u_\sigma^2 (4u_\Delta^2 + v^2)^2 \right]^{-1/2} \simeq \frac{1}{v^2 u_\sigma}, \tag{4.23}$$

where the last equality follows from the assumed hierarchy $u_\Delta \ll v \ll u_\sigma$ implied by the vev-seesaw relation. Using this, one can obtain

$$g_{J\nu_i\nu_j} = -\frac{m_i^\nu \delta_{ij}}{\sqrt{2}} \frac{1}{u_\sigma}, \tag{4.24}$$

leading to the decay width

$$\Gamma_{J\nu\nu} = \frac{m_J}{32\pi} \frac{\sum_i (m_i^\nu)^2}{2u_\sigma^2} . \tag{4.25}$$

It is worth mentioning that the sum $\sum_i (m_i^\nu)^2$ is in our framework constrained by the special form of the effective neutrino mass matrix shown in Eq. (4.16). In particular, there is a lower bound on the mass of the lightest neutrino: $m \gtrsim 0.03$ eV, as we saw in section 4.4.

### 4.5.2   Decay into photons

The Majoron also couples with photons (at the quantum level) and therefore the radiative decay $J \to \gamma\gamma$ is expected to occur with a photon energy $E_\gamma \simeq m_J/2$. Consequently, this decay exhibits a mono-energetic emission line which could be detected in a variety of X-ray observatories, see for example the discussion given in Refs. [15, 90].

The effective Majoron-photon interaction can be written as

$$\mathcal{L} = g_{J\gamma\gamma} \varepsilon^{\mu\nu\alpha\beta} F_{\mu\nu} F_{\alpha\beta}, \tag{4.26}$$

resulting from the one-loop diagrams shown in Fig. 4.3 (top diagrams). The effective coupling $g_{J\gamma\gamma}$ (bottom graph in Fig. 4.3) is

$$g^f_{J\gamma\gamma} \equiv \frac{N_f \alpha^2 g_{Jff} Q_f^2 X_f}{8\pi m_f}, \qquad (4.27)$$

with $X_f = -2m_f^2 C_0(0, 0, m_J^2, m_f^2, m_f^2, m_f^2) \simeq 1 + m_J^2/(12m_f^2)$ where $C_0$ is the invariant Passarino-Veltman loop function [91]. The last approximation is valid for $m_J \ll m_f$. $T_3^f$, $Q_f$ and $N_f$ denote the weak isospin, the electric charge and the colour factor of the corresponding charged fermion $f$, respectively. The coupling of the Majoron to the charged fermions $g_{Jff}$ is given by [90]

$$g_{Jff} = -\frac{2u_\Delta^2}{v^2 u_\sigma} m_f (-2T_3^f). \qquad (4.28)$$

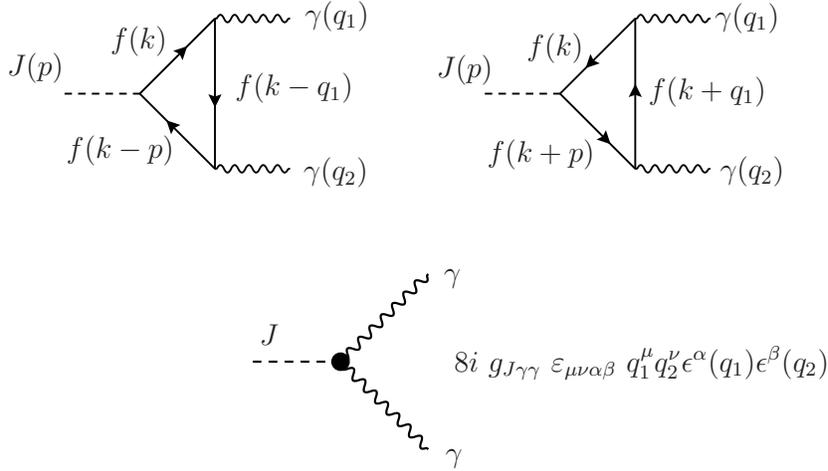

**Figure 4.3:** Top: One loop diagrams for the decay $J \to \gamma\gamma$. Bottom: Effective $J\gamma\gamma$ vertex.

We then get for the decay width,

$$\begin{aligned}
\Gamma_{J\gamma\gamma} &= \frac{m_J^3}{\pi} \left| \sum_f g^f_{J\gamma\gamma} \right|^2 = \frac{\alpha^2 m_J^3}{64\pi^3} \left| \sum_f \frac{N_f g_{Jff} Q_f^2 X_f}{m_f} \right|^2 = \\
&= \frac{\alpha^2 m_J^3}{64\pi^3} \left| \sum_f N_f Q_f^2 \frac{2u_\Delta^2}{v^2 u_\sigma} (-2T_3^f) \frac{m_J^2}{12 m_f^2} \right|^2,
\end{aligned} \qquad (4.29)$$

where the cancelation of the anomalous contribution has been taken into account. The above calculation was performed using the Mathematica package FeynCalc [92]. Here we give a more pedestrian derivation of (4.29). Take the diagrams in figure 4.4. The integral associated with the left diagram is

$$T_I^{\mu\nu} = (-1) \int \frac{d^4 q}{(2\pi)^4} \frac{\text{Tr}\left(\gamma_5(\not{p} + \not{q} + m_f)\gamma^\mu(\not{p}_2 + \not{q} + m_f)\gamma^\nu(\not{q} + m_f)\right)}{\left((p+q)^2 - m_f^2\right)\left((p_2+q)^2 - m_f^2\right)(q^2 - m_f^2)} \qquad (4.30)$$

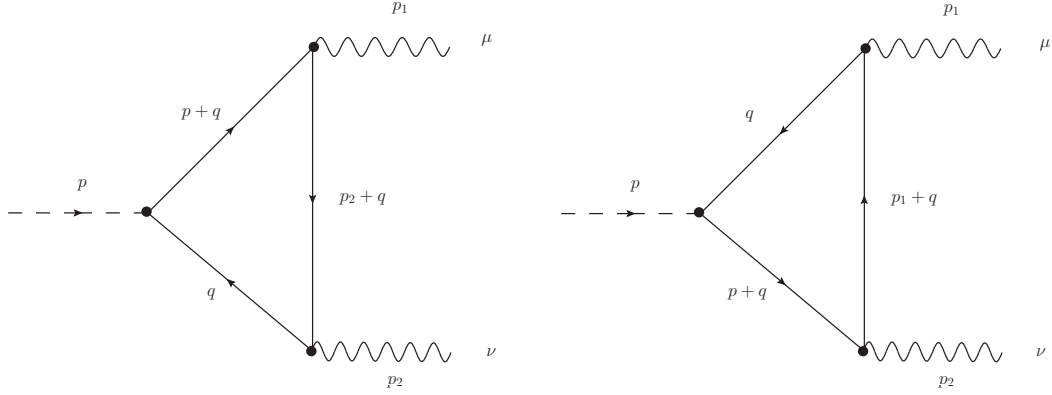

**Figure 4.4:** Radiative majoron decay into photons

and the right diagram gives an integral $T_{II}^{\mu\nu}$ that is obtained from this one with $p_2, (\mu, \nu) \longrightarrow p_1, (\nu, \mu)$.

The trace is easily computed:

$$
\begin{aligned}
&\text{Tr}\left(\gamma_5(\slashed{p}+\slashed{q}+m_f)\gamma^\mu(\slashed{p}_2+\slashed{q}+m_f)\gamma^\nu(\slashed{q}+m_f)\right) = \quad (4.31)\\
&m_f\text{Tr}\left(\gamma_5\gamma^\mu(\slashed{p}_2+\slashed{q})\gamma^\nu\slashed{q}\right) +\\
&m_f\text{Tr}\left(\gamma_5(\slashed{p}+\slashed{q})\gamma^\mu\gamma^\nu\slashed{q}\right) +\\
&m_f\text{Tr}\left(\gamma_5(\slashed{p}+\slashed{q})\gamma^\mu(\slashed{p}_2+\slashed{q})\gamma^\nu\right) =\\
&-4im_f\epsilon^{\mu\nu\rho\sigma}\left((p+q)_\rho q_\sigma - (p_2+q)_\rho q_\sigma - (p+q)_\rho(p_2+q)_\sigma\right) =\\
&= 4im_f\epsilon^{\mu\nu\rho\sigma}p_{1\rho}p_{2\sigma},
\end{aligned}
$$

which gives for $T_I^{\mu\nu}$

$$
T_I^{\mu\nu} = -4im_f\epsilon^{\mu\nu\rho\sigma}p_{1\rho}p_{2\sigma}\int\frac{d^4q}{(2\pi)^4}\frac{1}{\left((p+q)^2-m_f^2\right)\left((p_2+q)^2-m_f^2\right)(q^2-m_f^2)}. \quad (4.32)
$$

We see that the integral is in fact convergent, although naively we would expect it to be linearly divergent. Also, this expression is symmetric under the interchange $(p_1, p_2), (\mu, \nu) \longrightarrow (p_2, p_1), (\nu, \mu)$ and the change of variables $q \longrightarrow -q - p$ in the integration, which gives $T_{II}^{\mu\nu}$. So, we have

$$
\begin{aligned}
T^{\mu\nu} &= T_I^{\mu\nu} + T_{II}^{\mu\nu} \quad (4.33)\\
&= 2T_I^{\mu\nu}\\
&= -8im_f\epsilon^{\mu\nu\rho\sigma}p_{1\rho}p_{2\sigma}\int\frac{d^4q}{(2\pi)^4}\frac{1}{\left((p+q)^2-m_f^2\right)\left((p_2+q)^2-m_f^2\right)(q^2-m_f^2)}.
\end{aligned}
$$

Now we combine denominators with Feynman parameters to perform the integration: let us call $I$ the scalar integral

$$I = \int \frac{d^4q}{(2\pi)^4} \frac{1}{\left((p+q)^2 - m_f^2\right)\left((p_2+q)^2 - m_f^2\right)\left(q^2 - m_f^2\right)}. \tag{4.34}$$

We have

$$I = 2 \int \frac{d^4q}{(2\pi)^4} \int_0^1 dx \int_0^{1-x} dy \tag{4.35}$$

$$\frac{1}{\left(x\left((p+q)^2 - m_f^2\right) + y\left((p_2+q)^2 - m_f^2\right) + (1-x-y)(q^2 - m_f^2)\right)^3}.$$

Using $p^2 = m_J^2$ and $p_2^2 = 0$ the denominator can be cast in the form

$$\text{denom.} = q^2 - m_f^2 + x(1-x-y)m_J^2 = q^2 - C, \tag{4.36}$$

with $C = m_f^2 - x(1-x-y)m_J^2$. Then the scalar integral is

$$I = 2 \int_0^1 dx \int_0^{1-x} dy \int \frac{d^4q}{(2\pi)^4} \frac{1}{(q^2 - C)^3}.$$

Now we change to Euclidean momentum by a Wick rotation $q_0 \longrightarrow iq_{0E}$ to get

$$\begin{aligned}
I &= 2i \int_0^1 dx \int_0^{1-x} dy \int \frac{d^4q_E}{(2\pi)^4} \frac{1}{(q_E^2 + C)^3} \tag{4.37}\\
&= 2i \int_0^1 dx \int_0^{1-x} dy \int \frac{q_E^3 dq_E d\Omega_3}{(2\pi)^4} \frac{1}{(q_E^2 + C)^3}\\
&= 2i \int_0^1 dx \int_0^{1-x} dy \int d\Omega_3 \int_0^{\infty} du \frac{1}{2(2\pi)^4} \frac{u}{(u+C)^3}\\
&= 2i \int_0^1 dx \int_0^{1-x} dy \int d\Omega_3 \frac{1}{4(2\pi)^4 C}.
\end{aligned}$$

Using the result

$$\int d\Omega_3 = 2\pi^2 \tag{4.38}$$

we get

$$I = \frac{i}{16\pi^2} \int_0^1 dx \int_0^{1-x} dy \frac{1}{C}. \tag{4.39}$$

Remembering that $C = m_f^2 - x(1-x-y)m_J^2$ we then have

$$C^{-1} = m_f^{-2} \left( 1 - x(1-x-y)\frac{m_J^2}{m_f^2} \right)^{-1} \tag{4.40}$$

$$= m_f^{-2} \left( 1 + x(1-x-y)\frac{m_J^2}{m_f^2} + \dots \right) \tag{4.41}$$

and putting this in (4.39) we obtain

$$I = \frac{i}{16\pi^2 m_f^2} \int_0^1 dx \int_0^{1-x} dy \left( 1 + x(1-x-y)\frac{m_J^2}{m_f^2} + \dots \right) \tag{4.42}$$

$$= \frac{i}{32\pi^2 m_f^2} \left( 1 + \frac{m_J^2}{12 m_f^2} + \dots \right) \tag{4.43}$$

which gives for $T^{\mu\nu}$

$$T^{\mu\nu} = -8 i m_f \epsilon^{\mu\nu\rho\sigma} p_{1\rho} p_{2\sigma} I \tag{4.44}$$

$$= \epsilon^{\mu\nu\rho\sigma} \frac{p_{1\rho} p_{2\sigma}}{4\pi^2 m_f} \left( 1 + \frac{m_J^2}{12 m_f^2} + \dots \right). $$

Noting that the $J\bar{f}f$ coupling is $g_{Jff}$, that the electromagnetic one is proportional to $eQ_f$ and summing over all fermions we are able to write the $M$ matrix. The Feynman rules applied to the diagrams give

$$M = -\sum_f g_{Jff} N_f Q_f^2 e^2 T^{\mu\nu} \epsilon_\mu(p_1) \epsilon_\nu(p_2), \tag{4.45}$$

where $\epsilon_\mu(p)$ is the photon polarization vector and we can add the amplitudes for the two diagrams because the analytic expressions are the same. The coupling $g_{Jff}$ is

$$g_{Jff} = -\frac{2v_3^2}{v_2^2 v_1} m_f(-2T_{3f}) \tag{4.46}$$

which allows us to eliminate the anomalous like contribution, giving

$$M = \frac{2v_3^2 \alpha}{\pi v_2^2 v_1} \epsilon^{\mu\nu\rho\sigma} \epsilon_\mu(p_1) \epsilon_\nu(p_2) p_{1\rho} p_{2\sigma} \sum_f N_f Q_f^2 (-2T_{3f}) \left( \frac{m_J^2}{12 m_f^2} \right) \tag{4.47}$$

This gives for the square of $M$

$$|M|^2 = \frac{4\alpha^2}{\tilde{\Lambda}_\gamma^2 \pi^2} 2(p_1 \cdot p_2)^2 \tag{4.48}$$

$$= \frac{\alpha^2}{2\tilde{\Lambda}_\gamma^2} \frac{m_J^4}{\pi^2}$$

with

$$\tilde{\Lambda}_\gamma = \frac{1}{\sum_f N_f Q_f^2 (-2T_{3f})(1/12)(m_f^2/m_f^2)} \frac{2v_2^2 v_1}{v_3^2}. \tag{4.49}$$

In obtaining (4.48) we used the replacement rule

$$\sum_s \epsilon^{\mu*}(p)\epsilon^\nu(p) \longrightarrow -g^{\mu\nu} \tag{4.50}$$

and the identity

$$\epsilon^{\mu\nu\rho\sigma}\epsilon_{\mu\nu\alpha\beta} = -2\left(\delta_\alpha^\rho \delta_\beta^\sigma - \delta_\beta^\rho \delta_\alpha^\sigma\right). \tag{4.51}$$

The decay width is

$$\Gamma = \frac{1}{16\pi} \frac{1}{m_J} |M|^2 \frac{1}{2} \tag{4.52}$$

and so finally

$$\Gamma = \frac{\alpha^2}{64\pi^3} \frac{m_J^3}{\tilde{\Lambda}_\gamma^2}. \tag{4.53}$$

### 4.5.3 Numerical results

In this section we discuss some numerical results regarding the implementation of the decaying Majoron dark matter hypothesis in our scenario. In Ref. [90] it was shown that the experimental limit in the Majoron decay rate into photons is of the order of $10^{-30}$ s$^{-1}$. It was also shown that, in a generic seesaw model, a sizeable triplet vev plays a crucial role in bringing the decay rate close to this experimental bound. Here we have computed the width of the Majoron into neutrinos and photons in our extended seesaw model which incorporates the $A_4$ flavor symmetry, generalizing the models of Ref. [46]. The results are shown in Fig. 4.5. These take into account the current neutrino oscillation data, discussed in section 4.4. We chose five values for the triplet vev, $u_\Delta = 1$ eV (turquoise), 100 eV (dark green), 10 keV (magenta), 1 MeV (grey) and 10 MeV (dark blue) and 100 MeV (black). For the right panel we consider only points that satisfy the WMAP constraint (4.19) indicated by the red horizontal band on the top of the left plot.

In order to be able to probe our decaying Majoron dark matter scenario through the mono-energetic emission line one must be close to the present experimental limits on the photon decay channel, discussed in Ref. [90] and references therein. As mentioned, this requires the triplet vev to be sizeable, as shown on the right panel of Fig. 4.5 for the same choices of $u_\Delta$. In principle there is an additional lower bound on the Majoron mass coming from the Tremaine-Gunn argument [93], which, for fermionic dark matter would be around 500 eV. Under certain assumptions this bound could be extended to bosons, and is expected to be somewhat weaker [94]. The upper orange shaded region is the excluded region from X-ray observations given in Ref. [15]. One should point out that, in this model, because of the vev seesaw relation $u_\Delta u_\sigma \sim v^2$ one cannot arbitrarily take large values for $u_\Delta$ to enhance $\Gamma_{J\gamma\gamma}$ because then the singlet vev gets correspondingly smaller values, hence reducing the lifetime of the Majoron to values in conflict with the WMAP constraint. This interplay between the CMB bounds and the detectability of the gamma line is illustrated in Fig. 4.5, where the dark-blue points corresponding to $u_\Delta = 10$ MeV illustrate the experimental sensitivity to our signal.

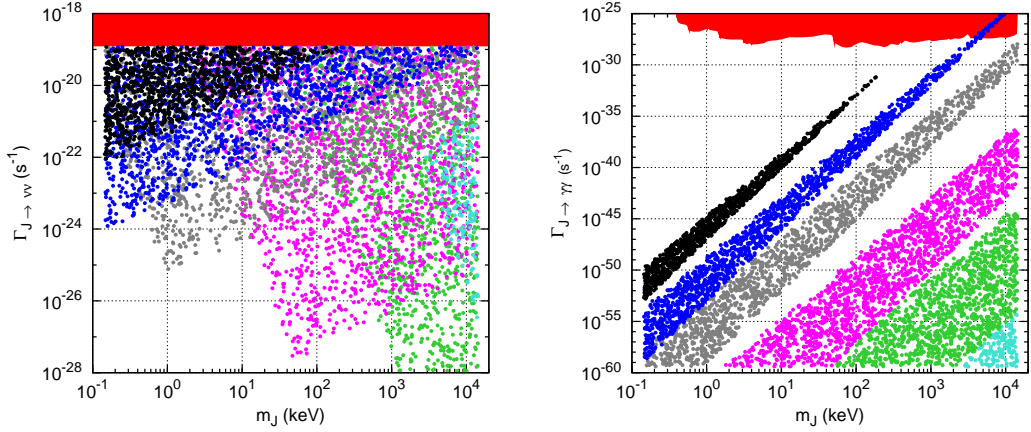

**Figure 4.5:** Left panel: $\Gamma_{J\nu\nu}$ as function of the Majoron mass respecting Eq. (4.19) for $u_\Delta =1$ eV (turquoise), 100 eV (dark green), 10keV (magenta), 1MeV (grey), 10MeV (dark blue) and 100 MeV (black). Right panel: $\Gamma_{J\gamma\gamma}$ as function of the Majoron mass for the same values of the triplet vev as in the left panel. The upper orange shaded region is the excluded region from X-ray observations taken from Ref.[15].

## 4.6 Conclusions

We have studied the possibility that the seesaw model with spontaneously broken ungauged lepton number may simultaneously account for the observed neutrino masses and mixing as well as the dark matter of the Universe. We have presented a two-texture structure for the neutrino mass which arises in a specific seesaw scheme implementing an $A_4$ flavor symmetry. A predictive pattern of neutrino masses emerges from the interplay of type-I and type-II seesaw contributions, with a lower bound on the neutrinoless double beta decay rate, which correlates with the deviation from maximality of the atmospheric mixing angle $\theta_{23}$, as well as nearly maximal CP violation, correlated with the reactor angle $\theta_{13}$.

On the other hand, assuming that associated Majoron picks up a mass due to explicit lepton number violating effects that may arise, say, from quantum gravity, we showed how it can constitute a viable candidate for decaying dark matter, consistent with cosmic microwave background lifetime constraints that follow from current WMAP observations. We have also shown how the Higgs boson triplet, the existence of which is required by the consistency of the model, plays a key role in providing a test of the decaying Majoron dark matter hypothesis, implying the existence of a mono-energetic emission line which arises from the sub-leading one-loop-induced decay of the Majoron into photons. We also discussed the possibility of probing its existence in future X-ray observations such as expected in NASA's Xenia mission [95]. The presence of the type-II seesaw Higgs triplet would also have other particle physics implications, such as lepton flavor violating decay rate enhancements due to the strong renormalization effects of the triplet, quite sizeable in a supersymmetric model.

# Chapter 5

# Seesaw Models and Lepton Flavour Violation

At "low" energies one cannot decide whether tree-level or loop physics generates the operator equation (1.176) responsible for neutrino masses, nor can any measurements of neutrino angles, phases or masses distinguish between the different tree-level seesaw realizations. Under the assumption of a pure type-I or pure type-II minimal supergravity seesaw mechanisms, we reconsider here the prospects for reconstructing the underlying high energy parameters from a combination of different measurements. Clearly, as mentioned at the end of section 1.2.3, observables outside the neutrino sector are needed in order to ultimately learn about the high energy parameters characterizing the seesaw. If the CERN LHC, due to take first data, finds signs of electroweak scale supersymmetry, indirect insight into the high-energy world might become possible through the search for flavour violation effects [96, 97].

Starting from flavour diagonal soft supersymmetry breaking terms at some high energy "unification" scale, flavour violation appears at lower energies due to the renormalization group evolution of the soft breaking parameters [98, 99]. If the seesaw mechanism is responsible for the observed neutrino masses, the neutrino Yukawa couplings leave their imprint in the slepton mass matrices as first shown in [100]. Potentially large LFV is then induced by the flavour off-diagonal structure in the Yukawa couplings required by the large mixing angles observed in oscillation experiments [28]. Expectations for LFV decays such as $l_i \rightarrow l_j + \gamma$ and $l_i \rightarrow 3l_j$ in the supersymmetric seesaw have been studied in [84, 101–105]. For the related process of $\mu - e$ conversion in nuclei see, for example [106, 107]. The potential of LHC experiments in probing the allowed seesaw parameters through measurements of masses and branching ratios of supersymmetric particles has also been discussed in Refs. [108–112].

In two previous studies [113, 114] it was pointed out that ratios of branching ratios are especially useful for learning about the unknown seesaw parameters. In [114] the case of type-I seesaw was discussed, whereas [113] addresses the case of seesaw type-II. For the type-I seesaw, there are in general too many unknown parameters that preclude making any definite predictions for LFV decays. In contrast, in the simplest type-II seesaw model (with only one triplet coupling to standard model leptons) neutrino mixing angles can be related to ratios such as $\mathrm{Br}(\tilde{\tau}_2 \to e\chi_1^0)/\mathrm{Br}(\tilde{\tau}_2 \to \mu\chi_1^0)$.

It has been shown that, to a good approximation, such ratios do not depend on the mSUGRA parameter values. However, from an experimental point of view, calculations of absolute event rates are needed, before ratios of different final state channels can be studied. In [113, 114] were taken as reference just a few benchmark mSUGRA points, for which we have made detailed studies. In the work [5], that is described here, we have calculated branching ratios and event rates over a large region of mSUGRA parameter space, in order to identify the maximal number of events one can expect in experiments at the LHC, while still respecting all low-energy constraints.

## 5.1 Theory setup

In order to fix the notation, we will briefly review the main features of the seesaw mechanism and mSUGRA. As already discussed in section 1.2.3, the type-I supersymmetric seesaw consists in extending the particle content of the MSSM by three gauge singlet "right-handed" neutrino superfields. The leptonic part of the superpotential is then

$$W = Y_e^{ji}\widehat{L}_i\widehat{H}_d\widehat{E}_j^c + Y_\nu^{ji}\widehat{L}_i\widehat{H}_u\widehat{N}_j^c + M_i\widehat{N}_i^c\widehat{N}_i^c, \quad i,j = 1,\dots,3, \qquad (5.1)$$

where $Y_e$ and $Y_\nu$ denote the charged lepton and neutrino Yukawa couplings, while $\widehat{N}_i^c$ are the "right-handed" neutrino superfields with $M_i$ Majorana mass terms. One can always choose a basis on which the Majorana mass matrix of the "right-handed" neutrinos is brought to diagonal form $\hat{M}_R = \mathrm{diag}(M_1, M_2, M_3)$. Without loss of generality we will also assume that equation (5.1) is written on the basis where the charged lepton Yukawa matrix is already diagonal. In this simple setup, the type-I seesaw model, as defined by equation (5.1), is characterized by a total of 21 parameters, from which only 12 are measurable in the low-energy theory, as we discuss below.

The effective mass matrix of the "left-handed" neutrinos at low energies is then given as (see section 1.2.3)

$$m_\nu = -\frac{v_u^2}{2}Y_\nu^T \cdot \hat{M}_R^{-1} \cdot Y_\nu, \qquad (5.2)$$

so that, for each "right-handed" neutrino, there is one non-zero eigenvalue in $m_\nu$. In equation (5.2) we use the notation $\langle H_{u,d}\rangle = \frac{v_{u,d}}{\sqrt{2}}$ for the vacuum expectation values of the neutral components of the Higgs boson doublets.

The parameters of equation(5.1) are defined at the Grand Unified Theory (GUT) scale, whereas the entries of equation (5.2) are measured at low energies. In order to connect these two scales we numerically solve the full set of renormalization group equations (RGE) [104, 115].

As already discussed on Chapter 1, the light Majorana neutrino mass matrix in equation (5.2), being complex symmetric, is diagonalized by a unitary $3 \times 3$ matrix $U$ [53] as shown in (1.184). Inverting the seesaw equation, equation (5.2), allows to express $Y_\nu$ as [21]

$$Y_\nu = \sqrt{2}\frac{i}{v_u}\sqrt{\hat{M}_R} \cdot R \cdot \sqrt{\hat{m}_\nu} \cdot U^\dagger, \qquad (5.3)$$

where $\hat{m}_\nu$ is the diagonal matrix with $m_i$ eigenvalues and in general $R$ is a complex orthogonal matrix. Note that, in the special case $R = 1$, $Y_\nu$ contains only "diagonal" products $\sqrt{M_i m_i}$. In this simplified case the 18 parameters in $Y_\nu$ are reduced to 12. Note also that in general type-I seesaw schemes, the unitary matrix diagonalizing the effective neutrino mass matrix differs from the lepton mixing matrix by terms of order $D/M_R$, where the $D = Y_\nu v_u$. For the high-scale schemes considered here one can safely neglect these deviations [1]. In this case we can set the diagonalization matrix as the lepton mixing matrix (partially) determined in neutrino oscillation measurements.

Implementing the type-II seesaw mechanism within supersymmetry requires at least two $SU(2)$ triplet states $T_{1,2}$ for the cancelation of anomalies and because of the superpotential holomorphy. A scalar triplet with mass below the GUT scale changes the running of $g_1$ and $g_2$ in an unwanted way and gauge coupling unification shown in section 1.1.4 is lost. If instead one adds only complete $SU(5)$ multiplets (or GUT multiplets which can be decomposed into complete $SU(5)$ multiplets) to the standard model particle content, the scale where couplings unify remains the same (at one loop level), only the value of the GUT coupling itself changes [116].

Our numerical calculation uses an $SU(5)$ inspired model [86, 87], which adds a pair of **15** and $\overline{\mathbf{15}}$ to the Minimal Supersymmetric Standard Model (MSSM) particle spectrum. This variant of the type-II seesaw mechanism, as discussed above, allows us to maintain gauge coupling unification even for $M_T \ll M_G$, $M_G$ being the unification scale. Under $SU(3) \times SU_L(2) \times U_Y(1)$ the **15** decomposes as

$$\begin{aligned}\mathbf{15} &= S + T + Z \\ S &\sim (6, 1, -\frac{2}{3}), \qquad T \sim (1, 3, 1), \qquad Z \sim (3, 2, \frac{1}{6}).\end{aligned} \qquad (5.4)$$

$T$ has the correct quantum numbers to generate the dimension-5 operator of equation (1.176). The $SU(5)$ invariant superpotential reads

$$\begin{aligned}W &= \frac{1}{\sqrt{2}}\mathbf{Y}_{ij}^{15}\bar{\mathbf{5}}_i \cdot \mathbf{15} \cdot \bar{\mathbf{5}}_j + \frac{1}{\sqrt{2}}\lambda_1\bar{\mathbf{5}}_H \cdot \mathbf{15} \cdot \bar{\mathbf{5}}_H + \frac{1}{\sqrt{2}}\lambda_2\mathbf{5}_H \cdot \overline{\mathbf{15}} \cdot \mathbf{5}_H + \mathbf{Y}_{ij}^5\mathbf{10}_i \cdot \bar{\mathbf{5}}_j \cdot \bar{\mathbf{5}}_H \\ &+ \mathbf{Y}_{ij}^{10}\mathbf{10}_i \cdot \mathbf{10}_j \cdot \mathbf{5}_H + M_{15}\mathbf{15} \cdot \overline{\mathbf{15}} + M_5\bar{\mathbf{5}}_H \cdot \mathbf{5}_H \ .\end{aligned} \qquad (5.5)$$

Here, $\bar{\mathbf{5}} = (d^c, L)$, $\mathbf{10} = (u^c, e^c, Q)$, $\mathbf{5}_H = (t, H_2)$ and $\bar{\mathbf{5}}_H = (\bar{t}, H_1)$. Below the GUT scale,

---

[1] However for other type-I schemes, like the inverse seesaw [84, 107] this approximation fails and leads to large LFV from right-handed neutrino exchange, even in the absence of supersymmetric contributions. For a systematic perturbative seesaw diagonalization method that covers all cases see Ref. [53].

in the $SU(5)$-broken phase, the superpotential contains the terms

$$\frac{1}{\sqrt{2}}(Y_T^{ij} L_i T_1 L_j + Y_S^{ij} d_i^c S d_j^c) + Y_Z^{ij} d_i^c Z L_j + Y_d^{ij} d_i^c Q_j H_d + Y_u^{ij} u_i^c Q_j H_u + Y_e^{ij} e_i^c L_j H_d$$

$$+\frac{1}{\sqrt{2}}(\lambda_1 H_d T_1 H_d + \lambda_2 H_u T_2 H_u) + M_T T_1 T_2 + M_Z Z_1 Z_2 + M_S S_1 S_2 + \mu H_d H_u \ . \quad (5.6)$$

As long as $M_Z \sim M_S \sim M_T$, gauge coupling unification will be preserved. Note that exact equality is not required for a successful unification. In our numerical studies we have taken into account the different running of these mass parameters.

Integrating out the heavy triplets at their mass scale, the dimension-5 operator of equation (1.176) is generated and after electroweak symmetry breaking the resulting neutrino mass matrix can be written as

$$m_\nu = \frac{v_u^2}{2} \frac{\lambda_2}{M_T} Y_T \ . \quad (5.7)$$

As in the case of the type-I seesaw, equation (5.7) depends on the energy scale. In order to compute the neutrino mass $m_\nu$ measured at low energies, one needs to know $\lambda_2$, $Y_T$ and $M_T$ as input parameters at the high energy scale. As will be discussed in section 5.2, one can use an iterative procedure in order to find the high scale parameters from the low energy measured quantities.

Note that, without loss of generality, we have the freedom to write eqs. (5.1) and (5.5) on the basis where the charged lepton mass matrix is diagonal, fitting the corresponding Yukawa couplings so as to reproduce the three measured charged lepton masses. However there are important differences between the type-I and type-II seesaw schemes. For example, in contrast to type-I, in a pure type-II seesaw scheme the unitary matrix $U$ that diagonalizes equation (5.7) coincides with the lepton mixing matrix studied in neutrino oscillations. Moreover, in sequential type-I seesaw for each "right-handed" neutrino added there is one non-zero light neutrino mass eigenstate [2]. In contrast, in type-II seesaw one can produce three neutrino masses with just one pair of triplet superfields, with only one triplet directly coupling to leptons. This implies that in the minimal type-II seesaw one has less parameters than in the sequential type-I seesaw. Indeed, as already mentioned in section 1.2.3, from the 12 parameters in the complex symmetric $Y_T$ matrix, one can remove 3 phases by redefining the charged leptons [53]. In addition, from the 3 complex parameters $\lambda_{1,2}$ and $M_T$, one does not enter, as only one of the triplets couples to leptons, and finally, two of the three phases can also be removed by field redefinitions. The net result is that there are only 11 physical parameters governing neutrino physics [113]. This number is substantially smaller than the 18 free parameters describing the simplest type-I

---

[2]We do not consider here the possibility of having just two right-handed neutrino states in the type-I seesaw, called (3,2) in Ref. [53]. This could well account for the current neutrino data with just 12 parameters, instead of the 18 characterizing the sequential (3,3) seesaw considered here.

seesaw scheme containing three "right-handed" neutrinos [117] [3].

At low energies a maximum of 9 neutrino parameters can be fixed by measuring lepton properties: 3 neutrino masses, 3 mixing angles and 3 CP phases. Thus from neutrino data only, neither type-I nor type-II seesaw schemes can be completely reconstructed, even in their simplest realizations. However, especially important in the following is the fact that low-energy neutrino angles are directly related to the high-energy Yukawa matrix in the type-II seesaw, whereas no such simple connection exists in the seesaw type-I (see also the discussion in [118]).

As already commented above and in Chapter 1, to a good approximation the lepton mixing matrix may be taken in unitary form, with three mixing angles $\theta_{ij}$, and three physical CP phases $\phi_{ij}$ [53]. Of these only the leptonic analogue of the Kobayashi-Maskawa phase $\delta$, taken as the invariant combination $\delta \equiv \phi_{12} - \phi_{13} + \phi_{23}$ would enter the class of LFV processes discussed in this paper, so that we get the standard form (1.186). Since no current experiment is sensitive enough to probe leptonic CP violation we take, for simplicity, $\delta = 0$. Neutrino oscillation experiments can be fitted with either a normal hierarchical spectrum (NH), or with an inverted hierarchy (IH) one. If one does not insist in ordering the neutrino mass eigenstates $m_{\nu_i}$, $i = 1, 2, 3$ with respect to increasing mass, the matrix $U$ can describe both possibilities without re-ordering of angles. In this convention, which we will use in the following, $m_{\nu_1} \simeq 0$ ($m_{\nu_3} \simeq 0$) corresponds to normal (inverse) hierarchy and $s_{12}$, $s_{13}$ and $s_{23}$ are the angles in both types of spectra. Basically $s_{12}$ is measured in solar + reactor experiments, $s_{23}$ in atmospheric + accelerator experiments and $s_{13}$ is constrained by reactor neutrino oscillation data.

In the general MSSM, LFV off-diagonal entries in the slepton mass matrices involve additional free parameters which arise from the mechanism of supersymmetry breaking. In order to relate LFV in the slepton sector with the LFV encoded in $Y_\nu$ or $Y_T$ one must assume some particular scheme for supersymmetry breaking. For simplicity and definiteness we will adopt mSUGRA boundary conditions, characterized by four continuous real and one discrete free parameter, usually denoted as

$$m_0, \ M_{1/2}, \ A_0, \ \tan\beta, \ \text{Sgn}(\mu) \ . \tag{5.8}$$

Here, $m_0$ is the common scalar mass, $M_{1/2}$ the gaugino mass and $A_0$ the common trilinear parameter, all defined at the grand unification scale, $M_G \simeq 2 \cdot 10^{16}$ GeV. The remaining two parameters are $\tan\beta = v_u/v_d$ and the sign of the Higgs mixing parameter $\mu$.

In order to have a qualitative understanding of the magnitudes of the LFV rates we first present approximate leading-log analytical solutions for the renormalization group equations [4]. For the case of type-I seesaw, the LFV elements induced in the charged left-

---

[3]We are treating the three charged lepton masses as experimentally determined parameters.

[4]Note that in the numerical code that leads to the results presented in our plots we have numerically solved the full set of RGEs.

slepton mass matrix by renormalization group evolution can be approximated as [104]

$$(\Delta M_{\tilde{L}}^2)_{ij} = -\frac{1}{8\pi^2}(3m_0^2 + A_0^2)(Y_\nu^\dagger L Y_\nu)_{ij}\,, \tag{5.9}$$

where $Y_\nu$ is given in terms of the neutrino parameters by equation (5.3) and the factor $L$ is defined as

$$L_{kl} = \log\Big(\frac{M_G}{M_k}\Big)\delta_{kl}\,. \tag{5.10}$$

Similarly, one can get an analogous approximate expression for the off-diagonal elements of the charged left-slepton mass matrix characterizing LFV in type-II seesaw schemes [86].

## 5.2 Numerical results

Due to the non-trivial structure of the neutrino Yukawa matrix $Y_\nu$ in equation (5.3) and of $Y_T$ in equation (5.7) for type-I and type-II seesaw, respectively, the slepton mass matrices contain calculable LFV entries [99, 100]. In order to determine their magnitude we solve the complete set of renormalization group equations, given in [86, 104, 115]. All results presented below have been obtained with the lepton flavour violating version of the program package SPheno [119], where the RGEs for the MSSM part are implemented at the 2-loop level. For definiteness we set neutrino mass squared differences to their current best fit values [28] and fix the angles to the Tri-Bi-Maximal (TBM) values [29].

Fixing the values of other mSUGRA parameters, we used SPheno to perform a numerical scan over the $m_0$-$M_{1/2}$ plane. For each point in this plane, we adjust the value of $M_R$ ($M_T$) in order to keep the low energy LFV observable BR($\mu \to e\gamma$) within its present experimental upper bound or within the expected sensitivity of the upcoming experiments [120].

For type-I seesaw our numerical procedure to fit these masses is as follows. As we have already commented, the large number of free parameters characterizing even the simplest type-I seesaw schemes forces us to make simplifying assumptions in inverting the seesaw equation, equation (5.3). As a first step we assume degenerate "right-handed" neutrinos and the simplest possible, flavourless, structure for the matrix $R$, i.e.

$$R = 1, \qquad \hat{M}_{Rij} = M_R\,\delta_{ij}\,. \tag{5.11}$$

Moreover, we fix the values of the light neutrino masses and Yukawa couplings to reproduce the TBM angle values. In order to determine the resulting LFV observables we numerically integrate the renormalization group equations taking into account the flavour structure of the $Y_\nu$ matrix. We integrate out every "right-handed" neutrino and its superpartner at the scale associated to its mass, and calculate the corresponding contribution to the dimension-five operator which is evolved to the electroweak scale. This way we obtain the exact neutrino masses and mixing angles for this first guess. The difference between the results numerically obtained and the input numbers is then minimized in an iterative procedure until convergence is achieved.

For the type-II seesaw the calculations are performed for the **15**-plet case, under the assumption $Y_Z = Y_T = Y_S$ at $M_G$, as discussed above, and including the one-loop RGEs for the new parameters in SPheno. For consistency, we have also included 1-loop threshold corrections for gauge couplings and gaugino mass parameters at the scale corresponding to the mass of the triplet, $M_T$. The MSSM part is implemented at the 2-loop level and, thus, in principle one should also consistently include the effect of the **15**-plets for all parameters at this level. However, as discussed in [113], the correct fit of the neutrino data requires that either the triplet (**15**-plet) Yukawa couplings are small and/or that $M_T$ is close to $M_G$, implying that the ratio $M_T/M_G$ is significantly smaller than $M_G/m_Z$ and thus one expects only small effects. Inverting the seesaw equation for any fixed value of $\lambda_2$ in equation (5.7), one can get a first guess of the Yukawa couplings for any fixed values of the light neutrino masses as a function of the corresponding triplet mass. This first guess will not give the correct Yukawa couplings, since the neutrino masses and mixing angles are measured at low energy, whereas for the calculation of $m_\nu$ we need to insert the parameters at the high-energy scale. However, we can use this first guess to numerically run the RGEs to obtain the exact neutrino parameters (at low energies) for these input values. The difference between the results obtained numerically and the input can then be minimized in a simple iterative procedure until convergence is achieved. As long as neutrino Yukawas are not large, convergence is reached in a few steps. However, in type-II seesaw schemes, the Yukawa couplings run stronger than in the type-I seesaw. Thus our initial guess can sizeable deviate from the exact Yukawa coupling values. Since neutrino oscillation data requires at least one neutrino mass to be larger than about 0.05 eV, we do not find any solutions for $M_T > 10^{15}$ GeV.

Finally, the calculation of cross sections for the production of supersymmetric particles was done using Prospino [121–125]. The input data was taken from SPheno using the SUSY Les Houches Accord standard format [126].

### 5.2.1 LFV stau decays

The eagerly awaited production of supersymmetric particles at the LHC would open new opportunities for the study of flavour violation in the supersymmetric sector [96]. Here we study how the LFV decays of staus may provide valuable cross-checks of neutrino properties determined at low energies as well as complementary information on the origin of neutrino mass.

The expected LFV branching ratios for $\tilde{\tau}_2 \rightarrow \mu + \chi_1^0$ and $\tilde{\tau}_2 \rightarrow e + \chi_1^0$ depend on the choice of the mSUGRA parameters. After a full scan over the mSUGRA parameter space we found that the dependence on $A_0$ and on the sign of $\mu$ is weaker, but that the rates decreased with increasing values of $\tan\beta$. Therefore, we chose our standard point with a relatively low value of $\tan\beta = 10$, and for definiteness took $\mu > 0$, and $A_0 = 0$. In Fig. 5.1 we show the contour plots for the LFV decays $\tilde{\tau}_2 \to \mu + \chi_1^0$ (left panel) and $\tilde{\tau}_2 \to e + \chi_1^0$ (right panel) in the $m_0, M_{1/2}$ plane for our standard choice of mSUGRA parameters for the simplest pure type-I seesaw scheme. One sees that there are regions in parameter space where the LFV decays of the $\tilde{\tau}_2$ can be as large as of order $10^{-1}$. In these plots the values of $M_R$ were chosen as to obtain the maximum LFV compatible with the present experimental limit of $Br(\mu \to e + \gamma) \leq 1.2 \cdot 10^{-11}$ [43]. Also shown in these plots are the exclusion regions coming from the LEP constraints on SUSY masses and also the exclusion obtained when the neutralino is not the LSP [5]. In Fig. 5.2 we show the same contour plots for $Br(\mu \to e + \gamma) \leq 10^{-13}$, which will be achievable in the coming experiments [120]. Also in this case one observes in Fig. 5.2 that the LFV stau decay rates may exceed the 10% level. Notice also that the nontrivial features present in both Figs. 5.1 and Fig. 5.2 reflect the well-known cancellations between chargino and neutralino contributions to $\mu \to e + \gamma$ already discussed above.

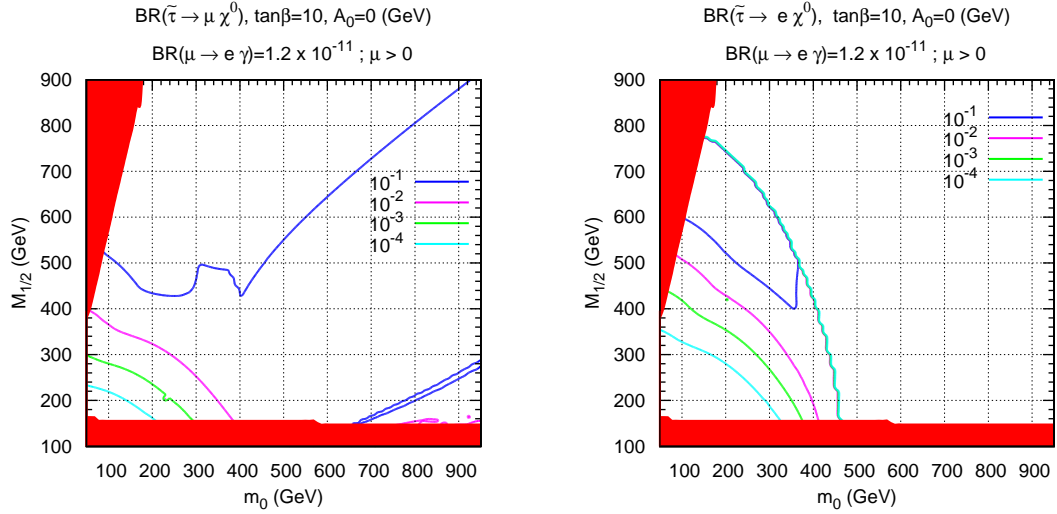

**Figure 5.1:** Br($\tilde{\tau}_2 \to \mu + \chi_1^0$) (left panel) and Br($\tilde{\tau}_2 \to e + \chi_1^0$) (right panel), in the $m_0, M_{1/2}$ plane for our standard choice of parameters: $\mu > 0$, $\tan\beta = 10$ and $A_0 = 0$ GeV, for type-I seesaw, imposing Br($\mu \to e + \gamma$) $\leq 1.2 \cdot 10^{-11}$.

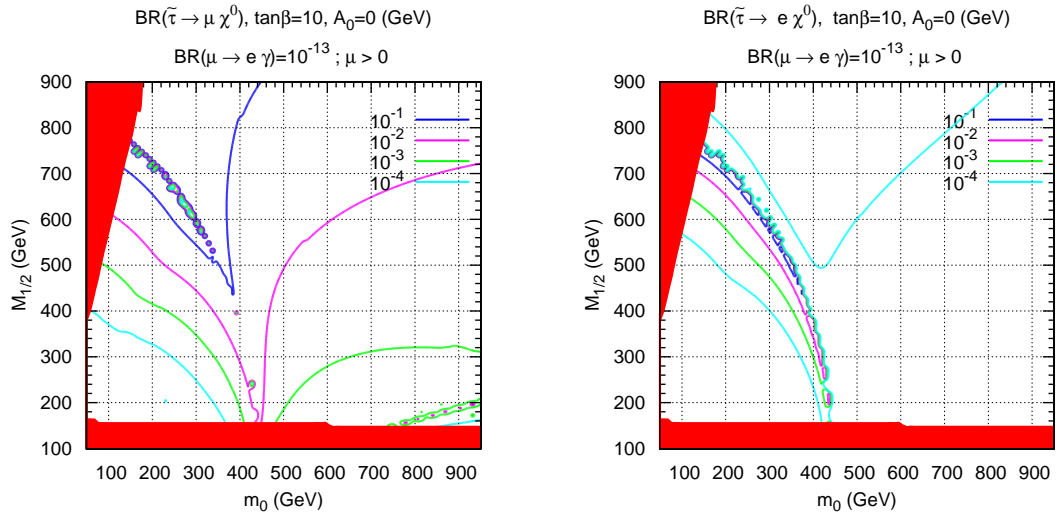

**Figure 5.2:** Br($\tilde{\tau}_2 \to \mu + \chi_1^0$) (left panel) and Br($\tilde{\tau}_2 \to e + \chi_1^0$) (right panel), in the $m_0, M_{1/2}$ plane for our standard choice of parameters: $\mu > 0$, $\tan\beta = 10$ and $A_0 = 0$ GeV, for type-I seesaw, imposing Br($\mu \to e + \gamma$) $\leq 10^{-13}$.

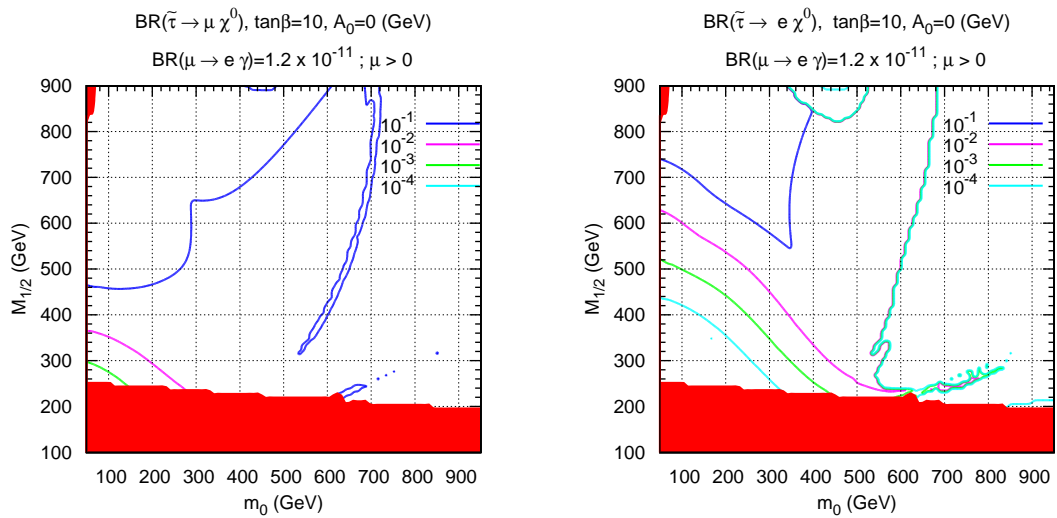

**Figure 5.3:** $Br(\tilde{\tau}_2 \to \mu + \chi_1^0)$ (left panel) and $Br(\tilde{\tau}_2 \to e + \chi_1^0)$ (right panel), in the $m_0$, $M_{1/2}$ plane for $\lambda_1 = 0.02$ and $\lambda_2 = 0.5$ and our standard choice of parameters: $\mu > 0$, $\tan \beta = 10$ and $A_0 = 0$ GeV, for type-II seesaw, imposing $Br(\mu \to e + \gamma) \leq 1.2 \cdot 10^{-11}$.

In Fig. 5.3 and Fig. 5.4 the same type of plots are shown for type-II seesaw. A comparison of these figures shows that, qualitatively, the behavior is very similar for the two types of seesaw. In both cases, the larger rates for $\tilde{\tau}_2 \to e + \chi_1^0$ are more constrained in parameter space than those for $\tilde{\tau}_2 \to \mu + \chi_1^0$. Notice however that there is an important difference between type-I and type-II seesaw, coming from the presence of the Higgs triplets that contribute sizeably to the running of the type-II beta functions. This gets reflected in the supersymmetric particle spectra and hence in the shapes of the red (shaded) regions in Fig. 5.3 and Fig. 5.4. One can observe, indeed, that the regions where the stau is the lightest supersymmetric particle, as well as the regions already excluded by LEP2 are substantially different for type-II seesaw, as compared to the corresponding ones for type-I. This follows from the modification in the beta functions introduced by the addition of the Higgs triplets, making $M_1$ and $M_2$ smaller in type-II than in type-I seesaw for the same value of $M_{1/2}$. The variation with the mSUGRA parameters is illustrated in Fig. 5.5 (type-I) and Fig. 5.6 (type-II) for the parameter $A_0$ and in Fig. 5.7 (type-I) and Fig. 5.8 (type-II) for $\tan\beta$. We can see that there is not much variation with $A_0$, while the rates decrease rapidly with increasing values of $\tan\beta$. The reason for this is that $\mathrm{BR}(\mu \to e + \gamma)$ increases along with $\tan^4\beta$, thus constraining more strongly the maximum attainable stau LFV rates. This effect is stronger for type-I as can be seen by noting the different values for the contour levels in Fig. 5.7 and Fig. 5.8. The variation with the sign of $\mu$ is weak and we do not show it here. So, in summary, large LFV rates prefer moderate values of $\tan\beta$ and this explains a posteriori the choice of our standard parameters.

---

[5]Note that we did not display the constraints coming from Dark Matter (DM) relic abundance.

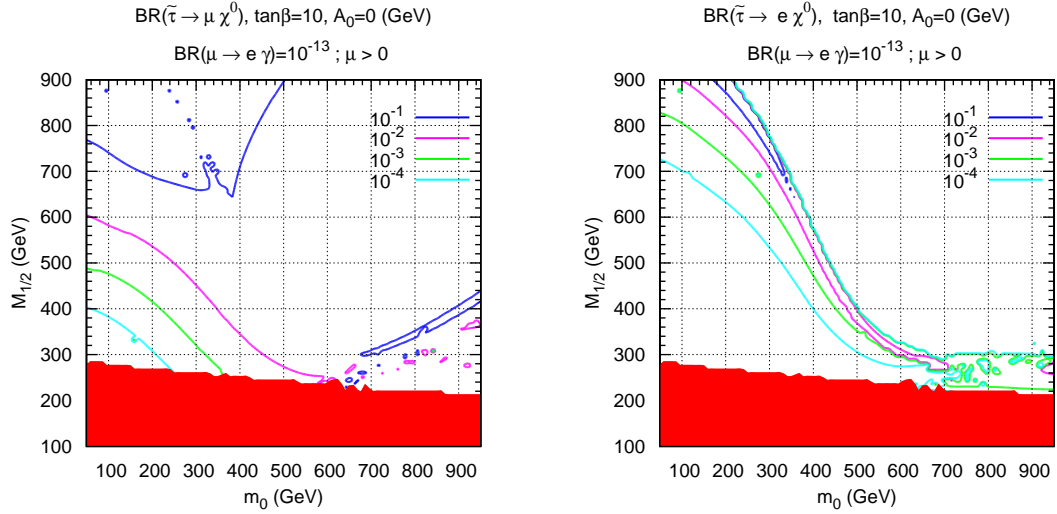

**Figure 5.4:** $\mathrm{Br}(\tilde{\tau}_2 \to \mu + \chi_1^0)$ (left panel) and $\mathrm{Br}(\tilde{\tau}_2 \to e + \chi_1^0)$ (right panel), in the $m_0, M_{1/2}$ plane, for $\lambda_1 = 0.02$ and $\lambda_2 = 0.5$ and our standard choice of parameters: $\mu > 0$, $\tan\beta = 10$ and $A_0 = 0$ GeV, for type-II seesaw, imposing $\mathrm{Br}(\mu \to e + \gamma) \leq 10^{-13}$.

### 5.2.2 Total production cross section of $\chi_2^0$

As important as having a large branching ratio into a LFV final state, it is to be able to produce a large enough event sample. In order to estimate the number of LFV events expected at the LHC, one notes that, from Figs. 5.1 - 5.8, in the regions where the LFV is sizeable, the direct production of staus at the LHC is negligible compared to that which arises from cascade decays of heavier neutralinos, mainly $\chi_2^0$. We focus on the $\chi_2^0$, because decays such as $\chi_2^0 \to \mu\tau\chi_1^0$ are sensitive to flavour violation, whereas in the corresponding chargino decays the flavour information is lost. Hence we first compute the total $\chi_2^0$ production cross section. In the left panel of Fig. 5.9 we show the results for the cross-section for $\chi_2^0$ production as a function of $M_{1/2}$, for different choices of $m_0$ and for our standard choice of mSUGRA parameters: $\mu > 0$, $\tan\beta = 10$ and $A_0 = 0$

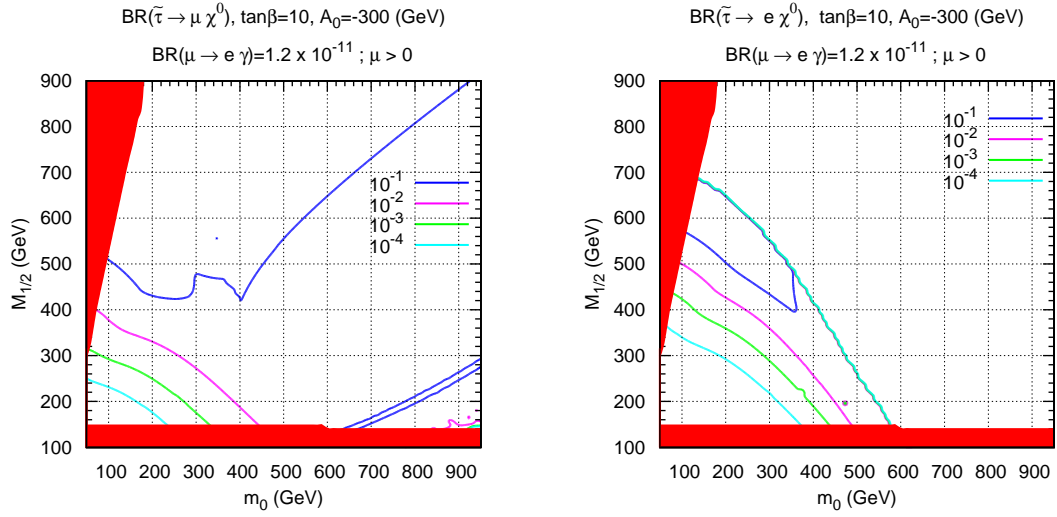

**Figure 5.5:** $\mathrm{Br}(\tilde{\tau}_2 \to \mu + \chi_1^0)$ (left panel) and $\mathrm{Br}(\tilde{\tau}_2 \to e + \chi_1^0)$ (right panel), in the $m_0, M_{1/2}$ plane for standard choice of parameters: $\mu > 0$, $\tan\beta = 10$ but different $A_0 = -300$ GeV, for type-I seesaw, imposing $\mathrm{Br}(\mu \to e + \gamma) \leq 1.2 \cdot 10^{-11}$.

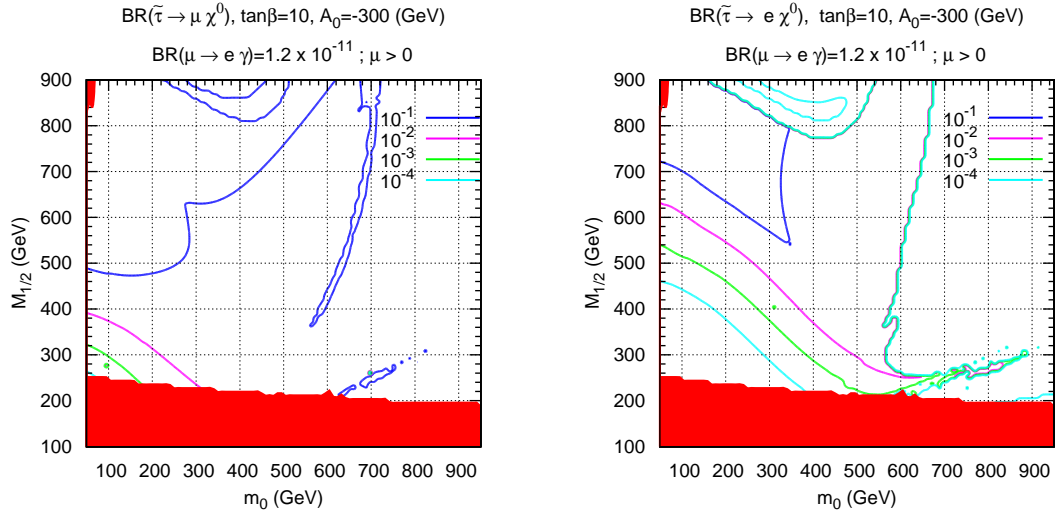

**Figure 5.6:** $\mathrm{Br}(\tilde{\tau}_2 \to \mu + \chi_1^0)$ (left panel) and $\mathrm{Br}(\tilde{\tau}_2 \to e + \chi_1^0)$ (right panel), in the $m_0, M_{1/2}$ plane for $\lambda_1 = 0.02$ and $\lambda_2 = 0.5$ and standard choice of parameters: $\mu > 0$, $\tan\beta = 10$ but different $A_0 = -300$ GeV, for type-II seesaw, imposing $\mathrm{Br}(\mu \to e + \gamma) \leq 1.2 \cdot 10^{-11}$.

GeV, for the pure type-I mSUGRA seesaw scheme. This choice of mSUGRA parameters corresponds, as will be discussed below, to the case where the branching ratios of the LFV stau decays are the largest. This result was obtained using the Prospino code [121–125] at Leading Order (LO) approximation. We have checked that the Next to Leading Order (NLO) calculation only changes the results slightly, due to an appropriate choice of the renormalization scale [121–125]. So, in all cross sections presented here, we only used the LO approximation. The corresponding results for type-II seesaw are shown in the right panel of Fig. 5.9, for the same choice of mSUGRA parameters and for $\lambda_1 = 0.02$ and $\lambda_2 = 0.5$.

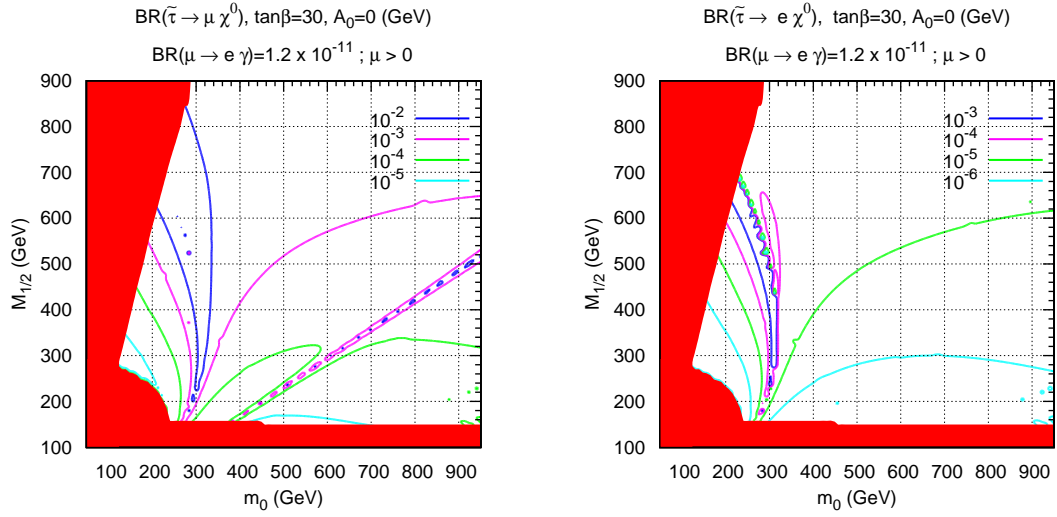

**Figure 5.7:** Br($\tilde{\tau}_2 \to \mu + \chi_1^0$) (left panel) and Br($\tilde{\tau}_2 \to e + \chi_1^0$) (right panel), in the $m_0, M_{1/2}$ plane for standard choice of parameters: $\mu > 0$, $A_0 = 0$ but different $\tan\beta = 30$, for type-I seesaw, imposing Br($\mu \to e + \gamma$) $\leq 1.2 \cdot 10^{-11}$.

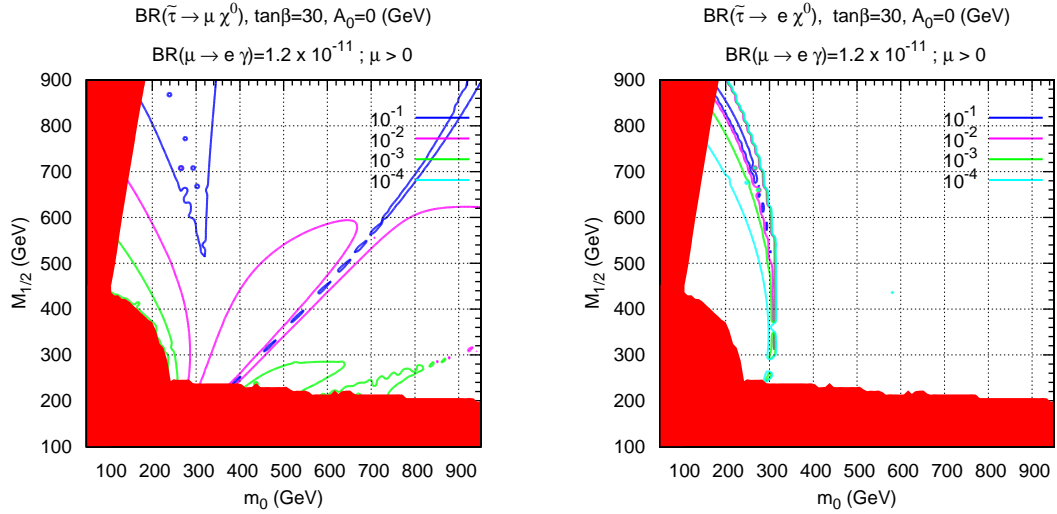

**Figure 5.8:** Br($\tilde{\tau}_2 \to \mu + \chi_1^0$) (left panel) and Br($\tilde{\tau}_2 \to e + \chi_1^0$) (right panel), for $\lambda_1 = 0.02$ and $\lambda_2 = 0.5$, in the $m_0, M_{1/2}$ plane for standard choice of parameters: $\mu > 0$, $A_0 = 0$, but different $\tan\beta = 30$, for type-II seesaw, imposing Br($\mu \to e + \gamma$) $\leq 1.2 \cdot 10^{-11}$.

### 5.2.3 Total production of $\chi_2^0$ times BR to $\mu$-$\tau$ lepton pair

In order to get an estimate of the expected number of LFV events at the LHC we now use a combination of the Prospino and SPheno codes to evaluate the product of the $\chi_2^0$ production cross section times the branching ratios into LFV processes. Once we know the luminosity at LHC we can multiply it by the above product to get the number of events.

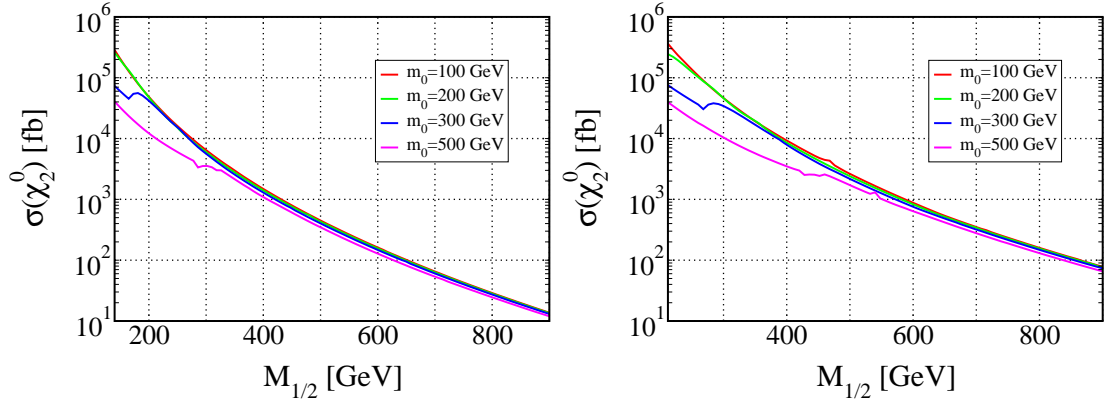

**Figure 5.9:** Production cross section (at leading order) of $\chi_2^0$ versus $M_{1/2}$ for varying $m_0$, and for our standard choice of parameters: $\mu > 0$, $\tan \beta = 10$ and $A_0 = 0$ GeV, in type-I seesaw (left panel) and type-II seesaw (right panel) for $\lambda_1 = 0.02$ and $\lambda_2 = 0.5$.

In Fig. 5.10, we have plotted, for type-I seesaw (left panel) and type-II (right panel), the production cross section at leading order of the second lightest neutralino $\sigma(\chi_2^0)$ times the BR of $\chi_2^0$ going to the opposite-sign dilepton signal $\chi_1^0 \mu \tau$ as a function of $M_{1/2}$, for different values of $m_0$. We have fixed the rest of the mSUGRA parameters to our standard mSUGRA point and imposed an upper limit on $\mathrm{Br}(\mu \to e + \gamma) \leq 1.2 \cdot 10^{-11}$. In type-I seesaw, the number of events of the opposite-sign dilepton signal $\chi_2^0 \to \chi_1^0 \mu \tau$ can be of the order of $10^3$ for $m_0 \sim 100$ GeV and $M_{1/2} \sim [450, 600]$ GeV, assuming a luminosity $\mathcal{L} = 100$ fb$^{-1}$ and $\sqrt{s} = 14$ TeV. In type-II seesaw, there can be a maximum number of events of the order of $10^3$ for $m_0 \sim 100$ GeV and $M_{1/2} \sim [600, 800]$ GeV.

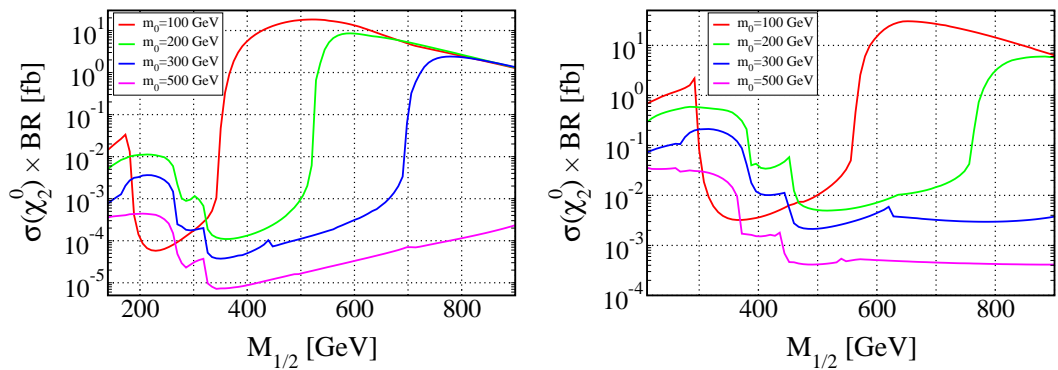

**Figure 5.10:** Production cross section (at leading order) of $\chi_2^0$ times BR of $\chi_2^0$ going to $\mu$-$\tau$ lepton pair versus $M_{1/2}$ for $m_0 = 100$ GeV (red), 200 GeV (green), 300 GeV (blue) and 500 GeV (magenta), and for our standard choice of parameters: $\mu > 0$, $\tan \beta = 10$ and $A_0 = 0$ GeV, for type-I (left panel) and for type-II seesaw (right panel) with $\lambda_1 = 0.02$ and $\lambda_2 = 0.5$, imposing $\mathrm{Br}(\mu \to e + \gamma) \leq 1.2 \cdot 10^{-11}$.

For type-II seesaw where we have less parameters, we can look at variations of the result with the values of the triplet Higgs boson coupling $\lambda_2$, a parameter that can not be determined from neutrino data alone as it appears only in the ratio $\lambda_2/M_T$, see equation (5.7). In Fig. 5.11 we show the dependence of the product of cross section times LFV branching ratios as function of $\lambda_2$ for our standard point. We should mention that the other Higgs boson triplet coupling $\lambda_1$, does not contribute to LFV decays, and hence is left undetermined by this analysis.

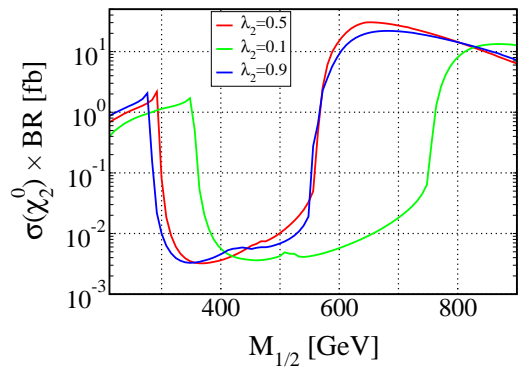

**Figure 5.11:** Production cross section (at leading order) of $\chi_2^0$ times BR of $\chi_2^0$ going to $\mu$-$\tau$ lepton pair versus $M_{1/2}$, for our standard choice of parameters: $\mu > 0$, $\tan\beta = 10$ and $A_0 = 0$ GeV, for type-II seesaw, imposing Br$(\mu \to e + \gamma) \leq 1.2 \cdot 10^{-11}$, for a fixed value of $m_0 = 100$ GeV and different values of $\lambda_2 = 0.1$ (green), 0.5 (red), 0.9 (blue).

As has been discussed in [127], the dominant standard model backgrounds for the process considered are expected to be $WW$ and $t\bar{t}$ production. The cuts necessary to reduce this background will depend on the details of the SUSY spectrum and a detailed investigation is beyond the scope of this paper. The results of [127] suggest that the signal should be visible for $\sigma(\chi_2^0)\times$BR of order $\mathcal{O}(10)$ fb.

### 5.3 Conclusions

Low energy neutrino experiments, including oscillation studies and neutrinoless double-beta decay searches, may optimistically determine at most 9 neutrino parameters: the 3 neutrino masses, the 3 mixing angles and potentially the 3 CP violating phases. This is insufficient to fully reconstruct the underlying mechanism of neutrino mass generation. Under the assumption that neutrino masses arise *a la seesaw*, we have considered the simplest pure type-I or pure type-II seesaw schemes in mSUGRA.

We have performed a full scan over the mSUGRA parameter space in order to identify regions where LFV decays of $\chi_2^0$ can be maximal, while still respecting low-energy constraints that follow from the upper bounds on Br$(\mu \to e\gamma)$. We have also estimated the expected number of events for $\chi_2^0 \to \chi_1^0 + \tau + \mu$, for a sample luminosity of $\mathcal{L} = 100$ fb$^{-1}$. The expected number of events for the other channel $\chi_2^0 \to \chi_1^0 + \tau + e$ is always smaller, as can be seen from the LVF branching ratios presented in section 5.2.1. We have found that the pure seesaw-II scheme is substantially simpler and comes closer to beéing fully reconstructable, provided additional LFV decays are detected and some supersymmetric particles are discovered at the Large Hadron Collider, providing the necessary parameter reconstruction information of the supersymmetric lagrangian.

Note that in what concerns the expected number of events both type-I and type-II schemes give similar results. However, as we have seen, given their smaller number of paramaters, type-II seesaw schemes are more likely to be reconstructable through a combination of low energy neutrino measurements, with the possible detection of supersymmetric states and lepton flavour violation at the LHC. This should encourage one to perform full-fledged dedicated simulations, in order to ascertain their feasibility within realistic experimental conditions [96].

Finally we note that we have not analyzed in detail the fact that LFV might induce new "edge variables", giving additional information [128]. We have focused here on LHC, but we should mention that a future ILC would be much more suited for measuring LFV SUSY processes [112, 129–134].

# Chapter 6

# Dark Matter in Seesaw Type II Model

Standard cosmology requires the existence of a non-baryonic dark matter (DM) contribution to the total energy budget of the universe [135, 136]. In the past few years estimates of the DM abundance have become increasingly precise. Indeed, the Particle Data Group now quotes at 1 $\sigma$ c.l. [43]

$$\Omega_{DM} h^2 = 0.105 \pm 0.008. \tag{6.1}$$

Since the data from the WMAP satellite [81, 137] and large scale structure formation [138] is best fitted if the DM is cold, weakly interacting mass particles (WIMP) are currently the preferred explanation, as discussed in section 2.4.3 - see formulas (2.186), (2.188) and the comment that follows. While there is certainly no shortage of WIMP candidates (lists can be found in many reviews, see for example [135, 136, 139, 140]), the literature is completely dominated by studies of the lightest neutralino. Neutrino oscillation experiments have shown that neutrinos have non-zero mass and mixing angles [10, 58, 59, 141, 142] (see the discussion in Chapter 1) and the most recent global fits to all data [28] confirm again that the mixing angles are surprisingly close to the so-called tri-bimaximal mixing (TBM) values [29], as in equation (1.228). In the minimal supersymmetric extension of the standard model (MSSM) with conserved R-parity neutrino masses are zero for the same reasons as in the SM. However, it was shown long ago that if neutrinos are Majorana particles, their mass is described by a unique dimension-5 operator [19] given by equation (1.176). All (Majorana) neutrino mass models reduce to this operator at low energies. If $f$ is a coefficient $\mathcal{O}(1)$, current neutrino data indicates $\Lambda \lesssim \mathcal{O}(10^{15})$ GeV. This is the essence of the "seesaw" mechanism. There are three different tree-level realizations of the seesaw, classified as type-I, type-II and type-III in [143]. As already discussed on Chapter 1, Type-I is the well-known case of the exchange of a heavy fermionic singlet [70, 144–146] and Type-II corresponds to the exchange of a scalar triplet [53, 77]. One could also add one (or more) fermionic triplets to the field content of the SM [147]. This is called seesaw type-III in [143].

Neutrino experiments at low energies measure only $f_{\alpha\beta}/\Lambda$, thus observables outside the neutrino sector will ultimately be needed to learn about the origin of equation (1.176).

Augmenting the SM with a high-scale seesaw mechanism does not lead to any conceivable phenomenology apart from neutrino masses, but if weak scale supersymmetry exists *indirect* probes into the high energy world might be possible. Two kind of measurements containing such indirect information exist in principle, lepton flavour violating (LFV) observables and sparticle masses.

Assuming complete flavour blindness in the soft supersymmetry breaking parameters at some large scale, the neutrino Yukawa matrices will, in general, lead to non-zero flavour violating entries in the slepton mass matrices, if the seesaw scale is lower than the scale at which SUSY is broken. This was first pointed out in [100]. The resulting LFV processes have been studied in many publications, for low-energy observables such as $\mu \to e\gamma$ and $\mu - e$ conversion in seesaw type-I see for example [101–106, 118], for seesaw type-II [86, 113]. LFV collider observables have also been studied in a number of papers, see for example [5, 96, 112–114, 127–130, 132–134, 148].

Mass measurements in the sparticle sector will not only be necessary to learn about the mechanism of SUSY breaking in general, but might also reveal indications about the scale of the seesaw mechanism. However, very precise knowledge of masses will be necessary before one can learn about the high scale parameters [108, 111]. Especially interesting in this context is the observation that from the different soft scalar and gaugino masses one can define certain combinations ("invariants") which are nearly constant over large parts of mSugra space. Adding a seesaw mechanism of type-II or type-III these invariants change in a characteristic way as a function of the seesaw scale and are thus especially suited to extract information about the high energy parameters [109]. Note, however, that the "invariants" are constants in mSugra space only in leading order and that quantitatively important 2-loop corrections exist [113].

In the work [6], that is reproduced here, we study neutralino dark matter [149–151] within a supersymmetric type-II seesaw model with mSugra boundary conditions. For definiteness, the model we consider consists of the MSSM particle spectrum to which we add a single pair of **15**- and $\overline{\mathbf{15}}$-plets. This is the simplest supersymmetric type-II setup, which allows one to maintain gauge coupling unification [86] and explains measured neutrino oscillation data.

In mSugra - assuming a standard thermal history of the early universe [1] - only four very specific regions in parameter space can correctly explain the most recent WMAP data [81]. As described in section 2.4.3, these are (i) the bulk region; (ii) the co-annihilation line; (iii) the "focus point" line and (iv) the "Higgs funnel" region. In the bulk region there are no specific relations among the sparticle masses. However, all sparticles are rather light in this region, so it is already very constrained from the view point of low-energy data [153]. In the co-annihilation line the lightest scalar tau is nearly degenerate with the lightest neutralino, thus reducing the neutralino relic density with respect to naive expectations [151, 154]. In the "focus point" line [154, 155] $\Omega_{\chi_1^0}h^2$ is small enough to explain $\Omega_{DM}h^2$ due to a rather small value of $\mu$ leading to an enhanced higgsino component in the lightest neutralino and thus an enhanced coupling to the $Z^0$ boson. Lastly, at large $\tan\beta$ an s-channel resonance pair annihilation of neutralinos through the CP-odd Higgs boson can become important. This is called the "Higgs funnel" region [149].

---

[1]In models with non-standard thermal history the relation between sparticle masses and relic density can be lost completely [152].

The addition of the **15** and $\overline{\bf 15}$ pair at the high scale does not, in general, lead to the appearance of new allowed regions. However, the deformed sparticle spectrum with respect to mSugra expectations leads to characteristic changes in the allowed regions as a function of the unknown seesaw scale. We discuss these changes in detail and compare the results to other indirect constraints, namely, the observed neutrino masses and upper limits on LFV processes. We concentrate on the seesaw type-II scheme, since for mSugra + seesaw type-I the changes in the DM allowed regions with respect to pure mSugra are, in general, expected to be tiny. [2]

## 6.1 Theory setup: mSugra and $SU(5)$ motivated type-II seesaw

In this section we summarize the main features of the model we will use in the numerical calculation. We will always refer to minimal Supergravity (mSugra) as the "standard" against which we compare all our results. The model consists in extending the MSSM particle spectrum by a pair of **15** and $\overline{\bf 15}$. It is the minimal supersymmetric seesaw type-II model which maintains gauge coupling unification [86].

mSugra is specified by 4 continuous and one discrete parameter [157]. These are usually chosen to be $m_0$, the common scalar mass, $M_{1/2}$, the gaugino mass parameter, $A_0$, the common trilinear parameter, $\tan\beta = \frac{v_2}{v_1}$ and the sign of $\mu$. $m_0$, $M_{1/2}$ and $A_0$ are defined at the GUT scale, the RGEs are known at the 2-loop level [158] [3].

Under $SU(3) \times SU_L(2) \times U(1)_Y$ the **15** decomposes as

$$\begin{aligned} {\bf 15} &= S + T + Z \\ S &\sim (6,1,-\frac{2}{3}), \qquad T \sim (1,3,1), \qquad Z \sim (3,2,\frac{1}{6}). \end{aligned} \tag{6.2}$$

The $SU(5)$ invariant superpotential reads as

$$\begin{aligned} W &= \frac{1}{\sqrt{2}}{\bf Y}_{15}\bar{5}\cdot 15\cdot\bar{5} + \frac{1}{\sqrt{2}}\lambda_1\bar{5}_H\cdot 15\cdot\bar{5}_H + \frac{1}{\sqrt{2}}\lambda_2 5_H\cdot\overline{15}\cdot 5_H + {\bf Y}_5 10\cdot\bar{5}\cdot\bar{5}_H \\ &+ {\bf Y}_{10}10\cdot 10\cdot 5_H + M_{15}15\cdot\overline{15} + M_5\bar{5}_H\cdot 5_H \end{aligned} \tag{6.3}$$

Here, $\bar{5} = (d^c, L)$, $10 = (u^c, e^c, Q)$, $5_H = (t, H_2)$ and $\bar{5}_H = (\bar{t}, H_1)$. Below the GUT scale

---

[2]We have confirmed this general expectation with some sample calculations. However, an exceptional case has been presented recently in [156], see the more detailed discussion in section (6.2).

[3]For reviews on mSugra and MSSM, see for example [157, 159, 160].

in the $SU(5)$-broken phase the potential contains the terms

$$\frac{1}{\sqrt{2}}(Y_T L T_1 L + Y_S d^c S d^c) + Y_Z d^c Z L + Y_d d^c Q H_1 + Y_u u^c Q H_2 + Y_e e^c L H_1 \quad (6.4)$$

$$+ \quad \frac{1}{\sqrt{2}}(\lambda_1 H_1 T_1 H_1 + \lambda_2 H_2 T_2 H_2) + M_T T_1 T_2 + M_Z Z_1 Z_2 + M_S S_1 S_2 + \mu H_1 H_2$$

$Y_d$, $Y_u$ and $Y_e$ generate quark and charged lepton masses in the usual manner. In addition there are the matrices $Y_T$, $Y_S$ and $Y_Z$. For the case of a complete **15**, apart from calculable threshold corrections, $Y_T = Y_S = Y_Z$ and $M_T$, $M_S$ and $M_Z$ are determined from $M_{15}$ by the RGEs. As long as $M_Z \sim M_S \sim M_T \sim M_{15}$ gauge coupling unification will be maintained. The equality need not be exact for successful unification.

The triplet $T_1$ has the correct quantum numbers to generate neutrino masses via the first term in equation (6.4). Integrating out the heavy triplets at their mass scale a dimension-5 operator of the form equation (1.176) is generated. This can be seen as follows: first we compute the $F$ term associated with $T_2$ which in abbreviated notation is

$$F_{T_2} = \frac{1}{\sqrt{2}} \lambda_2 H_2 H_2 + M_T T_1. \quad (6.5)$$

Next we compute the corresponding scalar potential term

$$V = \cdots + |F_{T_2}|^2 \quad (6.6)$$

$$= \cdots + \frac{1}{2}|\lambda_2|^2 |H_2|^2 + M_T^2 |T_1|^2 + \frac{2}{\sqrt{2}} \mathrm{Re}(\lambda_2 M_T T_1 H_2 H_2)$$

where in this expression $T_1, H_1$ and $H_2$ refer to the scalar components of the respective superfields. Then we assume that at this low energy scales the triplet is almost at rest such that we can ignore the kinetic term in Lagrange equations for $T_1$, leaving us with

$$-\frac{\partial \mathcal{L}}{\partial T_1} = \frac{\partial V}{\partial T_1} = 0 \quad (6.7)$$

which, solving it in order to $T_1$ gives

$$T_1 = -\frac{1}{\sqrt{2}} \frac{\lambda_2}{M_T} H_2 H_2. \quad (6.8)$$

Finally, the leptonic Yukawa couplings are computed from the superpotential term $W_L = \frac{1}{\sqrt{2}} Y_T L T_1 L$:

$$(m_L)_{ij} = -\frac{2}{\sqrt{2}} \frac{\partial^2 W_L}{\partial \tilde{\nu}_i \partial \tilde{\nu}_j} \quad (6.9)$$

$$= -\frac{2}{\sqrt{2}} (Y_T)_{ij} T_1$$

and after replacing (6.8) we end up with the dimension 5 operator

$$Lm_LL = \frac{\lambda_2 Y_T}{M_T} LH_2 H_2 L.$$ (6.10)

After electro-weak symmetry breaking the resulting neutrino mass matrix can be written as

$$m_\nu = \frac{v_2^2}{2} \frac{\lambda_2}{M_T} Y_T.$$ (6.11)

Here $v_2$ is the vacuum expectation value of Higgs doublet $H_2$ and we use the convention $\langle H_i \rangle = \frac{v_i}{\sqrt{2}}$. $m_\nu$ can be diagonalized in the standard way with a unitary matrix $U$, containing in general 3 angles and 3 phases. Note that $\hat{Y}_T = U^T \cdot Y_T \cdot U$ is diagonalized by *the same matrix as* $m_\nu$. This means that if all neutrino eigenvalues, angles and phases were known, $Y_T$ would be completely fixed up to an overall constant, which can be written as $\frac{M_T}{\lambda_2} \simeq 10^{15}$GeV $\left( \frac{0.05 \text{ eV}}{m_\nu} \right)$. Thus, current neutrino data requires $M_T$ to be lower than the GUT scale by (at least) an order or magnitude.

The full set of RGEs for the $\mathbf{15} + \overline{\mathbf{15}}$ can be found in [86] and in the numerical calculation, presented in the next section, we solve the exact RGEs. However, for a qualitative understanding of the results, the following approximative solutions are quite helpful.

For the gaugino masses one finds in leading order

$$M_i(m_{SUSY}) = \frac{\alpha_i(m_{SUSY})}{\alpha(M_G)} M_{1/2}.$$ (6.12)

Eq. (6.12) implies that the ratio $M_2/M_1$, which is measured at low-energies, has the usual mSugra value, but the relationship to $M_{1/2}$ is changed. Neglecting the Yukawa couplings $\mathbf{Y}_{15}$ (see below), for the soft mass parameters of the first two generations one gets

$$m_{\tilde{f}}^2 = M_0^2 + \sum_{i=1}^{3} c_i^{\tilde{f}} \left( \left( \frac{\alpha_i(M_T)}{\alpha(M_G)} \right)^2 f_i + f_i' \right) M_{1/2}^2,$$ (6.13)

$$f_i = \frac{1}{b_i} \left( 1 - \left[ 1 + \frac{\alpha_i(M_T)}{4\pi} b_i \log \frac{M_T^2}{m_Z^2} \right]^{-2} \right),$$

$$f_i' = \frac{1}{b_i + \Delta b_i} \left( 1 - \left[ 1 + \frac{\alpha(M_G)}{4\pi} (b_i + \Delta b_i) \log \frac{M_G^2}{M_T^2} \right]^{-2} \right).$$ (6.14)

The various coefficients $c_i^{\tilde{f}}$ can be found in [113]. The gauge couplings are given as

$$\alpha_1(m_Z) = \frac{5\alpha_{em}(m_Z)}{3\cos^2\theta_W}, \qquad \alpha_2(m_Z) = \frac{\alpha_{em}(m_Z)}{\sin^2\theta_W}, \qquad (6.15)$$

$$\alpha_i(m_{SUSY}) = \frac{\alpha_i(m_Z)}{1 - \frac{\alpha_i(m_Z)}{4\pi}b_i^{SM}\log\frac{m_{SUSY}^2}{m_Z^2}},$$

$$\alpha_i(M_T) = \frac{\alpha_i(m_{SUSY})}{1 - \frac{\alpha_i(m_{SUSY})}{4\pi}b_i\log\frac{M_T^2}{m_{SUSY}^2}},$$

$$\alpha_i(M_G) = \frac{\alpha_i(M_T)}{1 - \frac{\alpha_i(M_T)}{4\pi}(b_i + \Delta b_i)\log\frac{M_G^2}{M_T^2}}.$$

with $b_i^{SM}$ and $b_i^{MSSM}$ being the usual standard model and MSSM coefficients. $\Delta b_i = 7$ for all $i$ in case of a complete 15-plet.

We can estimate the soft mass parameters given the above formulas for a given choice of $m_0$, $M_{1/2}$ and $M_{15} = M_T$. We show some arbitrarily chosen examples in fig. (6.1). Note that the result shown is approximate, since we are (a) using the leading log approximation and (b) two loop effects are numerically important, especially for $m_Q$, but not included. The figure serves to show that for any $M_{15} < M_{GUT}$ the resulting mass parameters are always smaller than the mSugra expectations for the same choice of initial parameters $(m_0, M_{1/2})$. While the exact values depend on $(m_0, M_{1/2})$ and on the other mSugra parameters, this feature is quite generally true in all of the $(m_0, M_{1/2})$ plane. Note, that the running is different for the different scalar mass parameters, but the ratio of the gaugino mass parameters $M_1/M_2$ always stays close to the mSugra expectation, $M_1 \simeq \frac{5}{3}\tan^2\theta_W M_2$.

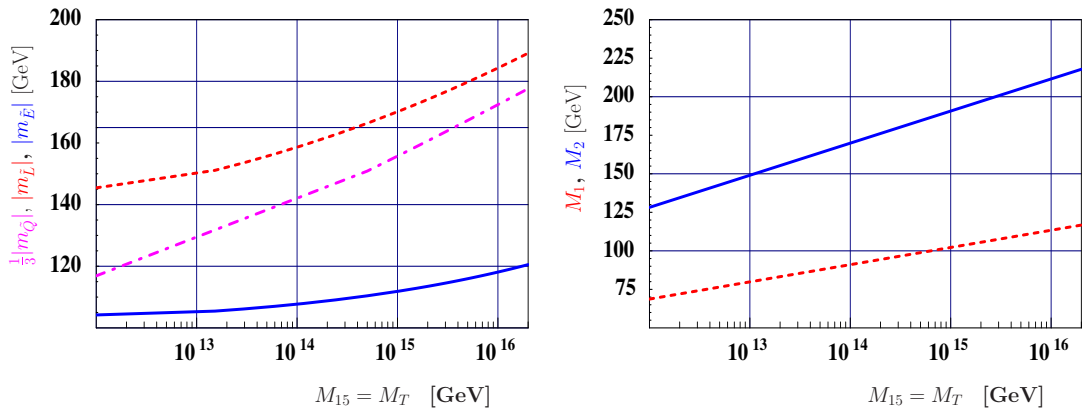

**Figure 6.1:** Analytically calculated running of scalar (to the left) and gaugino mass parameters (to the right), leading order only. The mass parameters are calculated as a function of $M_{15}$ for the mSugra parameters $m_0 = 70$ GeV and $M_{1/2} = 250$ GeV. For $M_{15} \simeq 2 \times 10^{16}$ GeV the mSugra values are recovered. Smaller $M_{15}$ lead to smaller soft masses in all cases. Note that the running is different for the different mass parameters with gaugino masses running faster than slepton mass parameters.

## 6.2  Numerical results

In this section we discuss our numerical results. All the plots shown below are based on the program packages SPheno [119] and micrOMEGAs [161, 162]. We use SPheno V3 [7], including the RGEs for the $\mathbf{15} + \overline{\mathbf{15}}$ case [86, 113] at the 2-loop level for gauge couplings and gaugino masses and at one-loop level for the remaining MSSM parameters and the 15-plet parameters, for a discussion see [113]. For any given set of mSugra and 15-plet parameters SPheno calculates the supersymmetric particle spectrum at the electro-weak scale, which is then interfaced with micrOMEGAs2.2 [163] to calculate the relic density of the lightest neutralino, $\Omega_{\chi_1^0} h^2$.

For the standard model parameters we use the PDG 2008 values [43], unless specified otherwise. As discussed below, especially important are the values (and errors) of the bottom and top quark masses, $m_b = 4.2 + 0.17 - 0.07$ GeV and $m_t = 171.2 \pm 2.1$ GeV. Note, the $m_t$ is understood to be the pole-mass and $m_b(m_b)$ is the $\overline{MS}$ mass. As the allowed range for $\Omega_{DM} h^2$ we always use the 3 $\sigma$ c.l. boundaries as given in [43], i.e. $\Omega_{DM} h^2 = [0.081, 0.129]$. Note, however, that the use of 1 $\sigma$ contours results in very similar plots, due to the small error bars.

In the "seesaw sector" we have the parameters connected with the 15-plets, i.e. $M_{\mathbf{15}}$, $Y_{\mathbf{15}}$, $\lambda_1$ and $\lambda_2$. For the calculation of the dark matter abundance the most important parameter is $M_{\mathbf{15}}$. It has turned out that the effects of $Y_{\mathbf{15}}$, $\lambda_1$ and $\lambda_2$ on the relic abundance of neutralinos are very minor. Note, however, that as discussed in the previous section, atmospheric neutrino oscillation data can not be explained in our setup, if the triplet mass is larger than approximately $M_{\mathbf{15}} = M_T = 10^{15}$ GeV. Also, the non-observation of lepton flavour violating (LFV) decays puts an upper bound on $M_{\mathbf{15}}$. The latter, however, is strongly dependent on $\tan\beta$ and depends also on $m_0$ and $M_{1/2}$. We will first show results using different values of $M_T$ as free parameter, without paying attention to neutrino masses and LFV. We will discuss how our results change for correctly fitted neutrino masses and angles towards the end of this section, where we also discuss and compare LFV excluded regions with DM allowed ones.

We define our "standard choice" of mSugra parameters as $\tan\beta = 10$, $A_0 = 0$ and $\mu > 0$ and use these values in all plots, unless specified otherwise. We then show our results in the plane of the remaining two free parameters, $(m_0, M_{1/2})$. Fig. (6.2) shows in the top panel contours of equal dark matter density, $\Omega_{\chi_1^0} h^2$. The lines are constant $\Omega_{\chi_1^0} h^2$ with $\Omega_{\chi_1^0} h^2 = 0.1, 0.2, 0.5, 1, 2$. In the bottom panel we show the range of parameters allowed by the DM constraint at 3 $\sigma$ c.l. In both cases, to the left a pure mSugra calculation, whereas the plot to the right shows mSugra + 15-plet with $M_T = 10^{14}$ GeV. In each plot the yellow regions are eluded either by the lighter scalar tau being the LSP (to the bottom right) or by the LEP limit on the mass of the lighter chargino (to the left), $m_{\chi_1^+} \geq 105$ GeV. In addition, we show two lines of constant lightest Higgs boson mass, $m_{h^0} = 110$

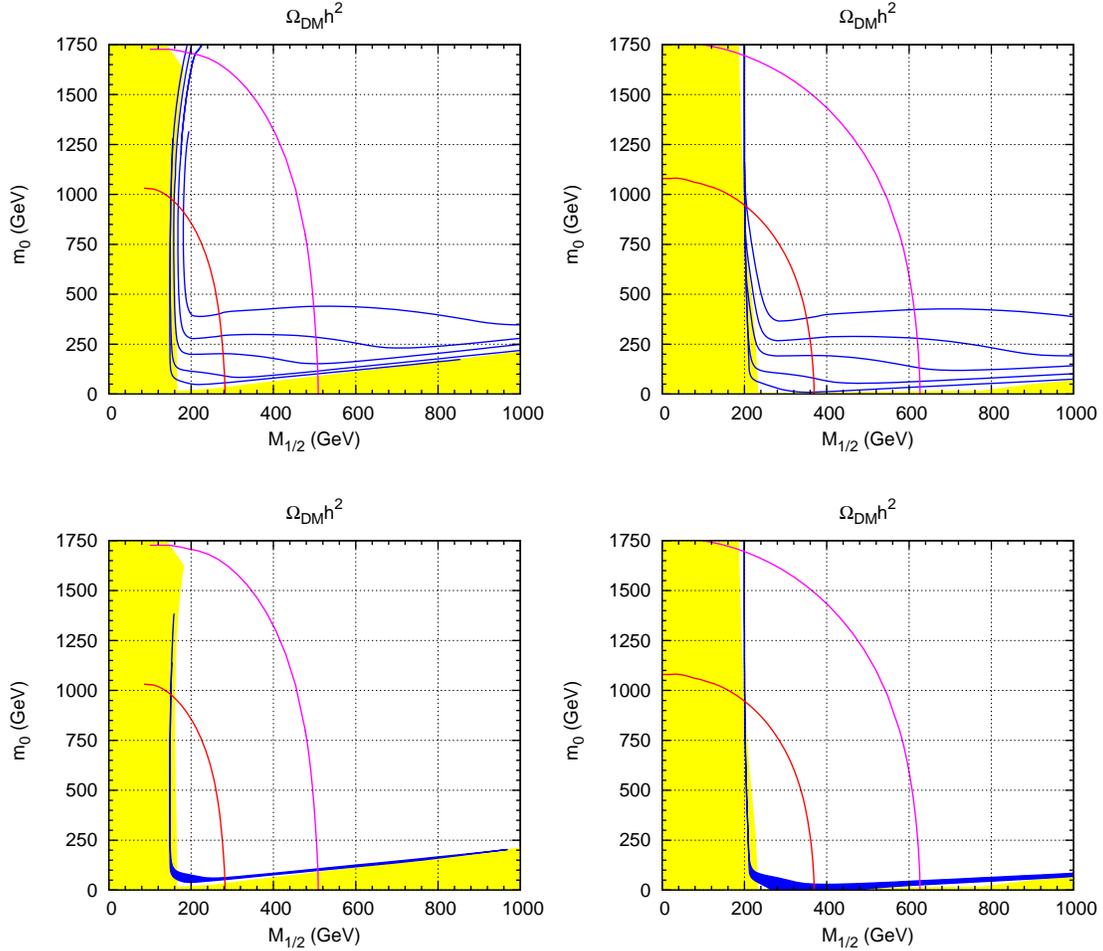

**Figure 6.2:** Top: Contours of equal dark matter density ($\Omega_{\chi_1^0} h^2$) in the ($m_0, M_{1/2}$) plane for the "standard choice" $\tan\beta = 10$, $A_0 = 0$ and $\mu \geq 0$, for mSugra (left panel) and type-II seesaw with $M_T = 10^{14}$ GeV (right panel). The lines are constant $\Omega_{\chi_1^0} h^2$ with $\Omega_{\chi_1^0} h^2 = 0.1, 0.2, 0.5, 1, 2$. Bottom: Range of parameters allowed by the DM constraint at $3\,\sigma$ c.l. To the left: mSugra; to the right: $M_T = 10^{14}$ GeV. For a discussion see text.

GeV (dotted) and $m_{h^0} = 114.4$ GeV (dashed), as calculated by SPheno, see the discussion below.

The plots show three of the different allowed regions discussed in the introduction. To the right the co-annihilation region, here the lightest neutralino and the lighter scalar tau are nearly degenerate in mass. The line going nearly vertically upwards at constant $M_{1/2}$ is the "focus point" line. The small region connecting the two lines are the remains of the bulk region, which has shrunk considerably due to the reduced error bars on $\Omega_{DM} h^2$ after the most recent WMAP data [81]. The focus point line is excluded by the LEP constraint on the lighter chargino mass at low and moderate values of $m_0$. It becomes allowed only at values of $m_0$ larger than (very roughly) 1-1.5 TeV. However, note that the exact value of $m_0$ at which the focus point line becomes allowed is extremely sensitive to errors in $m_{\chi_1^+}$, both from the experimental bound and the error in the theoretical calculation.

Comparing the results for the pure mSugra case to the mSugra+15-plet calculation, two differences are immediately visible in fig. (6.2). First, the focus point line is shifted towards larger values of $M_{1/2}$. This is due to the fact that for the 15-plet at $M_{15} = 10^{14}$ GeV the neutralino is lighter than in the mSugra case at the same value of $M_{1/2}$, compare to fig. (6.1). Maintaining the same relation between $M_1$ and $\mu$ as in the mSugra case requires a then a larger value of $M_{1/2}$. Note that for the same reason the excluded region from the LEP bound on the chargino mass is larger than in the mSugra case. Second one finds that the co-annihilation line is shifted towards smaller values of $m_0$. The latter can be understood from fig. (6.3).

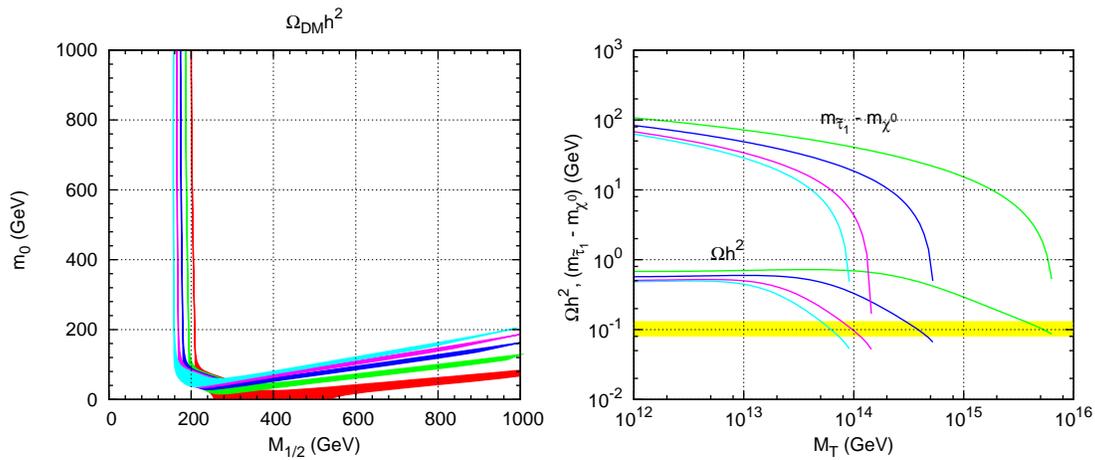

**Figure 6.3:** Allowed region for dark matter density ($0.081 < \Omega_{\chi_1^0} h^2 < 0.129$) in the $(m_0, M_{1/2})$ plane for the "standard choice" $\tan\beta = 10$, $A_0 = 0$ and $\mu \geq 0$, for five values from $M_T$, $M_T = 10^{14}$ GeV (red), to $M_T = 10^{16}$ GeV (cyan), to the left. To the right: Variation of the mass difference $m_{\tilde{\tau}_1} - m_{\chi^0}$ (top lines) and of $\Omega h^2$ (bottom lines), as a function of $M_T$ for four different values of $m_0$: 0 (cyan), 50 (magenta), 100 (blue) and 150 GeV (green) for one fixed value of $M_{1/2} = 800$ GeV. The yellow region corresponds to the experimentally allowed DM region.

Fig. (6.3) shows the allowed region for the dark matter density in the $(m_0, M_{1/2})$ plane for our "standard choice" of other mSugra parameters for a number of different $M_T$ (to the left). The plot shows how the co-annihilation line moves towards smaller values of $m_0$ for smaller values of $M_T$. The plot on the right in fig. (6.3) explains this behaviour. It shows the variation of the mass difference $m_{\tilde{\tau}_1} - m_{\chi^0}$ (top lines) and of $\Omega h^2$ (bottom lines), as a function of $M_T$ for four different values of $m_0$: 0 (cyan), 50 (magenta), 100 (blue) and 150 GeV (green) for one fixed value of $M_{1/2} = 800$ GeV. The yellow region corresponds to the experimentally allowed DM region. Co-annihilation requires a small value of $m_{\tilde{\tau}_1} - m_{\chi^0}$, typically smaller than a few GeV. With decreasing values of $M_T$ the gaugino masses run down to smaller values faster than the slepton masses, thus effectively increasing $m_{\tilde{\tau}_1} - m_{\chi^0}$ in these examples with respect to mSugra. To compensate for this effect at constant $M_{1/2}$ smaller values of $m_0$ are required to get the $m_{\tilde{\tau}_1} - m_{\chi^0}$ in the required range.

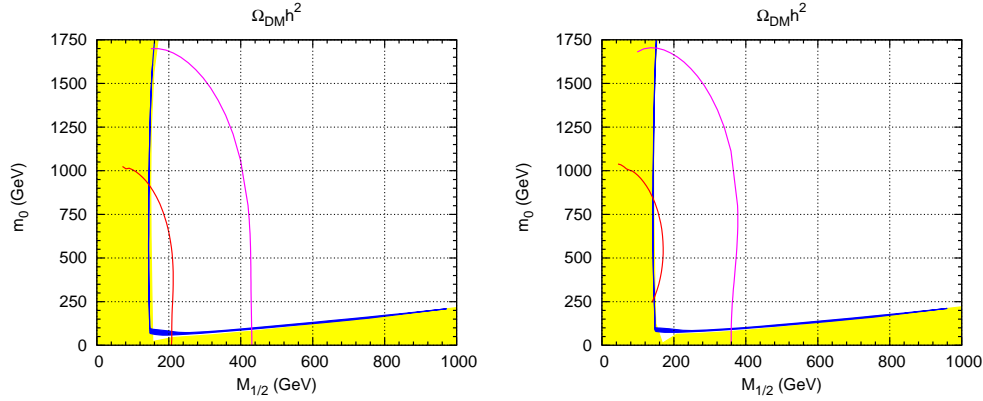

**Figure 6.4:** Limits for mSugra with $\tan\beta = 10$, and $\mu > 0$ for $A_0 = -300$ GeV (left panel) and $A_0 = -500$ GeV (right panel). The blue regions are allowed by the DM constraint, for the explanation of the bounds see fig. (6.2) and text.

At this point a short discussion of the Higgs boson mass bound might be in order. LEP excluded a light Higgs boson with SM couplings with masses below $m_h \leq 114.4$ GeV [43]. For reduced coupling of the Higgs boson to $b\bar{b}$ the bound is less severe, so this bound is not strictly valid in all of MSSM space. More important for us, however, is the *theoretical* uncertainty in the calculation of the lightest Higgs boson mass. SPheno calculates $m_{h^0}$ at two-loop level using $\overline{DR}$ renormalization. Expected errors for this kind of calculation, including a comparison of different public codes, have been discussed in [164]. As discussed in [164, 165] even at the 2-loop level uncertainties in the calculation of $m_{h^0}$ can be of the order of $3 - 5$ GeV. In this context it is interesting to note that FeynHiggs [166], which calculates the Higgs masses in a diagrammatic approach within the $\overline{OS}$ renormalization scheme tends to predict Higgs masses which are systematically larger by $3 - 4$ GeV, when compared with the $\overline{DR}$ calculation. We therefore showed in fig. (6.2) two lines of constant Higgs boson masses. The value of $m_{h^0} = 114.4$ GeV is taking the LEP bound at face value, while the lower value of $m_{h^0} = 110$ GeV estimates the parameter region which is excluded *conservatively*, including the theoretical error. Since the lightest Higgs boson mass varies slowly with $m_0$ and $M_{1/2}$, even a relatively tiny change in $m_{h^0}$ of, say 1 GeV, shifts the extreme values of the excluded region by $\sim 50$ GeV in $M_{1/2}$ (at small $m_0$) and by $\sim 150$ GeV in $m_0$ (at small $M_{1/2}$).

Moreover, it is well known that the calculated Higgs boson masses are strongly dependent on the mixing in the stop sector and thus, indirectly, on the value of $A_0$. This is shown for the case of a pure mSugra calculation in fig. (6.4). Here we show two examples for the DM allowed region and the regions disfavoured by the Higgs boson mass bound at $m_{h^0} = 114.4$ GeV and $m_{h^0} = 110$ GeV. Larger negative $A_0$ leads to a less stringent constraint (for $\mu > 0$). Note, that all of the bulk region becomes allowed at $A_0 = -500$ GeV, once the theoretical uncertainty in the Higgs boson mass calculation is taken into account. We have checked for a few values of $M_T$ that for the case of mSugra+**15** the resulting Higgs boson bounds are very similar. We thus do not repeat the corresponding plots here. Comparing the calculations shown in fig. (6.4) and the mSugra calculation in fig. (6.2) with each other, one finds that the DM allowed regions are actually affected very little by the choice of $A_0$. We have checked that this is also the case for mSugra + seesaw type-II.

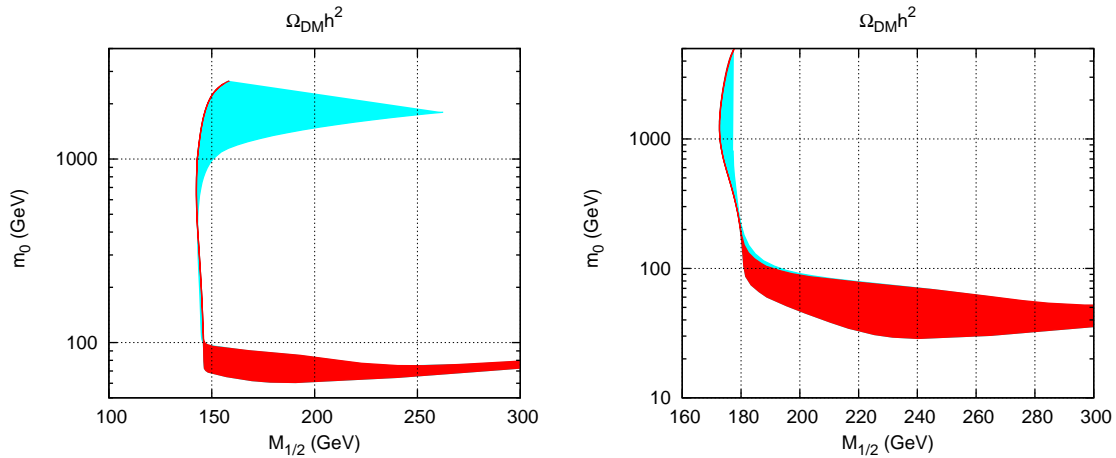

**Figure 6.5:** Logarithmically scaled zoom into the focus point region. In red the allowed region for $0.081 < \Omega h^2 < 0.129$ and in cyan the allowed region due the variation of $m_{\mathrm{top}} = 171.2 \pm 2.1$ GeV. The left panel is for mSugra case and the right panel for $M_T = 10^{15}$ GeV. The other parameters are taken at our "standard" values.

As mentioned above the uncertainty in the top mass is important for the calculation of the relic density. At low and moderate values of $\tan \beta$ the exact value of $m_t$ affects mainly the focus point region. As fig. (6.2) demonstrates near the focus point line the relic density changes very abruptly even for tiny changes of $M_{1/2}$. This is because a comparatively small value of $\mu$ is required to get a sufficiently enhanced coupling of the neutralino to the $Z^0$ boson. In mSugra the value of $\mu$ is determined from all other parameters by the condition of having correct electro-weak symmetry breaking (EWSB) and usually leads to $M_1, M_2 \ll \mu$. In the focus point region $\mu$ varies abruptly, points to the "left" of the focus point region are usually ruled out by the fact that EWSB can not be achieved. Since $m_t$ is the largest fermion mass, its exact value influences the value of $\mu$ required to achieve EWSB most. The change of $\mu$ with respect to a change of $m_t$ then can lead to a significant shift in the DM allowed region of parameter space. This is demonstrated in fig. (6.5), which shows a zoom into the focus point region for pure mSugra (to the left) and mSugra + **15** (to the right). The variation of the top mass shown corresponds to the current $1\ \sigma$ allowed range [43]. The pure mSugra is especially sensitive to a change of $m_t$. At large values of $m_0$ the uncertainty in "fixing" $M_{1/2}$ from the DM constraint can be larger than 100 GeV in the case of mSugra. Given this large uncertainty it would be impossible at present to distinguish the pure mSugra case from

mSugra + seesaw, if the focus point region is the correct explanation of the observed DM. Note, however, that in the future the top mass will be measured more precisely. At the LHC one expects an uncertainty of 1-2 GeV [167]. At a linear collider $m_t$ could be determined down to an uncertainty of 100 MeV [168].

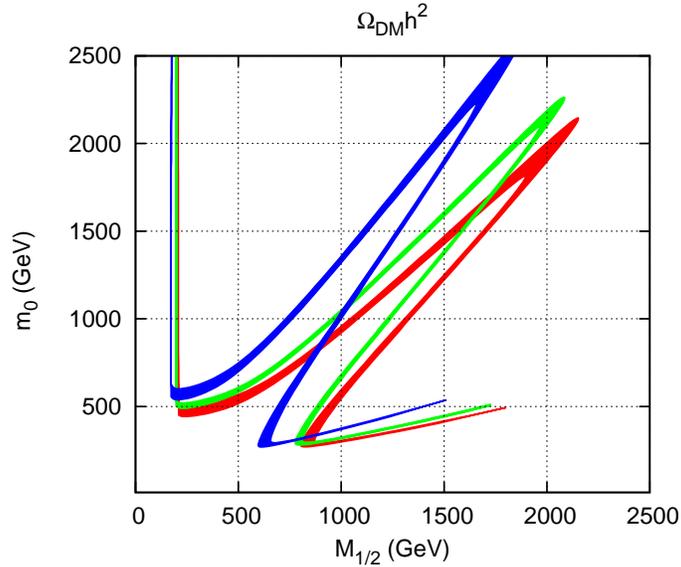

**Figure 6.6:** Allowed region for dark matter density in the $(m_0, M_{1/2})$ plane for $A_0 = 0$, $\mu \geq 0$ and $\tan \beta = 45$, for (from top to bottom) $M_T = 5 \times 10^{13}$ GeV (red), $M_T = 10^{14}$(green) and $M_T = 10^{15}$ GeV (blue).

We now turn to a discussion of large $\tan \beta$. At large values of $\tan \beta$ the width of the CP-odd Higgs boson $A$ becomes large, $\Gamma_A \sim M_A \tan^2 \beta (m_b^2 + m_t^2)$, and a wide s-channel resonance occurs in the region $m_{\chi_1^0} \simeq M_A/2$. The enhanced annihilation cross section reduces $\Omega_{\chi_1^0} h^2$ to acceptable levels, the resulting region is known as the "Higgs funnel" region. In fig. (6.6) we show the allowed range of parameters in the $(m_0, M_{1/2})$ plane for one specific value of $\tan \beta = 45$ and three different values of $M_T$. As demonstrated, the Higgs funnel region is very sensitive to the choice of $M_T$. It is fairly obvious that varying $M_T$ one can cover nearly all of the plane, even for fixed values of all other parameters. We have calculated the DM allowed region for various values of $\tan \beta$ and found that the funnel appears for all $\tan \beta \gtrsim 40$, approximately.

The strong dependence of the Higgs funnel region on $M_T$ unfortunately does not imply automatically that if large $\tan \beta$ is realized in nature one could get a very sensitive indirect "measurement" of the seesaw scale by determining $(m_0, M_{1/2})$. The reason is that the Higgs funnel is also very sensitive to the exact value of $\tan \beta$ and to the values (and errors) of the top and bottom quark mass. The latter is demonstrated in fig. (6.7), where we show the DM allowed range of parameters for a fixed choice of $\tan \beta$ and $M_T$ varying to the left (to the right) $m_t$ ($m_b$) within their current 1 $\sigma$ c.l. error band. The position of the funnel is especially sensitive to the exact value of $m_b$. Comparing fig. (6.7) with fig.

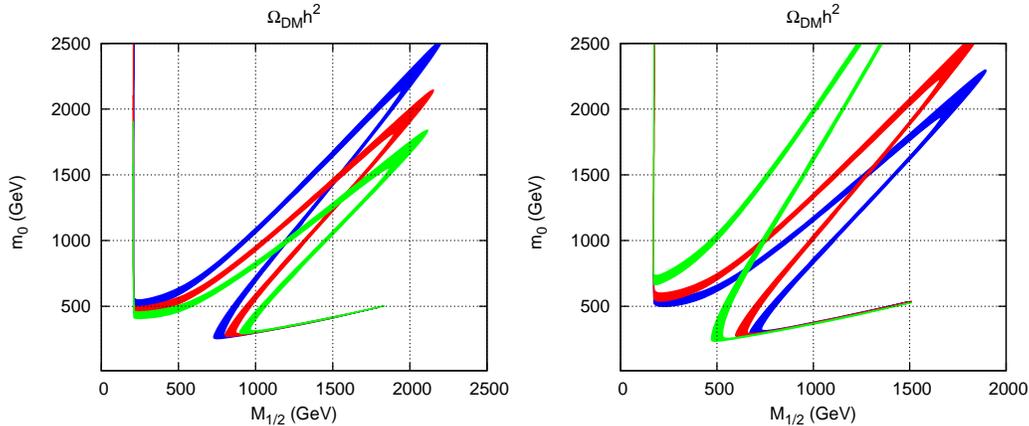

**Figure 6.7:** Allowed region for the dark matter density in the $(m_0, M_{1/2})$ plane for $A_0 = 0$, $\mu \geq 0$ and $\tan\beta = 45$, for $M_T = 5 \times 10^{13}$ GeV and (to the left) for three values of $m_{top} = 169.1$ GeV (blue), $m_{top} = 171.2$ GeV (red) and $m_{top} = 173.3$ GeV (green). To the right: The same, but varying $m_b$. $m_{bot} = 4.13$ GeV (blue), $m_{bot} = 4.2$ GeV (red) and $m_{bot} = 4.37$ GeV (green).

(6.6) one can see that the uncertainty in $m_b$ and $m_t$ currently severely limit any sensitivity one could get on $M_T$. However, future determinations of $m_b$ and $m_t$ could improve the situation considerably. For future uncertainties in $m_t$ see the discussion above for the focus point region. For $m_b$ reference [169] estimates that $m_b$ could be fixed to $4.17 \pm 0.05$ GeV, which might even be improved to an accuracy of $\Delta m_b \simeq 16$ MeV according to [170].

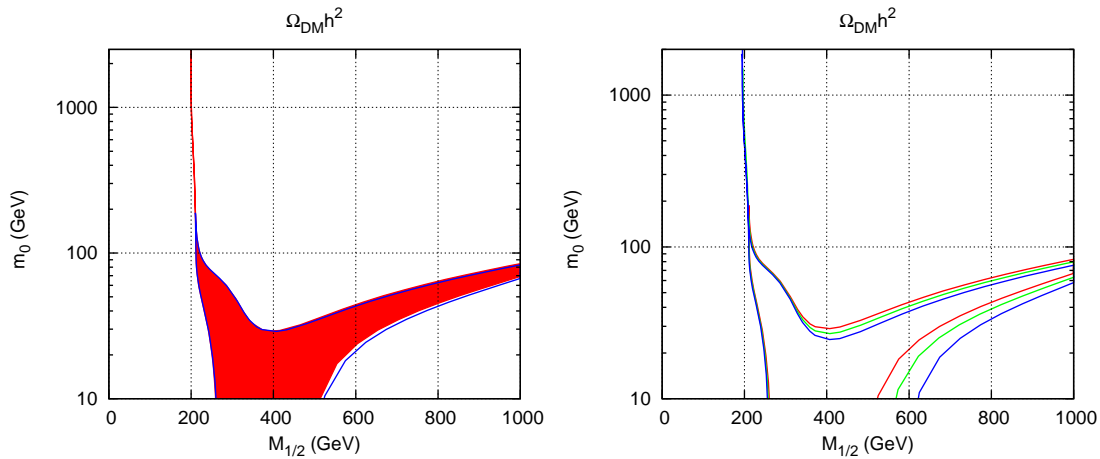

**Figure 6.8:** Allowed region for dark matter density in the $(m_0, M_{1/2})$ plane for the "standard choice" of mSugra parameters for $M_T = 10^{14}$ GeV. To the left: For one fixed value of $\lambda_2 = 0.5$ the allowed range for negligibly small neutrino Yukawa couplings (red) and $Y_T$ fitted to correctly explain solar and atmospheric neutrino data (blue lines). To the right: the DM allowed range of parameters for 3 different values of $\lambda_2$, $\lambda_2 = 0.5$ (red), $\lambda_2 = 0.75$ (green) and $\lambda_2 = 1$ (blue). Note the logarithmic scale.

All of the above figures have been calculated using fixed values for $\lambda_1$ and $\lambda_2$ and negligibly small Yukawa couplings $Y_T$. This choice in general does not affect the calculation of the DM allowed regions much. However, a fully consistent calculation can not vary $M_T$, $Y_T$ and $\lambda_2$ independently, since this will lead to neutrino masses and angles outside the experimentally allowed ranges. Since $Y_T$ is diagonalized by the same matrix as the effective neutrino mass matrix, $m_\nu$ (see the previous section) the measured neutrino angles provide constraints on the relative size of the entries in $Y_T$. The absolute size of $Y_T$ is then fixed for any fixed choice of $\lambda_2$ and $M_T$, once the neutrino spectrum is chosen to be hierarchical or quasi-degenerate. In the numerical calculation shown in fig. (6.8) we have chosen neutrino masses to be of the normal hierarchical type and fitted the neutrino angles to exact tri-bimaximal (TBM) values [29], i.e. $\tan^2 \theta_{\mathrm{Atm}} = 1$, $\tan^2 \theta_\odot = 1/2$ and $\sin^2 \theta_{\mathrm{R}} = 0$. This has to be done in a simple iterative procedure, since the triplet parameters are defined at the high scale, whereas neutrino masses and angles are measured at low scale. For more details on the fit procedure see [113].

In fig. (6.8) to the left we show two calculations of the DM allowed regions. The allowed range for negligibly small neutrino Yukawa couplings is shown by the filled (red) region, while the calculation with $Y_T$ fitted to correctly explain solar and atmospheric neutrino data is the one inside the (blue) lines. Note the logarithmic scale. As demonstrated, the exact values of $Y_T$ are of minor importance for the determination of the parameter region allowed by the DM constraint. Slightly larger differences between the fitted and unfitted calculations are found pushing $M_T$ to larger values (see, however, below). For smaller values of $M_T$, the entries in $Y_T$ needed to correctly explain neutrino data are smaller and, thus, $Y_T$ affects the DM allowed region even less for $M_T < 10^{14}$ GeV.

In fig. (6.8) to the right we compare three different calculations for $\lambda_2$, $\lambda_2 = 0.5$ (red), $\lambda_2 = 0.75$ (green) and $\lambda_2 = 1$ (blue), for fixed choice of other parameters. This plot serves to show that also the exact choice of $\lambda_2$ is of rather minor importance for the determination of the DM allowed region. Very similar results have been found for $\lambda_1$, we therefore do not repeat plots varying $\lambda_1$ here.

In fig. (6.9) we show the DM allowed parameter regions for $\tan\beta = 10$ and two values of $M_T$, $M_T = 5 \cdot 10^{13}$ GeV (to the left) and $M_T = 10^{14}$ GeV (to the right), for a fixed choice of all other parameters. Superimposed on this plot are lines of constant branching ratio for $Br(\mu \to e\gamma)$. The latter have been calculated requiring neutrino masses being hierarchical and fitted to solar and atmospheric neutrino mass squared differences and neutrino angles fitted to TBM values. Within the $(m_0, M_{1/2})$ region shown, $Br(\mu \to e\gamma)$ can vary by two orders of magnitude, depending on the exact combination of $(m_0, M_{1/2})$, even for all other parameters fixed. The most important parameter determining $Br(\mu \to e\gamma)$, once neutrino data is fixed, however, is $M_T$, as can be seen comparing the figure to the left with the plot on the right. While for $M_T = 10^{14}$ GeV about "half" of the plane is ruled out by the non-observation of $\mu \to e\gamma$, for $M_T = 5 \cdot 10^{13}$ GeV with the current upper limit nearly all of the plane becomes allowed. The strong dependence of $\mu \to e\gamma$ on $M_T$ can be understood from the analytical formulas presented in [113]. In this paper it was shown that $Br(\mu \to e\gamma)$ scales very roughly as $Br(\mu \to e\gamma) \propto M_T^4 \log(M_T)$, if neutrino masses are to be explained correctly. For $\tan\beta = 10$ one thus concludes that with present data values of $M_T$ larger than (few) $10^{13}$ GeV - (few) $10^{14}$ GeV are excluded by $Br(\mu \to e\gamma)$, to be compared with $M_T/\lambda_2 \lesssim 10^{15}$ GeV from the measured neutrino masses. Note, however, that (i) the constraint from neutrino masses is relatively independent of $\tan\beta$, $m_0$ and $M_{1/2}$, while $\mu \to e\gamma$ shows strong dependence on these parameters; and (ii) allowing the value of the reactor angle $\sin^2\theta_R$ to vary up to its experimental upper limit, $\sin^2\theta_R = 0.056$ [28], leads to larger values of $Br(\mu \to e\gamma)$ and thus to a tighter upper limit on $M_T$.

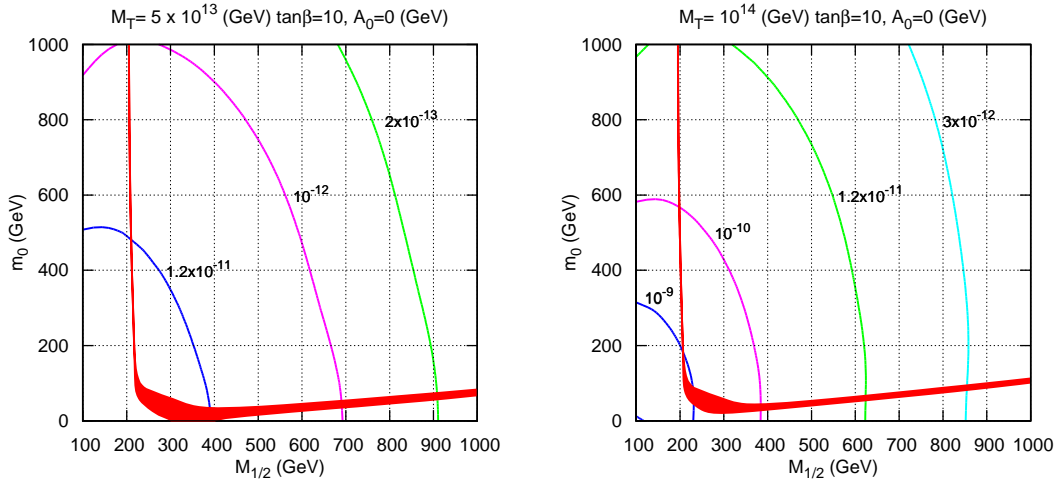

**Figure 6.9:** Allowed region for dark matter density in the $(m_0, M_{1/2})$ plane for our "standard choice" of mSugra parameters and for two values of $M_T$: $M_T = 5 \times 10^{13}$ (left panel) and for $M_T = 10^{14}$ (right panel). Superimposed are the contour lines for the $Br(\mu \to e\gamma)$.

## 6.3 Conclusions

In conclusion, we have calculated the neutralino relic density in a supersymmetric model with mSugra boundary conditions including a type-II seesaw mechanism to explain current neutrino data. We have discussed how the allowed ranges in mSugra parameter space change as a function of the seesaw scale. The stau co-annihilation region is shifted towards smaller $m_0$ for smaller values of the triplet mass $M_T$, while the bulk region and the focus point line are shifted towards larger values of $M_{1/2}$ for $M_T$ sufficiently below the GUT scale. The Higgs funnel, which appears at large values of $\tan\beta$ has turned out to be especially sensitive to the value of $M_T$. Determining $M_{1/2}$ from the mass of any gaugino and $m_0$ from a sparticle which is *not important for the DM calculation*, one could, therefore, get a constraint on $M_T$ from the requirement that the observed $\Omega_{DM}h^2$ is correctly explained by the calculated $\Omega_{\chi_1^0}h^2$.

On the positive side, we can remark that current data on neutrino masses put an upper bound on $M_T$ of the order of $\mathcal{O}(10^{15})$ GeV. Since this is at least one order of magnitude smaller than the GUT scale, the characteristic shifts in the DM regions are necessarily non-zero if our setup is the correct explanation of the observed neutrino oscillation data.

Even more stringent upper limits on $M_T$ follow, in principle, from the non-observation of LFV decays. A smaller $M_T$ implies larger shifts of the DM region. However, the "exact" upper limit on $M_T$ from LFV decays depends strongly on $\tan\beta$, $m_0$ and $M_{1/2}$, and thus can be quantified only once at least some information on these parameters is available.

On the down side, we need to add a word of caution. We have found that the DM calculation suffers from a number of uncertainties, even if we assume the soft masses to be perfectly known. The most important SM parameters turn out to be the bottom and the top quark mass. The focus point line depends extremely sensitively on the exact value of the top mass, the Higgs funnel shows a strong sensitivity on both, $m_b$ and $m_t$.

Finally, it is clear that quite accurate sparticle mass measurements will be necessary, before any quantitative conclusions can be taken from the effects we have discussed. Unfortunately, such accurate mass measurements might be very difficult to come by for different reasons. In the focus point region all scalars will be heavy, leading to small production cross section at the LHC. In the co-annihilation line with a nearly degenerate stau and a neutralino, the stau decays produce very soft taus, which are hard for the LHC to measure. And the Higgs funnel extends, depending on $\tan\beta$ and $M_T$, to very large values of $(m_0, M_{1/2})$, at least partially outside the LHC reach. Nevertheless, DM provides in principle an interesting constraint on the (supersymmetric) seesaw explanation of neutrino masses, if seesaw type-II is realized in nature, a fact which to our knowledge has not been discussed before in the literature.

## Chapter 7

# Lepton Flavour Violation and Dark Matter in Seesaw Type III Model

## 7.1 Introduction

This chapter is based on the work [171]. We study a supersymmetric version of the seesaw mechanism type-III. The model consists of the MSSM particle content plus three copies of **24** superfields. The fermionic part of the $SU(2)$ triplet contained in the **24** is responsible for the type-III seesaw, which is used to explain the observed neutrino masses and mixings. Complete copies of **24** are introduced to maintain gauge coupling unification. These additional states change the beta functions of the gauge couplings above the seesaw scale. Using mSUGRA boundary conditions we calculate the resulting supersymmetric mass spectra at the electro-weak scale using full 2-loop renormalization group equations. We show that the resulting spectrum can be quite different compared to the usual mSUGRA spectrum. We discuss how this might be used to obtain information on the seesaw scale from mass measurements. Constraints on the model space due to limits on lepton flavour violating decays are discussed. The main constraints come from the bounds on $\mu \to e\gamma$ but there are also regions where the decay $\tau \to \mu\gamma$ gives stronger constraints. We also calculate the regions allowed by the dark matter constraint. For the sake of completeness, we compare our results with those for the supersymmetric seesaw type-II and, to some extent, with type-I.

## 7.2 Supersymmetric seesaw type-III

In the case of a seesaw model type-III one needs new fermions $\Sigma$ at the high scale belonging to the adjoint representation of $SU(2)$. This has to be embedded in a **24**-plet to obtain a complete $SU(5)$ representation. The superpotential of the unbroken $SU(5)$ relevant for our discussion is

$$W = \sqrt{2}\,\bar{5}_M Y^5 10_M \bar{5}_H - \frac{1}{4}10_M Y^{10}10_M 5_H + 5_H 24_M Y_N^{III}\bar{5}_M + \frac{1}{2}24_M M_{24}24_M \ . \ (7.1)$$

We have not specified the Higgs sector responsible for the $SU(5)$ breaking. The new parts, which will give the seesaw mechanism, comes from the $24_M$. It decomposes under $SU(3) \times SU(2) \times U(1)$ as

$$
\begin{aligned}
24_M &= (1,1,0) + (8,1,0) + (1,3,0) + (3,2,-5/6) + (3^*,2,5/6) , \\
&= \widehat{B}_M + \widehat{G}_M + \widehat{W}_M + \widehat{X}_M + \widehat{\overline{X}}_M .
\end{aligned}
\tag{7.2}
$$

The fermionic components of $(1,1,0)$ and $(1,3,0)$ have exactly the same quantum numbers as $\widehat{N}^c$ and $\Sigma$. Thus, the $24_M$ always produces a combination of the type-I and type-III seesaw.

In the $SU(5)$ broken phase the superpotential becomes

$$
\begin{aligned}
W_{III} &= W_{MSSM} + \widehat{H}_u(\widehat{W}_M Y_N - \sqrt{\tfrac{3}{10}}\widehat{B}_M Y_B)\widehat{L} + \widehat{H}_u \widehat{\overline{X}}_M Y_X \widehat{D}^c \\
&\quad + \tfrac{1}{2}\widehat{B}_M M_B \widehat{B}_M + \tfrac{1}{2}\widehat{G}_M M_G \widehat{G}_M + \tfrac{1}{2}\widehat{W}_M M_W \widehat{W}_M + \widehat{X}_M M_X \widehat{\overline{X}}_M
\end{aligned}
\tag{7.3}
$$

As before we use at the GUT scale the boundary condition $Y_N = Y_B = Y_X$ and $M_B = M_G = M_W = M_X$. Integrating out the heavy fields yields the following formula for the neutrino masses at the low scale:

$$
m_\nu = -\frac{v_u^2}{2}\left(\frac{3}{10}Y_B^T M_B^{-1} Y_B + \frac{1}{2}Y_W^T M_W^{-1} Y_W\right) .
\tag{7.4}
$$

As mentioned above there are two contributions stemming from the gauge singlet as well as from the $SU(2)$ triplet. In this case the calculation of the Yukawa couplings in terms of a given high scale spectrum is more complicated than in the other two types of seesaw models. However, as we start from universal couplings and masses at $M_{GUT}$ we find that at the seesaw scale one still has $M_B \simeq M_W$ and $Y_B \simeq Y_W$ so that one can write in a good approximation

$$
m_\nu = -v_u^2 \frac{4}{10} Y_W^T M_W^{-1} Y_W
\tag{7.5}
$$

and one can use the corresponding decomposition for $Y_W$ as discussed in section 5.1 up to the overall factor $4/5$.

## 7.3  Effects of the heavy particles on the MSSM spectrum

The appearance of charged particles at scales between the electro-weak scale and the GUT scale leads to changes in the beta functions of the gauge couplings [86, 109]. In the MSSM the corresponding values at 1-loop level are $(b_1, b_2, b_3) = (33/5, 1, -3)$. In case of one **15**-plet the additional contribution is $\Delta b_i = 7/2$ whereas in case of **24**-plet it is $\Delta b_i = 5$. This results in case of type-II in a total shift of $\Delta b_i = 7$ for the minimal model and in case of type-III in $\Delta b_i = 15$ assuming 3 generations of **24**-plets.

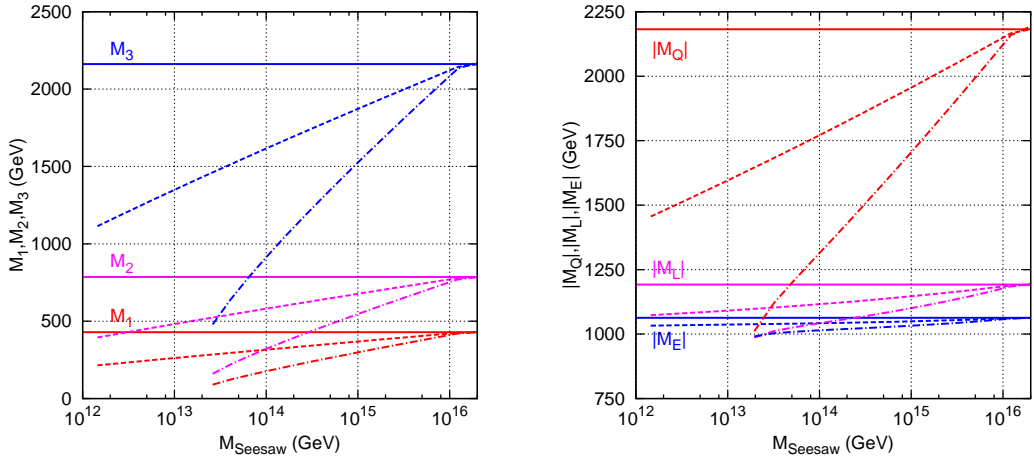

**Figure 7.1:** Mass parameters at $Q = 1$ TeV versus the seesaw scale for fixed high scale parameters $m_0 = M_{1/2} = 1$ TeV, $A_0 = 0$, $\tan\beta = 10$ and $\mu > 0$. The full lines correspond to seesaw type-I, the dashed ones to type-II and the dash-dotted ones to type-III. In all cases a degenerate spectrum of the seesaw particles has been assumed.

This does not only change the evolution of the gauge couplings but also the evolution of the gaugino and scalar mass parameters with profound implications on the spectrum [109, 113]. Additional effects on the spectrum of the scalars can be present if some of the Yukawa couplings get large [113, 172, 173]. In Fig. 7.1 we exemplify this by showing the values of selected mass parameters at $Q = 1$ TeV versus the seesaw scale for fixed high scale parameters $m_0 = M_{1/2} = 1$ TeV and we have set the additional Yukawa couplings to zero. As expected, the effects in case of models of type-II and III are larger the smaller the corresponding seesaw-scale is. The scalar mass parameters shown are of the first generation and, thus, the results are nearly independent of $\tan\beta$ and $A_0$. For illustration we show in Fig. 7.2 the corresponding spectrum where we have fixed $\tan\beta = 10$ and $A_0 = 0$.

We note that in all three model types the ratio of the gaugino mass parameters is nearly the same as in the usual mSUGRA scenarios but the ratios of the sfermion mass parameters change [109, 113]. One can form four 'invariants' for which at least at the 1-loop level the dependence on $M_{1/2}$ and $m_0$ is rather weak, e.g. $(m_L^2 - m_E^2)/M_1^2$, $(m_Q^2 - m_E^2)/M_1^2$, $(m_D^2 - m_L^2)/M_1^2$ and $(m_Q^2 - m_U^2)/M_1^2$. Here one could replace $M_1$ by any of the other two gaugino masses which simply would amount in an overall rescaling. In Fig. 7.3 we show these 'invariants' in the leading-log approximation at 1-loop order to demonstrate the principal behaviour for seesaw type-II with a pair of **15**-plets and seesaw type-III with three **24**-plets. From this one concludes that in principle one has a handle to obtain information on the seesaw scale for given assumptions on the underlying neutrino mass model, if universal boundary conditions are assumed. For the type-I, i.e. singlets only, of course $\Delta b_i = 0$ and no change with respect to mSUGRA are expected. If, for example, the seesaw III model would be realized in nature with three **24**-plets having similar masses around $10^{13}$ GeV one could e.g. show that the corresponding ratios cannot be obtained with one pair of **15**-plets in the seesaw II model, thus excluding this possibility. However, taking the seesaw II with two pairs of **15**-plets one would obtain similar ratios as in this case the corresponding additional beta-functions at 1-loop would be $\Delta b_i = 14$, e.g. nearly equal to our seesaw III model.

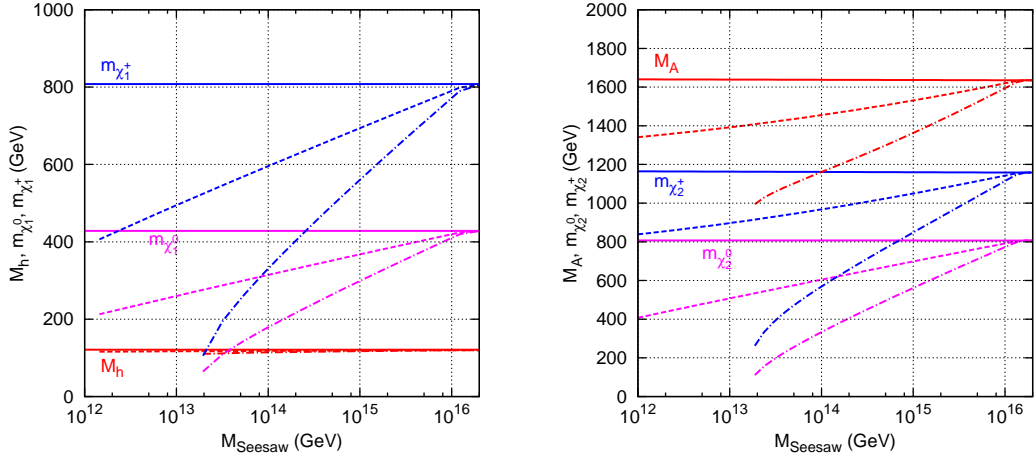

**Figure 7.2:** Example of spectra at $Q = 1$ TeV versus the seesaw scale for fixed high scale parameters $m_0 = M_{1/2} = 1$ TeV, $\tan\beta = 10$ and $\mu > 0$. On left panel $M_h, m_{\tilde\chi_1^0}, m_{\tilde\chi_1^+}$ while on the right panel we have $M_A, m_{\chi_2^0}, m_{\tilde\chi_2^+}$. The line codes are as in Fig. 7.1.

The leading-log approximation gives only the general trend, but there is an important dependence on the SUSY point chosen. In Fig. 7.4 we show as illustration $(m_L^2 - m_E^2)/M_1^2$ and $(m_Q^2 - m_E^2)/M_1^2$ for different mSUGRA points and at different loop orders: the dashed lines are at 1-loop level whereas the solid ones are at 2-loop level. The points considered are SPS3 [174] with $m_0 = 90$ GeV, $M_{1/2} = 400$ GeV, $A_0 = 0$, $\tan\beta = 10$, $\mu > 0$ and for the same values of $A_0$ and $\tan\beta$ two points with $M_{1/2} = 1$ TeV: $m_0 = 500$ GeV and $m_0 = 1$ TeV. The black line shows for comparison the leading-log approximation. We observe that usually the approximation gets worse for lower values of $M_{24}$ and this is even stronger at the 2-loop level which is a consequence of the large coefficient in the beta functions at the 2-loop level. Nevertheless, one sees that in general it gives the correct trend, but it might even fail completely, e.g. in the case of $M_{1/2} = m_0 = 1$ TeV. The reason for the drop around $M_{24} \simeq 3.5 \times 10^{13}$ is that the difference between the parameters goes to zero as can also be seen from the right of Fig. 7.1, see also discussion below.

Last but not least we note that the use of the 2-loop RGEs leads to a shift of $M_{GUT}$ from about $2 \times 10^{16}$ GeV for **24**-plet mass of $10^{16}$ GeV to about $4 \times 10^{16}$ GeV for **24**-plet mass of $10^{13}$ GeV, which is part of the differences between the 1-loop and 2-loop results in Fig. 7.4. Here $M_{GUT}$ is defined as the scale where the electro-weak couplings meet, e.g. $g_{U(1)} = g_{SU(2)}$. This implies also that there is some difference for the strong coupling which is, however, in the order of 5-10% which can easily be accounted for by threshold effects of the new GUT particles, e.g. the missing members of the gauge fields and the Higgs fields responsible for the breaking of the GUT group [175]. A second reason why the deviations between the leading log calculation, the case of 1-loop and 2-loop RGEs gets larger for smaller seesaw scale is that the increase of the beta coefficients implies larger values of the gauge couplings at the GUT scale. This implies that one reaches a Landau pole for sufficiently low values of the seesaw scale. As an example we show in Fig. 7.5 the value of the gauge coupling at $M_{GUT} = 2 \times 10^{16}$ GeV as a function of the seesaw scale for type-II with a pair of **15**-plets (black lines) and type-III with three degenerate **24**-plets (green lines). In both cases the 2-loop RGEs imply a larger gauge coupling for a fixed seesaw scale. One sees that in case of type-II (type-III) in principle one could reach a seesaw scale of about $10^8$ GeV ($10^{13}$ GeV). However, we believe that we can no longer trust even the 2-loop calculation for such large values of the $g_i$, as the neglected higher order terms become more and more important. Especially, we should not trust the "turn-over" of the invariants in Fig. 7.4 for very low values of the seesaw scale, since the numerical calculation at these points is already very close to breaking down.

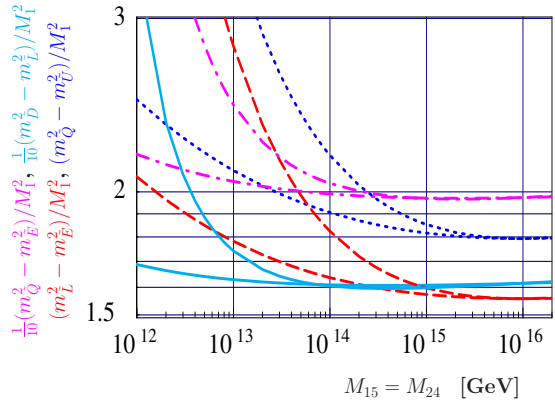

**Figure 7.3:** Four different "invariant" combinations of soft masses versus the mass of the **15**-plet *or* **24**-plet, $M_{15} = M_{24}$. The plot assumes that the Yukawa couplings are negligibly small. The calculation is at 1-loop order in the leading-log approximation. The lines running faster up towards smaller $M$ are for type-III seesaw, the values for type-II seesaw are shown for comparison.

We would also like to mention that, in the numerical calculation we find very often that one of the scalar masses squared, in particular staus and/or sbottoms, gets large negative values already for values of the seesaw scale larger than the Landau pole and thus we can not go to values of the seesaw scale as low as the examples shown in Fig. 7.4 in many SUSY points.

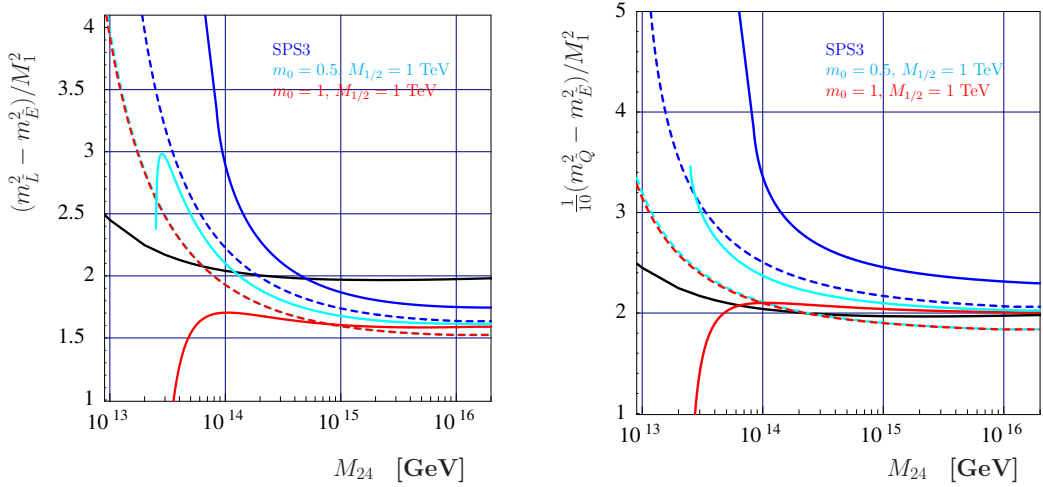

**Figure 7.4:** The limits of the invariants in seesaw type-III models. Left: $(m_L^2 - m_E^2)/M_1^2$, right $(m_Q^2 - m_E^2)/M_1^2$. The blue lines are for SPS3, the light blue one for $m_0 = 500$ GeV and $M_{1/2} = 1$ TeV, and the red one for $m_0 = M_{1/2} = 1$ TeV; full (dashed) lines are 2-loop (1-loop) results. The black line is the analytical approximation, for comparison.

## 7.4 Lepton flavour violation in the slepton sector

From a one-step integration of the RGEs one gets assuming mSUGRA boundary conditions a first rough estimate for the lepton flavour violating entries in the slepton mass parameters:

$$m_{L,ij}^2 \simeq -\frac{a_k}{8\pi^2}\left(3m_0^2 + A_0^2\right)\left(Y_N^{k,\dagger}LY_N^k\right)_{ij} , \qquad (7.6)$$

$$A_{l,ij} \simeq -a_k\frac{3}{16\pi^2}A_0\left(Y_eY_N^{k,\dagger}LY_N^k\right)_{ij} , \qquad (7.7)$$

for $i \neq j$ in the basis where $Y_e$ is diagonal, $L_{ij} = \ln(M_{GUT}/M_i)\delta_{ij}$ and $Y_N^k$ is the additional Yukawa coupling of the type-$k$ seesaw at $M_{GUT}$ $(k = I, II, III)$. We obtain

$$a_I = 1 , \; a_{II} = 6 \; \text{and} \; a_{III} = \frac{9}{5} . \qquad (7.8)$$

Note, that in case of the type-II the matrix $L$ is degenerate and thus can be factored out. All models have in common that they predict negligible flavour violation for the right-sleptons

$$m_{E,ij}^2 \simeq 0. \qquad (7.9)$$

We know that these approximations work well only in case of the type-I models. Nevertheless they give a rough idea on the relative size one has to expect for the rare lepton

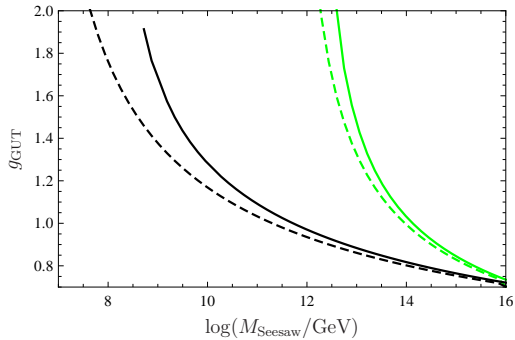

**Figure 7.5:** Values of the gauge coupling at $M_{GUT} = 2 \times 10^{16}$ GeV as a function of the seesaw scale, black lines seesaw type-II and green lines seesaw type-III with three **24**-plets with degenerate mass spectrum; full (dashed) lines are 2-loop (1-loop) results. For the calculation of the electroweak threshold the spectrum corresponds to $m_0 = M_{1/2} = 1$ TeV, $A_0 = 0$, $\tan\beta = 10$ and $\mu > 0$.

decays $l_i \to l_j \gamma$ which very roughly scale like

$$Br(l_i \to l_j \gamma) \propto \alpha^3 m_{l_i}^5 \frac{|m_{L,ij}^2|^2}{\tilde{m}^8} \tan^2\beta. \tag{7.10}$$

where $\tilde{m}$ is the average of the SUSY masses involved in the loops. Note, that for a given set of high scale parameters both, the different size of the flavour mixing entries and the changed mass spectrum, play a role.

## 7.5 Numerical results

In this section we present our numerical calculations. All results presented below have been obtained with the lepton flavour violating version of the program package SPheno [7, 119]. The RGEs of the seesaw II and seesaw III models have been calculated with SARAH [176–178]. All seesaw parameters are defined at $M_{GUT}$ and as mentioned in the previous section we require for models of type-II the boundary condition $Y_Z = Y_S = Y_T$ and $M_Z = M_S = M_T$ and in case of type-III models $Y_N = Y_B = Y_W$ and $M_B = M_G = M_W = M_X$. We evolve the RGEs to the scale(s) corresponding to the GUT scale values of the masses of the heavy particles. The RGE evolution implies also a splitting of the heavy masses. We therefore add at the corresponding scale the threshold effects due to the heavy particles to account for the different masses. In case of type-III models off-diagonal elements are induced in the mass matrices. This implies that one has to go the corresponding mass eigenbasis before calculating the threshold effects. We use 2-loop RGEs everywhere except stated otherwise. In the appendix we give the necessary ingredients on how to obtain them in the seesaw type-II and III models. The analogous anomalous dimensions for the type-I model can be found in [115].

Unless mentioned otherwise, we fit neutrino mass squared differences to their best fit values [14] and the angles to tri-bi-maximal (TBM) values [29]. Our numerical procedure is as follows. Inverting the seesaw equation, see eqs. (6.11) and (7.4), one can get a first guess of the Yukawa couplings for any fixed values of the light neutrino masses (and angles) as a function of the corresponding triplet mass for any fixed value of the couplings. This first guess will not give the correct Yukawa couplings, since the neutrino masses and mixing angles are measured at low energy, whereas for the calculation of $m_\nu$ we need to insert the parameters at the high energy scale. However, we can use this first guess to run numerically the RGEs to obtain the exact neutrino masses and angles (at low energies) for these input parameters. The difference between the results obtained numerically and the input numbers can then be minimized in a simple iterative procedure until convergence is achieved. As long as neutrino Yukawas are $\forall Y_{ij} < 1$ we reach convergence in a few steps. However, in seesaw type-II and type-III the Yukawas run stronger than in seesaw type-I, so our initial guess can deviate sizable from the correct Yukawas, implying in general also more iterations until full convergence is reached. Since neutrino data requires at least one neutrino mass to be larger than about 0.05 eV, we do not find any solutions for $M_T \gtrsim \lambda_2 \times 10^{15}$ GeV and $M_{24} \gtrsim 8 \times 10^{14}$ GeV, respectively. In the latter case we have assumed that all 24-plets have similar masses. For sake of completeness we note that one can also satisfy all neutrino data by giving one of the 24-plets a large mass in the order of $M_{GUT}$ or larger having a model with effectively only two 24-plets.

### 7.5.1 Lepton flavour violation

We have seen in eq. (7.10) that rates for the lepton flavour violating decays of $\mu$ and $\tau$ scale like the LFV entries in the slepton mass squared matrix squared and inverse to the overall SUSY mass to the power eight. From this one immediately concludes the rates for the rare lepton decays are in general larger in seesaw models of type-II and III than in type-I models for fixed SUSY masses and seesaw scales except if one arranges for special cancellations.

Comparing the type-II with the type-III model one finds that LFV decays are larger for type-III, as shown for the case of $\mu \to e\gamma$ in Fig. 7.6. From eqs. (7.7) and (7.8), however, one would expect that type-II should have larger LFV. Numerically we find the opposite for two reasons. (i) $Br(l_i \to l_j\gamma)$ strongly depends on the SUSY masses, see eq. (7.10) and type-III has a lighter spectrum than type-II (for the same mSUGRA input parameters). And (ii) 2-loop effects are very important in type-III, due to the large coefficients, in general leading to large flavor violating soft SUSY breaking parameters.

In Fig. 7.6 we compare $Br(l_i \to l_j\gamma)$ for the three seesaw models taking degenerate seesaw spectra in case of type-I and type-III. Note that in case of seesaw type-III we can only show a relatively short interval for the seesaw scale which is mainly due to two reasons: (i) for scales below approximately $10^{13}$ GeV the gauge couplings get large at $M_{GUT}$ as a consequence of the large beta functions and, thus, perturbation theory breaks down. (ii) One encounters negative mass squares for the scalars, in particular for the lighter stau and/or lighter sbottom. The latter point is also the reason why the possible range is larger in case of the larger soft SUSY breaking parameters.

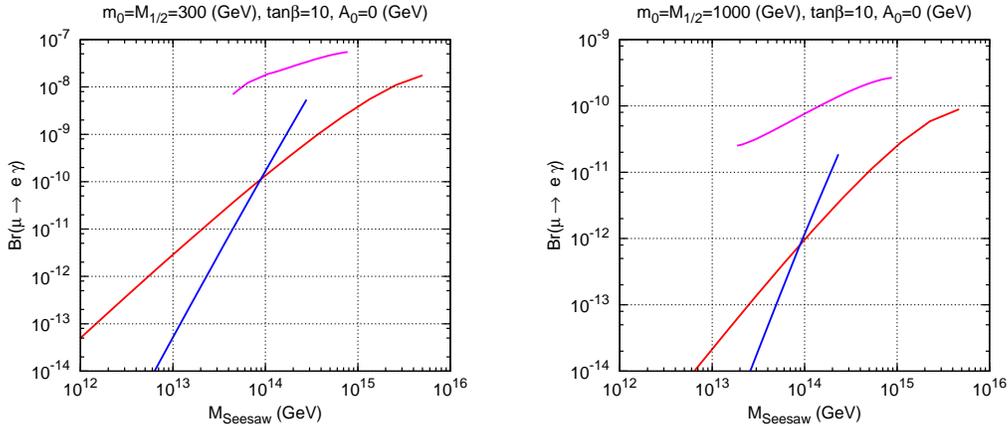

**Figure 7.6:** $Br(\mu \to e\gamma)$ as a function of the seesaw scale for seesaw type-I (red line), seesaw type-II (blue line) and seesaw type-III (magenta line). In case of type-I and type-III a degenerate spectrum has been assumed. On the left panel $m_0 = m_{1/2} = 300$ (GeV), on the right panel $m_0 = m_{1/2} = 1000$ (GeV). In both cases we take $\tan\beta = 10$, $A_0=0$ and $\mu > 0$.

| $m_0$ | $m_{\tilde{\chi}_1^0}$ | $m_{\tilde{\chi}_1^+}$ | $m_{\tilde{\chi}_2^+}$ | $m_{\tilde{g}}$ | $m_{\tilde{\tau}_1}$ | $m_{\tilde{e}_R}$ | $m_{\tilde{e}_L}$ | $m_{\tilde{t}_1}$ |
|---|---|---|---|---|---|---|---|---|
| 500 | 178 | 333 | 617 | 1029 | 535 | 543 | 600 | 772 |
| 1000 | 180 | 338 | 642 | 1057 | 1008 | 1020 | 1043 | 925 |

**Table 7.1:** Examples masses in GeV for $M_{1/2} = 1000$ GeV, $\tan\beta = 10$, $A_0 = 0$ GeV and $\mu > 0$, for seesaw type-III for a degenerate seesaw spectrum with $M_{24} = 10^{14}$ GeV.

The values for $Br(\mu \to e\gamma)$ in Fig. 7.6 are larger than the current experimental bound [56], so one might worry if in case of type-III models only SUSY spectra beyond the reach of the LHC are allowed. (Note, that even for the examples shown the masses of the sfermions are already in the range of several hundred GeVs as can be seen from table 7.1.) Indeed we find that by putting generic Yukawa couplings which are able to explain neutrino data one needs a heavy spectrum to be consistent with bounds on the rare lepton decays. However, this is strictly true only for the TBM angles and $R = \mathbf{1}$. Accidental cancelations due to different contributions to the flavor violating soft masses and thus to the rare lepton decays are possible in type-III (and in type-I). As an example we show in Fig. 7.7 $Br(\mu \to e\gamma)$ as a function of the reactor angle $s_{13}^2$ for different values of the Dirac phase $\delta$. For comparison we also show the calculation for a type-I model. For $\delta = \pi$ there is a range of $s_{13}^2$ where this branching ratio is below the experimental constraint.

At first glance this seem to require some fine-tuning of the underlying parameters. However, one can look at this from a different perspective: Assume that the MEG col-laboration has found a non-vanishing value for $Br(\mu \to e\gamma)$ and from LHC data one has found that the spectrum is consistent with the type-III seesaw model. For a fixed $R$-matrix, e.g. $R=\mathbf{1}$ one would obtain in this case a relation between $s_{13}^2$ and $M_{24}$. This can be exploited to put a bound on $M_{24}$ or even to determine it depending on the outcome of measurements of reactor angle and, thus, the model assumptions can be tested. In Fig. 7.8 we show the corresponding rare tau decays. Note that also for $\tau \to e\gamma$ such a cancelation exists in principle but the corresponding range is excluded by $\mu \to e\gamma$. In contrast $\tau \to \mu\gamma$ is insensitive to the reactor angle and should be measurable in the near future.

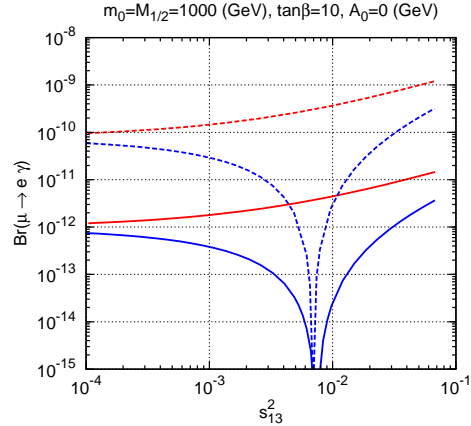

**Figure 7.7:** $Br(\mu \to e\gamma)$ versus $s_{13}^2$ for $m_0 = M_{1/2} = 1000$ GeV, $\tan\beta = 10$, $A_0 = 0$ GeV and $\mu > 0$, for seesaw type-I (solid lines) and seesaw type-III (dashed lines), for $M_{\text{Seesaw}} = 10^{14}$ GeV. The curves shown are for 2 values of the Dirac phase: $\delta = 0$ (red) and $\delta = \pi$ (blue), both for normal hierarchy.

Up to now we have assumed that the seesaw spectrum is nearly degenerate which is of course a strong assumption. We show in Fig. 7.9 two examples where we keep in each case two masses fixed and vary the third one. Note, that in contrast to SUSY particles the indices of the heavy particles are generation indices and do not correspond to a particular mass ordering, e.g. $M_{R_2}$ corresponds to the 'solar neutrino scale' and $M_{R_3}$ to the 'atmospheric neutrino scale'. In case that the mass of the first generation state is varied, e.g. the left plot of this figure, one finds a decrease of the branching ratios with increasing seesaw mass $M_{R_1}$. This is mainly caused by an increase of the SUSY spectrum while at the same time neutrino physics is only affected mildly requiring only a light increase of the corresponding Yukawa couplings to obtain the correct neutrino masses. If, on the other hand, the mass $M_{R_3}$ of the third generation seesaw particles is increased on needs also a sizable increase of the Yukawa couplings to obtain the correct neutrino mass difference squared for the atmospheric sector. This leads to the observed behaviour that the branching ratios for $\tau \to \mu\gamma$ and $\tau \to 3\mu$ increases while the other ones decrease.

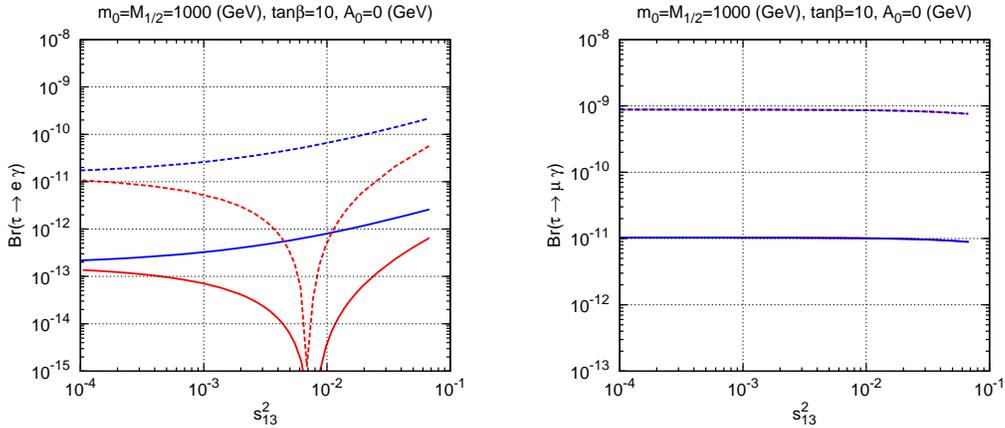

**Figure 7.8:** $Br(\tau \to e\gamma)$ versus $s_{13}^2$ (left) and $Br(\tau \to \mu\gamma)$ versus $s_{13}^2$ (right) for $m_0 = M_{1/2} = 1000$ GeV, $\tan\beta = 10$, $A_0 = 0$ GeV and $\mu > 0$, for seesaw type-I (solid lines) and seesaw type-III (dashed lines), for $M_{\text{Seesaw}} = 10^{14}$ GeV. The curves shown are for $\delta = 0$ (red) and $\delta = \pi$ (blue) for normal hierarchy.

### 7.5.2 Dark Matter

The changes in the spectrum induced by the new heavy states also impact on the predictions with respect to the relic density which we have calculated using the program `micrOMEGAs` [161]. As is well-known, within mSUGRA there are 4 regions in parameter space, in which the constraint from dark matter can be satisfied. These are (i) the bulk region; (ii) the stau co-annihilation region; (iii) the focus point line and (iv) the Higgs funnel. Below we will show usually the range of $\Omega h^2$ allowed at $3\,\sigma$ according to [56]

$$0.081 \leq \Omega h^2 \leq 0.129\,. \tag{7.11}$$

In particular, the co-annihilation region is very sensitive to the difference between the masses of the lightest stau and the lightest neutralino. In Fig. 7.10 we observe that this difference depends strongly on the seesaw scale in both models. For a fixed $M_{1/2}$ and $m_0$ lowering the seesaw scale increases this mass difference, which then leads to a larger calculated $\Omega h^2$. To compensate for this effect one needs to lower $m_0$, with the value depending on the seesaw scale chosen. For certain seesaw scales then $m_0$ needs to be lowered below $m_0 = 0$ and the co-annihilation region disappears. In this region of parameter space both models behave in a qualitatively similar way. However, recall that spectra run faster towards smaller masses in seesaw type-III.

Also the focus point region is very sensitive to the precise values of the input parameters. The focus point region appears in mSUGRA for large values of $m_0$ and small/moderate values of $M_{1/2}$ of the order of $\mathcal{O}(100)$ GeV, the exact value depending on $m_0$. This can be seen in figs. 7.11 and 7.12 where we show $m_{\tilde{\chi}_1^0}$, the higgsino content $|N_{13}|^2 + |N_{14}|^2$ and the corresponding $\Omega h^2$ as a function of $m_0$ for a fixed seesaw scale $M_{T,W} = 10^{14}$ GeV, $A_0 = 0$, $\tan\beta = 10$, $\mu > 0$ and various values of $M_{1/2}$. Note, that we take different values of $M_{1/2}$ for the two models in such a way that we obtain similar values for $m_{\tilde{\chi}_1^0}$. We find that both models behave differently in this region of parameter space, e.g. the higgsino content $|N_{13}|^2 + |N_{14}|^2$ decreases (increases) with increasing values $m_0$ for seesaw type-II (type-III). However, also for type-II the higgsino content increases for increasing $m_0$ once we reach the multi-TeV range but we did not get correct electroweak symmetry breaking in case of multi-TeV values for $m_0$ in case of type-III models. The increased higgsino content of the lightest neutralino leads to on increase (decrease) of its couplings to the $Z$-boson and the light Higgs boson (to sfermions) resulting in the observed dependence of $\Omega h^2$ for $m_0$ close to the 1-TeV region.

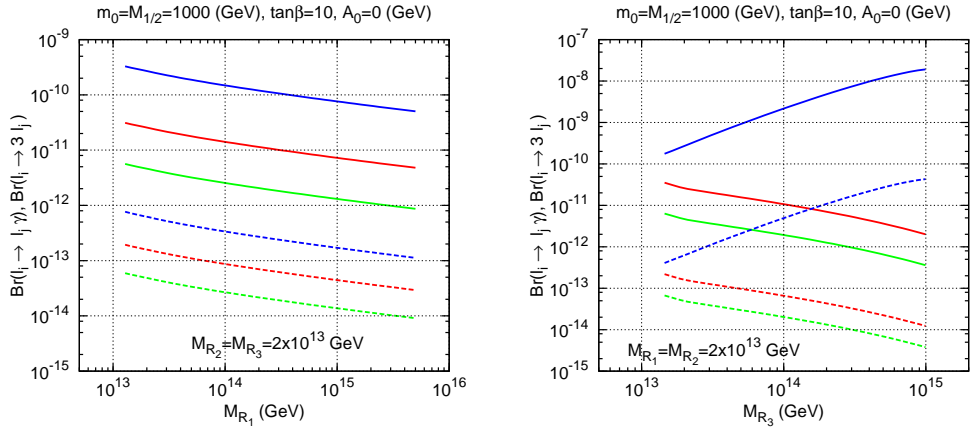

**Figure 7.9:** Branching ratios for $l_i \to l_j\gamma$ (solid lines) and $l_i \to 3l_j$ (dashed lines) versus the seesaw scale for $\tan\beta = 10$, $\mu > 0$, $A_O = 0$ GeV, $M_{1/2} = m_0 = 1000$ GeV. On the left panel we scan on $M_{R_1}$ with $M_{R_2} = M_{R_3} = 2 \times 10^{13}$ GeV while on the right panel we scan on $M_{R_3}$ with $M_{R_1} = M_{R_2} = 2 \times 10^{13}$ GeV. The color code is red for $\mu \to e\gamma$ or $\mu \to 3e$, blue for $\tau \to \mu\gamma$ or $\tau \to 3\mu$ and green for $\tau \to e\gamma$ or $\tau \to 3e$.

With these observations it is clear that the DM allowed regions will be shifted in the $m_0$-$M_{1/2}$ plane compared to the usual mSUGRA expectations. We fix in the following $m_{top} = 171.2$ GeV, $\tan\beta = 10$, $A_0 = 0$ and $\mu > 0$ as well as the seesaw scale to $10^{14}$ GeV. For comparison we show in Fig. 7.13 the usual mSUGRA case without any heavy intermediate particles (left plot) as well as the case of a seesaw type-I scenario (right plot). The blue bands show the $3\sigma$ range according to [56] and we see the three usual regions: the stau co-annihilation with a lighter stau mass close to the LSP mass for $M_{1/2} \lesssim 300$ GeV, the bulk region for moderate values of $M_{1/2}$ and $m_0$ resulting in small sfermion masses as well as the focus point region for $M_{1/2} \simeq 170$ GeV and large values of $m_0$. In addition, we show the lines corresponding to $M_h = 110$ GeV and 114 GeV. Note, that the theoretical uncertainty on $M_h$ is still of the order of 3-5 GeV [164, 165]. Moreover, the value of the Higgs boson mass also depends strongly on $A_0$ and in particular for negative values of $A_0$ one can easily increase the value of $M_h$ while the DM allowed regions hardly

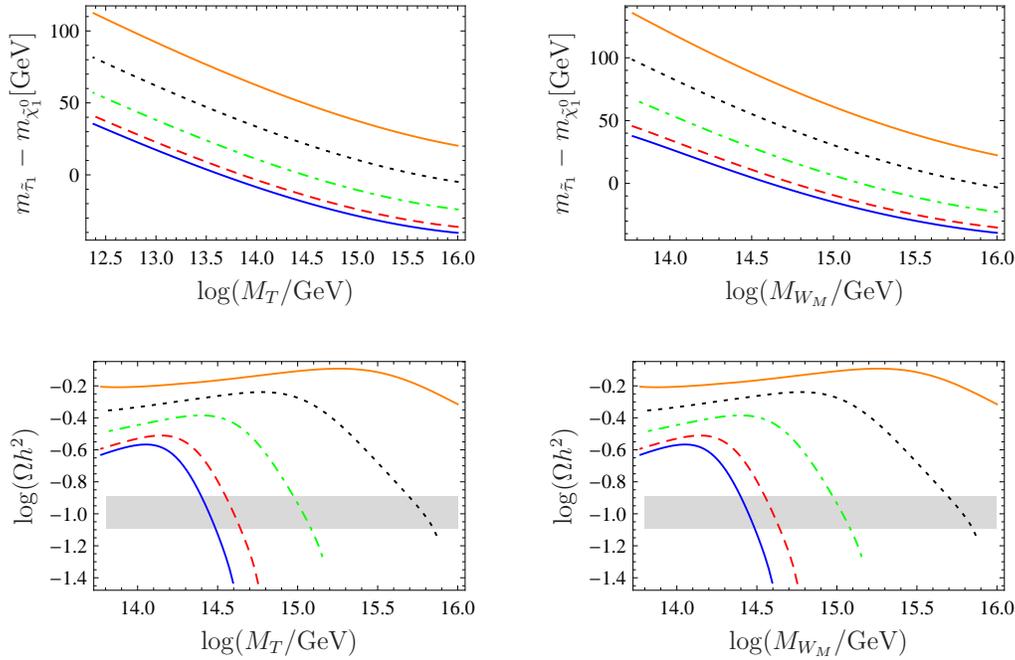

**Figure 7.10:** Difference between the masses and the lightest stau and the lightest neutralino (upper row) as well as the corresponding $\Omega h^2$ (lower row) as a function of the seesaw scale. The left (right) plots are for seesaw type-II (III). A degenerate seesaw spectrum has been assumed in case of seesaw type-III. $M_{1/2} = 800$ GeV, $A_0 = 0$, $\tan\beta = 10$ and $\mu > 0$. The lines correspond to full blue line $m_0 = 0$, red dashed line $m_0 = 50$ GeV, green dashed dotted line $m_0 = 100$ GeV, black dashed line $m_0 = 150$ GeV and orange full line $m_0 = 200$ GeV. The gray band shows the preferred range according to eq. (7.11).

change.

The part of parameter space most affected is the one at large $m_0$. Since in mSUGRA $\mu$ is calculated from the requirement of correct electroweak symmetry breaking, $\mu$ changes rapidly in this region. With the Higgsino content in the lightest neutralino changing rapidly as a function of $\mu$, this region is then very sensitive to any changes of parameters. Since the $Y_\nu$ also impacts on the running of the Higgs mass parameters and thus slightly affects the value predicted for $\mu$, some small changes are found relative to mSUGRA here. Note, however, that this region is highly constrained by the lower bound on the lightest chargino mass of the order of 103 GeV [179].

In case of the other two seesaw models the shift of the allowed regions is much more pronounced, as discussed above. In Figs. 7.14 and 7.15 we show to regions for type-II (left plot) and type-III (right plot) and two different values for $A_0$. As claimed above, the Higgs mass bounds gets shifted significantly while the DM allowed regions are hardly affected. As expected the effects are much more pronounced in case of type-III as the effects of the heavy particles on the spectrum is much stronger. Note, that in particular the bending of the allowed region for large $m_0$ is due to the changed higgsino content as discussed in case of figs. 7.11 and 7.12. Moreover, the case of stau co-annihilation is not viable anymore in case of the type-III model already for this value of the seesaw scale. For completeness we mention that for the type-II the stau co-annihilation region disappears (below $M_{1/2} = 1500$ GeV) for $M_T \lesssim 10^{13}$ GeV. For completeness we note that the results here differ slightly from the ones of our previous work [6] because (i) of the corrections of the 1-loop RGEs of ref. [86] by [180] and (ii) the complete set of 2-loop RGEs are now used.

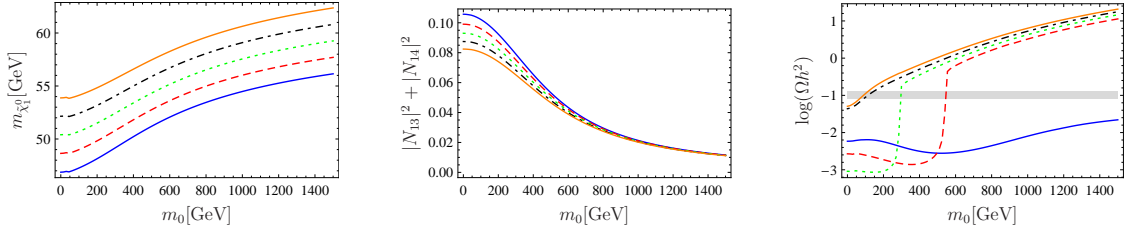

**Figure 7.11:** Mass of the lightest neutralino (left plot), its higgsino content (middle plot) and the corresponding $\Omega h^2$ (right plot) as a function of $m_0$ for a seesaw type-II model with $M_T = 10^{14}$ GeV, $m_{top} = 171.2$ GeV, $A_0 = 0$, $\tan \beta = 10$ and $\mu > 0$. The lines correspond to full blue line $M_{1/2} = 195$ GeV, red dashed line $M_{1/2} = 200$ GeV, green dashed dotted line $M_{1/2} = 205$ GeV, black dashed line $M_{1/2} = 210$ GeV and orange full line $M_{1/2} = 215$ GeV. The gray band shows the range eq. (7.11).

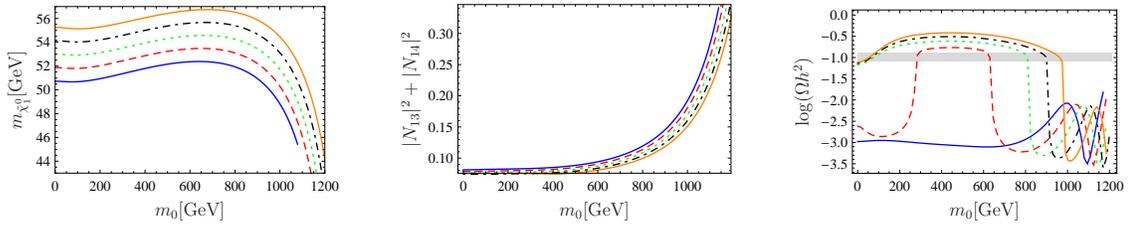

**Figure 7.12:** Mass of the lightest neutralino (left plot), its higgsino content (middle plot) and the corresponding $\Omega h^2$ (right plot) as a function of $m_0$ for a seesaw type-III model with a degenerate seesaw scale $M_W = 10^{14}$ GeV, $m_{top} = 171.2$ GeV, $A_0 = 0$, $\tan \beta = 10$ and $\mu > 0$. The lines correspond to full blue line $M_{1/2} = 400$ GeV, red dashed line $M_{1/2} = 405$ GeV, green dashed dotted line $M_{1/2} = 410$ GeV, black dashed line $M_{1/2} = 415$ GeV and orange full line $M_{1/2} = 420$ GeV. The gray band shows the range eq. (7.11).

In the case of large $\tan \beta$ an additional region, usually called the Higgs funnel, opens up. This region is characterized by $M_A \simeq 2m_{\tilde{\chi}_1^0}$. Also here the regions gets shifted compared to usual mSUGRA scenario. However, this region is very sensitive to higher order corrections and therefore it is quite important to use full 2-loop RGEs as can be seen in Fig. 7.16. We have again fixed $A_0 = 0$, $\mu > 0$, $m_{top} = 171.2$ GeV and the seesaw scale to $10^{14}$ GeV, with a degenerate spectrum in case of the type-III model. The main reason for the observed and rather surprisingly large differences between the different calculations is that the 2-loop contributions decrease the neutralino mass compared to the 1-loop case while at the same time increasing $M_A$. For example, in case of seesaw II and for fixed values of $m_0 = M_{1/2} = 1500$ GeV we get in case of 1-loop RGEs $m_{\tilde\chi_1^0} = 560$ GeV, $M_A = 1090$ GeV and in case of 2-loop RGEs $m_{\tilde\chi_1^0} = 498$ GeV, $M_A = 1100$ GeV. For completeness we note that this region is also very sensitive to input values for $m_t$ and $m_b$ [6].

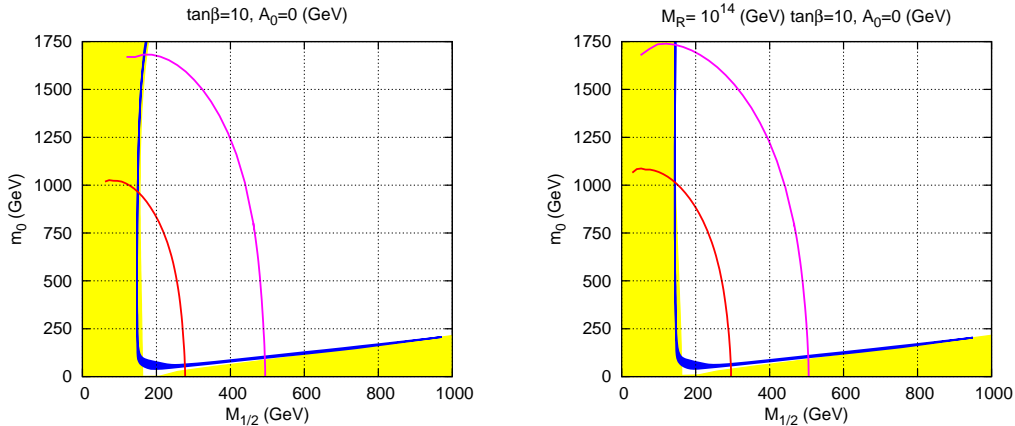

**Figure 7.13:** Dark matter allowed region (in blue) for mSUGRA (left panel) and for type-I seesaw (right panel). The parameters are $\tan\beta = 10$, $A_0 = 0$, $\mu > 0$ and $M_T = 10^{14}$ GeV for $m_{top} = 171.2$ GeV. Also shown (in yellow) are the regions excluded by LEP (small values of $M_{1/2}$), and by LSP constraint (small values of $m_0$). Also shown are the Higgs boson mass curves for $M_h = 110$ GeV (in red) and for $M_h = 114.4$ GeV (in magenta).

## 7.6 Conclusions

To summarize, we have investigated in detail a supersymmetric version of a seesaw model of type-III and compared it to seesaw models of type-I and type-II. In case of type-II and type-III models we have embedded the $SU(2)$ triplets in the corresponding $SU(5)$ representations to maintain gauge coupling unification, e.g. **15**-plets in case of type-II and **24**-plets in case of type-III models. For definiteness we have assumed mSUGRA boundary conditions for the soft SUSY breaking parameters.

The additional heavy charged states lead to changes in the beta-functions and, thus, also in the running of the SUSY mass parameters. We have calculated the soft masses as a function of the seesaw parameters. As discussed in some detail, there are certain combinations of soft masses, which are approximately constants over large regions of mSUGRA space. These "invariants" contain indirect information about the seesaw scale assuming the type of seesaw model. In certain parts of the parameter space, e.g. for low seesaw scales, one might even be able to exclude certain seesaw models by combining mass measurements at the LHC with the mSUGRA paradigm. We note, that using 2-loop RGEs will be crucial to obtain reliable results.

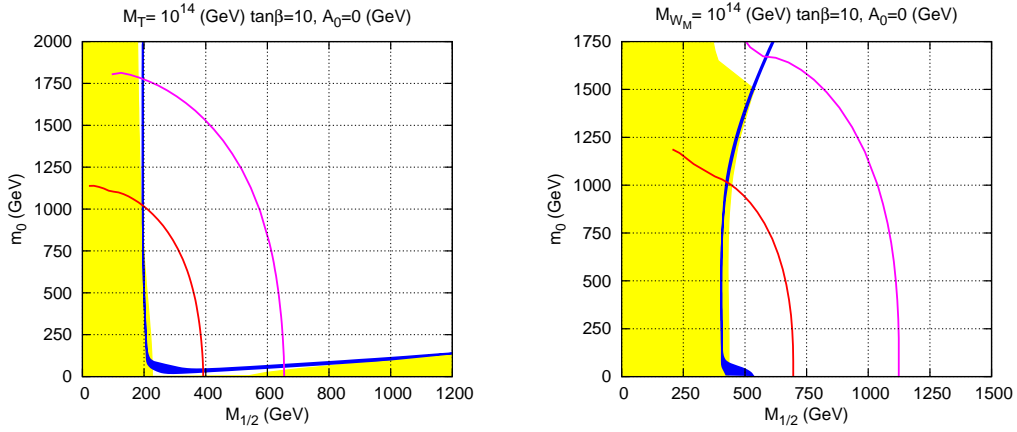

**Figure 7.14:** Like in Fig. 7.13 but for seesaw type-II (left panel) and type-III (right panel).

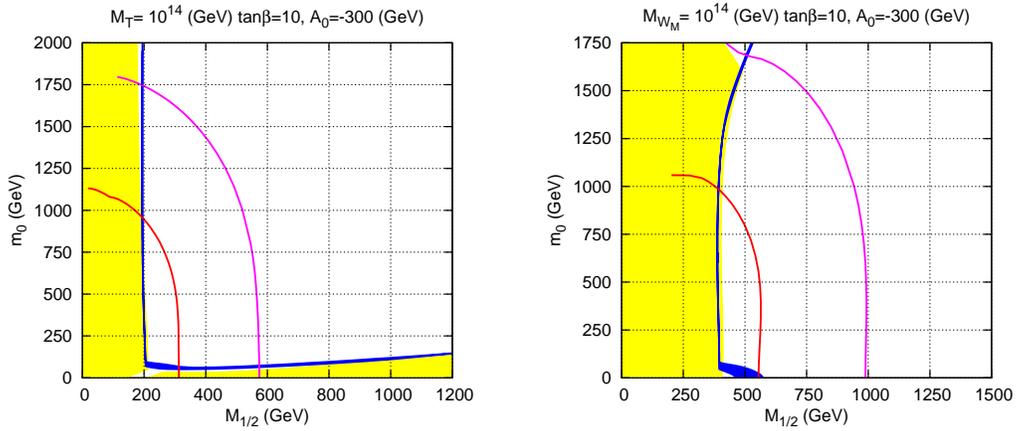

**Figure 7.15:** Like in Fig. 7.14 but for $A_0 = -300$. Seesaw type-II (left panel) and type-III (right panel).

The changes in the spectrum leads obviously to changes in the phenomenology. We have calculated lepton flavour violating observables, such as $Br(l_i \to l_j + \gamma)$. We find that for fixed (degenerate) seesaw scale these branching ratios are in general largest for type-III models followed by type-II and type-I. This is a consequence of the fact that for a given set of mSUGRA parameters the spectrum in type-III is lighter than for type-II models which is again lighter than in type-I models. However, the difference in the predictions of type-II and type-III is somewhat smaller than expected from these considerations because in type-II models the flavour violating entries are larger compared to the case of type-III models.

We also investigated the predictions for the relic density $\Omega h^2$ in the type-III model and compared them with the other models. We find the usual four regions in the mSUGRA parameter space but of course they are shifted due to the changes in the spectrum. It has been found that in particular in case of the Higgs-funnel the use of 2-loop RGEs is crucial to identify the correct allowed region. Last but not least we note, that for low seesaw scales the co-annihilation region vanishes for both, the type-II and the type-III models, as the required mass difference between the lightest neutralino and the stau cannot be obtained anymore.

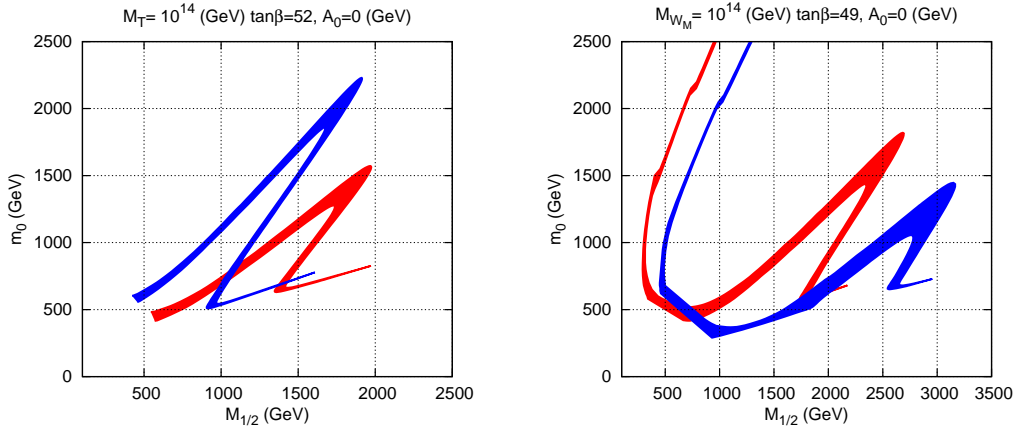

**Figure 7.16:** Comparison between using 1-loop (red) or 2-loop (blue) RGEs on the dark matter allowed region for type-II (left panel) and type-III (right panel). The parameters are: $A_0 = 0$, $\mu > 0$ and $M_{\text{Seesaw}} = 10^{14}$ GeV, $m_{top} = 171.2$ GeV and $\tan\beta = 52$ for type-II and $\tan\beta = 49$ for type-III.

# Chapter 8

# Conclusions

It is now an established fact that neutrinos have mass. This can have important consequences in the development of Physics in the near future. The quest for an explanation why these masses are so tiny puts some stress on the usual Higgs mechanism and asks for new Physical frameworks that, in turn, can have important phenomenological consequences beyond the neutrino sector. We have explored this in the present thesis, constraining the various predictions on Lepton Flavour Violation, Dark Matter and Baryon Asymmetry from the neutrino phenomenology.

In Chapter 3 we started by adopting the possibility of an $A_4$ symmetry in the leptonic and Higgs sector, as proposed in [46]. The Higgs sector is extended with an $SU(2)$ triplet and therefore a mixed Type-I/Type-II seesaw mechanism was proposed that could reproduce a zero texture classified as $B_1$ for the neutrino mass matrix and that is well motivated in explaining neutrino data. Then we explored the triplet decay into leptons and the consequent production of a lepton asymmetry to obtain the phenomenological baryonic asymmetry through the leptogenesis mechanism. To fully address the question of the possible washout of the leptonic asymmetry we have solved numerically the full set of Boltzmann equations and showed that it is possible to obtain the correct baryonic density with the assumptions of this model. Finally we discussed the problem of the Higgs potential minimization and substantiated with a full computation demonstrating that there is a limitation in obtaining the correct vacuum expectation values for the scalar fields within the $A_4$ symmetry and with the representation assignments made therein. We ended by proposing two solutions: breaking explicitly the symmetry group and adding a second $SU(2)$ triplet. This could find a natural framework in a supersymmetrization of the model, although it would make it more elaborate.

In Chapter 4, based on [22], we discussed an alternative model to the one proposed in the previous chapter. We have studied the possibility that the seesaw model with spontaneously broken ungauged lepton number may simultaneously account for the observed neutrino masses and mixing as well as the dark matter of the Universe. We have presented a two-texture structure for the neutrino mass which arises in a inverse seesaw mechanism implementing an $A_4$ flavor symmetry. A predictive pattern of neutrino masses emerges from the interplay of type-I and type-II seesaw contributions, with a lower bound on the neutrinoless double beta decay rate, which correlates with the deviation from maximality of the atmospheric mixing angle $\theta_{23}$, as well as nearly maximal CP violation, correlated with the reactor angle $\theta_{13}$.

On the other hand, assuming that associated Majoron picks up a mass due to explicit lepton number violating effects that may arise, say, from quantum gravity, we showed how it can constitute a viable candidate for decaying dark matter, consistent with cosmic microwave background lifetime constraints that follow from current WMAP observations. We have also shown how the Higgs boson triplet, whose existence is required by the consistency of the model, plays a key role in providing a test of the decaying Majoron dark matter hypothesis, implying the existence of a mono-energetic emission line which arises from the sub-leading one-loop-induced decay of the Majoron into photons. We also discussed the possibility of probing its existence in future X-ray observations such as expected in NASA's Xenia mission [95]. The presence of the type-II seesaw Higgs triplet would also have other particle physics implications, such as lepton flavor violating decay rate enhancements due to the strong renormalization effects of the triplet, quite sizeable in a supersymmetric model.

In Chapter 5, based on [5], we started by stressing that low energy neutrino experiments, including oscillation studies and neutrinoless double-beta decay searches may, optimistically, determine at most 9 neutrino parameters: the 3 neutrino masses, the 3 mixing angles and potentially the 3 CP violating phases. This is insufficient to fully reconstruct the underlying mechanism of neutrino mass generation. Then, under the assumption that neutrino masses arise *a la seesaw*, we have considered the simplest pure type-I or pure type-II seesaw schemes in mSugra. We performed a full scan over the mSugra parameter space in order to identify regions where LFV decays of $\chi_2^0$ can be maximal, while still respecting low-energy constraints that follow from the upper bounds on $\mathrm{Br}(\mu \to e\gamma)$. We have also estimated the expected number of events for $\chi_2^0 \to \chi_1^0 + \tau + \mu$, for a sample luminosity of $\mathcal{L} = 100 \; \mathrm{fb}^{-1}$. The expected number of events for the other channel $\chi_2^0 \to \chi_1^0 + \tau + e$ is always smaller, as can be seen from the LFV branching ratios presented in section 5.2.1. We have found that the pure seesaw-II scheme is substantially simpler and comes closer to being fully reconstructible, provided additional LFV decays are detected and some supersymmetric particles are discovered at the Large Hadron Collider, providing the necessary parameter reconstruction information of the supersymmetric lagrangian.

Note that in what concerns the expected number of events both type-I and type-II schemes give similar results. However, as we have seen, given their smaller number of parameters, type-II seesaw schemes are more likely to be reconstructible through a combination of low energy neutrino measurements, with the possible detection of supersymmetric states and lepton flavour violation at the LHC. This should encourage one to perform full-fledged dedicated simulations, in order to ascertain their feasibility within realistic experimental conditions [96].

We note that we have not done a thorough analysis of the fact that the LFV might induce new "edge variables", giving additional information [128]. We have focused here on LHC, but we should mention that a future ILC would be much more suited for measuring LFV SUSY processes [112, 129–134].

In Chapter 6, based on [6], we have calculated the neutralino relic density in a supersymmetric model with mSugra boundary conditions including a type-II seesaw mechanism to explain current neutrino data. We have discussed how the allowed ranges in mSugra parameter space change as a function of the seesaw scale. The stau co-annihilation region is shifted towards smaller $m_0$ for smaller values of the triplet mass $M_T$, while the bulk region and the focus point line are shifted towards larger values of $M_{1/2}$ for $M_T$ sufficiently below the GUT scale. The Higgs funnel, which appears at large values of $\tan\beta$ has turned out to be especially sensitive to the value of $M_T$. Determining $M_{1/2}$ from the mass of any gaugino and $m_0$ from a sparticle which is *not important for the DM calculation*, one could, therefore, get a constraint on $M_T$ from the requirement that the observed $\Omega_{DM}h^2$ is correctly explained by the calculated $\Omega_{\chi_1^0}h^2$.

On the positive side, we can remark that current data on neutrino masses put an upper bound on $M_T$ of the order of $\mathcal{O}(10^{15})$ GeV. Since this is at least one order of magnitude smaller than the GUT scale, the characteristic shifts in the DM regions are necessarily non-zero if our setup is the correct explanation of the observed neutrino oscillation data. Even more stringent upper limits on $M_T$ follow, in principle, from the non-observation of LFV decays. A smaller $M_T$ implies larger shifts of the DM region. However, the "exact" upper limit on $M_T$ from LFV decays depends strongly on $\tan\beta$, $m_0$ and $M_{1/2}$, and thus can be quantified only when at least some information about these parameters is available.

On the down side, we need to add a word of caution. We have found that the DM calculation suffers from a number of uncertainties, even if we assume the soft masses to be perfectly known. The most important SM parameters turn out to be the bottom and the top quark mass. The focus point line depends extremely sensitively on the exact value of the top mass, the Higgs funnel shows a strong sensitivity on both, $m_b$ and $m_t$.

It is clear that quite accurate sparticle mass measurements will be necessary, before any quantitative conclusions can be taken from the effects we have discussed. Unfortunately, such accurate mass measurements might be very difficult to come by for different reasons. In the focus point region all scalars will be heavy, leading to small production cross section at the LHC. In the co-annihilation line with a nearly degenerate stau and a neutralino, the stau decays produce very soft taus, which are hard for the LHC to measure. And the Higgs funnel extends, depending on $\tan\beta$ and $M_T$, to very large values of $(m_0, M_{1/2})$, at least partially outside the LHC reach. Nevertheless, DM provides in principle an interesting constraint on the (supersymmetric) seesaw explanation of neutrino masses, if seesaw type-II is realized in nature, a fact which to our knowledge has not yet been discussed in the literature.

Finally, in Chapter 7, based on [171], we have studied LFV processes and Dark Matter in the framework of a seesaw Type III model and compare it to Type II and in some extent to Type I. In case of type-II and type-III models we have embedded the $SU(2)$ triplets in the corresponding $SU(5)$ representations to maintain gauge coupling unification, e.g. **15**-plets in case of type-II and **24**-plets in case of type-III models. For definiteness we have assumed mSUGRA boundary conditions for the soft SUSY breaking parameters. The additional heavy charged states lead to changes in the beta-functions and, thus, also in the running of the SUSY mass parameters. We have calculated the soft masses as a function of the seesaw parameters. As discussed in some detail, there are certain combinations of soft masses, which are approximately constants over large regions of mSUGRA space. These "invariants" contain indirect information about the seesaw scale assuming the type of seesaw model. In certain parts of the parameter space, e.g. for low seesaw scales, one might even be able to exclude certain seesaw models by combining mass measurements at the LHC with the mSUGRA paradigm. We note, that using 2-loop RGEs will be crucial to obtain reliable results. he changes in the spectrum leads obviously to changes in the phenomenology. We have calculated lepton flavour violating observables, such as $Br(l_i \rightarrow l_j + \gamma)$. We find that for fixed (degenerate) seesaw scale these branching ratios are in general largest for type-III models followed by type-II and type-I. This is a consequence of the fact that for a given set of mSUGRA parameters the spectrum in type-III is lighter than for type-II models which is again lighter than in type-I models. However, the difference in the predictions of type-II and type-III is somewhat smaller than expected from these considerations because in type-II models the flavour violating entries are larger compared to the case of type-III models. We also investigated the predictions for the relic density $\Omega h^2$ in the type-III model and compared them with the other models. We find the usual four regions in the mSUGRA parameter space but of course they are shifted due to the changes in the spectrum. It has been found that in particular in case of the Higgs-funnel the use of 2-loop RGEs is crucial to identify the correct allowed region. Last but not least we note, that for low seesaw scales the co-annihilation region vanishes for both, the type-II and the type-III models, as the required mass difference between the lightest neutralino and the stau cannot be obtained anymore.

In summary we have considered some direct and indirect consequences of models leading to massive neutrinos, with a two-way perspective. From current phenomenological data we tried to obtain a hint on the actual high scale framework, adopting some previous models [46, 75, 86], and have constrained its parameters, using a couple of simplifying assumptions, like setting to the identity the Casas-Ibarra $R$-matrix [21] - see (1.188) - assuming the Tri-Bimaximal mixing *ansatz* [29] - formula (1.228), the Normal Hierarchy for neutrino masses and setting to zero the Dirac and Majorana CP-phases in Chapters 5 and 6. Then, with the model defined at GUT scale and within these assumptions, we did a study of some of its phenomenological implications at the electroweak scale and below for the present baryonic density in the Universe in Chapter 3, for lepton flavour violation processes in Chapter 5 and for Dark Matter in Chapters 4 and 6. There are still many open questions regarding these matters, like what is the magnitude of $CP$-violation, if any, on the leptonic sector, or if it is seesaw and which type the actual mechanism that leads to such tiny masses for neutrinos, or what is the magnitude of lepton flavour violation (does it remains within the predictions of the Standard Model extended minimally to include neutrino masses or is it there some new Physics), or else why there is an asymmetry between matter and anti-matter in the Universe and what is the mechanism that generates it... LHC surely will give answers, some partial, some definitive, as much as there are definitive answers in Physics.

# Bibliography

[1] S. Weinberg, *The Quantum Theory of Fields - Volume II, Modern Applications* (Cambridge University Press, Cambridge, 2000).

[2] S. Weinberg, *The Quantum Theory of Fields - Volume III, Supersymmetry* (Cambridge University Press, Cambridge, 2000).

[3] C. Giunti and C. W. Kim, *Fundamentals of Neutrino Physics and Astrophysics* (Oxford University Press, Oxford, 2007).

[4] M. Drees, R. M. Godbole, and P. Roy, *Theory and Phenomenology of Sparticles* (World Scientific, Singapore, 2005).

[5] J. N. Esteves *et al.*, JHEP **05**, 003 (2009), arXiv:0903.1408 hep-ph.

[6] J. N. Esteves, S. Kaneko, J. C. Romao, M. Hirsch, and W. Porod, Phys. Rev. D **80**, 095003 (2009), arXiv:0907.5090 hep-ph.

[7] The latest version of SPheno can be obtained from: http://theorie.physik.uni-wuerzburg.de/˜porod/SPheno.html.

[8] V. Castellani, S. Degl'Innocenti, G. Fiorentini, M. Lissia, and B. Ricci, Phys. Rep. **281**, 309 (1997), astro-ph/9606180.

[9] G. Barr, T. Gaisser, P. Lipari, S. Robbins, and T. Stanev, Phys. Rev. D **70**, 023006 (2004), astro-ph/0403630.

[10] KamLAND collaboration, K. Eguchi *et al.*, Phys. Rev. Lett. **90**, 021802 (2003), hep-ex/0212021.

[11] CHOOZ, M. Apollonio *et al.*, Eur. Phys. J. **C27**, 331 (2003), arXiv:hep-ex/0301017.

[12] KamLAND collaboration, T. Araki *et al.*, Phys. Rev. Lett. **94**, 081801 (2005).

[13] K2K, E. Aliu *et al.*, Phys. Rev. Lett. **94**, 081802 (2005), arXiv:hep-ex/0411038.

[14] T. Schwetz, M. Tortola, and J. W. F. Valle, New J. Phys. **10**, 113011 (2008), arXiv:0808.2016.

[15] J. W. d. Herder *et al.*, 0906.1788 [astro-ph].

[16] J. N. Bahcall, M. H. Pinsonneault, and S. Basu, Astrophys. J. **555**, 990 (2001), astro-ph/0010346.

[17] M. E. Peskin and D. Schroeder, *An Introduction to Quantum Field Theory* (Perseus Book, Cambridge, 1995).

[18] G. 't Hooft and M. Veltman, Nucl. Phys. B **44**, 189 (1972).

[19] S. Weinberg, Phys. Rev. D **22**, 1694 (1980).

[20] J. Schechter and J. W. F. Valle, Phys. Rev. **D25**, 2951 (1982).

[21] J. A. Casas and A. Ibarra, Nucl. Phys. B **618**, 171 (2001), hep-ph/0103065.

[22] J. N. Esteves *et al.*, (2010), arXiv:1007.0898v1 hep-ph.

[23] GNO, M. Altmann *et al.*, Phys. Lett. **B616**, 174 (2005), arXiv:hep-ex/0504037.

[24] SAGE, J. N. Abdurashitov *et al.*, J. Exp. Theor. Phys. **95**, 181 (2002), astro-ph/0204245.

[25] Kamiokande, Y. Fukuda *et al.*, Phys. Rev. Lett. **77**, 1683 (1996).

[26] Super-Kamiokande, J. Hosaka *et al.*, Phys. Rev. **D73**, 112001 (2006), arXiv:hep-ex/0508053.

[27] SNO, B. Aharmim *et al.*, Phys. Rev. **C72**, 055502 (2005), nucl-ex/0502021.

[28] M. Maltoni, T. Schwetz, M. A. Tortola, and J. W. F. Valle, New J. Phys.**6** (122 (2004)).

[29] P. F. Harrison, D. H. Perkins, and W. G. Scott, Phys. Lett. B **530**, 167 (2002), hep-ph/0202074.

[30] K. Fujikawa, Phys. Rev. Lett. **42**, 1195 (1972).

[31] M. F. Atiyah and I. M. Singer, Proc. Nat. Acad. Sci. **81**, 2597 (1972).

[32] A. A. Belavin, A. M. Polyakov, A. S. Schwarz, and Y. S. Tyupkin, Phys. Lett. B **59**, 85 (1975).

[33] F. R. Klinkhammer and N. S. Manton, Phys. Rev. D **30**, 2212 (1984).

[34] N. S. Manton, Phys. Rev. D **28**, 2019 (1983).

[35] A. Riotto, (1998), hep-ph/9807454.

[36] V. A. Kuzmin, V. A. Rubakov, and M. E. Shaposhnikov, Phys. Lett. B **155**, 36 (1985).

[37] M. Fukugita and T. Yanagida, Phys. Lett. B **174**, 45 (1986).

[38] J. A. Harvey and M. S. Turner, Phys. Rev. D **42**, 3344 (1990).

[39] S. Weinberg, *Cosmology* (Oxford University Press, Oxford, 2008).

[40] E. W. Kolb and M. S. Turner, *The Early universe* (Westview Press, Colorado, 1990).

[41] J. Bernstein, *Kinetic theory in the expanding universe* (Cambridge University Press, Cambridge, 2004).

[42] J. Bernstein, L. S. Brown, and G. Feinberg, Phys. Rev. D **32**, 3261 (1985).

[43] C. Amsler *et al.*, Physics Letters B **667**, 1 (2008), http://pdg.lbl.gov/.

[44] B. W. Lee and S. Weinberg, Phys. Rev. Lett. **39**, 165 (1977).

[45] J. Schechter and J. W. F. Valle, Phys. Rev. **D25**, 774 (1982).

[46] M. Hirsch, A. S. Joshipura, S. Kaneko, and J. W. F. Valle, Phys. Rev. Lett. **99**, 151802 (2007), arXiv:hep-ph/0703046.

[47] P. Ramond, *Group Theory - A Physicist's Survey* (Cambridge University Press, United Kingdom, 2010).

[48] E. Ma and G. Rajasekaran, Phys. Rev. **D64**, 113012 (2001), hep-ph/0106291.

[49] P. H. Frampton, S. L. Glashow, and D. Marfatia, Phys. Lett. **B536**, 79 (2002), hep-ph/0201008.

[50] J. Lesgourgues and S. Pastor, Phys. Rep. **429**, 307 (2006), astro-ph/0603494.

[51] KATRIN collaboration, G. Drexlin, Nucl. Phys. Proc. Suppl. **145**, 263 (2005).

[52] M. Hirsch, (2006), hep-ph/0609146, invited talk at Neutrino 2006, Santa Fe, New Mexico,13-19 Jun 2006.

[53] J. Schechter and J. W. F. Valle, Phys. Rev. **D22**, 2227 (1980).

[54] T. Hambye and G. Senjanovic, Phys. Lett. **B582**, 73 (2004), arXiv:hep-ph/0307237.

[55] T. Hambye, M. Raidal, and A. Strumia, Phys. Lett. **B632**, 667 (2006), arXiv:hep-ph/0510008.

[56] Particle Data Group, C. Amsler *et al.*, Phys. Lett. **B667**, 1 (2008).

[57] S. Fukuda *et al.*, Phys. Lett. **B539**, 179 (2002), hep-ex/0205075.

[58] SNO collaboration, Q. R. Ahmad *et al.*, Phys. Rev. Lett. **89**, 011301 (2002), nucl-ex/0204008.

[59] Super-Kamiokande collaboration, Y. Fukuda *et al.*, Phys. Rev. Lett. **81**, 1562 (1998), hep-ex/9807003.

[60] T. Kajita, New J. Phys. **6**, 194 (2004).

[61] M. H. Ahn *et al.*, Phys. Rev. **D74**, 072003 (2006), hep-ex/0606032.

[62] P. Adamson, Phys. Rev. Lett. **101**, 131802 (2008).

[63] S. Abe *et al.*, Phys. Rev. Lett. **100**, 221803 (2008).

[64] G. Altarelli and F. Feruglio, New J. Phys. **6**, 106 (2004), hep-ph/0405048.

[65] H. Nunokawa, S. J. Parke, and J. W. F. Valle, Prog. Part. Nucl. Phys. **60**, 338 (2008).

[66] J. W. F. Valle, J. Phys. Conf. Ser. **53**, 473 (2006), hep-ph/0608101, Review lectures at Corfu.

[67] K. S. Babu, E. Ma, and J. W. F. Valle, Phys. Lett. **B552**, 207 (2003), hep-ph/0206292.

[68] A. Zee, Phys. Lett. **B630**, 58 (2005), arXiv:hep-ph/0508278.

[69] G. Altarelli and F. Feruglio, Nucl. Phys. **B741**, 215 (2006), hep-ph/0512103.

[70] P. Minkowski, Phys. Lett. B **67**, 421 (1977).

[71] M. Gell-Mann, P. Ramond, and R. Slansky, (1979), Print-80-0576 (CERN).

[72] T. Yanagida, (KEK lectures, 1979), ed. O. Sawada and A. Sugamoto (KEK, 1979).

[73] S. Glashow, (1980), ed. M. Levy et al. (Plenum, New York), p. 707.

[74] R. N. Mohapatra and G. Senjanovic, Phys. Rev. Lett. **44**, 912 (1980).

[75] J. Schechter and J. W. F. Valle, Phys. Rev. **D25**, 774 (1982).

[76] Y. Chikashige, R. N. Mohapatra, and R. D. Peccei, Phys. Lett. **B98**, 265 (1981).

[77] T. P. Cheng and L. F. Li, Phys. Rev. D **22**, 2860 (1980).

[78] M. Magg and C. Wetterich, Phys. Lett. **B94**, 61 (1980).

[79] G. Lazarides, Q. Shafi, and C. Wetterich, Nucl. Phys. **B181**, 287 (1981).

[80] R. N. Mohapatra and G. Senjanovic, Phys. Rev. **D23**, 165 (1981).

[81] WMAP collaboration, E. Komatsu *et al.*, Astrophys. J. Suppl. **180**, 1330 (2009), arXiv:0803.0547 astro-ph.

[82] R. N. Mohapatra and J. W. F. Valle, Phys. Rev. **D34**, 1642 (1986).

[83] M. C. Gonzalez-Garcia and J. W. F. Valle, Phys. Lett. **B216**, 360 (1989).

[84] F. Deppisch and J. W. F. Valle, Phys. Rev. D **72**, 036001 (2005), hep-ph/0406040.

[85] S. Dev, S. Kumar, S. Verma, and S. Gupta, (2006), hep-ph/0612102.

[86] A. Rossi, Phys. Rev. D **66**, 075003 (2002), hep-ph/0207006.

[87] F. R. Joaquim and A. Rossi, Phys. Rev. Lett. **97**, 181801 (2006), hep-ph/0604083.

[88] V. Berezinsky and J. W. F. Valle, Phys. Lett. **B318**, 360 (1993), hep-ph/9309214.

[89] M. Lattanzi and J. W. F. Valle, Phys. Rev. Lett. **99**, 121301 (2007), arXiv:0705.2406 [astro-ph].

[90] F. Bazzocchi, M. Lattanzi, S. Riemer-Sorensen, and J. W. F. Valle, JCAP **0808**, 013 (2008), arXiv:0805.2372.

[91] G. Passarino and M. J. G. Veltman, Nucl. Phys. **B160**, 151 (1979).

[92] R. Mertig, http://www.feyncalc.org .

[93] S. Tremaine and J. E. Gunn, Phys. Rev. Lett. **42**, 407 (1979).

[94] J. Madsen, Phys. Rev. Lett. **64**, 2744 (1990).

[95] D. Hartmann *et al.*, Decadal Review white paper for the Xenia mission, http://sms.msfc.nasa.gov/xenia/pdf/xenia_whitepaper.pdf.

[96] F. del Aguila *et al.*, Eur. Phys. J. C **57**, 183 (2008), arXiv:0801.1800 hep-ph.

[97] M. Raidal *et al.*, Eur. Phys. J. C **57**, 13 (2008).

[98] J. F. Donoghue, H. P. Nilles, and D. Wyler, Phys. Lett. B **128**, 55 (1983).

[99] L. J. Hall, V. A. Kostelecky, and S. Raby, Nucl. Phys. B **267**, 415 (1986).

[100] F. Borzumati and A. Masiero, Phys. Rev. Lett. **57**, 961 (1986).

[101] S. Antusch, E. Arganda, M. J. Herrero, and A. M. Teixeira, JHEP **11**, 090 (2006), hep-ph/0607263.

[102] E. Arganda and M. J. Herrero, Phys. Rev. D **73**, 055003 (2006), hep-ph/0510405.

[103] F. Deppisch, H. Päs, A. Redelbach, R. Rückl, and Y. Shimizu, Eur. Phys. J. C **28**, 65 (2003), hep-ph/0206122.

[104] J. Hisano, T. Moroi, K. Tobe, and M. Yamaguchi, Phys. Rev. D **53**, 2442 (1996), hep-ph/9510309.

[105] J. Hisano, T. Moroi, K. Tobe, M. Yamaguchi, and T. Yanagida, Phys. Lett. B **357**, 579 (1995), hep-ph/9501407.

[106] E. Arganda, M. J. Herrero, and A. M. Teixeira, JHEP **10**, 104 (2007).

[107] F. Deppisch, T. S. Kosmas, and J. W. F. Valle, Nucl. Phys. B **752**, 80 (2006), hep-ph/0512360.

[108] G. A. Blair, W. Porod, and P. M. Zerwas, Eur. Phys. J. C **27**, 263 (2003), hep-ph/0210058.

[109] M. R. Buckley and H. Murayamas, Phys. Rev. Lett. **97**, 231801 (2006), hep-ph/0606088.

[110] F. Deppisch, A. Freitas, W. Porod, and P. M. Zerwas, Phys. Rev. D **77**, 075009 (2008), arXiv:0712.0361 hep-ph.

[111] A. Freitas, W. Porod, and P. M. Zerwas, Phys. Rev. D **72**, 115002 (2005), hep-ph/050905.

[112] J. Hisano, M. M. Nojiri, Y. Shimizu, and M. Tanaka, Phys. Rev. D **60**, 055008 (1999), hep-ph/9808410.

[113] M. Hirsch, S. Kaneko, and W. Porod, Phys. Rev. D **78**, 093004 (2008), arXiv:0806.3361 hep-ph.

[114] M. Hirsch, J. W. F. Valle, W. Porod, J. C. Romao, and A. V. del Moral, Phys. Rev. D **78**, 013006 (2008), arXiv:0804.4072 hep-ph.

[115] S. Antusch and M. Ratz, JHEP **07**, 059 (2002), hep-ph/0203027.

[116] P. Langacker, Phys. Rept. **72**, 185 (1981).

[117] A. Santamaria, Phys. Lett. B **305**, 90 (1993), hep-ph/9302301.

[118] J. R. Ellis, J. Hisano, M. Raidal, and Y. Shimizu, Phys. Rev. D **66**, 115013 (2002), hep-ph/0206110.

[119] W. Porod, Comput. Phys. Commun. **153**, 275 (2003), hep-ph/0301101.

[120] See, for example, http://meg.icepp.s.u-tokyo.ac.jp/.

[121] W. Beenakker, R. Hopker, and M. Spira, hep-ph/9611232.

[122] W. Beenakker *et al.*, Phys. Rev. Lett. **83**, 3780 (1999), hep-ph/9906298 Erratum-ibid. **100**, 029901 (2008).

[123] W. Beenakker, M. Kramer, T. Plehn, M. Spira, and P. M. Zerwas, Nucl. Phys. B **515**, 3 (1998), hep-ph/9710451.

[124] T. Plehn, Czech. J. Phys. **55**, B213 (2005), hep-ph/0410063.

[125] M. Spira, hep-ph/0211145.

[126] P. Skands *et al.*, JHEP **0407**, 036 (2004), hep-ph/0311123.

[127] E. Carquin, J. Ellis, M. E. Gomez, S. Lola, and J. Rodriguez-Quintero, arXiv:0812.4243 hep-ph.

[128] A. Bartl *et al.*, Eur. Phys. J. C **46**, 783 (2006), hep-ph/0510074.

[129] N. V. Krasnikov, Phys. Lett. B **388**, 783 (1996), hep-ph/9511464.

[130] N. Arkani-Hamed, H. C. Cheng, J. L. Feng, and L. J. Hall, Phys. Rev. Lett. **77**, 1937 (1996), hep-ph/9603431.

[131] N. Arkani-Hamed, J. L. Feng, L. J. Hall, and H.-C. Cheng, Nucl. Phys. **B505**, 3 (1997), hep-ph/9704205.

[132] D. Nomura, Phys. Rev. D **64**, 075001 (2001), hep-ph/0004256.

[133] W. Porod and W. Majerotto, Phys. Rev. D **66**, 015003 (2002), hep-ph/0201284.

[134] F. Deppisch, H. Pas, A. Redelbach, R. Ruckl, and Y. Shimizu, Phys. Rev. D **69**, 054014 (2004), hep-ph/0310053.

[135] G. Bertone, D. Hooper, and J. Silk, Phys. Rept. **405**, 279 (2005), hep-ph/0404175.

[136] G. Jungmanm, M. Kamionkowski, and K. Griest, Phys. Rept. **267**, 195 (1996), hep-ph/9506380.

[137] WMAP collaboration, D. N. Spergel *et al.*, Astrophys. J. Suppl. **148**, 175 (2003), astro-ph/0302209.

[138] SDSS collaboration, M. Tegmark *et al.*, Phys. Rev. D **74**, 123507 (2006), astro-ph/0608632.

[139] H. Baer, E. K. Park, and X. Tata, arXiv:0903.0555 hep-ph.

[140] L. Bergstrom, arXiv:0903.4849 hep-ph.

[141] MINOS collaboration, arXiv:0708.1495 hep-ex.

[142] KamLAND collaboration, K. Eguchi *et al.*, arXiv:0801.4589 hep-ex.

[143] E. Ma, Phys. Rev. Lett. **81**, 1171 (1998), hep-ph/9805219.

[144] M. Gell-Mann, P. Ramond, and R. Slansky, in *Supergravity*, edited by P. van Niewenhuizen and D. Freedman, North Holland, 1979.

[145] R. Mohapatra and G. Senjanovic, Phys. Rev. Lett. **44**, 912 (1980).

[146] T. Yanagida, in *KEK lectures*, edited by O. Sawada and A. Sugamoto, KEK, 1979.

[147] R. Foot, H. Lew, X. G. He, and G. C. Josh, Z. Phys. C **44**, 441 (1989).

[148] F. Deppisch, H. Pas, A. Redelbach, and R. Ruckl, Phys. Rev. D **73**, 033004 (2006), hep-ph/0511062.

[149] M. Drees and M. M. Nojiri, Phys. Rev. D **47**, 376 (1993), hep-ph/9207234.

[150] J. R. Ellis, J. S. Hagelin, D. V. Nanopoulos, K. A. Olive, and M. Srednicki, Nucl. Phys. B **238**, 453 (1984).

[151] K. Griest and D. Seckel, Phys. Rev. D **43**, 3191 (1991).

[152] G. B. Gelmini and P. Gondolo, Phys. Rev. D **74**, 023510 (2006), hep-ph/0602230.

[153] B. C. Allanach, G. Belanger, F. Boudjema, and A. Pukhov, JHEP **0412**, 020 (2002), hep-ph/0410091.

[154] H. Baer, C. Balazs, and A. Belyaev, JHEP **0203**, 042 (2002), hep-ph/0202076.

[155] J. L. Feng, K. T. Matchev, and T. Moroi, Phys. Rev. D **61**, 075005 (2000), hep-ph/9909334.

[156] K. Kadota, K. A. Olive, and L. Velasco-Sevilla, arXiv:0902.2510 hep-ph.

[157] H. P. Nilles, Phys. Rept. **110**, 1 (1984).

[158] S. P. Martin and M. T. Vaughn, Phys. Rev. D **50**, 2282 (1994), hep-ph/9311340, Erratum-ibid. **78**, 039903 (2008).

[159] H. E. Haber and G. L. Kane, Phys. Rept. **117**, 75 (1985).

[160] S. P. Martin, arXiv:hep-ph/9709356.

[161] G. Belanger, F. Boudjema, A. Pukhov, and A. Semenov, Comput. Phys. Commun. **176**, 367 (2007), arXiv:hep-ph/0607059.

[162] G. Belanger, F. Boudjema, A. Pukhov, and A. Semenov, Comput. Phys. Commun. **174**, 577 (2006), hep-ph/0405253.

[163] http://wwwlapp.in2p3.fr/lapth/micromegas/.

[164] B. C. Allanach, A. Djouadi, J. L. Kneur, W. Porod, and P. Slavich, JHEP **0409**, 044 (2004), hep-ph/0406166.

[165] S. Heinemeyer, W. Hollik, H. Rzehak, and G. Weiglein, Eur. Phys. J. C **39**, 465 (2005).

[166] M. Frank *et al.*, JHEP **0702**, 047 (2007), hep-ph/0611326, The FeynHiggs code can be found at http://www.feynhiggs.de/.

[167] T. A. Collaboration, arXiv:0901.0512 hep-ex.

[168] J. A. Aguilar-Saavedra *et al.*, hep-ph/0106315.

[169] Quarkonium Working Group, N. Brambilla *et al.*, hep-ph/0412158.

[170] K. G. Chetyrkin *et al.*, arXiv:0907.2110 hep-ph.

[171] J. N. Esteves, M. Hirsch, J. C. Romao, W. Porod, and F. Staub, arXiv:1010.6000 [hep-ph].

[172] L. Calibbi, M. Frigerio, S. Lavignac, and A. Romanino, JHEP **0912**, 057 (2009), arXiv:0910.0377 [hep-ph].

[173] C. Biggio and L. Calibbi, arXiv:1007.3750 [hep-ph].

[174] B. C. Allanach *et al.*, Eur. Phys. J. C **25**, 113 (2002), arXiv:hep-ph/0202233.

[175] W. Martens, L. Mihaila, J. Salomon, and M. Steinhauser, arXiv:1008.3070 [hep-ph].

[176] F. Staub, arXiv:0806.0538 [hep-ph].

[177] F. Staub, Comput. Phys. Commun. **181**, 1077 (2010), arXiv:0909.2863 [hep-ph].

[178] F. Staub, arXiv:1002.0840 [Unknown].

[179] Lepsusywg, aleph, delphi, l3 and opal experiments, note lepsusywg/01-03.1, http://lepsusy.web.cern.ch/lepsusy/Welcome.html.

[180] F. Borzumati and T. Yamashita, arXiv:0903.2793 [hep-ph].